\documentclass{PASAFreshNewFlavour}%

\usepackage[utf8]{inputenc}
\usepackage[T1]{fontenc}  

\usepackage{graphicx}
\usepackage{xcolor}

\usepackage{aas_macros}
\usepackage{hyperref}
\definecolor{dodgerblue}{RGB}{30, 144, 255}
\definecolor{crimson}{RGB}{220, 20, 60}
\definecolor{darkerblue}{RGB}{0, 0, 139}
\hypersetup{colorlinks,citecolor=dodgerblue,linkcolor=blue,urlcolor=darkerblue}

\usepackage{amsmath}

\usepackage{booktabs}
\usepackage{threeparttable,longtable}
\usepackage{etoolbox}

\AtBeginEnvironment{tablenotes}{\sffamily\sansmath}
\AtBeginEnvironment{caption}{\sffamily\sansmath}
\documentstatus{\textcolor{acceptcolorPASA}{accepted} for publication in PASA}

\usepackage[skip=0.5ex]{subcaption}
\usepackage{cprotect}
\usepackage{multirow}

\usepackage{pdflscape}
\usepackage{afterpage}
\usepackage{capt-of}
\usepackage{sidecap}

\usepackage{enumitem}

\usepackage{newtxtext,newtxmath}
\usepackage{beramono}
\usepackage{sansmath}

\usepackage[capitalise]{cleveref}

\newcommand{\tabft}[1]{(#1)}
\newcounter{ft}
\newcommand{\ft}[1]{\noindent%
  ~\refstepcounter{ft}\tabft{\alph{ft}\label{#1}}}

\newcommand\setItemnumber[1]{\setcounter{enumi}{\numexpr#1-1\relax}}

\usepackage[nohyperlinks,nolist]{acronym}
\begin{acronym}
\acro{MWA}{Murchison Widefield Array \citep{tgb+13}}
\acro{MWA2}[MWA-2]{\ac{MWA} Phase II `extended' \citep{wtt+18}}
\acro{ASKAP}{Australian Square Kilometre Array Pathfinder}
\acro{ATCA}{Australia Telescope Compact Array}
\acro{POSSUM}{Polarisation Sky Survey of the Universe's Magnetism}
\acro{EMU}{Evolutionary Map of the Universe}
\acro{ICM}{intra-cluster medium}
\acro{RM}{rotation measure}
\acro{GLEAM}{GaLactic and Extragalactic All-sky \ac{MWA} survey \citep{wlb+15,gleamegc}}
\end{acronym}

\def\LLS{\ensuremath{\text{LLS}}}
\def\LAS{\ensuremath{\text{LAS}}}

\def\xmm{XMM-\textit{Newton}}

\usepackage{siunitx}
\DeclareSIUnit{\gauss}{G}
\DeclareSIUnit{\jansky}{Jy}

\newcommand{\corrs}[1]{{#1}}
\newcommand{\CORRS}[1]{{#1}}

\title[\sffamily Digging for fossils in the Outback]{\sffamily Low-frequency integrated radio spectra of diffuse, steep-spectrum sources in galaxy clusters: palaeontology with the MWA and ASKAP}

\author[S.~W. Duchesne et al.]{S.~W. Duchesne $^\mathsf{1}$\thanks{\texttt{\href{mailto:stefan.duchesne.astro@gmail.com}{stefan.duchesne.astro@gmail.com}}}, M.~Johnston-Hollitt$^{1,2}$, and I.~Bartalucci$^{3}$
\affil{$^\text{1}$International Centre for Radio Astronomy Research (ICRAR), Curtin University, Bentley, WA 6102, Australia}
\affil{$^2$Curtin Institute for Computation, Curtin University,
GPO Box U1987, Perth, WA 6845, Australia}%
\affil{$^\text{3}$INAF - Istituto di Astrofisica Spaziale e Fisica Cosmica di Milano, Via A. Corti 12, 20133 Milano, Italy}
}%

\jid{PASA}
\doi{10.1017/pas.\the\year.xxx}
\jyear{\the\year}

\begin{document}


\begin{frontmatter}
\maketitle

\rule{\linewidth}{0.75pt}\vspace{11.5pt}
\begin{abstract}
Galaxy clusters have been found to host a range of diffuse, non-thermal emission components, generally with steep, power law spectra. In this work we report on the detection and follow-up of radio halos, relics, remnant radio galaxies, and other fossil radio plasmas in Southern Sky galaxy clusters using the Murchison Widefield Array \corrs{and the} Australian Square Kilometre Array Pathfinder. We make use of the frequency coverage between the two radio interferometers---from 88 to $\sim 900$~MHz---to characterise the integrated spectra of these sources within this frequency range. Highlights from the sample include the detection of a double relic system in Abell~3186, a mini-halo in RXC~J0137.2$-$0912, a candidate halo and relic in Abell~3399, and a complex multi-episodic head-tail radio galaxy in Abell~3164. We compare this selection of sources and candidates to the literature sample, finding sources consistent with established radio power--cluster mass scaling relations. Finally, we use the low-frequency integrated spectral index, $\alpha$ ($S_\nu \propto \nu^\alpha$), of the detected sample of cluster remnants and fossil sources to compare with samples of known halos, relics, remnants and fossils to investigate a possible link between their electron populations. We find the distributions of $\alpha$ to be consistent with relic and halo emission generated by seed electrons that originated in fossil or remnant sources. However, the present sample sizes are insufficient to rule out other scenarios.
\end{abstract}

\rule{\linewidth}{0.75pt}\vspace{11.5pt}
\end{frontmatter}

\section{Introduction}
\label{sec:intro}

\begin{figure*}[!h]
\centering
\includegraphics[width=0.8\linewidth]{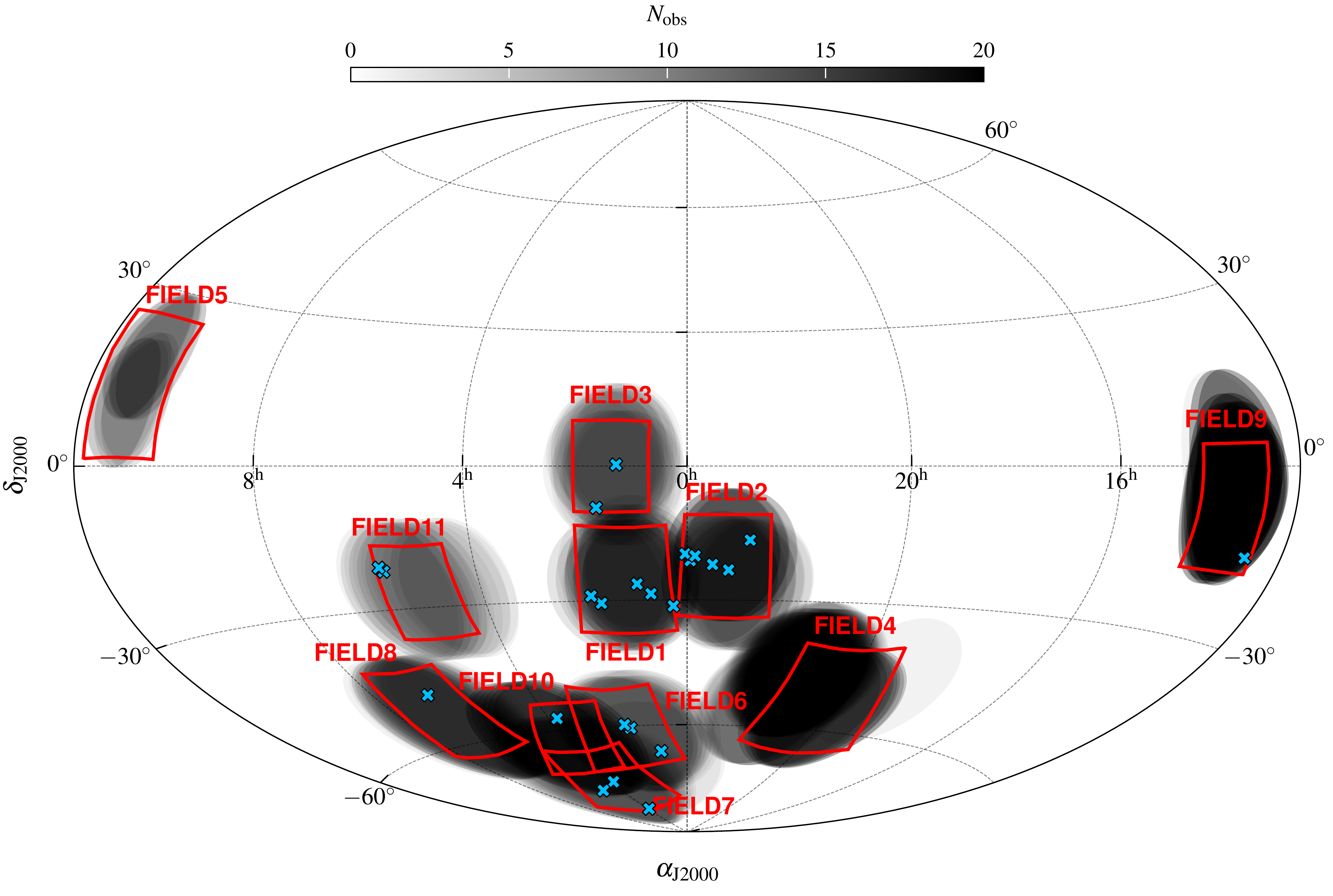}
\caption{\label{fig:survey} The sky coverage of the MWA-2 diffuse source follow-up survey, with named fields labelled and cluster targets reported in this work noted as blue `x' marks. Actual MWA-2 pointings at 154~MHz are \corrs{shown} as transparent black circles, indicating relative sensitivity of fields. While no sources from either \texttt{FIELD4} or \texttt{FIELD5} are reported in this work (as discussed in the text) we show their locations for completeness.}
\end{figure*}

Clusters of galaxies are formed through often highly energetic merger events and accretion from filaments of the Cosmic Web. Clusters are comprised of constituent galaxies, X-ray emitting plasmas, and $\sim$~\si{\micro\gauss}-level magnetic fields \citep{Clarke2001,mj-h}. In a fraction of clusters, large-scale ($\sim 1$~Mpc) steep-spectrum ($\alpha \lesssim -1$ \footnote{Where $\alpha$ is defined via $S_\nu \propto \nu^\alpha$.}), diffuse radio emission is observed as centrally-located radio \textit{halos} and peripherally-located radio \textit{relics} \citep[see][and references therein]{vda+19}. These large-scale synchtrotron-emitting sources are not thought to be presently fuelled by active galactic nuclei (AGN), rather they are assumed to be generated through in situ (re-)acceleration of particles \citep[e.g.][]{Jaffe1977,Ensslin1998}. Such sources are observed in predominantly merging, or otherwise morphologically disturbed clusters \citep[e.g.][]{Buote2001,Brunetti2009,Cassano2010,Botteon2018,Golovich2019}.

Radio \textit{halos} are generally \corrs{spatially} correlated with the thermal, X-ray--emitting core of the cluster \corrs{and are observed with morhpologies ranging from circular \citep[e.g.][]{Orru2007,Murgia2009} to more complex and elongated structures \citep[e.g.][]{vanWeeren2012}.} Radio halos are generally observed to have power law spectra, though some halos with significant spectral coverage show steepening beyond GHz frequencies \citep{tkw03,Xie2020,Rajpurohit2021a}. \corrs{Merger-driven turbulence in the ICM may provide a mechanism for in situ (re-)acceleration of seed particles from either the thermal pool of electrons or from a pre-accelerated population of mildly-relativistic `fossil' electrons throughout the cluster volume \citep[e.g.][]{bsfg01,Buote2001,bj14}.}

Radio \textit{relics} occur in the low-density cluster outskirts, where strong shocks in the intra-cluster medium (ICM) are thought to (re-)accelerate electrons through diffusive-shock acceleration (DSA; e.g. \citealt{Ensslin1998}, and similar processes; e.g. \citealt{Kang2015}) either originating a \corrs{`fossil' electron population} \citep[e.g.][]{Markevitch2005,Kang2011,Kang2016} or accelerated from the thermal pool of electrons in the cluster \citep[e.g.][]{Ensslin1998,Hoeft2007}. Unlike radio halos, relics are often observed with highly ordered, linearly polarized emission \citep[e.g.][]{mj-h,vanWeeren2010,Pearce2017}. The integrated spectra of relics are generally power laws \citep[e.g.][]{Hindson2014,Loi2017,Rajpurohit2020,Duchesne2020b}, though few examples exist with curvature beyond GHz frequencies \citep[e.g.][]{Trasatti2015}. In some cases, radio relics have been observed to be located co-spatially with X-ray shocks/\corrs{surface brightness} discontinuities \citep[e.g.][]{Finoguenov2010,botteon_shock_2016,botteon_mathcal_2016}.

Along with the large-scale radio halos and relics, other diffuse, non-thermal sources have been observed in clusters with many observational and physical similarities \citep[see e.g.][for a review of source types and nomenclature]{vda+19}. Radio \textit{mini-halos} are $\lesssim 500$~kpc synchrotron-emitting regions surrounding AGN in the centres of some cool-core (CC) clusters \citep[see e.g.][]{bgb16,Giacintucci2019}. Observationally, they appear as small radio halos with similar spectral and morphological properties but are thought to form via re-acceleration of AGN outflow from small-scale turbulence and sloshing within the cluster core \citep[e.g.][]{Gitti2002}. Mini-halos are not typically associated with major mergers. 

Beyond the cluster core, smaller-scale relic-like sources of various types are also found: radio \textit{phoenices} or otherwise revived \textit{fossil} sources have been observed \citep[e.g.][]{srm+01,cc11,Giacintucci2020,Hodgson2021}. These sources are typically on the order of a few hundred kpc in size and vary morphologically. They range from ultra-steep spectrum fossil plasmas that have possibly been revived via shock-driven adiabatic compression \citep[e.g.][]{eg01,eb02}, radio galaxies with shocks passing through an outer lobe/tail, re-energising the radio plasma (e.g. gentle re-energisation; \citealt{deGasperin2017a}, or less-gentle processes; \citealt{Bonafede2014,vanWeeren2017}), to true remnant radio galaxies with no evidence of re-energisation and are simply fading from normal energy losses after their AGN have switched off or have entered a low-power state \citep[e.g.][]{Parma2007,mpm+11}. Distinguishing between what are effectively radio galaxies at various stages through their life-cycle is difficult and in the case of radio phoenices often the hosting cluster does not show evidence of merger-driven shocks. Finally, in rare cases synchrotron-emitting bridges have been observed between cluster pairs \citep[e.g.][]{Govoni2019,Botteon2020c}, likely formed through turbulence in the inter-cluster region \citep{Brunetti2020}.

It is not yet clear whether the seed electrons responsible for radio halos and relics are from the thermal pool or from fossils that have diffused into the surrounding ICM---observations of spectra of such sources are beginning to provide answers \citep[e.g.][]{Rajpurohit2021a,Rajpurohit2021b,Rajpurohit2021c}. The current generation of radio interferometers, including the Murchison Widefield Array \citep[MWA;][]{tgb+13,wtt+18}, the Australian Square Kilometre Array Pathfinder \citep[ASKAP;][]{Hotan2021}, the LOw Frequency ARray \citep[LOFAR;][]{lofar}, and MeerKAT \footnote{Karoo Array Telescope} \citep{Jonas2016} are beginning to uncover diffuse cluster sources at higher rates \citep[e.g.][]{Wilber2020,Duchesne2020a,DiGennaro2020,Bruggen2020,vanWeeren2020,Knowles2020,Duchesne2020b,Hodgson2021,Duchesne2017,Duchesne2021a}, providing unprecedented insight into cluster diffuse source populations, paving the way for future observations with the Square Kilometre Array (SKA). In this work we detail a targeted campaign to follow-up diffuse radio emission in clusters originally detected in MWA surveys, leveraging the wide bandwidth of the MWA to investigate the low-frequency integrated spectra of these sources.

Throughout this paper, we assume a standard $\Lambda$ Cold Dark Matter cosmology with $H_0 = 70$~km\,s$^{-1}$\,Mpc$^{-1}$, $\Omega_\text{M} = 0.3$, and $\Omega_\Lambda = 1-\Omega_\text{M}$. Unless otherwise stated, frequency subscripts and superscripts on quantities are in units of MHz.

\section{Data \& methods}\label{sec:data}

\subsection{Cluster sample}\label{section:sample}
\defcitealias{Duchesne2017}{D21}

\begin{table*}[t]
    \centering
    \begin{threeparttable}
    \caption{\label{tab:clusters} Clusters and sources discussed in this work.}
    \begin{tabular}{r c c c l c c l}\toprule
         Cluster & \texttt{FIELD} & $\alpha_\text{J2000}$ \tnote{1} & $\delta_\text{J2000}$ \tnote{1} & $z$ & $M_{500}$ \tnote{2} & Type \tnote{3} & Ref. \tnote{4} \\
         & & (J2000) & (J2000) & & ($\times 10^{14}$~M$_\odot$) & & \\\midrule
          Abell~0122 & 1 &00:57:24 & $-$26:16:50 & 0.113 & 1.73 & r/F & \tabft{\ref{t1:abell}}/\tabft{\ref{t1:zgz06}}/\tabft{\ref{t1:pap+11}}/\tabft{\ref{t1:Duchesne2017}} \\
         Abell~2751 & 1 & 00:16:20 & $-$31:21:55 & 0.107 & 1.26 & r/F & \tabft{\ref{t1:abell}}/\tabft{\ref{t1:sr99}}/\tabft{\ref{t1:pap+11}}/\tabft{\ref{t1:Duchesne2017}} \\
         Abell~2811 & 1 & 00:42:09 & $-$28:32:09 & 0.108 & $3.67_{-0.37}^{+0.35}$ & cH & \tabft{\ref{t1:abell}}/\tabft{\ref{t1:zgz06}}/\tabft{\ref{t1:planck15}}/\tabft{\ref{t1:Duchesne2017}} \\ 
         Abell~2496 & 2 & 22:51:00 & $-$16:24:24 & 0.123 & $3.36_{-0.32}^{+0.30}$ & r & \tabft{\ref{t1:abell}}/\tabft{\ref{t1:sr99}}/\tabft{\ref{t1:planck16}}/\tabft{\ref{t1:Duchesne2017}}\\
         Abell~2680 & 2 & 23:56:28 & $-$21:02:18 & 0.190 & $3.2_{-1.0}^{+0.8}$ & cH & \tabft{\ref{t1:abell}}/\tabft{\ref{t1:cac+09}}/\tabft{\ref{t1:Wen15}}/\tabft{\ref{t1:Duchesne2017}} \\
         Abell~2693 & 2 & 00:02:10 & $-$19:33:18 & 0.173 & $2.1_{-0.6}^{+0.5}$ & cH/p & \tabft{\ref{t1:abell}}/\tabft{\ref{t1:cac+09}}/-/\tabft{\ref{t1:Duchesne2017}} \\
         Abell~S1099 & 2 & 23:13:16 & $-$23:08:40 & 0.110 & - & r & \tabft{\ref{t1:aco89}}/\tabft{\ref{t1:cmkw02}}/-/\tabft{\ref{t1:Duchesne2017}} \\
         AqrCC~087 & 2 & 23:31:30 & $-$21:55:00 & - & - & F & \tabft{\ref{t1:cmkw02}}/-/-/- \\
         RXC~J2351.0$-$1954 & 2 & 23:51:07 & $-$19:58:52 & 0.248 & $5.60_{-0.62}^{+0.59}$ & p, r/F, U & \tabft{\ref{t1:cb12}}/\tabft{\ref{t1:cb12}}/\tabft{\ref{t1:planck15}}/\tabft{\ref{t1:Duchesne2017}} \\
         Abell~0168 & 3 & 01:15:12 & $+$00:19:48 & 0.045 & $1.87_{-0.31}^{+0.29}$ & R & \tabft{\ref{t1:abell}}/\tabft{\ref{t1:sr99}}/\tabft{\ref{t1:planck15}}/\tabft{\ref{t1:Dwarakanath2018}} \\
         RXC~J0137.2$-$0912 & 3 & 01:37:15 & $-$09:12:10 & 0.039 & 0.95 & mH, RG & \tabft{\ref{t1:Cruddace2002}}/\tabft{\ref{t1:Cruddace2002}}/\tabft{\ref{t1:pap+11}}/- \\
         Abell~S0112 & 6 & 00:57:48 & $-$66:48:44 & 0.067 & 1.62 & F & \tabft{\ref{t1:aco89}}/\tabft{\ref{t1:Garilli1993}}/\tabft{\ref{t1:pap+11}}/- \\
         MCXC~J0145.2$-$6033 & 6 & 01:45:12 & $-$60:33:45 & 0.180 & $3.55_{-0.44}^{+0.42}$ & cmH & \tabft{\ref{t1:shl+00}}/\tabft{\ref{t1:shl+00}}/\tabft{\ref{t1:planck15}}/- \\
         MCXC~J0154.2$-$5937 & 6 & 01:54:15 & $-$59:39:38 & 0.360 & 1.41 & cGRG & \tabft{\ref{t1:Vikhlinin1998}}/\tabft{\ref{t1:Vikhlinin1998}}/\tabft{\ref{t1:pap+11}}/- \\
         Abell~3186 & 7 &  03:52:30 & $-$74:01:51 & 0.127 & $6.44_{-0.24}^{+0.24}$ & R, R, cH & \tabft{\ref{t1:aco89}}/\tabft{\ref{t1:sr99}}/\tabft{\ref{t1:planck15}}/- \\
         Abell~S0405 & 7 & 03:51:09 & $-$82:13:00 & 0.061 & $2.51_{-0.21}^{+0.20}$ & r & \tabft{\ref{t1:aco89}}/\tabft{\ref{t1:DeGrandi1999}}/\tabft{\ref{t1:planck15}}/- \\ 
         PSZ1~G287.95$-$32.98 & 7 &  04:59:38 & $-$75:47:48 &  0.250 & $5.88_{-0.41}^{+0.40}$ & cH & \tabft{\ref{t1:planck15}}/\tabft{\ref{t1:planck14}}/\tabft{\ref{t1:planck15}}/- \\
         Abell~3399 & 8 & 06:37:19 & $-$48:28:42 & 0.203 & $4.81_{-0.39}^{+0.37}$ & cR, cH & \tabft{\ref{t1:aco89}}/\tabft{\ref{t1:bsg+04}}/\tabft{\ref{t1:planck15}}/- \\
         MCXC~J1253.2$-$1522 & 9 & 12:53:14 & $-$15:22:48 & 0.046 \tnote{5} & 0.98 & F & \tabft{\ref{t1:pap+11}}/\tabft{\ref{t1:pap+11}}/\tabft{\ref{t1:pap+11}}/- \\
         Abell~3164 & 10 & 03:46:10 & $-$57:03:00 &  0.059 & $1.62_{-0.28}^{+0.26}$ & F, F, F & \tabft{\ref{t1:aco89}}/\tabft{\ref{t1:Fleenor2006}}/\tabft{\ref{t1:planck15}}/- \\
         Abell~3365 & 11 & 05:48:50 & $-$21:54:43 & 0.093 & 1.66 & R, U & \tabft{\ref{t1:aco89}}/\tabft{\ref{t1:sr99}}/\tabft{\ref{t1:pap+11}}/\tabft{\ref{t1:vanWeeren2011b}} \\
         Abell~0550 & 11 & 05:52:52 & $-$21:03:25 & 0.099 & $3.87_{-0.27}^{+0.25}$ & r & \tabft{\ref{t1:abell}}/\tabft{\ref{t1:DeGrandi1999}}/\tabft{\ref{t1:planck16}}/- \\
        \bottomrule
    \end{tabular}
    \begin{tablenotes}[flushleft]
    \footnotesize \item[1] Coordinates are shown in units of hours, minutes, seconds, and degrees, arcminutes, arcseconds. \item[2] Mass within $R_{500}$, the radius within which the mean density of the cluster is 500 times the critical density of the Universe. \item[3] Detected diffuse source types (either as reported in the literature or as determined in this work): relic (R), halo (H), mini-halo (mH), remnant radio galaxy/AGN (r), miscellaneous fossil plasma/re-accelerated fossil plasma source (e.g. phoenix) (F), candidate (c), point source (p), normal radio galaxy (RG), giant radio galaxy (GRG), unclassified (U). \item[4] References for position/$z$/$M_{500}$/previously detected diffuse emission: \ft{t1:abell} \citet{abell}. \ft{t1:sr99} \citet{sr99}. \ft{t1:pap+11} ($M_{\text{X},500}$) \citet{pap+11}. \ft{t1:Duchesne2017} \citet{Duchesne2017}. \ft{t1:zgz06} \citet{zgz06}. \ft{t1:planck15} ($M_{\text{SZ},500}$) \citet{planck15}. \ft{t1:Cavagnolo2008} \citet{Cavagnolo2008}. \ft{t1:cac+09} \citet{cac+09}. \ft{t1:Wen15} \citet{Wen2015}. \ft{t1:aco89} \citet{aco89}. \ft{t1:cmkw02} \citet{cmkw02}. \ft{t1:cb12} \citet{cb12}. \ft{t1:Dwarakanath2018} \citet{Dwarakanath2018}. \ft{t1:Cruddace2002} \citet{Cruddace2002}. \ft{t1:Garilli1993} \citet{Garilli1993}. \ft{t1:shl+00} \citet{shl+00}. \ft{t1:Vikhlinin1998} \citet{Vikhlinin1998}. \ft{t1:DeGrandi1999} \citet{DeGrandi1999}. \ft{t1:planck14} \citet{planck14}. \ft{t1:bsg+04} \citet{bsg+04}. \ft{t1:Fleenor2006} \citet{Fleenor2006}. \ft{t1:vanWeeren2011b} \citet{vanWeeren2011b}. \ft{t1:planck16} \citet{psz2}. \item[5] A second system (Abell~1631) is detected at $z=0.014$ \citep{cac+09}---see cluster entry in \cref{sec:sources} for details. 
    \end{tablenotes}
    \end{threeparttable}
\end{table*}

\citet[][hereafter \citetalias{Duchesne2017}]{Duchesne2017} report a number of candidate diffuse cluster sources detected in a large, deep $45^\circ \times 45^\circ$ MWA image created for foreground modelling of the Epoch of Re-ionization 0-h field \citep{oth+16}. Due to the low resolution of the MWA, many of these sources had an uncertain nature. With the upgrade to the Phase 2 `extended' MWA \citep[][hereafter MWA-2]{wtt+18} and the allure of an increase in resolution by a factor of two, we carried out re-observation of a selection of these sources as part of MWA project G0045 with Director's Time observations of two additional fields and the addition of overlapping archival observations. \par
At the same time, a candidate list of diffuse cluster sources had been prepared based on visual searches of the GaLactic and Extragalactic MWA (GLEAM) survey \citep[][]{wlb+15,gleamegc}. These searches focused on clusters from the Meta-Catalogue of X-ray detected Clusters \citep[MCXC;][]{pap+11}, the Abell catalogues \citep[][]{abell,aco89}, \emph{Planck} Sunyaev--Zel'dovich clusters \citep{planck15,psz2}, and a handful of miscellaneous clusters serendipitously found to host candidate diffuse emission that are nearby other clusters from the aforementioned catalogues. While the full sample is not within the scope of this work (it would be prohibitive to perform targeted follow-up of close to 200 sources), we present here 31 sources across 9 fields \footnote{A total of 11 fields were observed, 1 is presented in \citet{Duchesne2020a}, and one was not able to be processed.}. Due to the large field of view of the MWA ($\sim 20$~deg at 216~MHz and $\sim 60$~deg at 88~MHz) we planned MWA-2 observations to cover a total 22 clusters (\cref{tab:clusters}). While the 200-MHz wideband GLEAM image is usually sufficient to detect and measure flux density of these sources, the lower-frequency bands become prohibitively confused for use here. The fields observed (labelled \texttt{FIELD1}--\texttt{11}) are shown on \cref{fig:survey}. \texttt{FIELD5} and our target source within Abell~1127 were presented in \citet{Duchesne2020a}, while two sources from the current survey have already been reported: Abell~141 in \texttt{FIELD1} and Abell~3404 in \texttt{FIELD8} \corrs{\citep{Duchesne2021a}}. While generally we will not report non-detections (or more accurately, non-confirmations) from the non-public candidate list, we will, where available, report any such sources from \citetalias{Duchesne2017} in \cref{app:nondetections}.\par

\subsection{Observations with the MWA-2}\label{sec:data:mwa}

\begin{table*}[h!]
    \centering
    \begin{threeparttable}
    \caption{\label{tab:mwa2} MWA-2 and ASKAP observation and image details. Note due to the large number of separate images produced, there is a large range of values and here we report the minimum and maximum values for each quantity for each field. Exact PSF values used in measurements are provided as part of the online table described in \cref{sec:app:table}.}

    \begin{tabular}{l c r c c c c}\toprule
    Field/Name/Project & $\nu$ & Weighting  & $\tau$ \tnote{a} & PSF \tnote{b} & min($\sigma_\text{rms}$) & $\theta_\text{max}$ \\
             &  (MHz) & & (min)  & (${}^{\prime\prime} \times{}^{\prime\prime}$) &  (mJy\,beam$^{-1}$) & ($^\prime$) \\\midrule
    \multicolumn{6}{l}{MWA-2 \tnote{f}} \\\midrule
    \texttt{FIELD1} \tnote{c} & \multirow{14}{*}{88--216} & $+2.0$ & 68(56) & $91 \times 65$($230 \times 160$) &  2.0(8.3) & 90 \\
    \multirow{2}{*}{\texttt{FIELD2}} & & $+2.0$ & 62(88) & $94 \times 68$($222 \times 165$) &  1.4(7.9) & 120 \\
    & & $+0.5$ & 60(86) & $60 \times 60$($150 \times 148$) &  0.83(5.4) & 90 \\
    \multirow{2}{*}{\texttt{FIELD3}} & & $+2.0$ & 22(66) & $100 \times 66$($248 \times 157$) &  3.0(10.9) & 120 \\
    & & $0.0$ & 30(46) & $60 \times 53$($152 \times 126$) &  2.3(7.1) & 120 \\
    \multirow{2}{*}{\texttt{FIELD6}} & & $+2.0$ & 24(42) & $63 \times 51$($229 \times145$) &  2.3(10.3) & 90 \\
    & & $0.0$ & 24(42) & $63 \times 51$($145 \times 118$) &  2.4(9.7) & 90 \\
    \multirow{2}{*}{\texttt{FIELD7} \tnote{c}} & & $+2.0$ & 64(94) & $128 \times 67$($321\times 161$) &  4.0(12.2) & 120 \\
    & & $0.0$ & 56(86) & $67 \times 52$($162 \times 125$) &  2.3(7.0) & 120 \\
    \texttt{FIELD8} \tnote{e} & & $+2.0$ & 64(94) & $128 \times 67$($321 \times 161$) &  3.1(16.1) & 120 \\
    \multirow{2}{*}{\texttt{FIELD9}} & & $+2.0$ & 124(198) & $96 \times 67$($235 \times163$) &  2.0(10.0) & 90 \\
    & & $0.0$ & 200(216) & $59 \times 53$($140 \times 126$) & 1.2(6.1) & 120 \\
    \texttt{FIELD10} \tnote{c} & & $+2.0$ & 64(94) & $103 \times 67$($254 \times 161$) &  2.6(9.9) & 120 \\
    \texttt{FIELD11} & & $+1.0$ & 36(38) & $79 \times 61$($194 \times 148$) &  1.5(9.6) & 90 \\\midrule
    \multicolumn{6}{l}{ASKAP \tnote{g}} \\\midrule
    \multirow{3}{*}{RACS} & \multirow{3}{*}{887} & $+0.25$ & 15 & $14.5 \times 14.5$($22 \times 22$) & 0.12(0.18) & 10 \\
    & & $0.0$ & 15 & $12 \times 12$($16 \times 16$) & 0.16 & 10 \\
    & & $+0.25$, tapered & 15 & $39 \times 39$($90 \times 90$) & 0.31(0.60) & 10 \\\midrule
    \multirow{3}{*}{VAST} & \multirow{3}{*}{887}  & $+0.25$ & 60 & $16 \times 16$ & 0.10 & 20 \\ 
    & & $0.0$ & 60 & $13\times 13$ & 0.075 &  20 \\
    & & $+0.5$, taper & 60 & $50 \times 50$ & 0.16 & 20 \\\midrule
    \multirow{2}{*}{SB25035} & \multirow{2}{*}{887}  & $+0.25$ & 780 & $21 \times 21$ & 0.25 & 20 \\
    &  & $+0.25$, taper & 780 & $60 \times 60$ & 0.70 & 20 \\\midrule
    SB15191 & 943 & $+0.25$ & 595 & $12.8 \times 12.8$ & 0.026 & 20 \\\bottomrule
    \end{tabular}
    \begin{tablenotes}[flushleft]
    \footnotesize \item[a] Range of total stacked times for MWA snapshots, though note that effective sensitivity varies over the map due to mixed primary beam pointings/patterns. For ASKAP observations, this is simply integration time. \item[b] Range of major and minor axes of the PSF at the centre of the stacked images for the various images/frequencies. \item[c] Alternate imaging published in \citet{Duchesne2021a}. \item[d] \texttt{FIELD7} and \texttt{FIELD10} have significant enough overlap that they are combined for a joint \texttt{FIELD7+FIELD10} for increased sensitivity, though two individual maps are made centered on each field. \item[e] Alternate imaging for this field published in \citet{Bruggen2020} and \citet{Duchesne2021a}. \item{f} For MWA-2 observations, all fields are observed at 88, 118, 154, 185, and 216~MHz, and in general resolution increases with frequency, sensitivity peaks at 154~MHz except for zenith fields where sensitivity peaks at 216~MHz, and integration time varies across frequencies due to difference in data lost to ionospheric problems or other calibration problems. As discussed in the text, \texttt{FIELD4} and \texttt{FIELD5} are not presented in this work. \item[g] All ASKAP data are re-imaged. 
    \end{tablenotes}
    \end{threeparttable}
\end{table*}

For all fields, we observed a range of frequencies, mirroring the GLEAM survey frequency selections: 30-MHz instantaneous bandwidth observations centered on 88, 118, 154, 185, and 216~MHz. Observations are performed in the MWA-standard `snapshot' observing mode, with 2-min drift-scan snapshots. Each snapshot is calibrated and imaged independently prior to stacking/mosaicking. \par
Processing of the MWA-2 data follow the recipe described in detail by \citet{Duchesne2020a} making use of the purpose-built Phase II Pipeline ( \href{https://gitlab.com/Sunmish/piip}{\texttt{piip}} \footnote{\url{https://gitlab.com/Sunmish/piip}}) with constituent software which will be briefly described. Individual snapshots are retrieved from the Pawsey Supercomputing Centre \footnote{\url{https://pawsey.org.au/}} archive using the MWA component of the All-Sky Virtual Observatory \footnote{\url{https://asvo.org.au/}} which performs general pre-processing and initial RFI flagging with \texttt{AOFlagger} \footnote{\url{https://gitlab.com/aroffringa/aoflagger}} \citep{owh+15}. After snapshots are retrieved and pre-processed, they are calibrated using an implementation of the \texttt{Mitchcal} algorithm \citep{oth+16} using a global sky model as described in \citet{Duchesne2020a}. Imaging per snapshot is performed with \texttt{WSClean} \footnote{\url{https://gitlab.com/aroffringa/wsclean}} \citep[\corrs{version 2.9.0;}][]{wsclean1,wsclean2} using multi-scale CLEANing. 

Final images are corrected for astrometry using \texttt{fits\_warp.py} \citep[\corrs{version 2.0;}][]{hth18} and the flux scale \corrs{is} set using \texttt{flux\_warp} \footnote{\url{https://gitlab.com/Sunmish/flux_warp}} \corrs{(version 1.14)}. \corrs{Both of these tools take an input sky model generated by cross-matching and spectral modelling of GLEAM, the NRAO \footnote{National Radio Astronomy Observatory} VLA \footnote{Very Large Array} Sky Survey \citep[NVSS;][]{ccg+98} and/or the Sydney University Molonglo Sky Survey \citep[SUMSS;][]{bls99,mmb+03,mmg+07} using the Positional Update and Matching Algorithm \citep[PUMA;][]{puma}. This sky model is in turn cross-matched to point sources in the snapshot image catalogues to calculate astrometric offsets and flux density discrepancies. Corrections are applied over the snapshots via interpolation between cross-matched sources. Finally, snapshot images are stacked to create mosaics as described in \citet{Duchesne2020a}.} \corrs{Flux density scale uncertainties are derived by comparing point source flux densities with the PUMA-generated sky model finding $\sim 2$--10~per cent standard deviation across the observed fields and frequencies. An additional 8~per cent is added in quadrature as inherited from the GLEAM survey, which dominates the flux densities in the sky model.} Bulk image details are presented in \cref{tab:mwa2}. \par

The `extended' configuration \corrs{of the MWA} was created with the same \corrs{number of tiles} (i.e. 128) as the Phase I MWA due to limitations of the current correlator. \corrs{Creating the longer baselines of the MWA-2 therefore required removing a significant number of short baselines, reducing the sensitivity to larger angular scales compared to the Phase I MWA \citep{Hodgson2020}.} While the loss of sensitivity for this work is comparatively minimal, we still find that \corrs{images weighted with a} `Briggs' \citep{bri95} robust parameter \corrs{of} $\lesssim +0.5$ begin to significantly lose large-scale flux. Therefore, for flux density measurements we create at set of robust $+2.0$ images for all fields except \texttt{FIELD11} for which we use robust $+1.0$ \corrs{\footnote{\corrs{There is no functional difference between the robust $+1.0$ and $+2.0$ weighting for the MWA-2 data with respect to image resolution and sensitivity, however, the \texttt{FIELD11} data were processed at an earlier date for a separate project while more recent data-processing is done at $+2.0$.}}}. We also create images at 0.0 and $+0.5$ to leverage the resolution increase, though these images are typically used for morphological reference only, unless otherwise noted. \cref{fig:dirty:f1}--\subref{fig:dirty:f422} in \cref{app:dirty} highlight the `dirty flux' bias introduced due to the snapshot stacking method used which is corrected as described in Section 2.1.2 of \citet{Duchesne2021a}. Final imaging details are collected in \cref{tab:mwa2}. Note \texttt{FIELD5} was published in \citet{Duchesne2020a} and no further sources have been detected in that field so is not discussed here. \texttt{FIELD4} suffered from significant sidelobe contamination from Cygnus~A with the 185- and 216-MHz bands rendered unusable and will not be considered until future observations can be made when Cygnus~A is not present in the primary beam sidelobe \footnote{Note that Cygnus~A appeared with an apparent flux of $\sim 800$~Jy\,beam$^{-1}$ at 185~MHz on the horizon for these observations so proved particularly resistant to subtraction/peeling techniques. While images were eventually made, residual errors remain which made the images unusable for this work.}. 

\subsection{ASKAP survey data}

\subsubsection{Data and re-processing}

The Rapid ASKAP Continuum Survey \citep{racs1} at 887~MHz covers the entire sky below $\delta_\text{J2000} \sim +30^\circ$ and covers all clusters in our sample. The survey has a resolution of $\sim 15$~arcsec and noise of $\sim 250$--$400$~$\mu$Jy\,beam$^{-1}$. This imaging is sufficient in most cases to detect discrete source populations within the emission regions in the MWA data. ASKAP data (images and calibrated visibilities) are publicly available through the CSIRO \footnote{Commonwealth Scientific and Industrial Research Organisation} ASKAP Science Data Archive \citep[CASDA;][]{casda,Huynh2020}. RACS data products are available under project AS110 \citep{askap:racs}. \par

We are able to obtain slightly higher sensitivity in the RACS images by re-imaging with a robust $+0.25$ weighting using \texttt{WSClean} which has the added benefit of enhancing any detected diffuse emission with only a minor loss in resolution. For clusters where discrete sources are strong enough to be subtracted using a suitable $u,v$ cut (ranging from 1700--3000$\lambda$, additionally see \citealt{Knowles2020} for some discussion of this problem), we subtract discrete sources and re-image with additional tapering---dependent on the scale of the emission---at a robust $+0.25$ image weighting. For a selection of observations where point sources are either too faint or non-existent, a low-resolution image is made without additional subtraction and intervening source contributions (if any) are subtracted from the flux density measurements. As a quick quality assurance check, we compare any re-processed maps to the RACS survey images and find no significant discrepancies in astrometry or flux scale. 

Two clusters in our sample also benefit from being within archival ASKAP observations performed for the ASKAP survey for Variability And Slow Transients \citep[VAST;][]{Murphy2013} under pilot project AS107 \citep{askap:vast}. The set up for these observations is similar to RACS, except they have 5--6 $\sim 15$~min identical pointings which we combine and image as above. These data have some overlap in $u,v$ coverage, so the additional $u,v$ coverage is typically only equivalent to 2--3 additional 15-min observations. Source-subtraction is done in the combined visibilities and flux densities of points sources are equivalent to within a few per cent of RACS data at the location of the VAST observations.

Abell~0122 features at the centre of a beam in a deep observation, SB25035 \citep{askap:a141} \footnote{Additional observations are also available, however, the sensitivity in the single observation used here is sufficient and sensitivity improves only moderately with the additional observations.}. These data are processed identically to images presented of Abell~0141 by \citet{Duchesne2021a} and no flux scale discrepancy is observed. Due to the smaller size of the emission, no low-resolution image is made. 

Finally, a single cluster, Abell~3186, is present outside of the full width at half maximum (FWHM) of some beams in a deep, 12-h observation near the Large Magellanic Cloud \citep[SB25035;][]{askap:sb25035}. As the primary beam is not well modelled by a simple 2-d Gaussian $\sim2$~deg away from the beam centre, we instead cross-match sources in the image to a catalogue derived from the RACS image in the region, and create a pseudo primary beam correction using \texttt{flux\_warp} with a linear radial basis function interpolation scheme. This results in flux densities of the surrounding point sources that do not different by more than $\sim 10$~per cent from RACS. While the point source sensitivity of this image is comparable to the 15-min RACS image, the inner $u,v$ sampling is denser due to the longer synthesis rotation allowing better recovery of extended emission.

While the deep ASKAP observations have a well-sampled $u,v$ plane, as discussed by \citet{racs1}, the short $\sim15$-min observations performed for RACS do not allow significant sampling of the inner $u,v$ plane due to lack of significant Earth-rotation synthesis (see e.g. their Figure 4 for an example of the $u,v$ coverage, and see e.g. Figure~2 from \citealt{Duchesne2020b} for an example of the $u,v$ coverage for a 10-h ASKAP observation). While in principle structures up to $\sim10$~arcmin can be recovered, the lower sensitivity at this large angular scale only allows the brightest large-scale objects to be recovered fully. Generally the sources we will discuss in this work are sufficiently small to not be heavily affected (with some exceptions, noted where appropriate) and measurements typically agree with spectra obtained from MWA-2 data alone.

General ASKAP imaging details are presented in \cref{tab:mwa2}, and as with the MWA-2 data a range of imaging properties are reported for the various RACS images made. For non-RACS images, we report the exact properties.

\subsection{\corrs{Spectral properties}}

\subsubsection{Intervening source contributions}
Due to the low resolution of the MWA (even in its extended configuration) we have to carefully consider contamination from confusing sources. The two main scenarios we consider are case (1) brighter sources blended with the diffuse emission, and/or case (2) faint underlying/intervening sources within the detected MWA emission. Case (1) is simple in the sense that bright sources are easily detected with low resolution surveys such as the \corrs{NVSS or SUMSS}, both with $\sim 45$~arcsec resolution, or the TIFR \footnote{Tata Institute for Fundamental Research} GMRT \footnote{Giant Metrewave Radio Telescope} Sky Survey \citep[TGSS;][]{ijmf16} with $\sim 25$~arcsec resolution. The RACS survey data are suitable for this purpose also, and the MWA-2 and GLEAM data can also be useful in this case. 

In case (1), we can generally detect these brighter sources across multiple frequencies and model their spectra to remove their contribution in the MWA images, fitting a normal power law model of the form \corrs{\begin{equation}\label{eq:powerlaw}
    S_{\nu,\text{discrete}} = S_{0,\text{discrete}} \left( \nu / \nu_0 \right)^\alpha \, ,
\end{equation}}
for extrapolation to $S_{\nu,\text{\corrs{discrete}}}$ from a measured flux density $S_{0,\text{\corrs{discrete}}}$. For sources with only two measurements we derive a two-point spectral index rearranging \cref{eq:powerlaw}. We did not encounter any intervening discrete sources that required more complex spectral energy distribution (SED) modelling. \corrs{Uncertainty in the initial discrete source measurements and spectral index are propagated to the extrapolated value.}

Case (2) typically involves sources that are only detected in RACS or other higher-resolution data due to the relative sensitivities of the various low-resolution surveys. If multiple data sets are available, we model the SED as above to extrapolate discrete source flux densities at MWA frequencies. For sources without spectral coverage, we assume a spectral index. Typically this is assumed to be $\langle\alpha\rangle = -0.7$, though for some sources we note a non-detection in some MWA-2 bands/TGSS imply flatter spectra and modify the assumed spectral index appropriately. We use a range of $\alpha$ to estimate additional uncertainty in the unknown spectral index, via:
\corrs{\begin{equation}
    S_{\nu,\text{discrete}} = S_{0,\text{discrete}} \nu^{\langle\alpha\rangle} \pm \sigma_{S_{\nu,\text{discrete}}} \quad [\text{Jy}] \, ,
\end{equation}
and 
\begin{equation}
    \sigma_{S_{\nu,\text{discrete}}} = S_{0,\text{discrete}}\| \nu^{\alpha_\text{min}} - \nu^{\alpha_\text{max}} \| \quad [\text{Jy}] \, ,
\end{equation}}
where $\alpha_\text{min} = -1.0$ and $\alpha_\text{max} = -0.5$, typically, though may be chosen to reflect limits on point source contributions \corrs{as seen in TGSS or MWA images}. For each source, we report the total confusing flux density contributions that are subtracted, along with associated uncertainty in the online table (see \cref{sec:app:table} for details of the online table). 

\subsubsection{\corrs{Flux density measurements}}

\corrs{Flux density measurements are predominantly made using the lower-resolution robust $+2.0$/$+1.0$ images along with the GLEAM 200-MHz image and select ASKAP images. For certain sources/fields MWA-2 robust $0.0$/$+0.5$ images are used to maximise the signal-to-noise ratio (SNR) for smaller sources. Flux density measurements are performed using in-house code, \texttt{fluxtools.py} \footnote{\url{https://gist.github.com/Sunmish/198ef88e1815d9ba66c0f3ef3b18f74c},} by integration over a bespoke polygon region enclosing the source at all frequencies. This means the region is large enough to enclose the emission seen in the lowest-resolution images (usually the 88-MHz maps).}

\corrs{As the MWA-2 images are only CLEANed to the noise level in the individual 2-min snapshots, additional consideration is made for the un-deconvolved/`dirty' flux density contribution in the final stacked images. As described in \citet{Duchesne2021a}, the measurement of flux density may not be consistent before and after CLEANing, and the measurement process has the added complexity of normalising the residual, `dirty' flux density to the CLEAN flux density. \cref{fig:dirty:f1}--\subref{fig:dirty:f422} in \cref{app:dirty} show this effect for simulated Gaussian sources of varying size, highlighting the dependence on source size. }

\corrs{Flux density measurements, $S_\nu$, can therefore be described by \begin{equation}
    S_\nu = 
    \begin{cases} 
    S_\text{CLEAN} + S_\text{dirty}/f - S_{\nu,\text{discrete}}, & \text{for MWA-2} \\
    S_\text{image} - S_{\nu,\text{discrete}}, & \text{otherwise} \\
    \end{cases}
\end{equation}
where $S_\text{CLEAN}$ is the contribution from the stacked CLEAN component model, $S_\text{dirty}$ is contribution from the stacked residual map, $f$ is the model ratio $\overline{S_\text{dirty}/S_\text{CLEAN}}$ determined from simulated Gaussian sources, dependent on source size (\cref{fig:dirty:f1}--\subref{fig:dirty:f422}), and $S_\text{discrete}$ is the contribution from intervening discrete sources. For the non-MWA-2 images, $S_\text{image}$ is measured directly from the restored images.}

\corrs{The uncertainty on the flux density measurement, $\sigma_{S_\nu}$, is estimated as the quadrature sum of the various sources of uncertainty following 
\begin{multline}
    \sigma_{S_\nu} = \Big[ \left(\sigma_\text{scale} S_\nu \right)^2 +  \left(\sigma_\text{discrete}\right)^2 + N_\text{beam} \left(\sigma_\text{rms}\right)^2 + \\ \left(\sigma_{\text{std},f} S_\text{dirty} \right)^2\Big]^{0.5} \, ,
\end{multline}
where $\sigma_\text{scale}$ is the flux scale uncertainty for the image, $\sigma_{\text{std},f}$ is the standard deviation in values of $f$ over all snapshots for a given stacked MWA-2 image, $\sigma_\text{discrete}$ is the uncertainty in the subtracted discrete source contribution, and $N_\text{beam}$ is the number of independent restoring beams that cover the polygon region used for measurement. Typically the $\sigma_\text{scale}$ term dominates, as this is $\sim 8$--$10$~per cent for all MWA and ASKAP images. The $\left(\sigma_{\text{std},f}S_\text{dirty}\right)^2$ term is only included for MWA-2 images.}

\subsubsection{\corrs{Spectra and spectral indices}}
\corrs{The measured flux densities and uncertainties are used for modelling the integrated spectra within the observed frequency range. For sources with only MWA-2 data, we find a normal power law (as in \autoref{eq:powerlaw}) describes the data sufficiently well \footnote{\corrs{Though note in low-SNR cases curvature could be hidden in the noise.}} and provides a spectral index for the source. For sources where additional flux density measurements are available, we find a mixture of power law and curved power law models can be used to describe the observed spectra. We use a generic curved power law model of the form
 \citep{db12} 
\begin{equation}\label{eq:cpowerlaw}
    S_\nu \propto \nu^\alpha \exp \left[ q \left( \ln \nu \right)^2\right] \, ,
\end{equation}
where $q$ gives an indication of curvature in the spectrum. For each source we provide a fitted power law model or a curved power law model if appropriate, with the combined MWA-2 and supplementary data. An additional power law model is fit solely to the MWA-2 measurements providing a low-frequency spectral index. Model parameters and uncertainties are estimated via non-linear weighted least-squares curve fitting with the Levenberg--Marquardt algorithm and we report $1\sigma$ uncertainties. }

\begin{figure*}[tp]
    \centering
    \begin{minipage}[b]{0.33\linewidth}
    \begin{subfigure}[b]{1\linewidth}
    \includegraphics[width=1\linewidth]{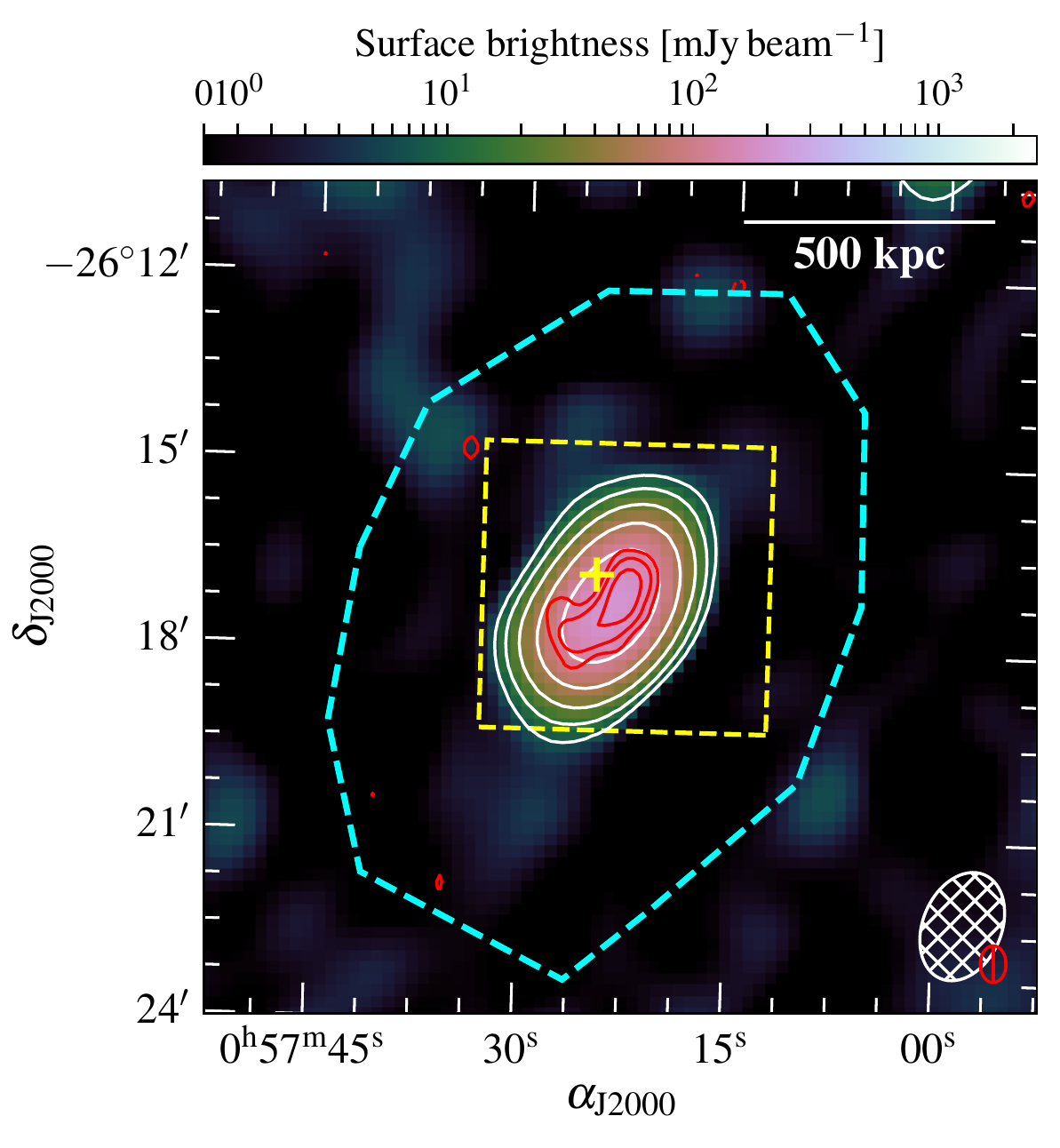}
    \caption{\label{fig:a0122:radio}}
    \end{subfigure}%
    \end{minipage}%
    \begin{minipage}[b]{0.33\linewidth}
    \begin{subfigure}[b]{1\linewidth}
    \includegraphics[width=1\linewidth]{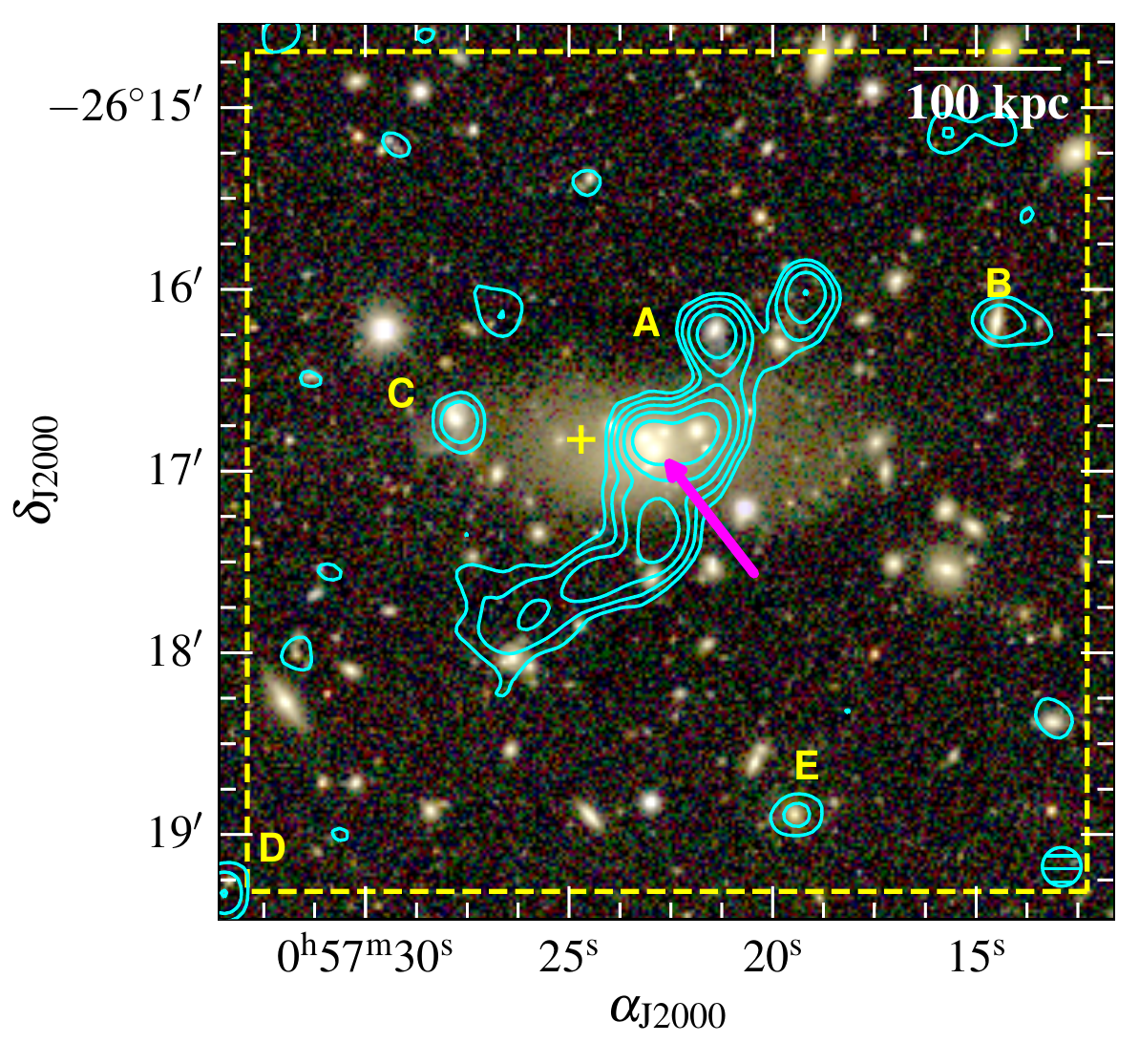}
    \caption{\label{fig:a0122:optical}}
    \end{subfigure}%
    \end{minipage}\hfill%
    \begin{minipage}[b]{0.30\linewidth}
    \caption{\label{fig:a0122} \hyperref[para:a0122]{Abell~0122}. \subref{fig:a0122:radio}. Background: MWA-2, 185~MHz, robust $+2.0$ image. \subref{fig:a0122:optical}. Background: RGB DES image ($i$, $r$, $g$). Where relevant, the white contours are from the background image in \subref{fig:a0122:radio}, in levels of $[\pm 3, 6, 12, 24, 48] \times \sigma_\text{rms}$ ($\sigma_\text{rms} = 2.5$~mJy\,beam$^{-1}$). Red contours: TGSS image, in levels of $[\pm 3, 6, 12, 24, 48] \times \sigma_\text{rms}$ ($\sigma_\text{rms} = 4.5$~mJy\,beam$^{-1}$). Cyan contours: deep ASKAP robust $+0.25$ image, in levels of $[\pm 3, 6, 12, 24, 48] \times \sigma_\text{rms}$ ($\sigma_\text{rms} = 0.026$~mJy\,beam$^{-1}$). The dashed, yellow box is identical in both panels. The ellipses in the lower corners correspond to the respective beams. Sources discussed in the text are labelled. Linear scale bars are at the redshift of the cluster. The magenta arrow points towards the brightest cluster galaxy (BCG). \corrs{The yellow cross indicates the reported cluster centre}\CORRS{, and the dashed cyan polygon indicates the region used for flux density measurement.}}
    \end{minipage}%
\end{figure*}

\begin{figure*}[tp]
    \centering
    \begin{subfigure}[b]{0.33\linewidth}
    \includegraphics[width=1\linewidth]{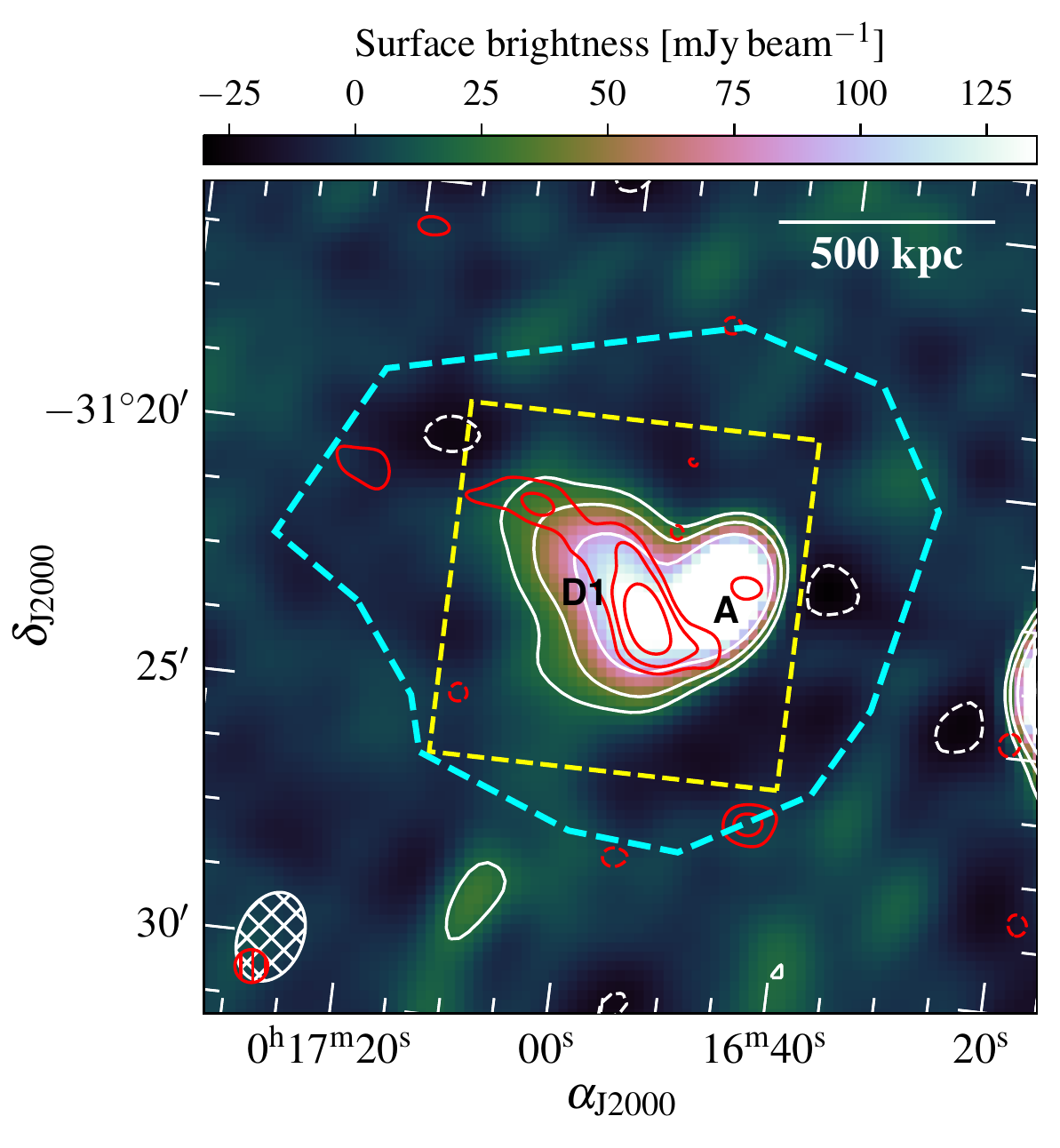}
    \caption{\label{fig:a2751:radio}}
    \end{subfigure}\hfill%
    \begin{subfigure}[b]{0.33\linewidth}
    \includegraphics[width=1\linewidth]{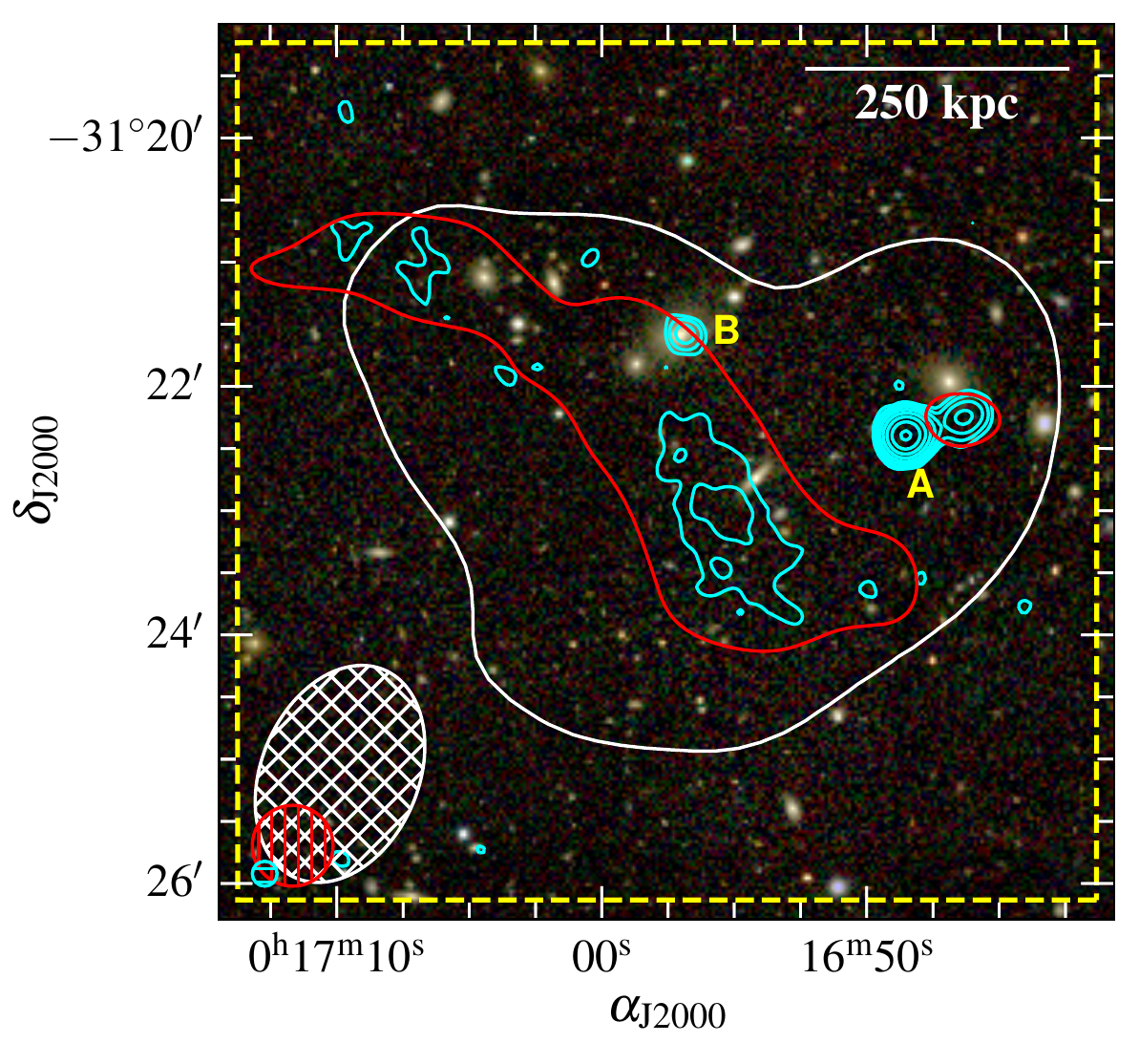}
    \caption{\label{fig:a2751:optical}}
    \end{subfigure}%
    \begin{subfigure}[b]{0.33\linewidth}
    \includegraphics[width=1\linewidth]{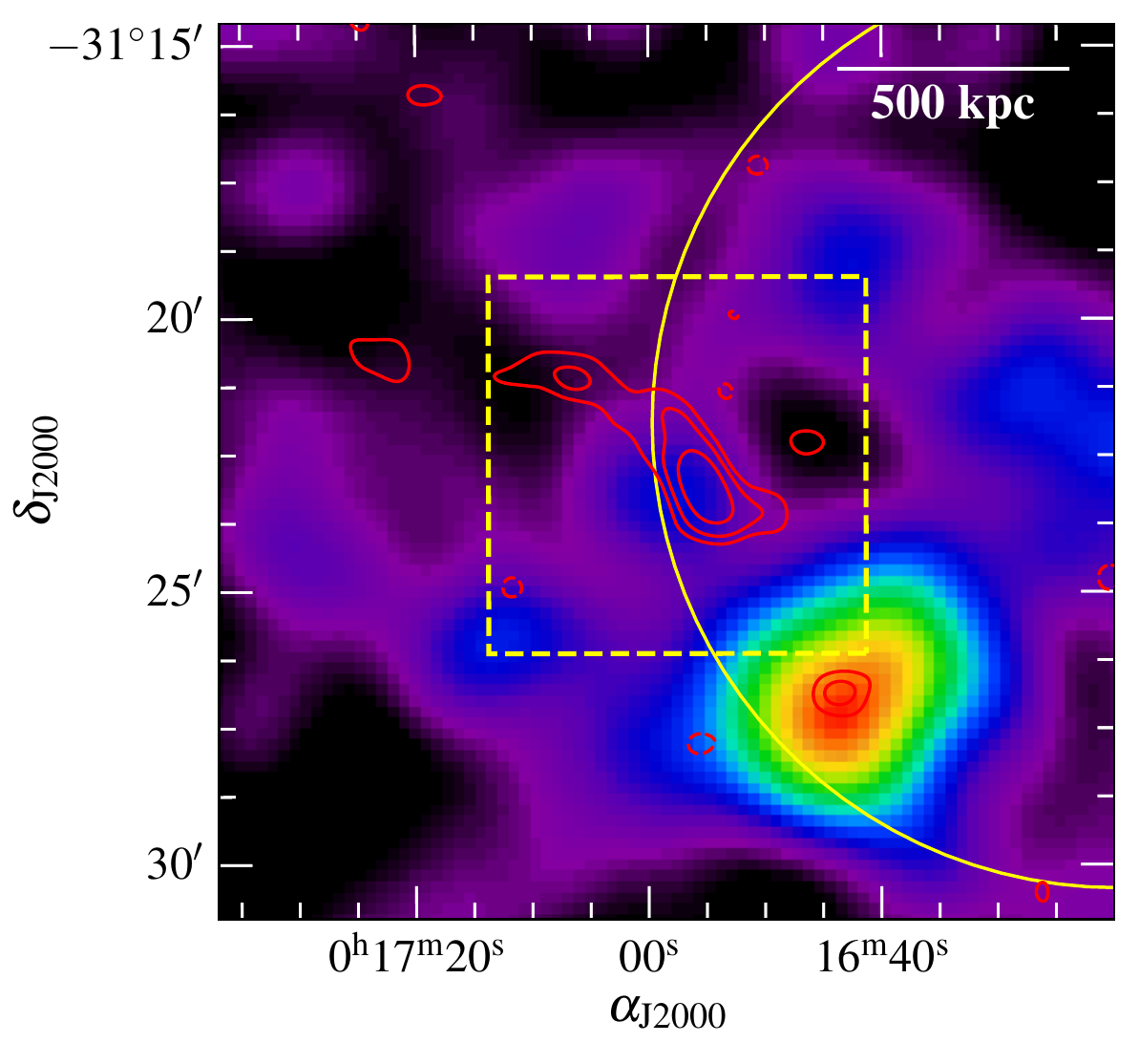}
    \caption{\label{fig:a2751:xray}}
    \end{subfigure}%
    \caption{\label{fig:a2751} \hyperref[para:a2751]{Abell~2751.} \subref{fig:a2751:radio}. Background: MWA-2, 185~MHz, robust $+2.0$ image. \subref{fig:a2751:optical}. Background: RGB DES image ($i$, $r$, $g$). \subref{fig:a2751:xray}. Background: Smoothed RASS image. The white contours are as in \cref{fig:a0122:radio} for the background of \corrs{\subref{fig:a2751:radio}} ($\sigma_\text{rms} = 7$~mJy\,beam$^{-1}$), except in \subref{fig:a2751:optical} with a single contour at $3\sigma_\text{rms}$. Red contours: RACS \corrs{discrete} source-subtracted image, $[\pm 3, 6, 12, 24, 48] \times \sigma_\text{rms}$ ($\sigma_\text{rms} = 0.44$~mJy\,beam$^{-1}$), except in \subref{fig:a2751:optical} with a single contour at $3\sigma_\text{rms}$. Cyan contours: RACS robust $+0.25$ image, $[\pm 3, 6, 12, 24, 48] \times \sigma_\text{rms}$ ($\sigma_\text{rms} = 0.2$~mJy\,beam$^{-1}$). \corrs{The yellow circle in \subref{fig:a2751:xray} has a 1~Mpc radius centered on the reported cluster coordinates.} Other features are as in \cref{fig:a0122}.}
\end{figure*}

\begin{figure*}[tp]
    \centering
    \begin{minipage}[b]{0.33\linewidth}
    \begin{subfigure}[b]{1\linewidth}
    \includegraphics[width=1\linewidth]{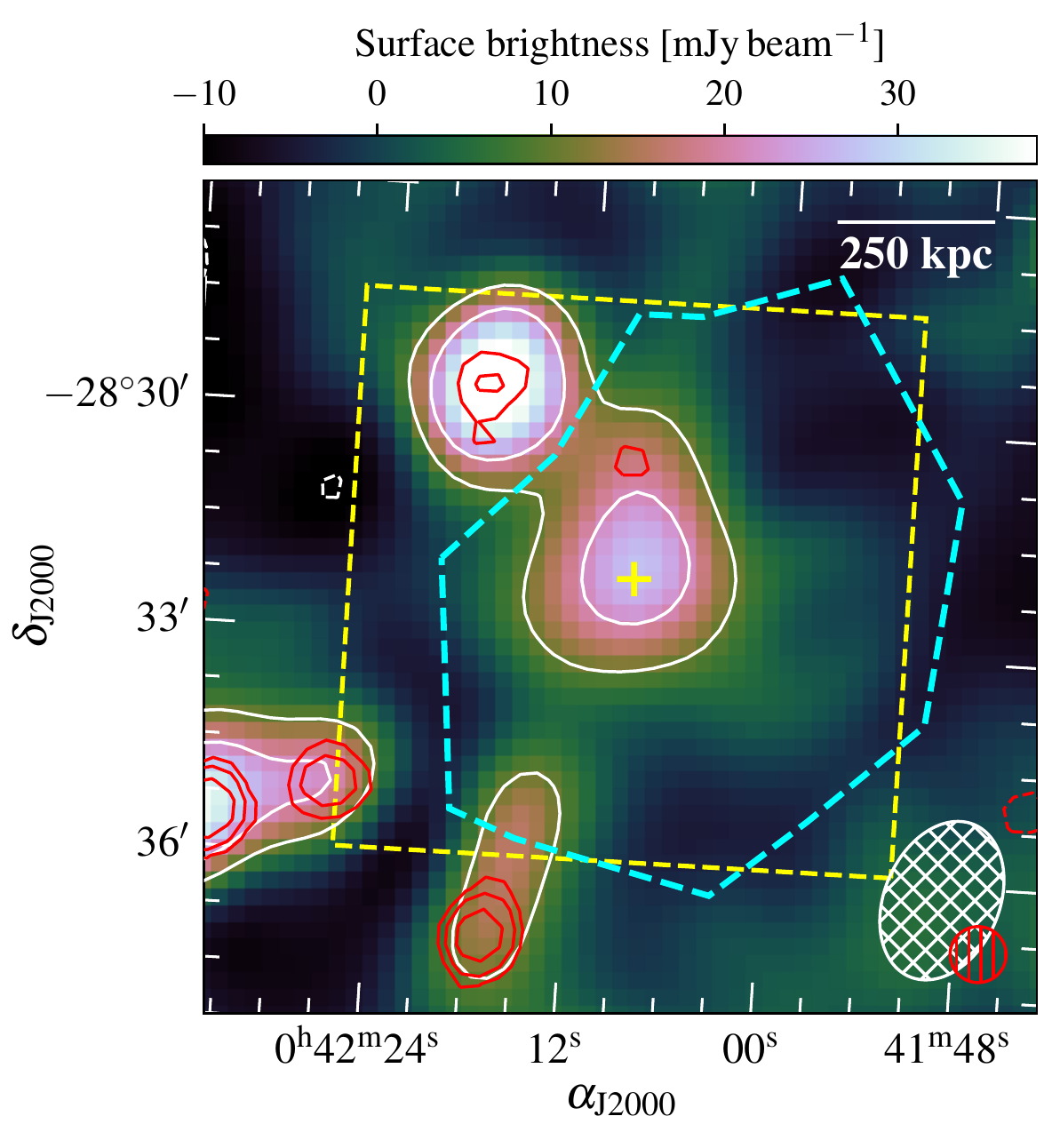}
    \caption{\label{fig:a2811:radio}}
    \end{subfigure}%
    \end{minipage}%
    \begin{minipage}[b]{0.33\linewidth}
    \begin{subfigure}[b]{1\linewidth}
    \includegraphics[width=1\linewidth]{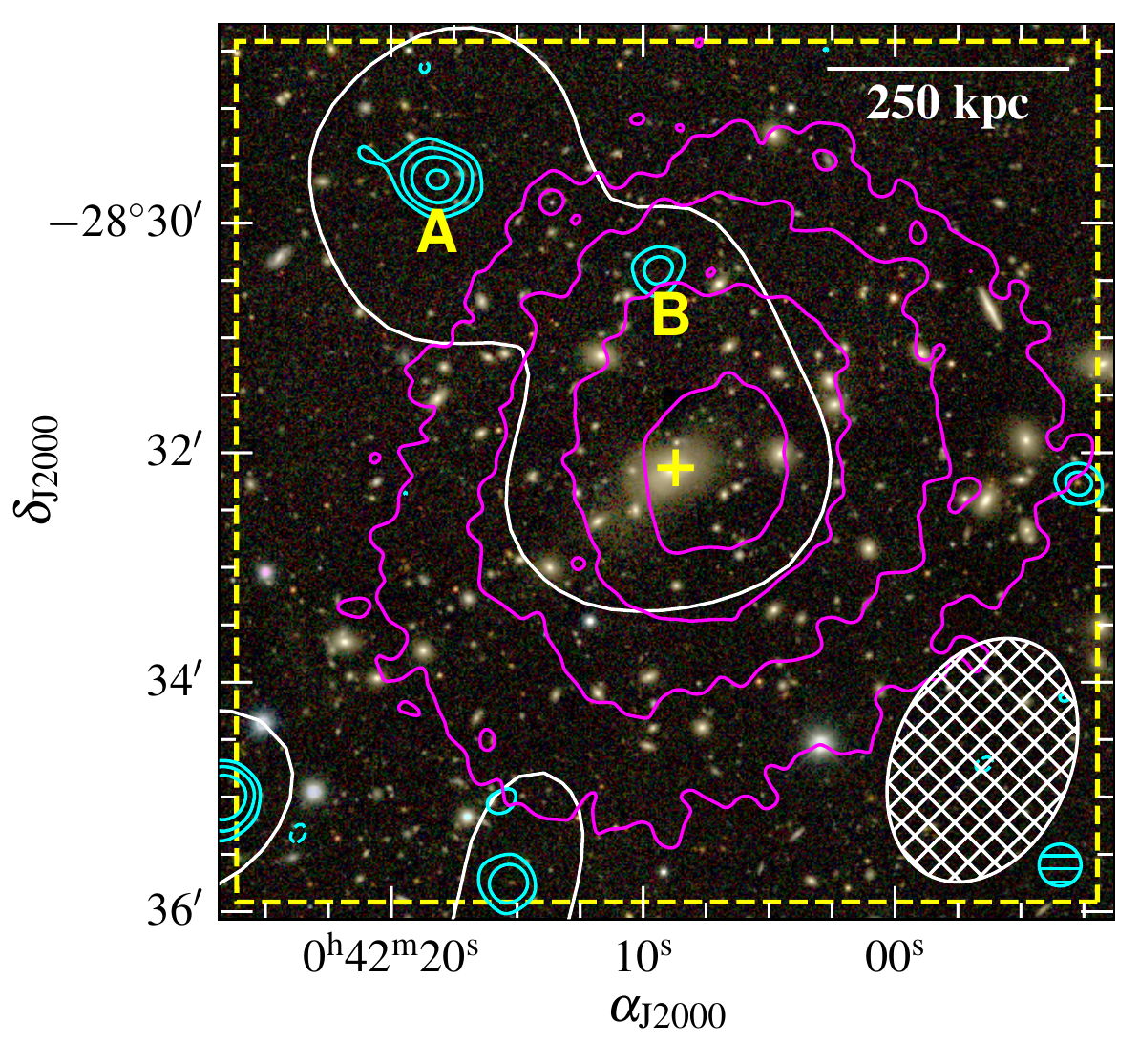}
    \caption{\label{fig:a2811:optical}}
    \end{subfigure}%
    \end{minipage}\hfill%
    \begin{minipage}[b]{0.30\linewidth}
    \caption{\label{fig:a2811} \hyperref[para:a2811]{Abell~2811}. \subref{fig:a2811:radio}. Background: MWA-2, 154~MHz, robust $+2.0$ image. \subref{fig:a2811:optical}. Background: RGB DES image ($i$, $r$, $g$). The white contours are as in \cref{fig:a0122:radio} for the background of \subref{fig:a2811:radio} (for $\sigma_\text{rms} = 3.5$~mJy\,beam$^{-1}$). Red contours: NVSS image, in levels of $[\pm 3, 6, 12, 24, 48] \times \sigma_\text{rms}$ ($\sigma_\text{rms} = 0.45$~mJy\,beam$^{-1}$. Cyan contours: RACS robust $+0.25$ image, in levels of $[\pm 3, 6, 12, 24, 48] \times \sigma_\text{rms}$ ($\sigma_\text{rms} = 0.17$~mJy\,beam$^{-1}$). \corrs{Magenta contours: exposure-corrected, background-subtracted \xmm{} data as presented in \citetalias{Duchesne2017}}. Other image features are as in \cref{fig:a0122}.}
    \end{minipage}%
\end{figure*}

\subsection{Archival X-ray observations}\label{sec:data:xray}

\begin{table}
\centering
\begin{threeparttable}
\caption{X-ray observation properties. \label{tab:xray_prop}}
 \begin{tabular}{lcc}
\hline\toprule
Cluster name            & Obs. ID                   & Exp. time \tnote{a} \\
                        &                           &     (ks)      \\\midrule
RXC~J0137.2$-$0912      & 0765001101                &  14.4\\
Abell~S0112             & 0653880201                & 42.3 \\
MCXC~J0154.2$-$5937     & 0109460201                & 6.3 \\
\multirow{2}{*}{Abell~3186}              & 0692931401   & \multirow{2}{*}{22} \\
& 0723161201    &\\
Abell~S0405             & 0720250601                & 8\\ 
PSZ1~G287.95$-$32.98    & 0762800101                & 15\\
Abell~3399              & 0692933101 \tnote{b}      & 25 \\
\multirow{2}{*}{Abell~0550}              & 0675470101 & \multirow{2}{*}{17} \\
& 0720250101    & \\
\bottomrule
\end{tabular}
\begin{tablenotes}[flushleft]
\footnotesize{\item[a] Exposure time after the cleaning procedures described in \cref{sec:data:xray}. \item[b] \textit{Chandra} dataset.}
\end{tablenotes}
\end{threeparttable}
\end{table}

X-ray datasets used in this work were taken using the XMM-\textit{Newton} European Photon Imaging Camera (EPIC, \citealt{turner2001} and \citealt{struder2001}) \corrs{ except for the observation of Abell~3399 which} was taken using the Advanced CCD Imaging Spectrometer (ACIS, \citealt{garmire2003}) on board of the \textit{Chandra} observatory. The details of data reduction can be found in the Appendix A of \cite{bartalucci2017}. We used the same reduction and cleaning procedures but updated versions of the \textit{Chandra} and XMM-\textit{Newton} analysis software \texttt{CIAO} \citep{Fuscione2006} ver.~4.11 with CALDB 4.8.5 and \texttt{SAS} (ver.~15.0) with CCF updated up to March 2021, respectively. The useful exposure times after the cleaning procedures and the observations used are reported in \cref{tab:xray_prop}.
The datasets were then arranged in data-cubes and corresponding exposure and background maps were calculated as detailed in \citealt{bourdin08}, \citealt{bourdin13} and \citealt{bogdan13}. Point sources were detected using the technique described in \citealt{bogdan13}, visually inspected for false positives or missed sources and then removed from the analysis. Exposure-corrected and background subtracted images are produced in the [0.5--2.5]~keV band.

\subsection{Additional survey data}\label{sec:data:survey}
In addition to the already discussed radio survey images (NVSS, TGSS, SUMSS, and GLEAM), we make use of images from the \emph{ROSAT} \footnote{\emph{R\"{O}entgen SATellite}} All Sky Survey \citep[RASS;][]{vab+99} for select clusters without deep \textit{Chandra} or \xmm\ observations and optical data from the SuperCOSMOS Sky Survey \citep[SSS;][]{supercosmos1,supercosmos2,supercosmos3}, the first Pan-STARRS \footnote{Panoramic Survey Telescope And Rapid Response System} survey \citep[PS1;][]{tsl+12,cmm+16}, and the Dark Energy Survey Data Release 2 \citep[DES DR2;][ hereafter DES]{des1,des2,decam}.

\section{Results}
\subsection{Individual clusters}\label{sec:sources}
In this section we will describe the individual clusters ordered by observed field. Individual plots of source SEDs are shown in \cref{app:seds} and measurements for cluster sources are provided as an online table described in \cref{sec:app:table}.

\begin{figure*}[tp]
    \centering
    \begin{minipage}[b]{0.33\linewidth}
    \begin{subfigure}[b]{1\linewidth}
    \includegraphics[width=1\linewidth]{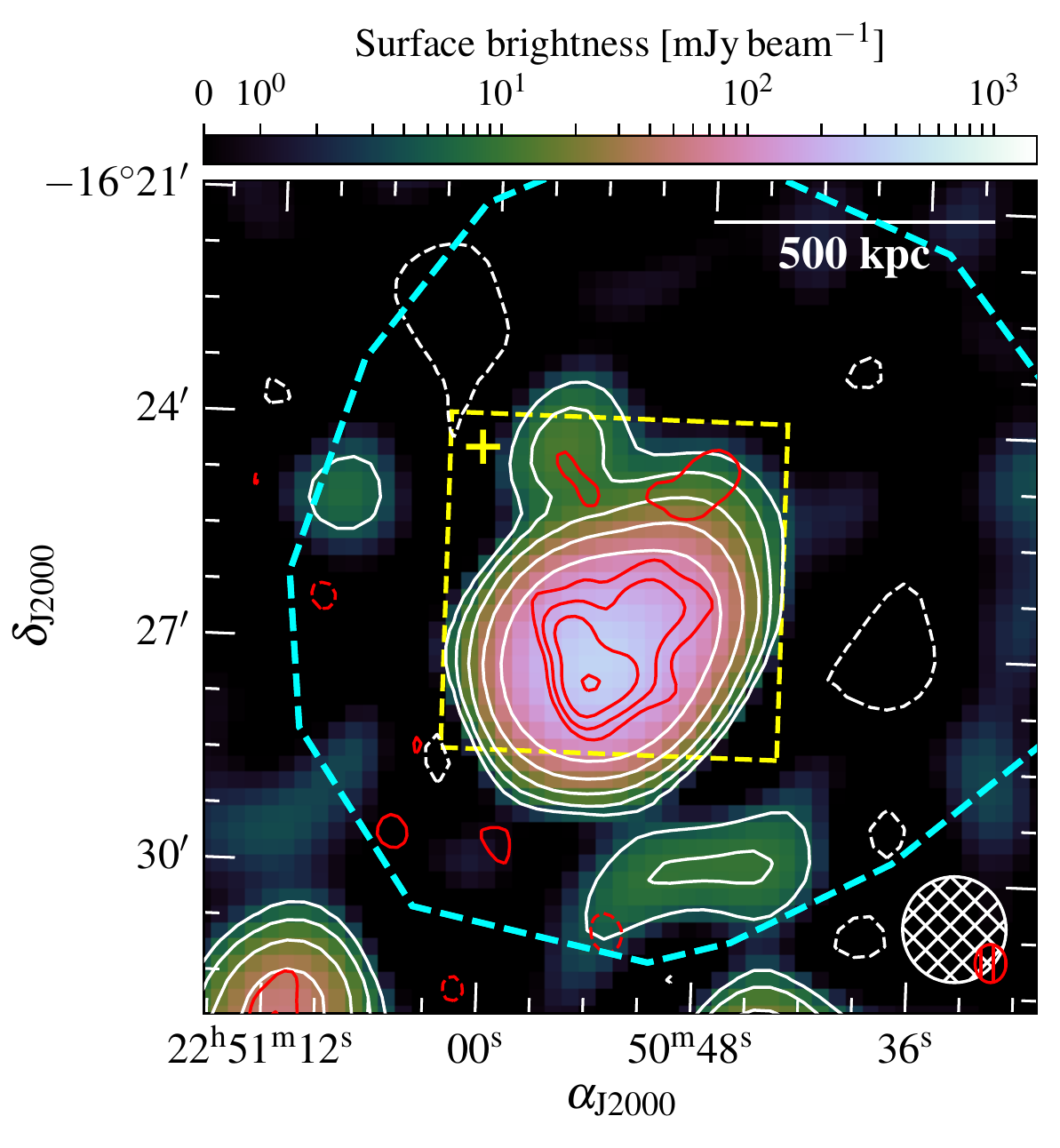}
    \caption{\label{fig:a2496:radio} }
    \end{subfigure}%
    \end{minipage}%
    \begin{minipage}[b]{0.33\linewidth}
    \begin{subfigure}[b]{1\linewidth}
    \includegraphics[width=1\linewidth]{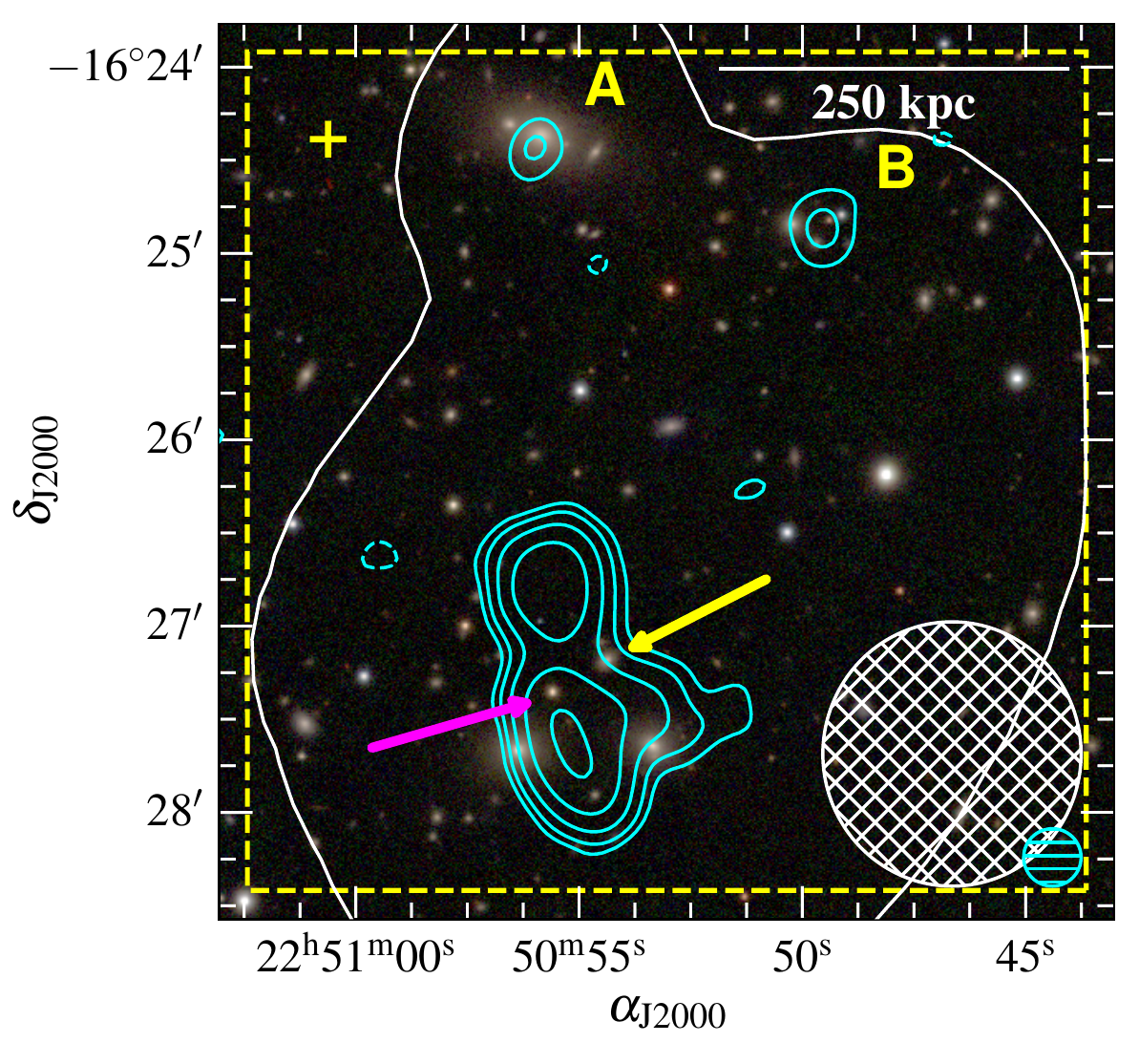}
    \caption{\label{fig:a2496:optical}}
    \end{subfigure}%
    \end{minipage}\hfill%
    \begin{minipage}[b]{0.30\linewidth}
    \caption{\label{fig:a2496} \hyperref[para:a2496]{Abell~2496}. \subref{fig:a2496:radio}. Background: MWA-2, 185~MHz, robust $+0.5$ image. \subref{fig:a2496:optical}. Background: RGB PS1 image ($i$, $r$, $g$). The white contours are as in \cref{fig:a0122:radio} for the background of \subref{fig:a2496:radio} (for $\sigma_\text{rms} = 1.5$~mJy\,beam$^{-1}$). Red contours: TGSS image, in levels of $[\pm 3, 6, 12, 24, 48] \times \sigma_\text{rms}$ ($\sigma_\text{rms} = 4$~mJy\,beam$^{-1}$). Cyan contours: RACS robust $+0.25$ image, in levels of $[\pm 3, 6, 12, 24, 48] \times \sigma_\text{rms}$ ($\sigma_\text{rms} = 0.23$~mJy\,beam$^{-1}$). Other image features are as in \cref{fig:a0122}.}
    \end{minipage}%
\end{figure*}

\subsubsection{\texttt{FIELD1}}

\paragraph{\hyperref[fig:a0122]{Abell~0122}} \label{para:a0122} (\cref{fig:a0122}). Reported by \citetalias{Duchesne2017} as an unclassified steep spectrum source. The source is detected in the MWA-2, TGSS, and deep ASKAP data, shown in \cref{fig:a0122:radio} and \cref{fig:a0122:optical}. The deep ASKAP data show a complex source with additional point source contributions (labelled in \cref{fig:a0122:optical}) and with contribution from what may be the core of the emission, the brightest cluster galaxy (BCG) (6dF~J0057228$-$261653; \citealt{Jones2009}) indicated by a magenta arrow in \cref{fig:a0122:optical}. The projected extent of the source is $\sim 2.6$~arcmin (corresponding to $\sim 310$~kpc), including the protrusion to the West of Source A. This is slightly smaller than \corrs{that} reported by \citetalias{Duchesne2017} due to less source blending. The SED between 88--943~MHz is shown in \cref{fig:sed:a0122}, finding curvature between the MWA and ASKAP data after subtraction of the labelled sources, and with a spectral index from 88--216~MHz of $\alpha_{88}^{216} = -1.6 \pm 0.1$. We consider this a remnant radio galaxy, likely associated with the BCG, or otherwise fossil plasma originally from the BCG.

\paragraph{\hyperref[fig:a2751]{Abell~2751}} \label{para:a2751} (\cref{fig:a2751}). \citetalias{Duchesne2017} report a relic source on the outskirts of Abell~2751 (D1 in \cref{fig:a2751:radio}). We show the MWA-2 and RACS \corrs{discrete} source-subtracted \corrs{images} in \cref{fig:a2751:radio}, and the higher-resolution RACS \corrs{image} in \cref{fig:a2751:optical} showing the embedded compact source labelled B. {The largest angular size (LAS) is 4.7~arcmin corresponding to a largest linear size (LLS) of 580~kpc, slightly smaller than \corrs{that} reported by \citetalias{Duchesne2017} due to the less confused images.} Sources A and B are subtracted from MWA-2 measurements, and we subtract the contribution of B from the measurements presented in \citetalias{Duchesne2017}. A plot of the SED between 88--1400~MHz is shown in \cref{fig:sed:a2751} in \cref{app:seds}, and we find a well-fit power law distribution with $\alpha_{88}^{1400} = -1.23 \pm 0.06$, consistent with $\alpha$ reported by \citetalias{Duchesne2017}. RASS data shown in the \cref{fig:a2751:xray} indicates the bulk ICM sits to the southwest, with D1 oriented almost perpendicular, which is abnormal for large-scale relics (with the exception of the relic source in MACS~J1149.5$+$2223, though the nature of that source is unclear; \citealt{Bonafede2012,Bruno2021}). With no evidence of shocks (and an absence of more sensitive X-ray data) we cannot differentiate from relic or fossil electrons/remnant radio galaxy. The reported cluster centre by \citet{aco89} is offset from the RASS X-ray peak by $\sim 2$~arcmin ($\sim 230$~kpc); the optical concentration of galaxies is also elongated \citep[][see their Figure 15]{Duchesne2017}---we suggest the system is merging based on these observations, and significant shocks may be present in the cluster volume. We consider this an ambiguous fossil source or remnant.

\paragraph{\hyperref[fig:a2811]{Abell~2811}} \label{para:a2811} (\cref{fig:a2811}). Halo/mini-halo candidate reported by \citetalias{Duchesne2017}, detected in MWA-2 data up to 185~MHz, with only partial detection at 216~MHz, and no detection in the RACS data (\cref{fig:a2811}). We measure the integrated flux density across the MWA-2 band, including the 200-MHz GLEAM image, and fit a power law model to the SED (\cref{fig:sed:a2811}), finding $\alpha_{88}^{200} = -2.5 \pm 0.4$ ($\alpha_{\text{MWA-2}} = -3.1 \pm 0.5$ for the MWA-2 data only), after subtraction of the contribution of Source B. The 168-MHz measurement reported by \citetalias{Duchesne2017} is slightly higher than expected due to additional blending with Source A. {Additionally, the LAS is $2.7$~arcmin (with an LLS of 320~kpc), slightly smaller again due to less blending with Source A.}
We fit the exposure-corrected and background-subtracted \xmm\ data presented in \corrs{\citetalias{Duchesne2017}} with a single-$\beta$ model \citep{Cavaliere1976}, and estimate the X-ray morphological parameters, the centroid shift, $w$ \citep{Poole2006}, with an outer radius set to $R_{500} = 1.035$~Mpc \citep{pap+11}. We find $w = 0.072R_{500}$, consistent with disturbed systems \citep{pcab09}. Additionally, the surface brightness concentration, $c_{100/500}$ is found to be 0.21, placing it right on the border of merging, halo-hosting clusters \citep{Cassano2010}. Similarly, the centroid shift within 500~kpc is found to be $w_{500} = 0.07$, placing it outside of halo-hosting quadrant, near Abell~697 which has been reported to host a radio halo \citep[but see also \citealt{Kempner2001}][]{vgd+08} with an ultra-steep spectrum \citep[$\alpha = -1.5$;][]{Macario2013}, though not as steep as the spectrum for Abell~2811. We also note the concentration parameter, $c_{40/400} = 0.048$, is below what is typically seen in cool-core clusters (CC; \citealt{Santos2008}, with $c_{40/400} \gtrsim 0.075$). Many of the properties are consistent with a radio halo, however, such a steep spectrum is rare for radio halos: while we consider this an extreme case of an ultra-steep--spectrum radio halo (USSRH) it may be a fossil plasma source projected onto the cluster centre.

\subsubsection{\texttt{FIELD2}}

\paragraph{\hyperref[fig:a2496]{Abell~2496}} \label{para:a2496} (\cref{fig:a2496}). Reported by \citetalias{Duchesne2017} as an unclassifed diffuse cluster source. The MWA-2 and TGSS data in \cref{fig:a2496:radio} show an extended source, with the RACS data in \cref{fig:a2496:optical} showing a clear double-lobed morphology. A small extension is seen in the RACS data in the direction of the larger extension seen in the TGSS and MWA-2 images, tracing an older plasma component. {The angular and linear extend of the source is the same as reported in \citetalias{Duchesne2017}.} The overall emission is modelled with a normal power law with $\alpha_{88}^{1400} = -1.23 \pm 0.05$ (\cref{fig:sed:a0122}). The PS1 data show possible hosts between the lobes: WISEA~J225055.58$-$162721.0; indicated by a yellow arrow in \cref{fig:a2496:optical}, and WISEA J225054.36-162710.7; indicated by a magenta arrow, neither with known redshifts. No distinct radio core is seen. We suggest this is a remnant radio galaxy. 

\begin{figure*}[tp]
    \centering
    \begin{subfigure}[b]{0.33\linewidth}
    \includegraphics[width=1\linewidth]{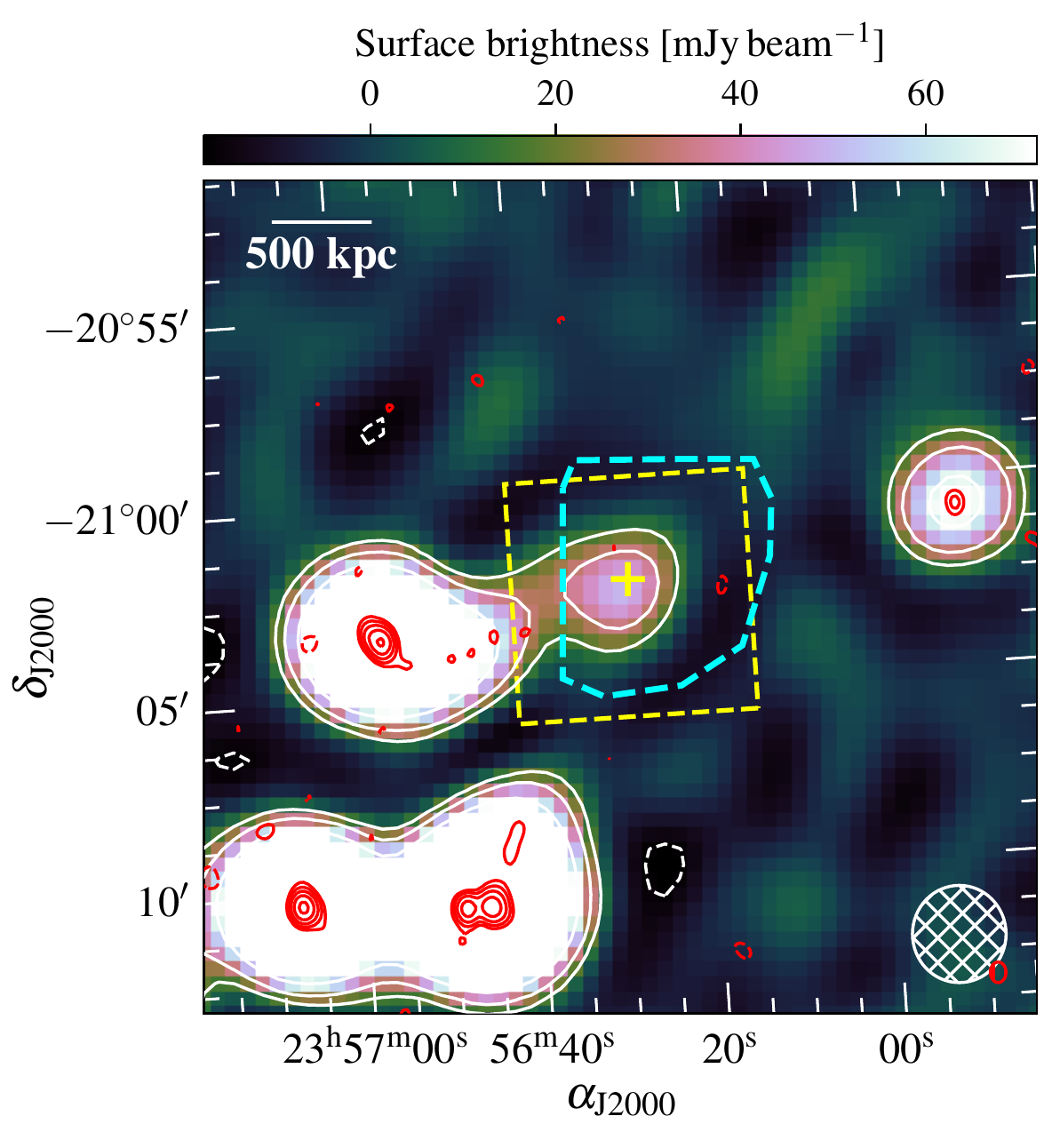}
    \caption{\label{fig:a2680:radio}}
    \end{subfigure}%
    \begin{subfigure}[b]{0.33\linewidth}
    \includegraphics[width=1\linewidth]{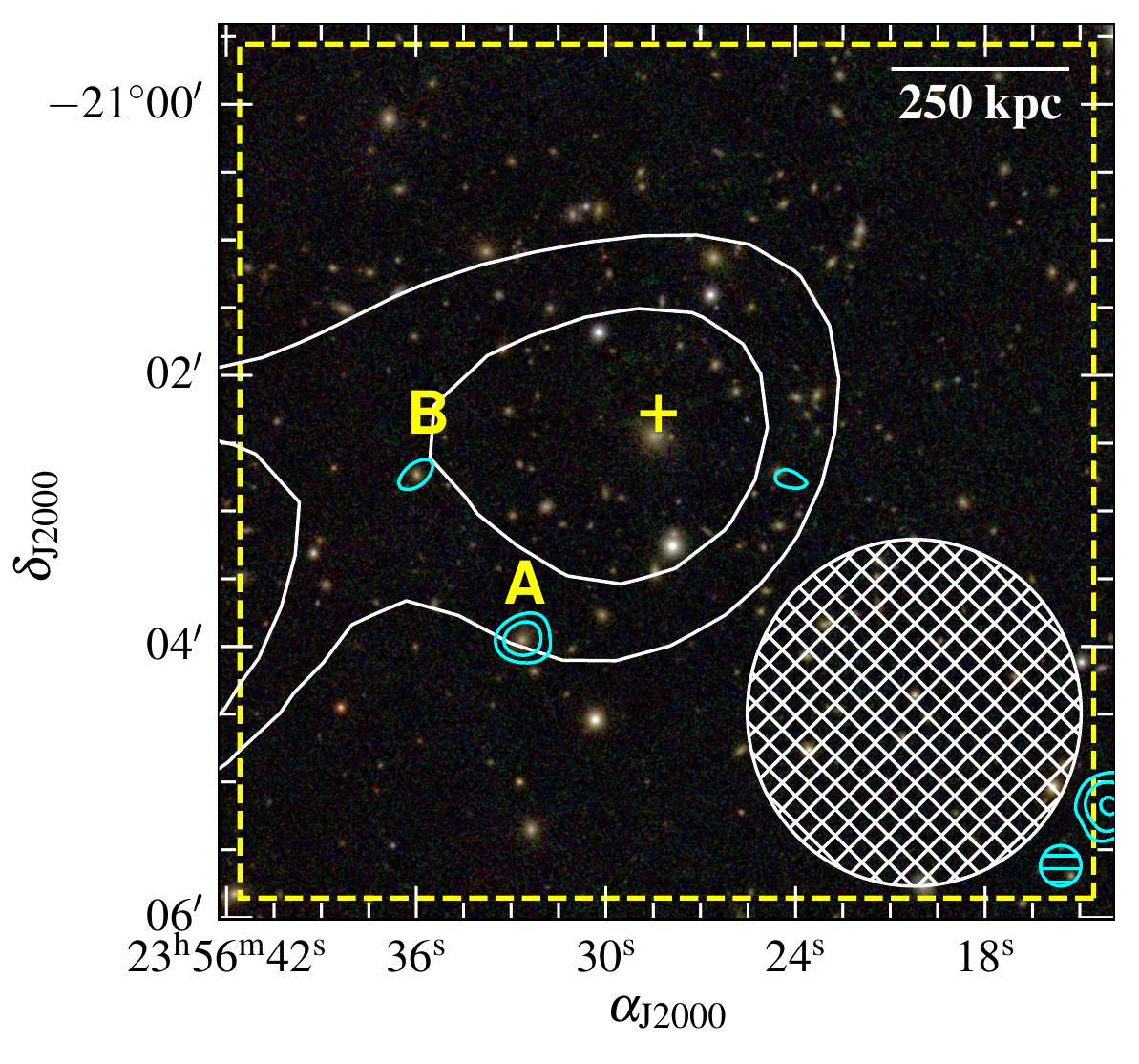}
    \caption{\label{fig:a2680:optical}}
    \end{subfigure}%
    \begin{subfigure}[b]{0.33\linewidth}
    \includegraphics[width=1\linewidth]{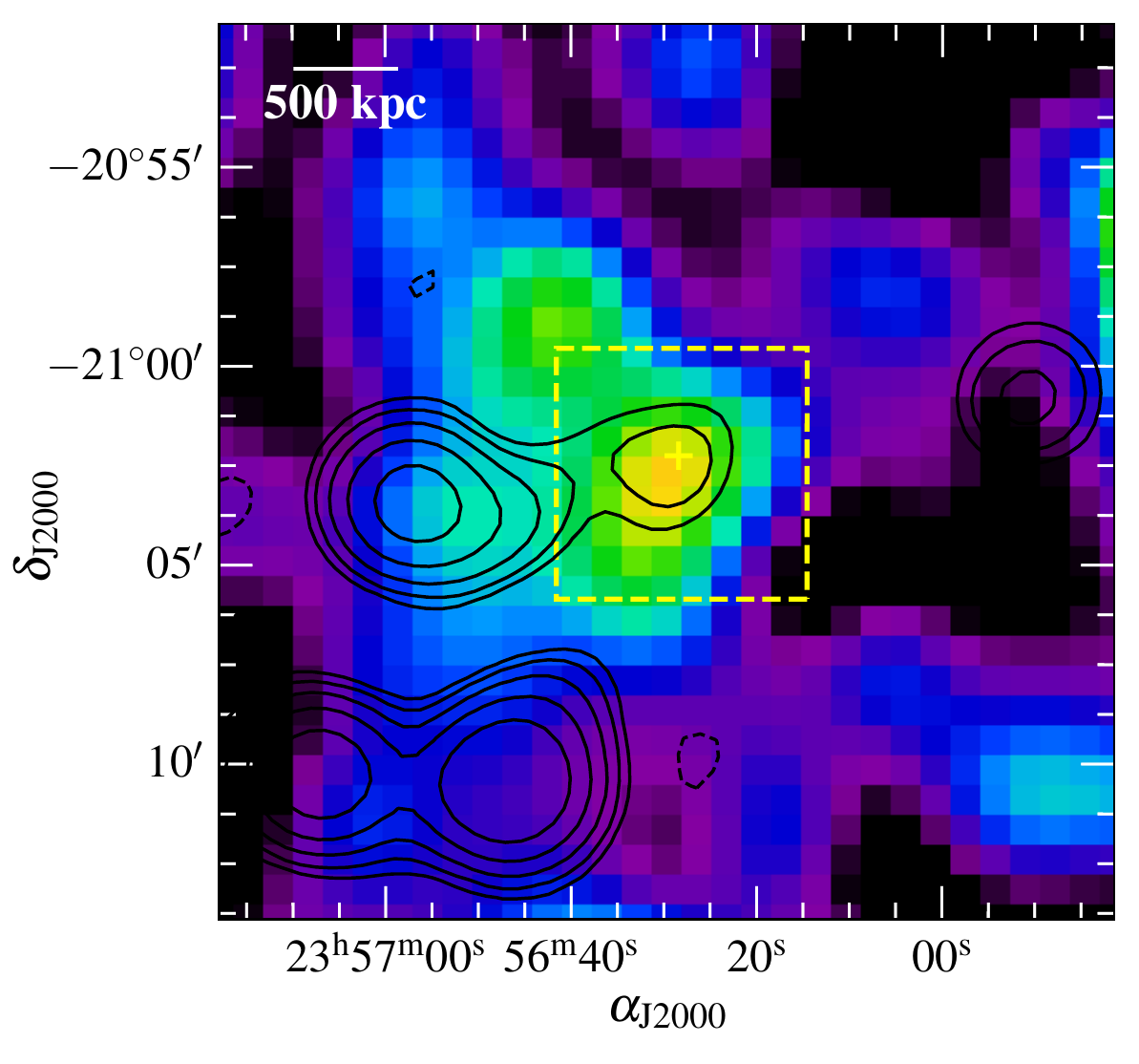}
    \caption{\label{fig:a2680:xray}}
    \end{subfigure}%
    \caption{\label{fig:a2680} \hyperref[para:a2680]{Abell~2680}. \subref{fig:a2680:radio}. Background: MWA-2, 88~MHz, robust $+0.5$ image. \subref{fig:a2680:optical}. Background: RGB PS1 image ($i$, $r$, $g$). \subref{fig:a2680:xray}. Background: Smoothed RASS image. The white (black) contours are as in \cref{fig:a0122:radio} for the background of \subref{fig:a2680:radio} (with $\sigma_\text{rms} = 5.5$~mJy\,beam$^{-1}$). Red contours: TGSS image, in levels of $[\pm 3, 6, 12, 24, 48] \times \sigma_\text{rms}$ ($\sigma_\text{rms} = 3.7$~mJy\,beam$^{-1}$). Cyan contours: RACS robust $+0.25$ image, in levels of $[\pm 3, 6, 12, 24, 48] \times \sigma_\text{rms}$ ($\sigma_\text{rms} = 0.25$~mJy\,beam$^{-1}$). Other image features are as in \cref{fig:a0122}.}
\end{figure*}

\paragraph{\hyperref[fig:a2680]{Abell~2680}} \label{para:a2680} (\cref{fig:a2680}). Reported by \citetalias{Duchesne2017}. \cref{fig:a2680:radio} shows the MWA-2 and TGSS radio data, and \cref{fig:a2680:optical} the PS1 data with MWA-2 and RACS data overlaid. {The LAS of the source is 3.0~arcmin (with an LLS of 580~kpc), slightly larger than reported by \citetalias{Duchesne2017} and the reduced confusion enables a better estimate of the size}. A single compact source is detected within the emission with RACS (Source A) and is subtracted from subsequent flux density measurements. We are only able to provide measurements in the 88-, 118-, and 154-MHz MWA-2 bands as the cluster is towards the edge of \texttt{FIELD2} with lessened sensitivity in the higher frequency images. We find $\alpha_{88}^{200} = -1.7 \pm 0.7$ (\cref{fig:sed:a2680}), consistent with the limited reported by \citetalias{Duchesne2017}. Smoothed RASS data is shown in \cref{fig:a2680:xray} highlighting the location of the radio emission relative to the thermal ICM though noting that the RASS data provide limited insight to the morphology of the ICM. From an optical analysis, \citet[][but see also \citealt{whl12}]{Wen2015} report an $R_{500}$ \footnote{Radius within which the mean density of the cluster is 500 times the critical density of the Universe.} of 1.26~Mpc which corresponds to mass of $M_{500} = 2.4 \times 10^{14}$~M$_\odot$ \corrs{following Equation~1 from \citet{Wen2015}}. As the cluster is detected in the RASS data, a mass is estimated following the procedures described by \citet{Tarrio2016,Tarrio2018}, resulting in $M_{500} = 3.2_{-1.0}^{+0.8} \times 10^{14}$~M$_\odot$, somewhat consistent with the mass derived from the optical radius. We consider this a candidate halo. 

\begin{figure*}[tp]
    \centering
    \begin{subfigure}[b]{0.33\linewidth}
    \includegraphics[width=1\linewidth]{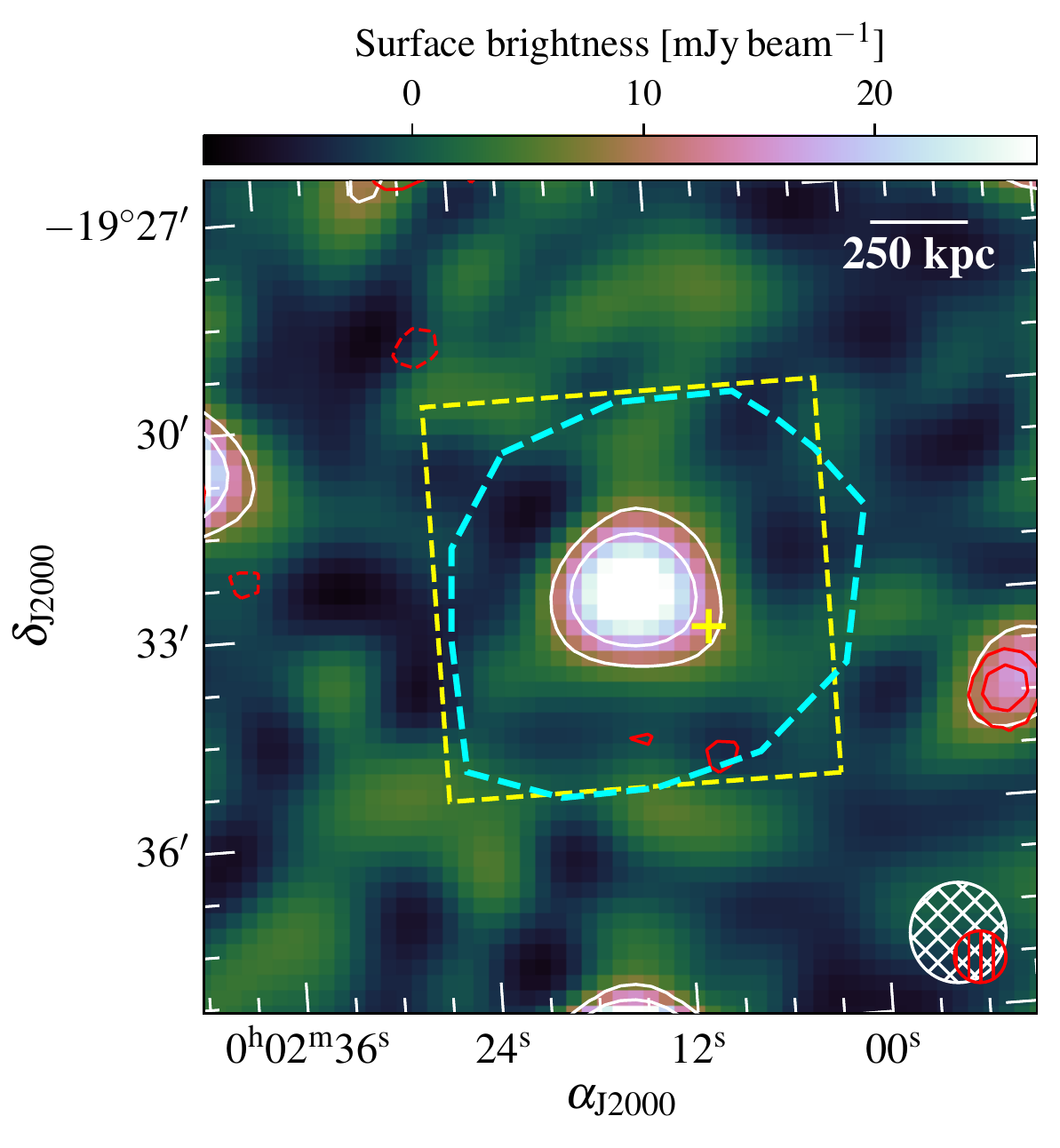}
    \caption{\label{fig:a2693:radio}}
    \end{subfigure}%
    \begin{subfigure}[b]{0.33\linewidth}
    \includegraphics[width=1\linewidth]{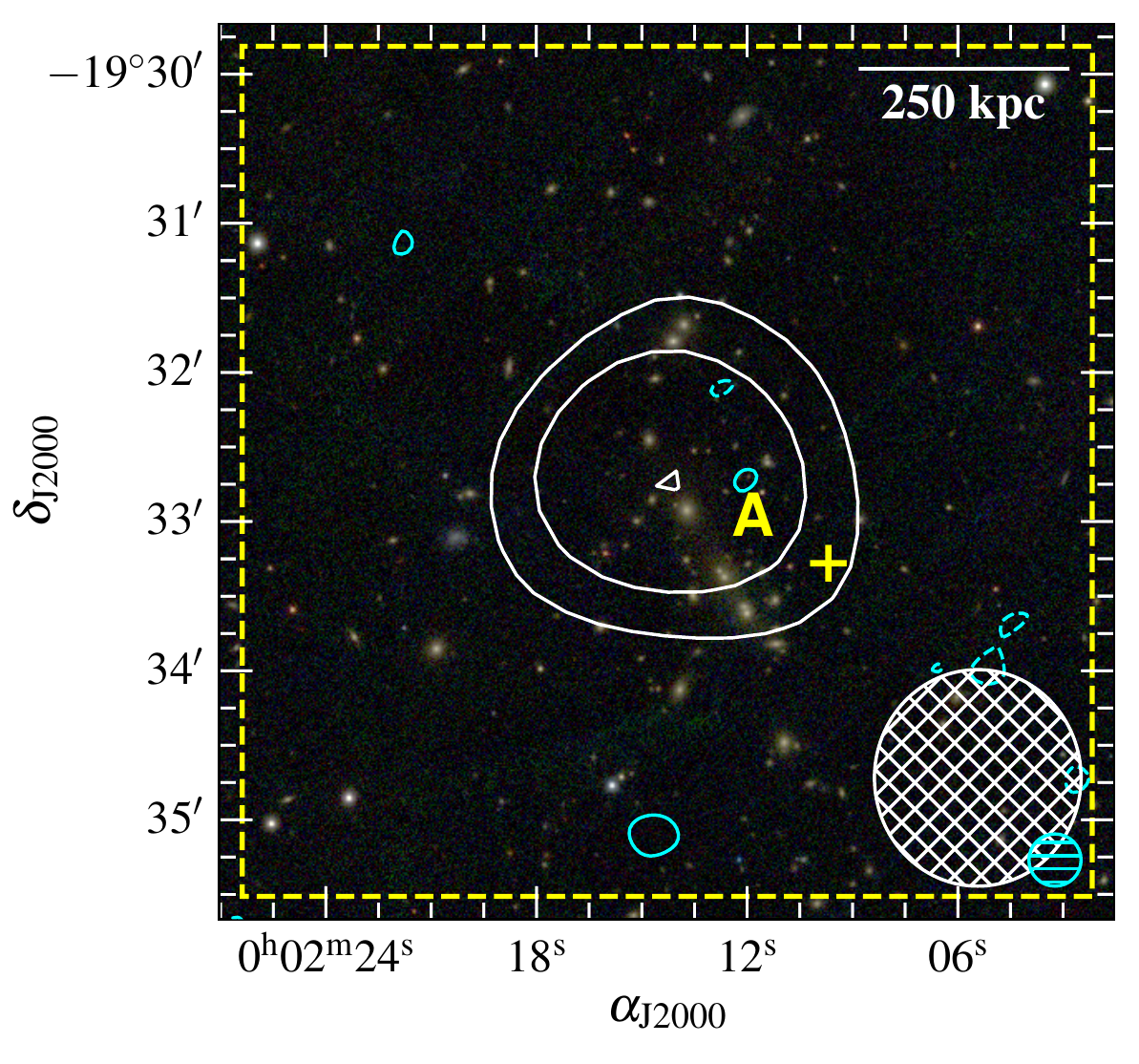}
    \caption{\label{fig:a2693:optical}}
    \end{subfigure}%
    \begin{subfigure}[b]{0.33\linewidth}
    \includegraphics[width=1\linewidth]{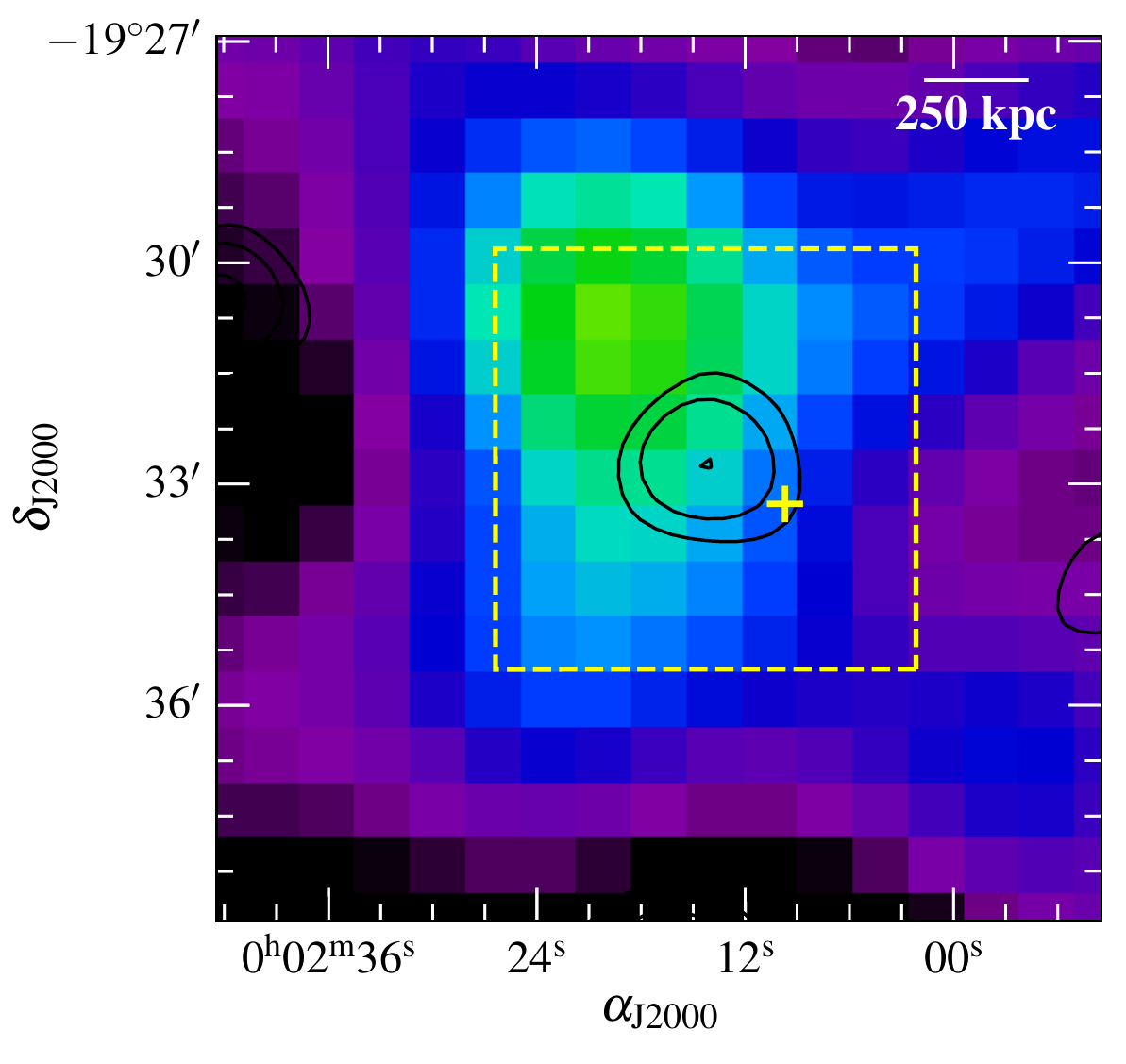}
    \caption{\label{fig:a2693:xray}}
    \end{subfigure}%
    \caption{\label{fig:a2693} \hyperref[para:a2693]{Abell~2693}. \subref{fig:a2693:radio} Background: MWA-2, 154~MHz, robust $+0.5$ image. \subref{fig:a2693:optical}. Background: RGB PS1 image ($i$, $r$, $g$). \subref{fig:a2693:xray}. Background: Smoothed RASS image. The white (black) contours are as in \cref{fig:a0122:radio} for the background of \subref{fig:a2693:radio} (with $\sigma_\text{rms} = 2.8$~mJy\,beam$^{-1}$). Red contours: NVSS image as in \cref{fig:a2811:radio}. Cyan contours: RACS robust $+0.25$ image, in levels of $[\pm 3, 6, 12, 24, 48] \times \sigma_\text{rms}$ ($\sigma_\text{rms} = 0.15$~mJy\,beam$^{-1}$). Other image features are as in \cref{fig:a0122}.}
\end{figure*}

\paragraph{\hyperref[fig:a2693]{Abell~2693}} \label{para:a2693} (\cref{fig:a2693}). Reported by \citetalias{Duchesne2017}. The candidate radio halo in Abell~2693 is largely similar to that in Abell~2680, with only a faint discrete source detected in the RACS data (Source A) with $S_{\text{A},887} \sim 0.8$~mJy. We provide additional flux density measurements, subtracting the contribution of Source A, to obtain a spectral index of $\alpha_{88}^{200} = -1.5 \pm 0.2$ (\cref{fig:sed:a2693}). {A mass is derived from the RASS data: $M_{500} = 2.1_{-0.6}^{+0.5} \times 10^{14}$~M$_\odot$, placing Abell~2693 as one of the least massive halo-hosting clusters if confirmed \citep[surpassed only by the `Ant' cluster;][]{Botteon2021b}. {The LAS for the source is 2.0~arcmin ($\text{LLS}=370$~kpc), marginally smaller than \corrs{that} reported by \citetalias{Duchesne2017}.} Alternatively, this may be a point source with $\alpha_{88}^{887} = - 2.2 \pm 0.2$.} 

\begin{figure*}[tp]
    \centering
    \begin{minipage}[b]{0.33\linewidth}
    \begin{subfigure}[b]{1\linewidth}
    \includegraphics[width=1\linewidth]{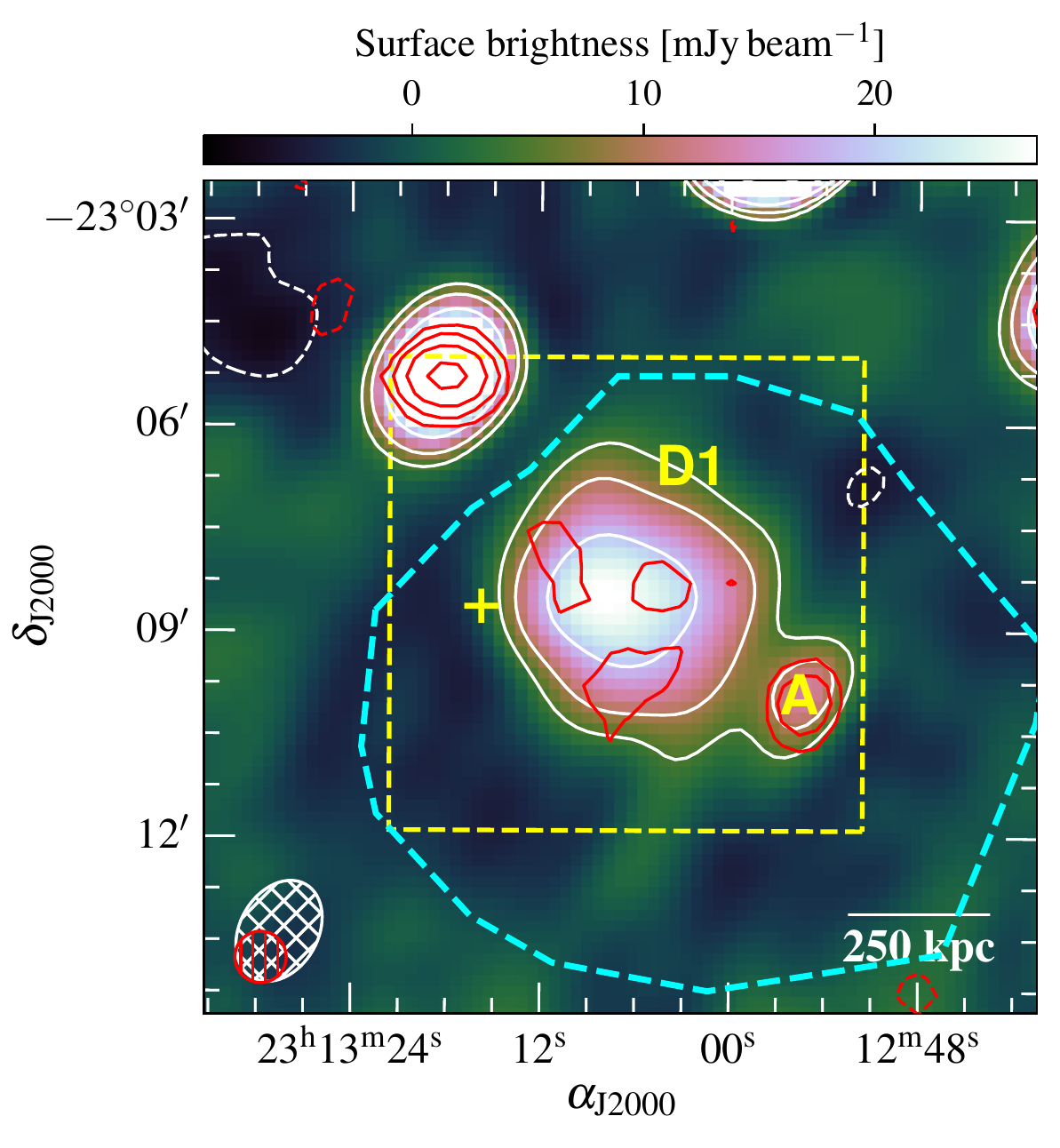}
    \caption{\label{fig:as1099:radio}}
    \end{subfigure}%
    \end{minipage}%
    \begin{minipage}[b]{0.33\linewidth}
    \begin{subfigure}[b]{1\linewidth}
    \includegraphics[width=1\linewidth]{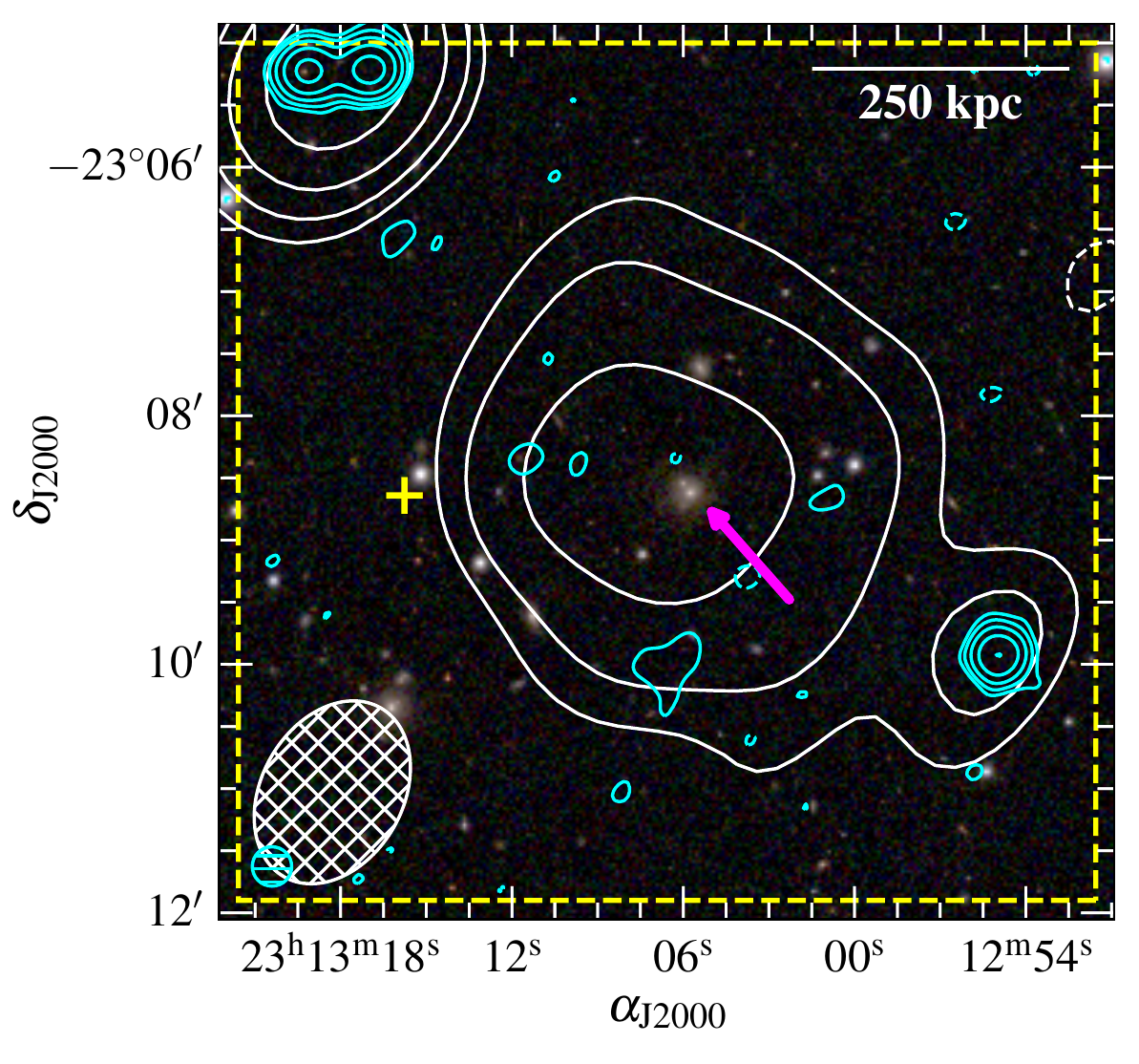}
    \caption{\label{fig:as1099:optical}}
    \end{subfigure}%
    \end{minipage}\hfill%
    \begin{minipage}[b]{0.3\linewidth}
    \caption{\label{fig:as1099} \hyperref[para:as1099]{Abell~S1099}. \subref{fig:as1099:radio} Background: MWA-2, 216-MHz, robust $+2.0$ image. \subref{fig:as1099:optical} Background: RGB PS1 image ($i$, $r$, $g$). The white contours are as in \cref{fig:a0122:radio} for the background of \subref{fig:as1099:radio} (with $\sigma_\text{rms} = 1.6$~mJy\,beam$^{-1}$). Red contours: NVSS image as in \cref{fig:a2811:radio}. Cyan contours: RACS robust $+0.25$ image, in levels of $[\pm 3, 6, 12, 24, 48] \times \sigma_\text{rms}$ ($\sigma_\text{rms} = 0.19$~mJy\,beam$^{-1}$). Other image features are as in \cref{fig:a0122}.}
    \end{minipage}%
\end{figure*}

\paragraph{\hyperref[fig:as1099]{Abell~S1099}} \label{para:as1099} (\cref{fig:as1099}). Reported by \citetalias{Duchesne2017}. MWA-2 radio data shown in \cref{fig:as1099:radio} and PS1 optical data shown in \cref{fig:as1099:optical}. RACS data shows no additional discrete sources beyond Source A, which is subtracted from flux density measurements where appropriate with $\alpha_{\text{A},216}^{1400} = -0.5 \pm 0.1$. The resulting spectral index of the diffuse source D1 is found to be $\alpha_{88}^{1400} = -0.87\pm0.11$ (\cref{fig:sed:as1099}). The lack of obvious core or any clear lobes/hot spots suggests a remnant radio source that has diffused into the surrounding medium. No deep X-ray observations are available, no cluster or source is detected in the RASS image, and there is no detection as a \emph{Planck}-SZ source. We consider this a remnant radio galaxy with the putative host (LEDA~195207) indicated by a magenta arrow on \cref{fig:as1099:optical}.
\par

\begin{figure*}[t]
    \centering
    \begin{minipage}[b]{0.33\linewidth}
    \begin{subfigure}[b]{1\linewidth}
    \includegraphics[width=1\linewidth]{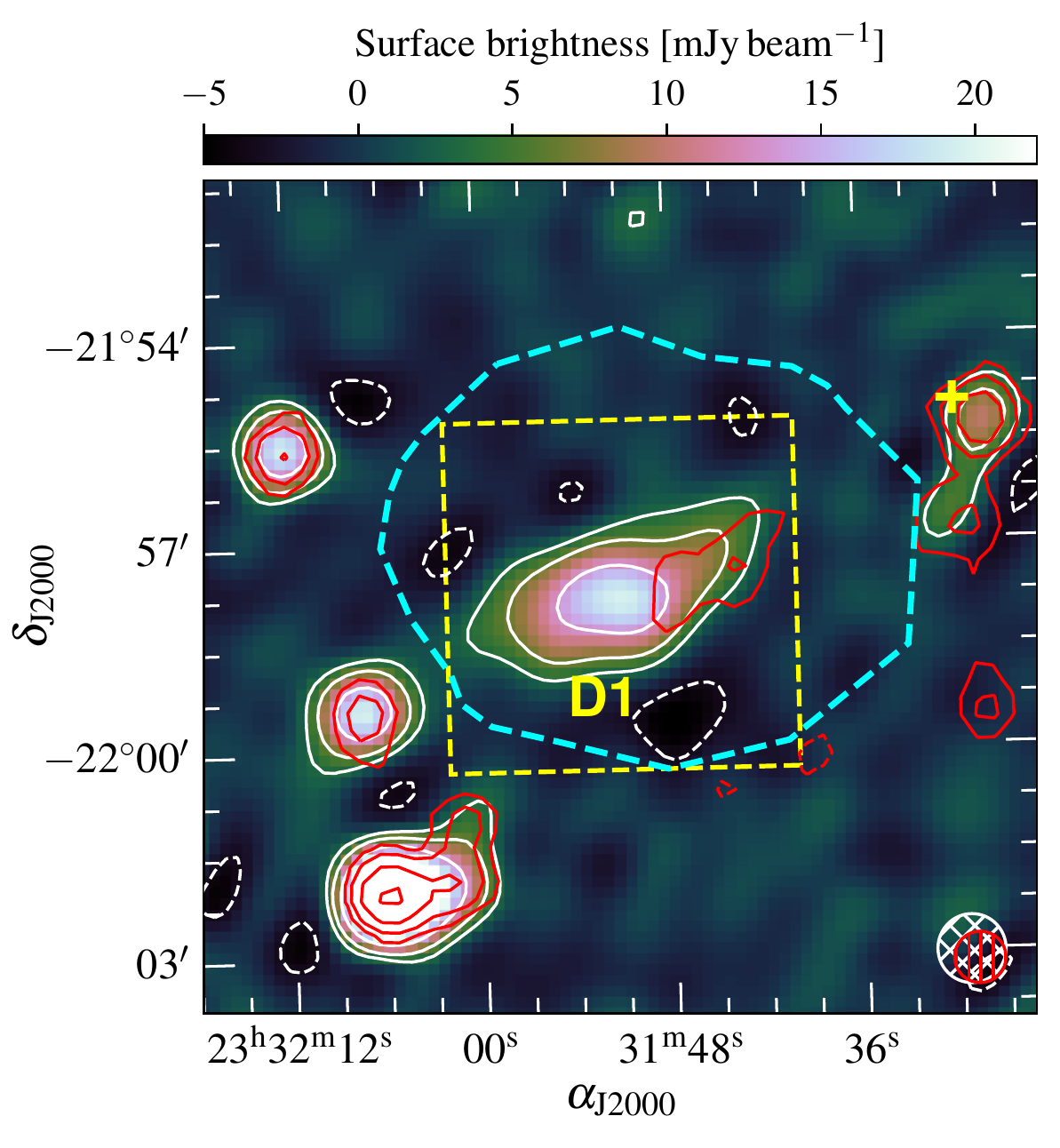}
    \caption{\label{fig:aqrcc:radio}}
    \end{subfigure}%
    \end{minipage}%
    \begin{minipage}[b]{0.33\linewidth}
    \begin{subfigure}[b]{1\linewidth}
    \includegraphics[width=1\linewidth]{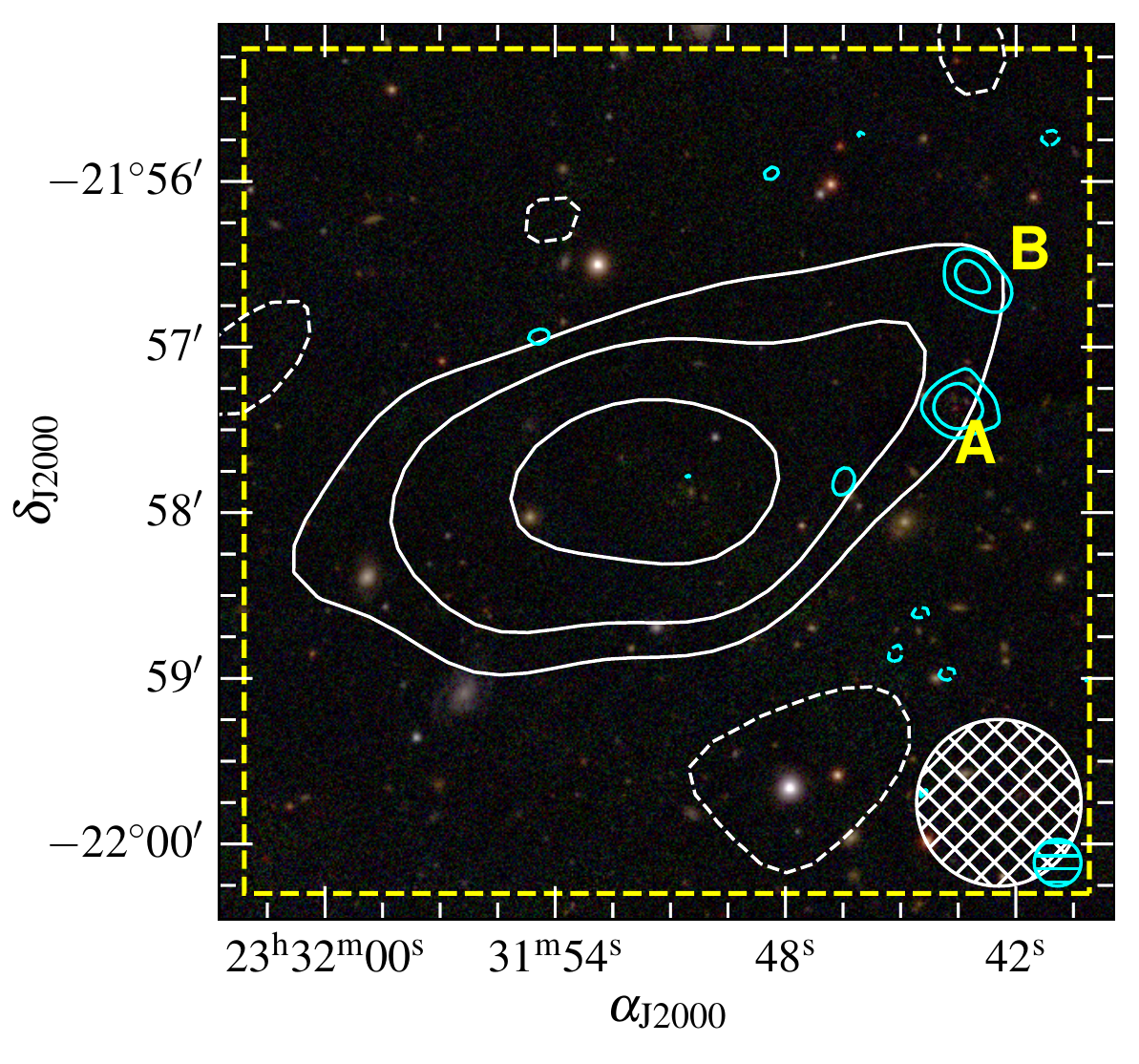}
    \caption{\label{fig:aqrcc:optical}}
    \end{subfigure}%
    \end{minipage}\hfill%
    \begin{minipage}[b]{0.3\linewidth}
    \caption{\label{fig:aqrcc} \hyperref[para:aqrcc]{AqrCC~087}. \subref{fig:aqrcc:radio} Background: MWA-2, 216-MHz, robust $+0.5$ image. \subref{fig:aqrcc:optical} Background: RGB PS1 image ($i$, $r$, $g$). The white contours are as in \cref{fig:a0122:radio} for the background of \subref{fig:aqrcc:radio} (with $\sigma_\text{rms} = 1.1$~mJy\,beam$^{-1}$). Red contours: NVSS image as in \cref{fig:a2811:radio}. Cyan contours: RACS robust $+0.25$ image, in levels of $[\pm 3, 6, 12, 24, 48] \times \sigma_\text{rms}$ ($\sigma_\text{rms} = 0.15$~mJy\,beam$^{-1}$). Other image features are as in \cref{fig:a0122}, though note no scalebar is given as no redshift is available for the reported cluster.}
    \end{minipage}%
\end{figure*}

\paragraph{\hyperref[fig:aqrcc]{AqrCC~087}} \label{para:aqrcc} (\cref{fig:aqrcc}). The cluster is reported in the Aquarius cluster catalogue \citep{cmkw02}, though no redshift is available. Additionally, there is no cross-identification with other cluster catalogues, and as with Abell~S1099 no X-ray or SZ observations provide detections. \corrs{We suggest this is a poor cluster or group.} The redshift distribution of galaxies within 1~deg around AqrCC~087 peaks around $z \approx 0.1$. We detect an elongated radio source $\sim 5.6$~arcmin from the reported cluster centre (\cref{fig:aqrcc:radio}, \corrs{$\sim 620$~kpc at $z=0.1$}) with no obvious optical host (\cref{fig:aqrcc:optical}). The angular size is $\sim 4.5$~arcmin, which if at $z=0.1$ is a linear projected extent of $\sim 500$~kpc. The source is not detected in RACS, with a partial detection in the NVSS image (though note there is confusion with the discrete Source A). Source A is subtracted from subsequent MWA-2 flux density measurements, and we obtain a spectral index of $\alpha_{88}^{216} = -1.7 \pm 0.1$ (\cref{fig:sed:aqrcc}). As with Abell~S1099, we consider this likely to be a remnant radio galaxy.

\begin{figure*}[h!]
    \centering
    \begin{subfigure}[b]{0.33\linewidth}
    \includegraphics[width=1\linewidth]{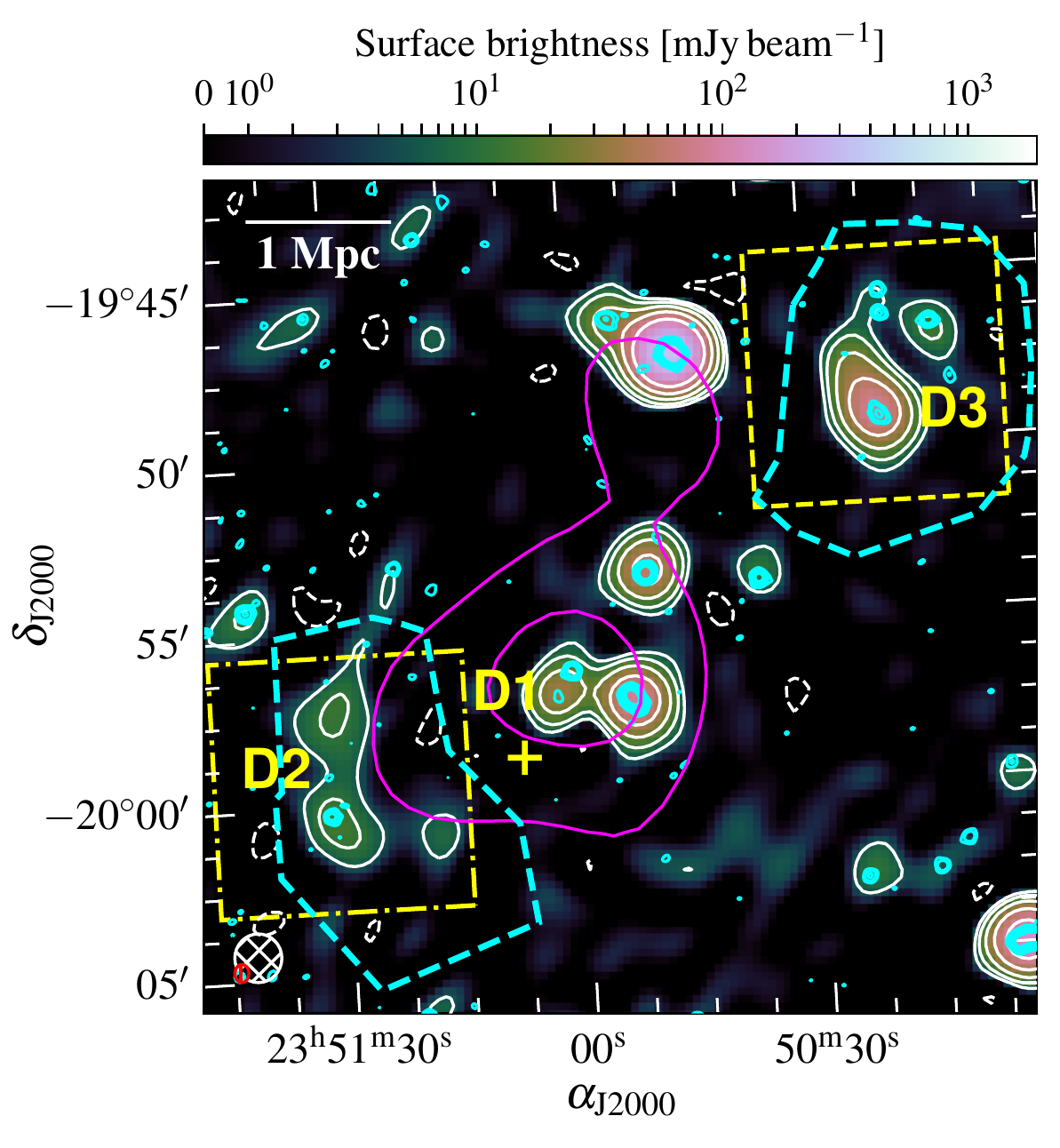}
    \caption{\label{fig:rxcj2351:radio}}
    \end{subfigure}%
    \begin{subfigure}[b]{0.33\linewidth}
    \includegraphics[width=1\linewidth]{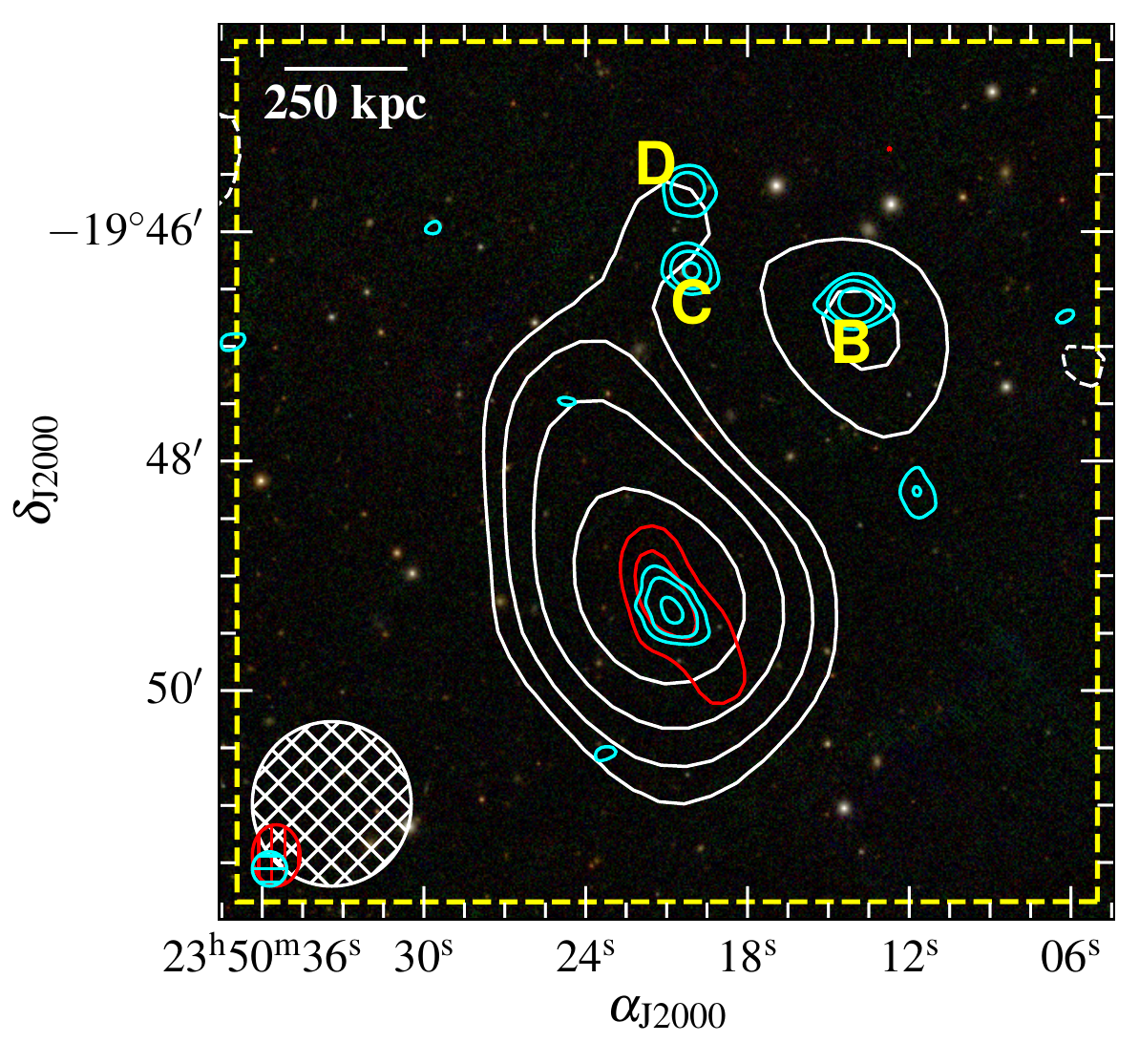}
    \caption{\label{fig:rxcj2351:optical}}
    \end{subfigure}%
    \begin{subfigure}[b]{0.33\linewidth}
    \includegraphics[width=1\linewidth]{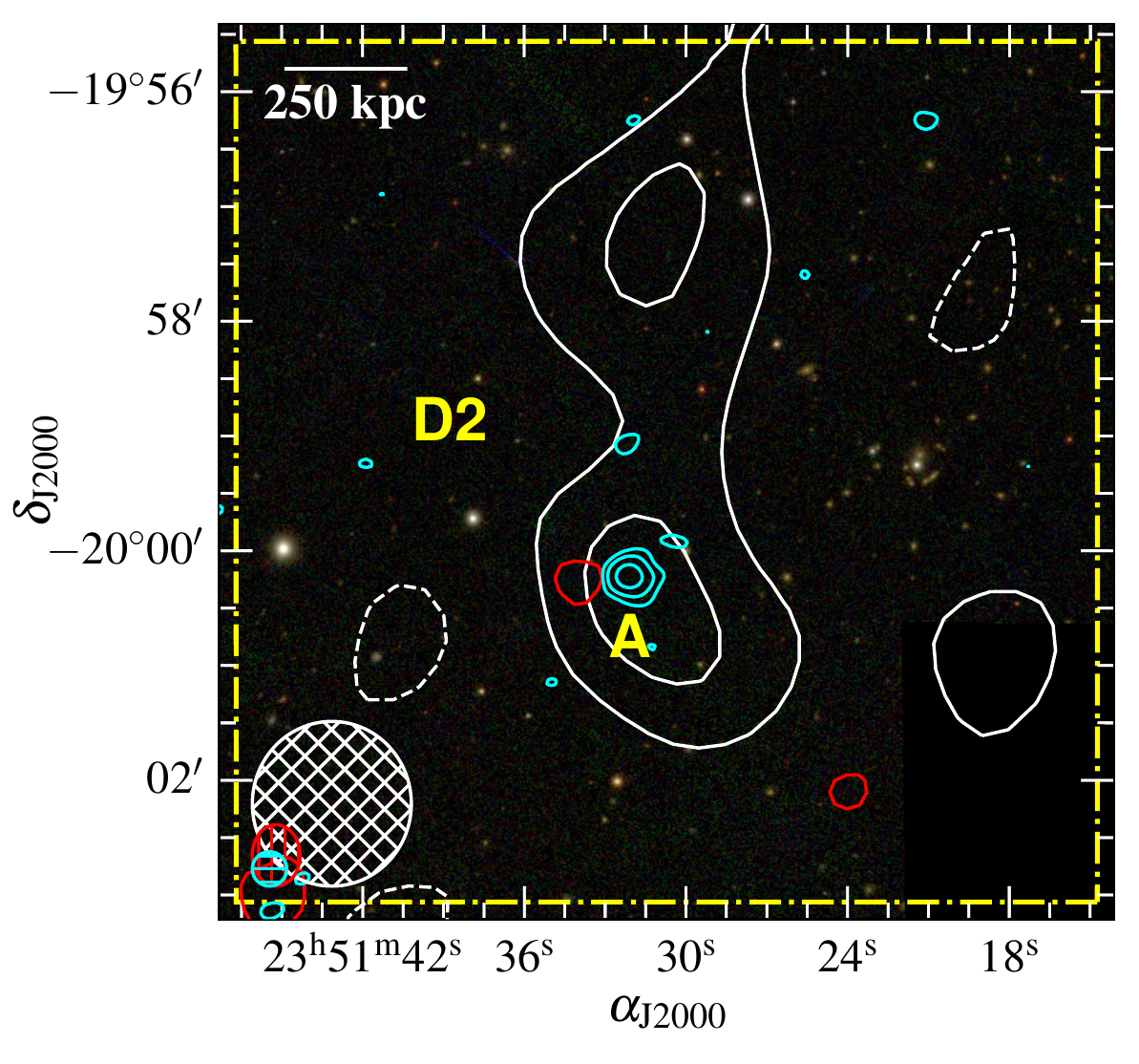}
    \caption{\label{fig:rxcj2351:optical2}}
    \end{subfigure}%
    \caption{\label{fig:rxcj2351} \hyperref[para:rxcj]{RXC~J2351.0$-$1954}. \subref{fig:rxcj2351:radio} Background: MWA-2, 154-MHz, robust $+0.5$ image. \subref{fig:rxcj2351:optical} and \subref{fig:rxcj2351:optical2} Background: RGB PS1 image ($i$, $r$, $g$). The white contours are as in \cref{fig:a0122:radio} for the background of \subref{fig:rxcj2351:radio} (with $\sigma_\text{rms} = 4.9$~mJy\,beam$^{-1}$). Red contours: TGSS image, in levels of $[\pm 3, 6, 12, 24, 48] \times \sigma_\text{rms}$ ($\sigma_\text{rms} = 3.6$~mJy\,beam$^{-1}$). Cyan contours: RACS robust $+0.25$ image, in levels of $[3, 6, 12, 24, 48] \times \sigma_\text{rms}$ ($\sigma_\text{rms} = 0.20$~mJy\,beam$^{-1}$). Magenta contours: smoothed RASS image, increasing with factors of $\sqrt{2}$. Other image features are as in \cref{fig:a0122}.}
\end{figure*}

\paragraph{\hyperref[fig:rxcj2351]{RXC~J2351.0$-$1954} (PSZ1~G057.09-74.45)} \label{para:rxcj} (\cref{fig:rxcj2351}). Originally \citetalias{Duchesne2017} reported three candidate sources: a halo at the centre (labelled D1 in \cref{fig:rxcj2351:radio}), and two relics: SE (labelled D2, inset \cref{fig:rxcj2351:optical2}), and NW, (labelled D3, inset \cref{fig:rxcj2351:optical}). The candidate halo is shown by the RACS data to be blended sources. The SE candidate relic detected at low significance ($\sim 3\sigma$), though the RACS data show a single compact source within the southern portion of the emission (Source A). The NW candidate relic is partially detected in RACS, with other compact blended sources (B--D in \cref{fig:rxcj2351:optical}). {Excluding point sources, the LAS of D2 and D3 are 1.9~arcmin ($\text{LLS}=230$) and 3.5~arcmin ($\text{LLS}=430$~kpc), respectively.} All labelled compact sources are subtracted from MWA measurements where appropriate. The spectral index of D2 is measured to be $\alpha_{\text{D2},88}^{216} = -1.3 \pm 0.3$ (\cref{fig:sed:rxcja}). D3 shows significant curvature within the MWA band and to 887.5~MHz and is fit with a generic curved model, with a power law model fit across the MWA-2 band: $\alpha_{\text{D3},88}^{216} = -1.2 \pm 0.1$ ( \cref{fig:sed:rxcjb}).The smoothed RASS image is shown as contours on \cref{fig:rxcj2351:radio} showing elongation hinting at an un-relaxed ICM, though we consider that D3 is likely a remnant with the classical spectral steepening and D2 is still unconfirmed.

\subsubsection{\texttt{FIELD3}}

\begin{figure}[h!]
    \centering
    \includegraphics[width=1\linewidth]{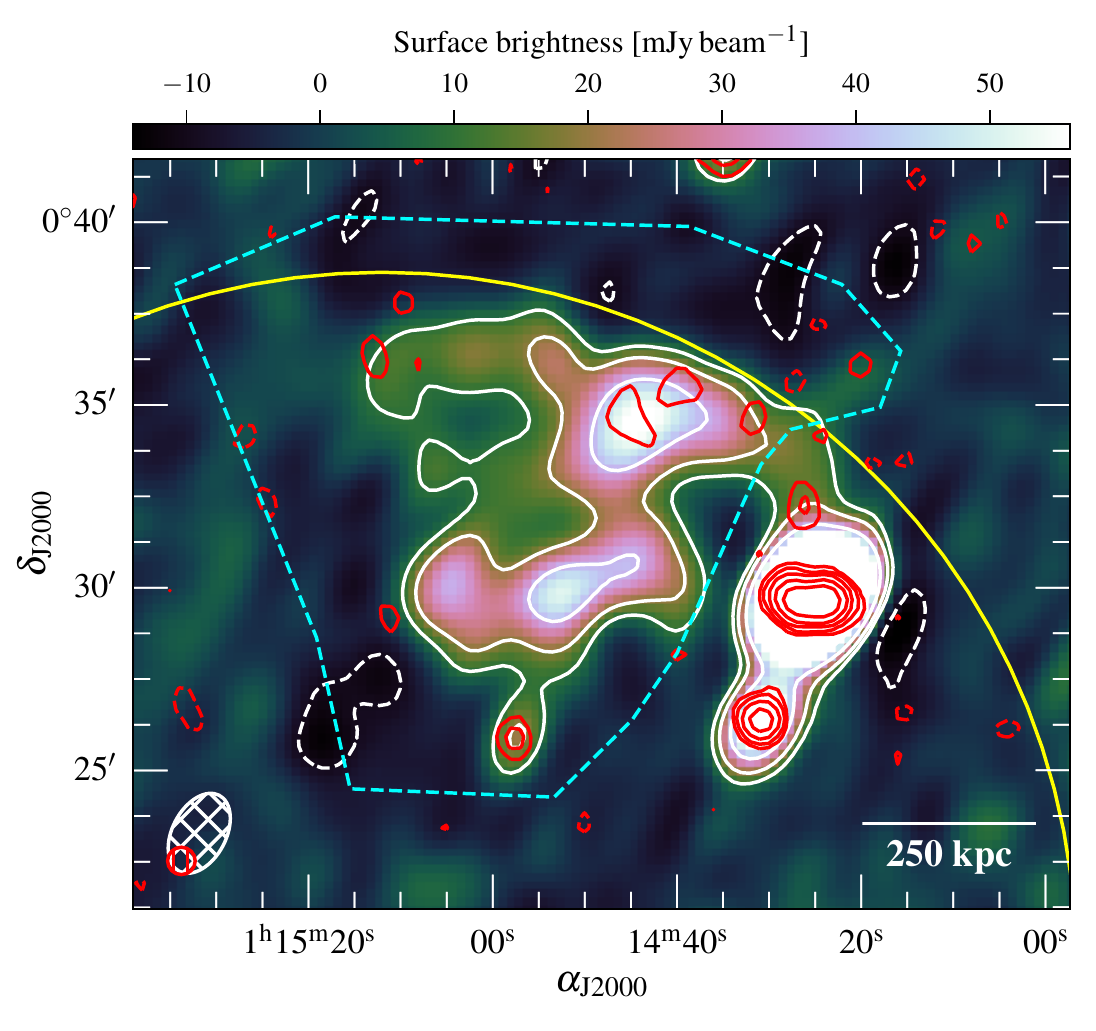}
    \caption{\label{fig:a168} \hyperref[para:a168]{Abell~0168}. Background: MWA-2, 154-MHz, robust $+2.0$ image. The white contours are as in \cref{fig:a0122:radio} for the background image (with $\sigma_\text{rms} = 3.2$~mJy\,beam$^{-1}$). Red contours: NVSS, $[\pm 3, 6, 12, 24, 48] \times \sigma_\text{rms}$ ($\sigma_\text{rms} = 0.45$~mJy\,beam$^{-1}$). Other image features are as in \cref{fig:a0122}, with a yellow circle with a 1~Mpc radius centered on the cluster.}
\end{figure}

\paragraph{\hyperref[fig:a168]{Abell~0168}} \label{para:a168} \cref{fig:a168} shows the radio relic \corrs{that was} reported by \citet{Dwarakanath2018} \corrs{and is} detected in the MWA-2 data. \citet{Dwarakanath2018} split the total relic source into two distinct components\corrs{--a large exterior component and a smaller interior component with a steeper spectrum. Here we consider it a single emission region due in part to the limitation of resolution but there also appears to be a fainter diffuse component connecting the two regions}. While emission is detected in all MWA-2 bands (see e.g. 118-MHz in \cref{fig:a168}), due to the large size ($\sim 11.5$~arcmin) we note that flux recovery diminishes significantly in the 154-, 185-, and 216-MHz bands. With supplemental GLEAM data and flux densities reported by \citet{Dwarakanath2018}, we find the spectral index of the whole relic to be $\alpha_{88}^{608} = -1.50 \pm 0.08$. 

\begin{figure*}[h!]
    \centering
    \begin{subfigure}[b]{0.33\linewidth}
    \includegraphics[width=1\linewidth]{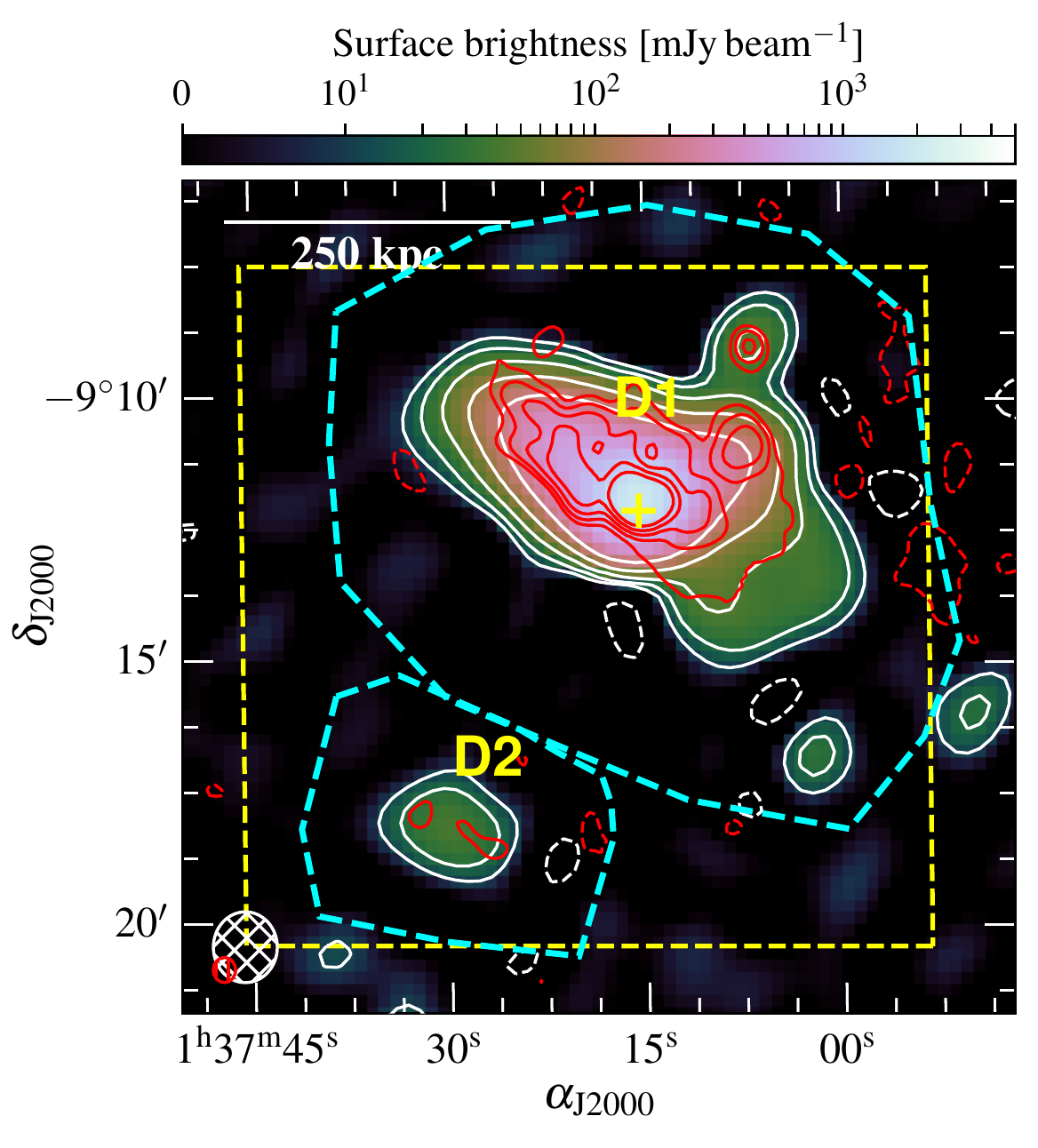}
    \caption{\label{fig:mcxc137:radio}}
    \end{subfigure}%
    \begin{subfigure}[b]{0.33\linewidth}
    \includegraphics[width=1\linewidth]{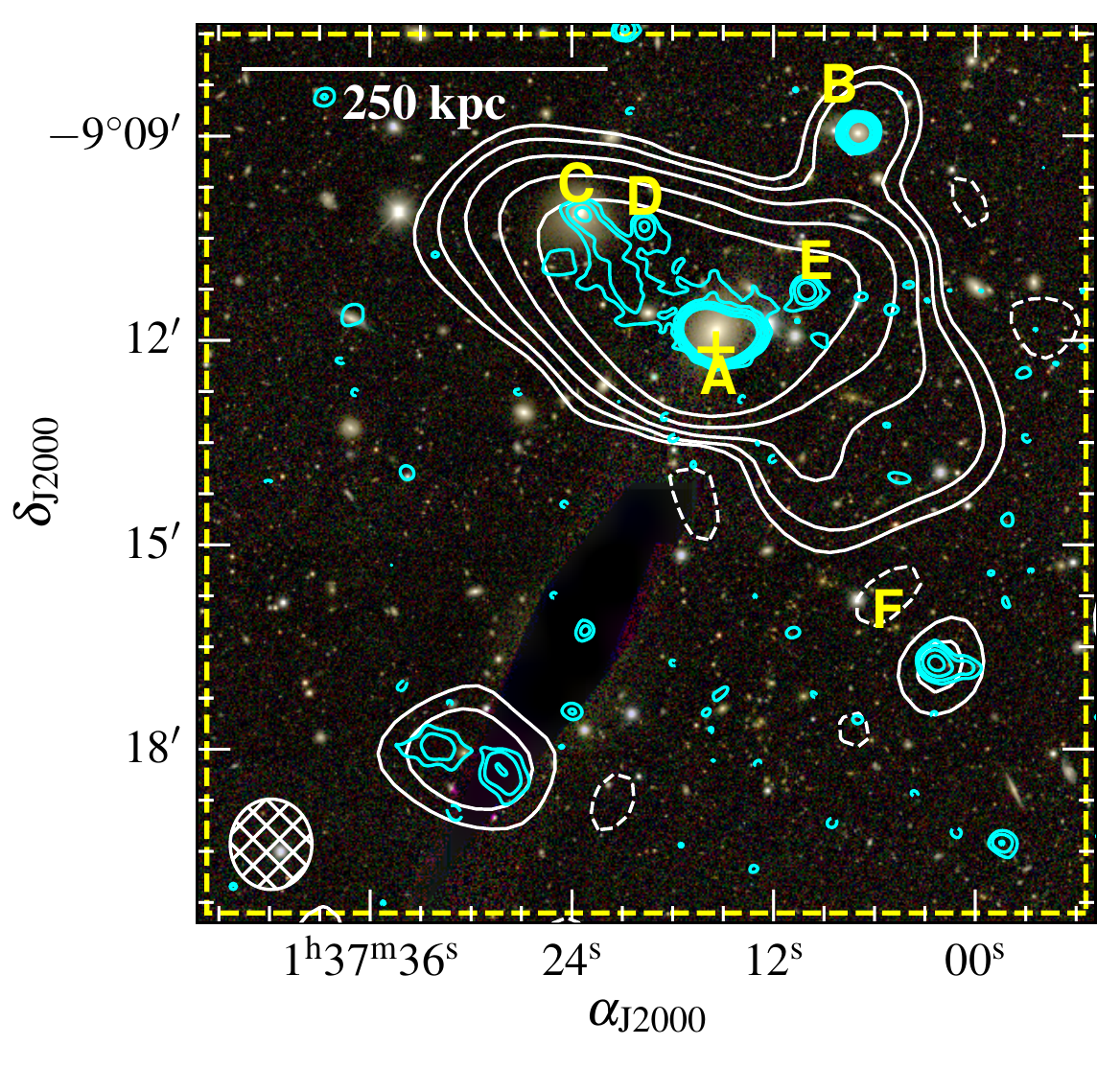}
    \caption{\label{fig:mcxc137:optical}}
    \end{subfigure}%
    \begin{subfigure}[b]{0.33\linewidth}
    \includegraphics[width=1\linewidth]{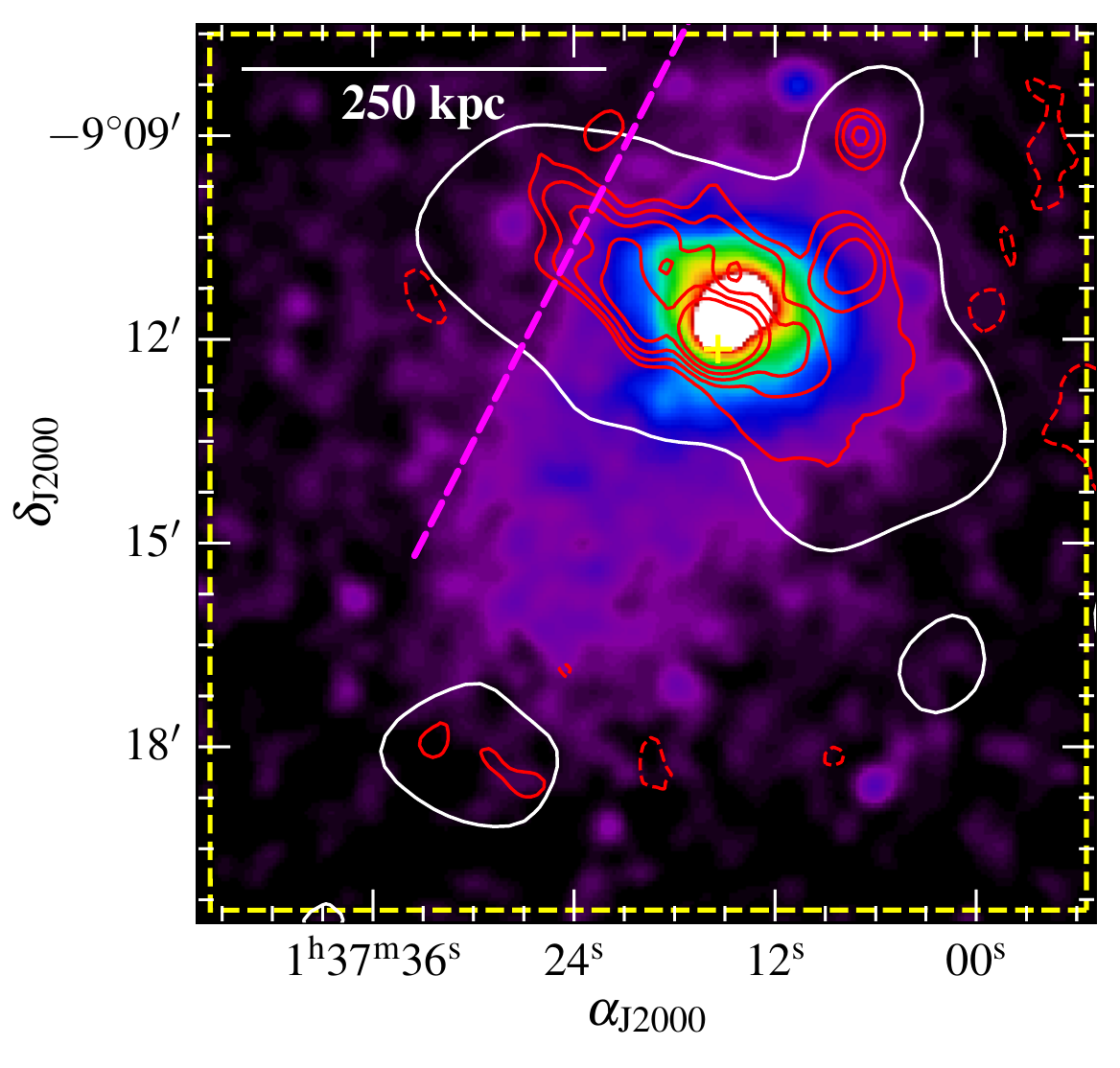}
    \caption{\label{fig:mcxc137:xray}}
    \end{subfigure}%
    \caption{\label{fig:mcxc137} \hyperref[para:mcxc137]{RXC~J0137.2$-$0912}. \subref{fig:mcxc137:radio} Background: MWA-2, 154-MHz, robust $0.0$ image. \subref{fig:mcxc137:optical} Background: RGB DES image ($i$, $r$, $g$). \subref{fig:mcxc137:xray}. Background: smoothed \xmm\ EPIC image. The white contours are as in \cref{fig:a0122:radio} for the background of \subref{fig:mcxc137:radio} (with $\sigma_\text{rms} = 3.8$~mJy\,beam$^{-1}$). Red contours: TGSS image, in levels of $[\pm 3, 6, 12, 24, 48] \times \sigma_\text{rms}$ ($\sigma_\text{rms} = 2.8$~mJy\,beam$^{-1}$). Cyan contours: VAST $+0.25$ image, in levels of $[\pm 3, 6, 12, 24, 48] \times \sigma_\text{rms}$ ($\sigma_\text{rms} = 0.14$~mJy\,beam$^{-1}$). \corrs{The dashed, magenta line on \subref{fig:mcxc137:xray} indicates the location of an \xmm{} chip gap with lessened sensitivity.} Other image features are as in \cref{fig:a0122}.}
\end{figure*}

\paragraph{\hyperref[fig:mcxc137]{RXC~J0137.2$-$0912}} \label{para:mcxc137} (\cref{fig:mcxc137}). We report the detection of steep-spectrum emission within RXC~J0137.2$-$0912, shown in \cref{fig:mcxc137:radio} in MWA and TGSS data. \cref{fig:mcxc137:optical} shows the optical host of the central compact emission with contours from re-processed VAST data (cyan) overlaid. \cref{fig:mcxc137:xray} shows the extent of the cluster's X-ray emission with archival \xmm\ data, noting some elongation perpendicular to the orientation of the radio emission. The compact emission from Sources A and B is subtracted after extrapolating to MWA-2 frequencies using the VAST and TGSS measurements, and C--E are subtracted assuming $\alpha = -0.7$. A power law SED is modelled, with a spectral index of $\alpha_{88}^{887} = -1.62 \pm 0.07$ (\cref{fig:sed:mcxcj137:1}, or $\alpha_{88}^{216} = -1.5 \pm 0.1$ across the MWA band alone). \corrs{The total angular size of the source is $7.8$~arcmin, corresponding to $410$~kpc.} From the archival \xmm\ data, we find a concentration parameter $c_{40/400} = 0.19$, consistent with cool-core (CC) clusters (where non-CC clusters are found to have $c_{40/400} \lesssim 0.075$; \citealt{Santos2008}). Based on the likelihood of a CC, the prominent BCG with significant AGN emission, and steep-spectrum diffuse emission surrounding the BCG we suggest the emission is a mini-halo. Some structure in the centre of the cluster gives some evidence for sloshing, and with a centroid shift within $R_{500} = 684$~kpc \citep{pap+11} of $w = 0.037R_{500}$ suggesting some disturbance \citep{pcab09,Bohringer2010}. \corrs{While the orientation of the radio emission appears largely perpendicular to the X-ray emission, the extension to the SW in the radio is traced by an extension in the X-ray as well. The NE direction is ambiguous as the \xmm{} exposure drops significantly due to a chip gap at that location. This is indicated on \autoref{fig:mcxc137:xray} as a dashed, magenta line.}

We see also an additional diffuse source resembling a remnant radio galaxy, D2; however, its SED shows significant curvature uncharacteristic of remnants (\cref{fig:sed:mcxcj137:2}).

\subsubsection{\texttt{FIELD6}}

\begin{figure*}[h!]
    \centering
    \begin{subfigure}[b]{0.33\linewidth}
    \includegraphics[width=1\linewidth]{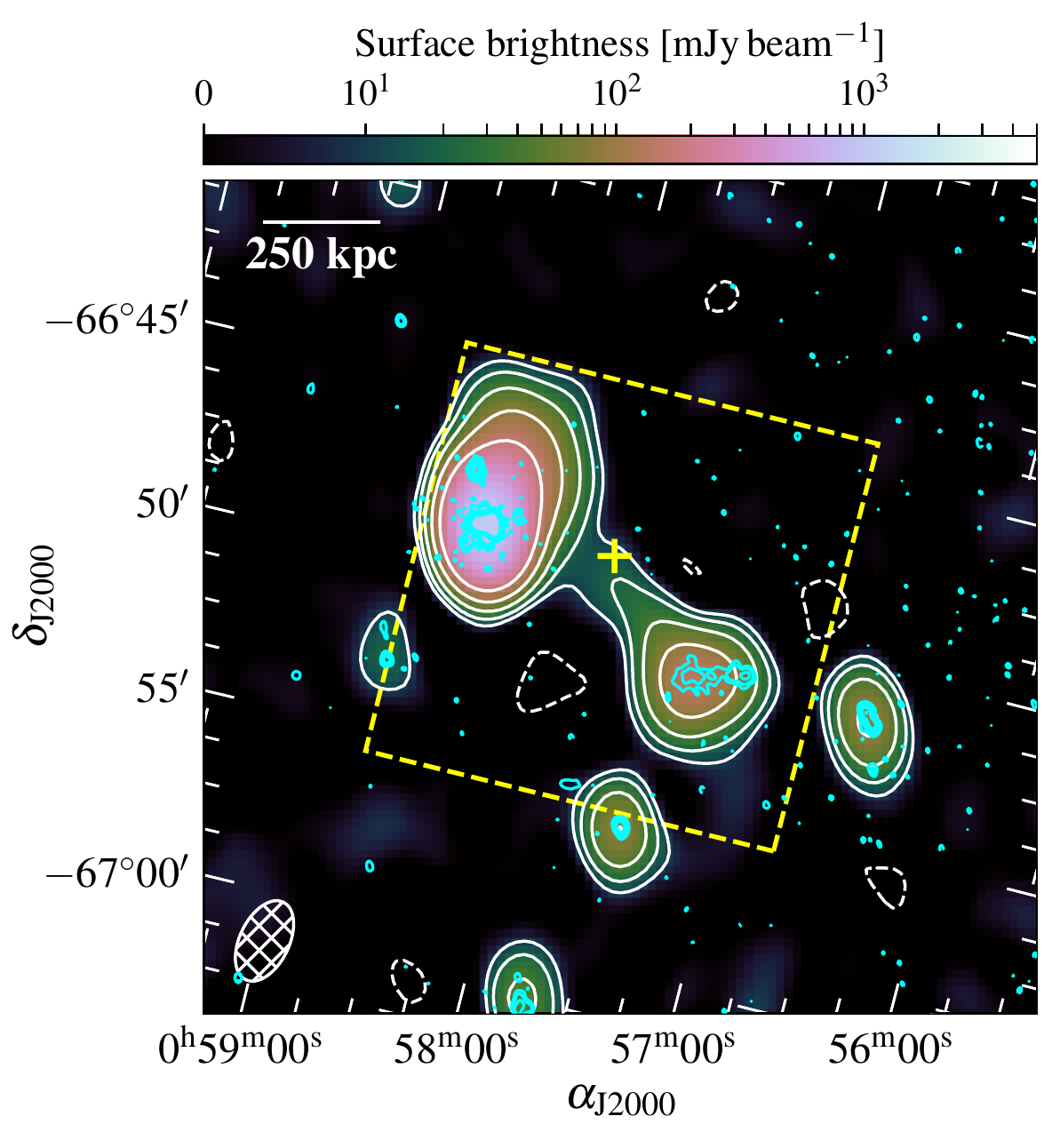}
    \caption{\label{fig:as112:radio}}
    \end{subfigure}%
    \begin{subfigure}[b]{0.33\linewidth}
    \includegraphics[width=1\linewidth]{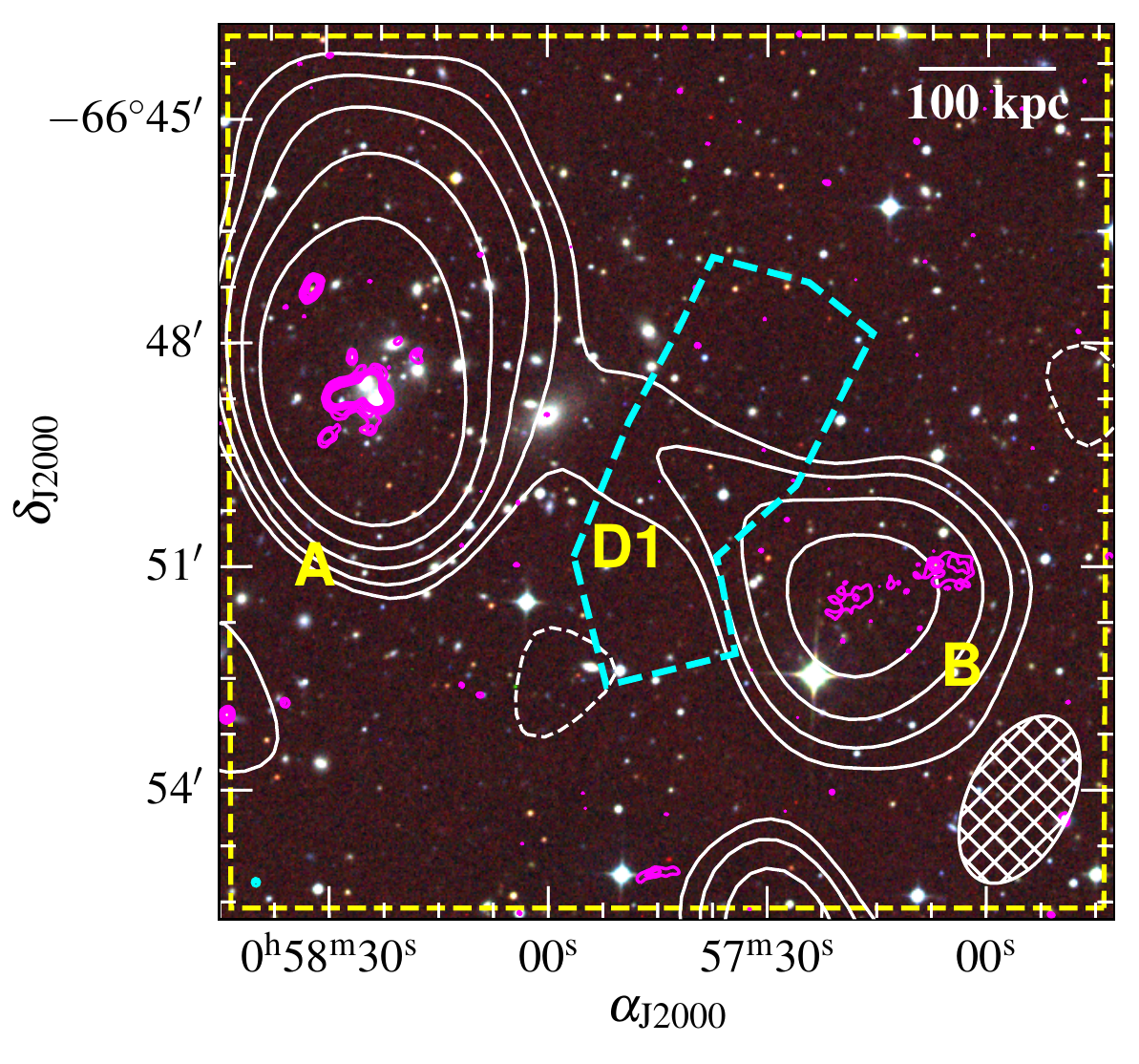}
    \caption{\label{fig:as112:optical}}
    \end{subfigure}%
    \begin{subfigure}[b]{0.33\linewidth}
    \includegraphics[width=1\linewidth]{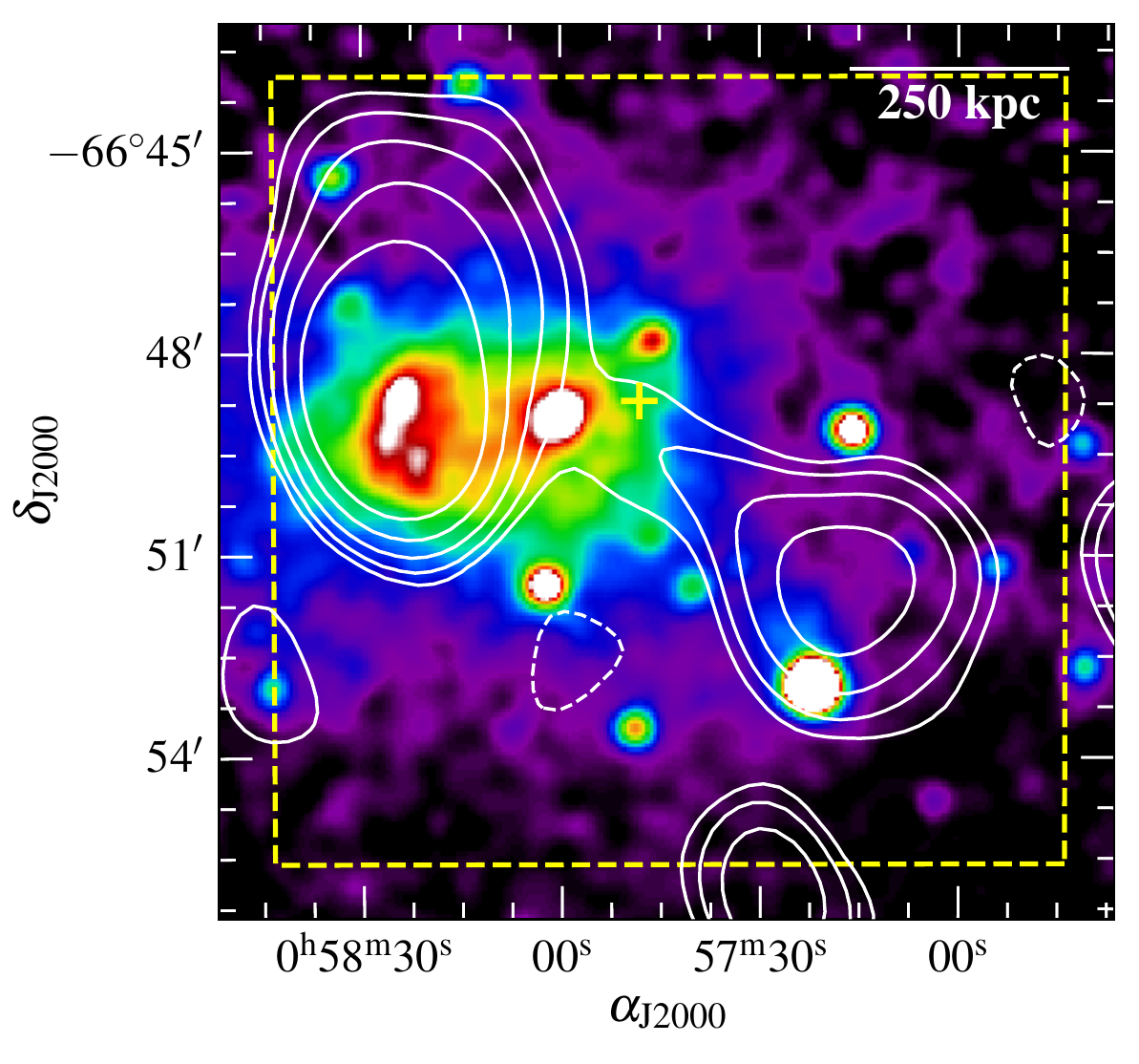}
    \caption{\label{fig:as112:xray}}
    \end{subfigure}%
    \caption{\label{fig:as112} \hyperref[para:as112]{Abell~S0112}. \subref{fig:as112:radio} Background: MWA-2, 154-MHz, robust $+2.0$ image. \subref{fig:as112:optical} Background: RGB SSS image ($i$, $r$, $b$). \subref{fig:as112:xray}. Background: smoothed \xmm\ EPIC image. The white contours are as in \cref{fig:a0122:radio} for the background of \subref{fig:as112:radio} (with $\sigma_\text{rms} = 3.5$~mJy\,beam$^{-1}$). Magenta contours: ATLBS high-resolution image, in levels of $[\pm 3, 6, 12, 24, 48] \times \sigma_\text{rms}$ ($\sigma_\text{rms} = 0.08$~mJy\,beam$^{-1}$. Cyan contours: RACS survey image, in levels of $[\pm 3, 6, 12, 24, 48] \times \sigma_\text{rms}$ ($\sigma_\text{rms} = 0.29$~mJy\,beam$^{-1}$). Other image features are as in \cref{fig:a0122}.}
\end{figure*}

\begin{figure*}[h!]
    \centering
    \begin{subfigure}[b]{0.33\linewidth}
    \includegraphics[width=1\linewidth]{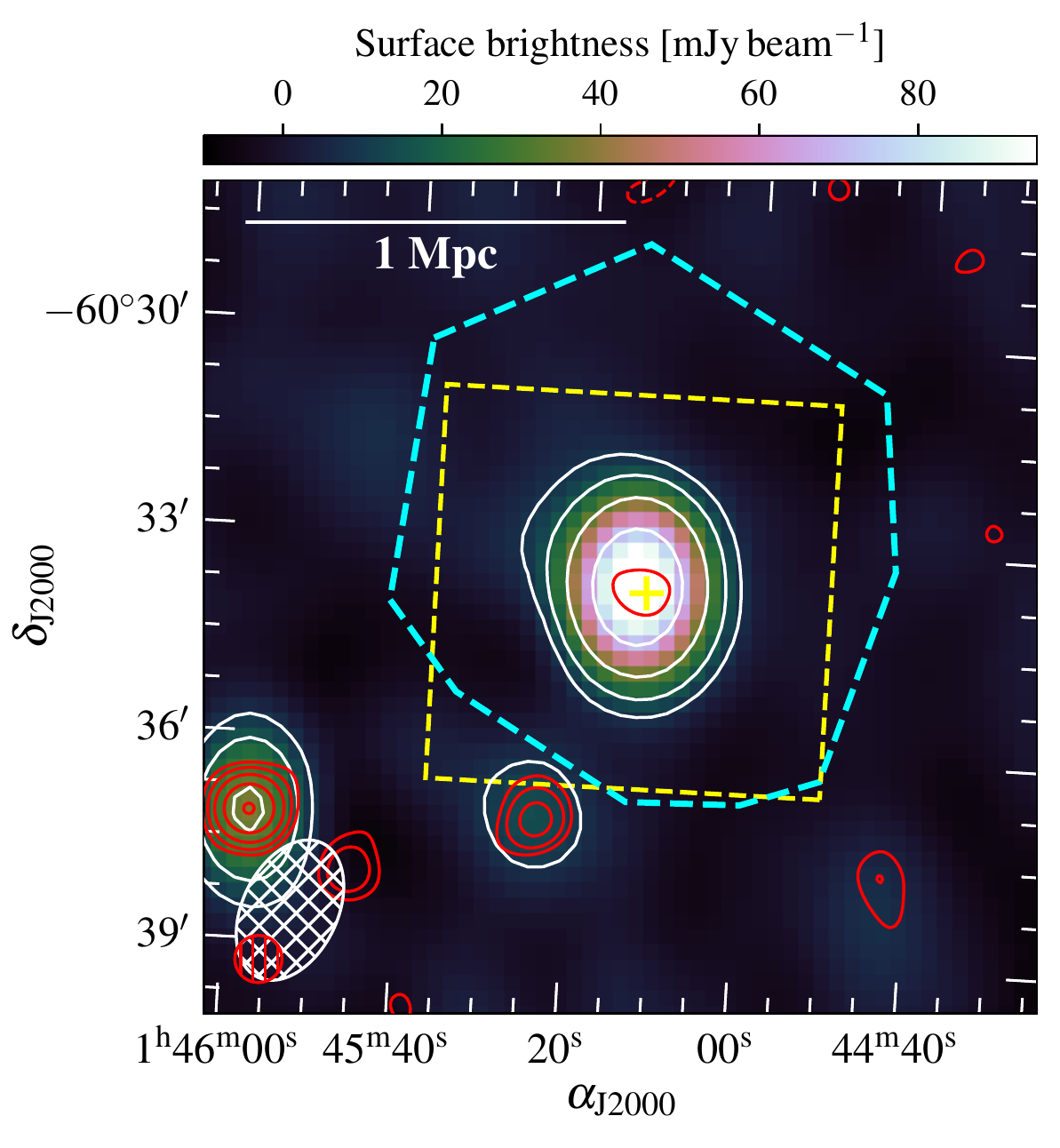}
    \caption{\label{fig:mcxc145:radio}}
    \end{subfigure}%
    \begin{subfigure}[b]{0.33\linewidth}
    \includegraphics[width=1\linewidth]{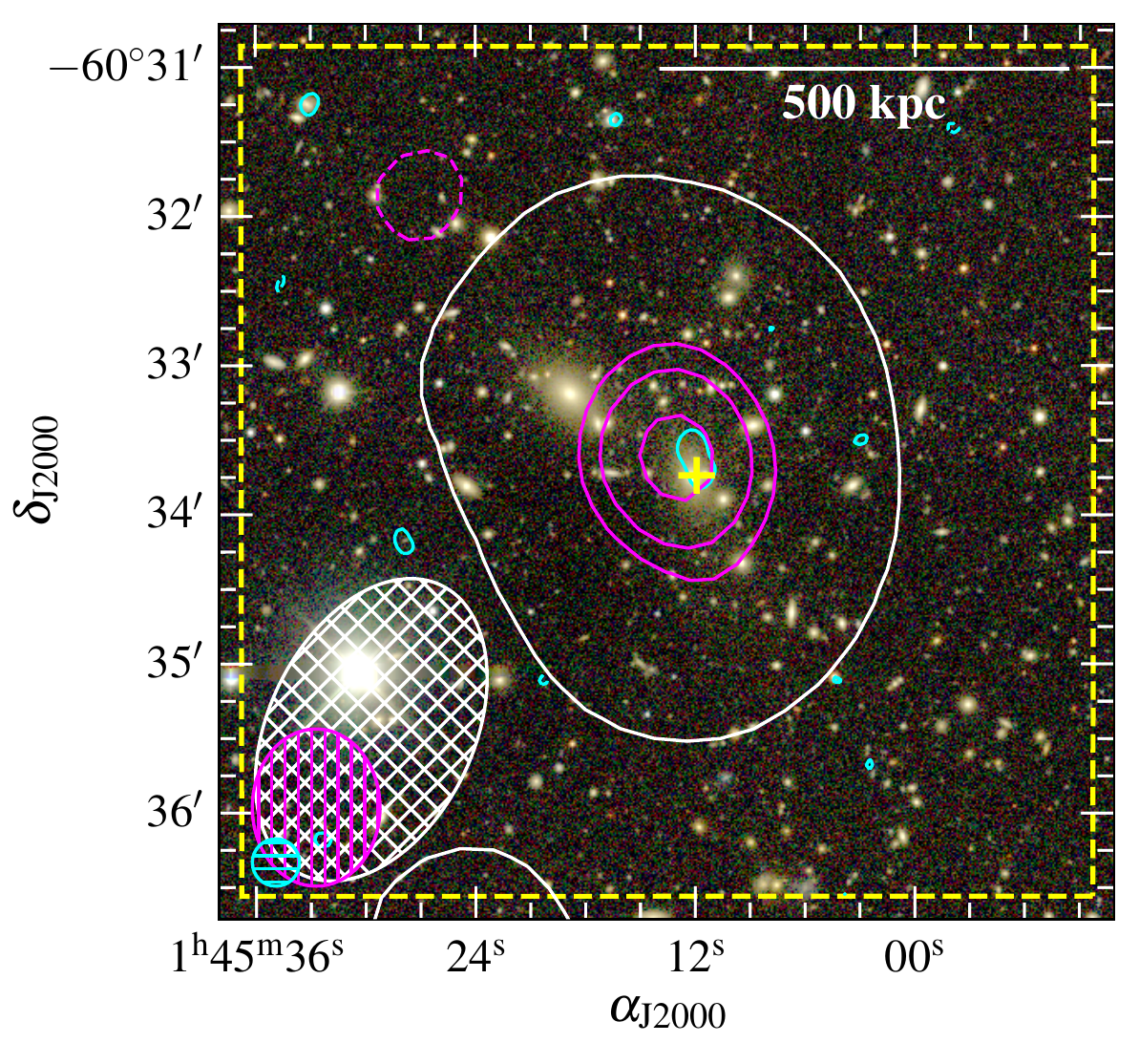}
    \caption{\label{fig:mcxc145:optical}}
    \end{subfigure}%
    \begin{subfigure}[b]{0.33\linewidth}
    \includegraphics[width=1\linewidth]{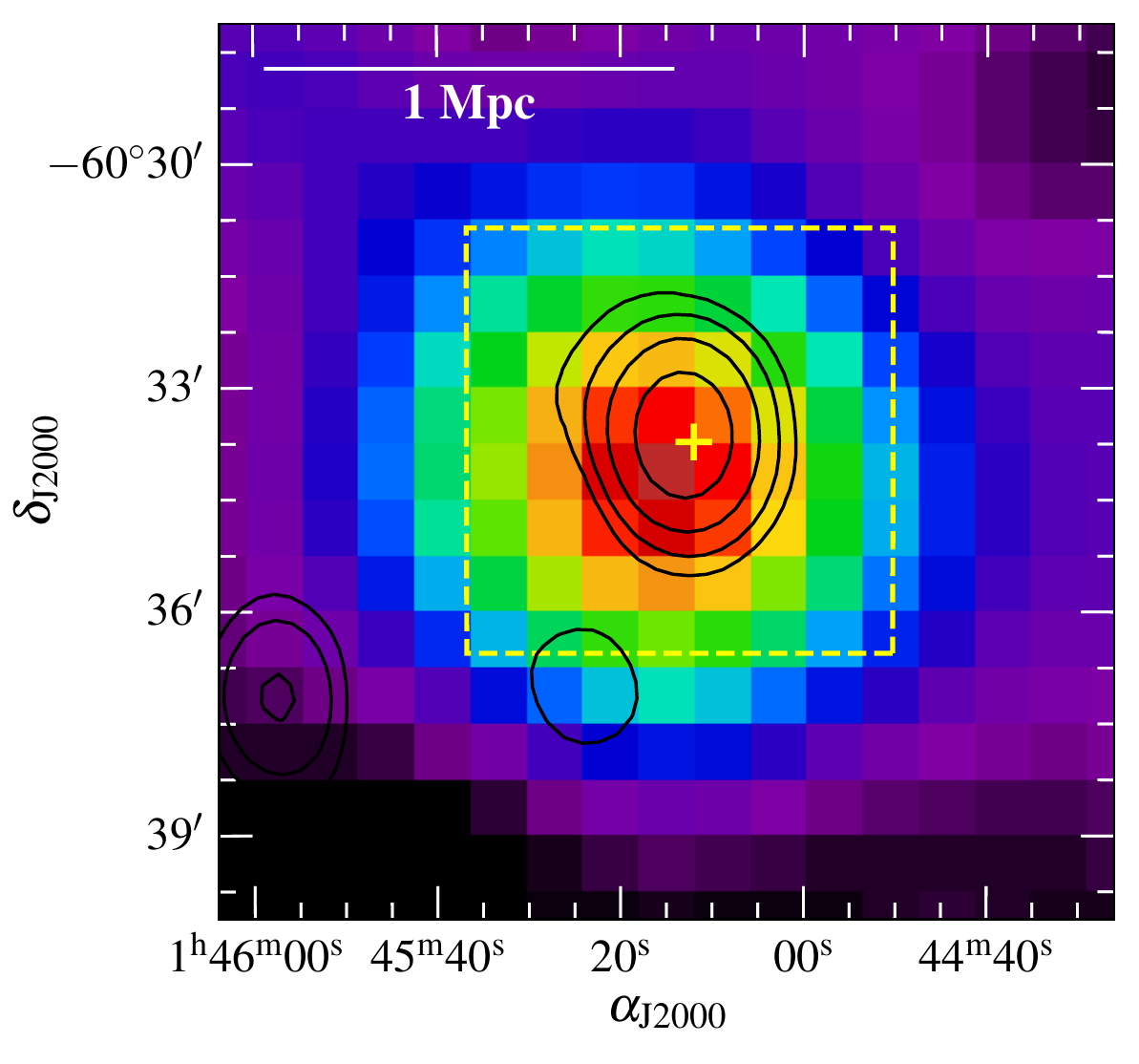}
    \caption{\label{fig:mcxc145:xray}}
    \end{subfigure}%
    \caption{\label{fig:mcxc145} \hyperref[para:mcxc145]{MCXC~J0145.2$-$6033}. \subref{fig:mcxc145:radio} Background: MWA-2, 154-MHz, robust $+2.0$ image. \subref{fig:mcxc145:optical} Background: RGB DES image ($i$, $r$, $g$). \subref{fig:mcxc145:xray}. Background: smoothed RASS image. The white (black) contours are as in \cref{fig:a0122:radio} for the background of \subref{fig:mcxc145:radio} (with $\sigma_\text{rms} = 2.9$~mJy\,beam$^{-1}$). Red contours: RACS low-resolution image, $[\pm 3, 6, 12, 24, 48] \times \sigma_\text{rms}$ ($\sigma_\text{rms} = 0.34$~mJy\,beam$^{-1}$). Cyan contours: RACS robust $+0.25$ image, $[\pm 3, 6, 12, 24, 48] \times \sigma_\text{rms}$ ($\sigma_\text{rms} = 0.17$~mJy\,beam$^{-1}$). Magenta contours: MWA-2, 216-MHz, robust $0.0$ image, $[\pm 3, 6, 12, 24, 48] \times \sigma_\text{rms}$ ($\sigma_\text{rms} = 3.4$~mJy\,beam$^{-1}$). Other image features are as in \cref{fig:a0122}.}
\end{figure*}

\begin{figure*}[h!]
    \centering
    \begin{subfigure}[b]{0.33\linewidth}
    \includegraphics[width=1\linewidth]{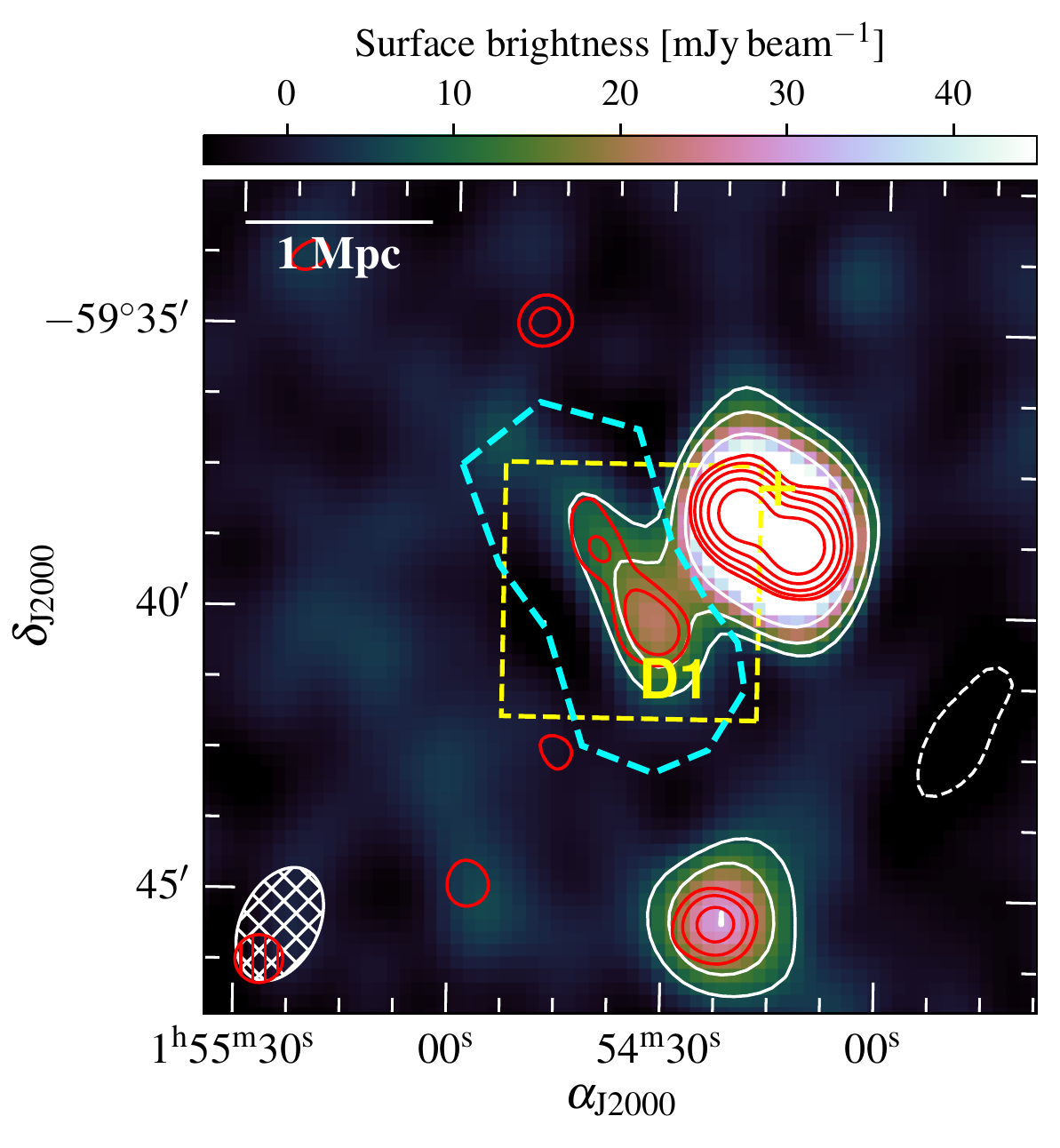}
    \caption{\label{fig:mcxc154:radio}}
    \end{subfigure}%
    \begin{subfigure}[b]{0.33\linewidth}
    \includegraphics[width=1\linewidth]{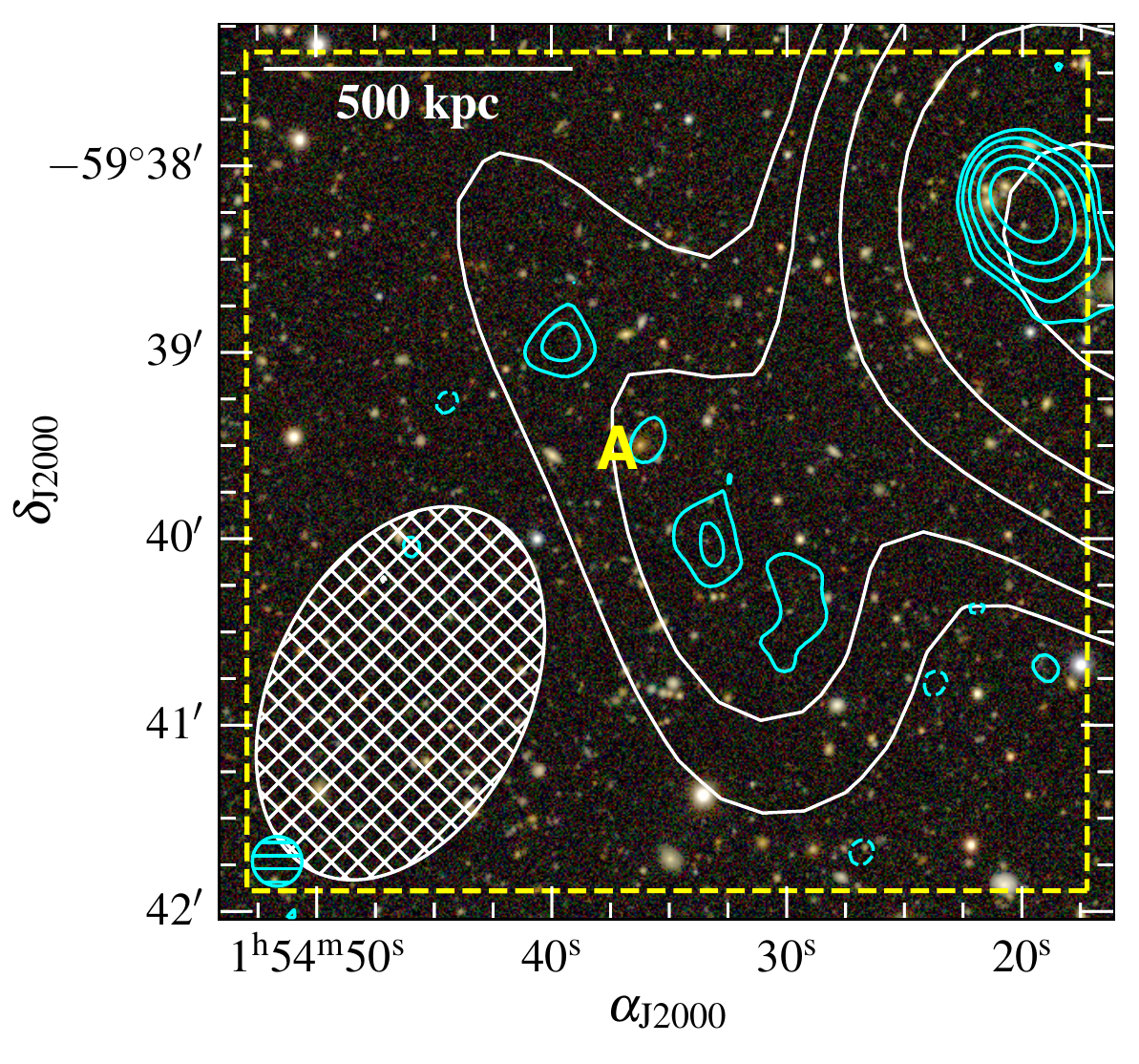}
    \caption{\label{fig:mcxc154:optical}}
    \end{subfigure}%
    \begin{subfigure}[b]{0.33\linewidth}
    \includegraphics[width=1\linewidth]{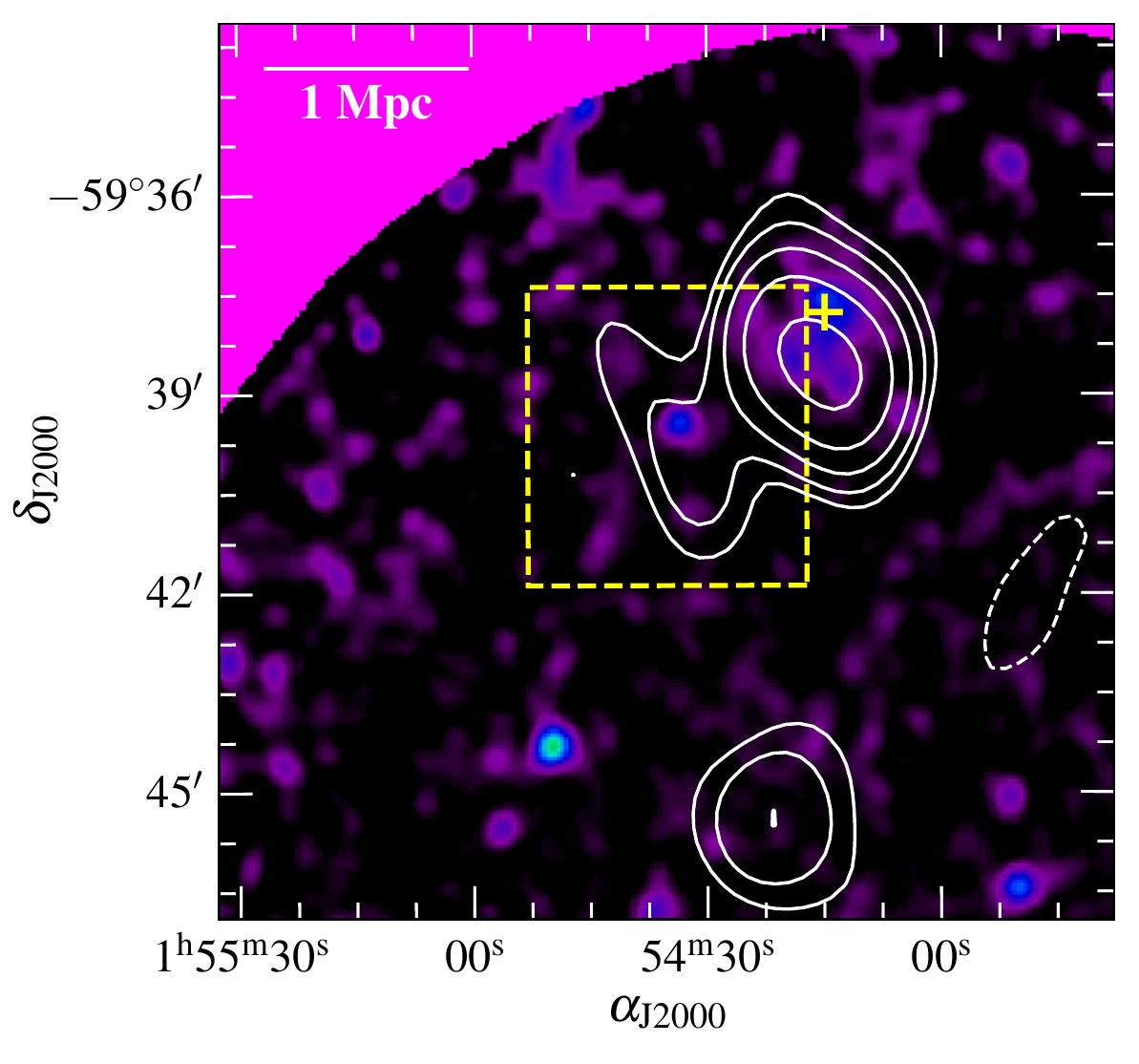}
    \caption{\label{fig:mcxc154:xray}}
    \end{subfigure}%
    \caption{\label{fig:mcxc154} \hyperref[para:mcxc154]{MCXC~J0154.2$-$5937}. \subref{fig:mcxc154:radio} Background: MWA-2, 154-MHz, robust $+2.0$ image. \subref{fig:mcxc154:optical} Background: RGB DES image ($i$, $r$, $g$). \subref{fig:mcxc154:xray}. Background: smoothed \xmm\ EPIC image. The white contours are as in \cref{fig:a0122:radio} for the background of \subref{fig:mcxc154:radio} (with $\sigma_\text{rms} = 2.5$~mJy\,beam$^{-1}$). Red contours: RACS low-resolution image, $[\pm 3, 6, 12, 24, 48] \times \sigma_\text{rms}$ ($\sigma_\text{rms} = 0.40$~mJy\,beam$^{-1}$). Cyan contours: RACS robust $+0.0$ image, $[\pm 3, 6, 12, 24, 48] \times \sigma_\text{rms}$ ($\sigma_\text{rms} = 0.20$~mJy\,beam$^{-1}$). Other image features are as in \cref{fig:a0122}.}
\end{figure*}

\paragraph{\hyperref[fig:as112]{Abell~S0112}} \label{para:as112} (\cref{fig:as112}). Abell~S0112 is covered by the Australia Telescope Low Brightness Survey \citep[ATLBS, region B;][]{Subrahmanyan2010,Thorat2013} which reveals two extended radio galaxies (A and B) within the cluster. The MWA data reveal an additional emission component (D1) between the two bright radio galaxies, creating an asymmetric dumbbell shape. The emission is only detected across the MWA-2 band, from 88--185~MHz and in the 200-MHz GLEAM data. {The extent of the emission between the two radio galaxies is $\sim 3$~arcmin ($\text{LLS}\approx 230$~kpc).} We measure integrated flux density within the D1 region between A and B, finding a spectral index of $\alpha_{88}^{185} = -1.9 \pm 0.5$ \cref{fig:sed:as112}). In \cref{fig:as112:optical} we show optical data which reveals a lack of obvious optical host for the emission, and \cref{fig:as112:xray} shows the \xmm\ image, highlighting the offset of the emission from the main component of the X-ray--emitting ICM. Source A is an active radio galaxy with a normal a radio spectrum ($\alpha \sim -0.8$), and D1 may be associated with an older episode of outflow. This could be true for B as well, with the extension of B to the north in the MWA-2 data also suggesting additional emission components not detected in higher-frequency/resolution images. We consider this emission fossil plasma associated with either A or B, with potential for some re-acceleration due to the dynamic nature of the cluster.

\paragraph{\hyperref[fig:mcxc145]{MCXC~J0145.2$-$6033}}\label{para:mcxc145} (\cref{fig:mcxc145}). We report the detection of a candidate mini-halo in MWA-2 and RACS data, shown in \cref{fig:mcxc145:radio}. The detection is marginal in the RACS data, and $2\sigma$ contours of the low resolution image are shown in \cref{fig:mcxc145:radio} to highlight the extent of the emission. \cref{fig:mcxc145:optical} shows the robust $+0.25$ RACS data which stems from the BCG and extends northwards. We measure the flux densities across the available images and find $\alpha_{88}^{887} = -2.1 \pm 0.1$ (\cref{fig:sed:mcxcj145}, with the same value across only the MWA-2 band). The extent of the source is $\LAS = 1.9$~arcmin; $\LLS = 350$~kpc. The optical data is shown in \cref{fig:mcxc145:optical}, with a BCG clear near the centre of the emission. RASS data shown in \cref{fig:mcxc145:xray} does not show any significant offset from the BCG, and we suggest the cluster is reasonably relaxed. {With an LAS of 1.9~arcmin ($\text{LLS}=350$~kpc),} the source is only barely extended in the robust $+2.0$ MWA-2 images, and the robust $0.0$ images show no significant extension. There is a $\sim 20$~mJy difference between the 0.0 and $+2.0$ measurements at 154~MHz ($S_{\text{r0}} = 110 \pm 10$ and $S_\text{r2} = 130 \pm 10$) which may be allowable within uncertainties. Given the low SNR in the RACS data, it is difficult to confirm if this is an ultra-steep spectrum point source.

\begin{figure*}[h!]
    \centering
    \begin{subfigure}[b]{0.5\linewidth}
    \includegraphics[width=1\linewidth]{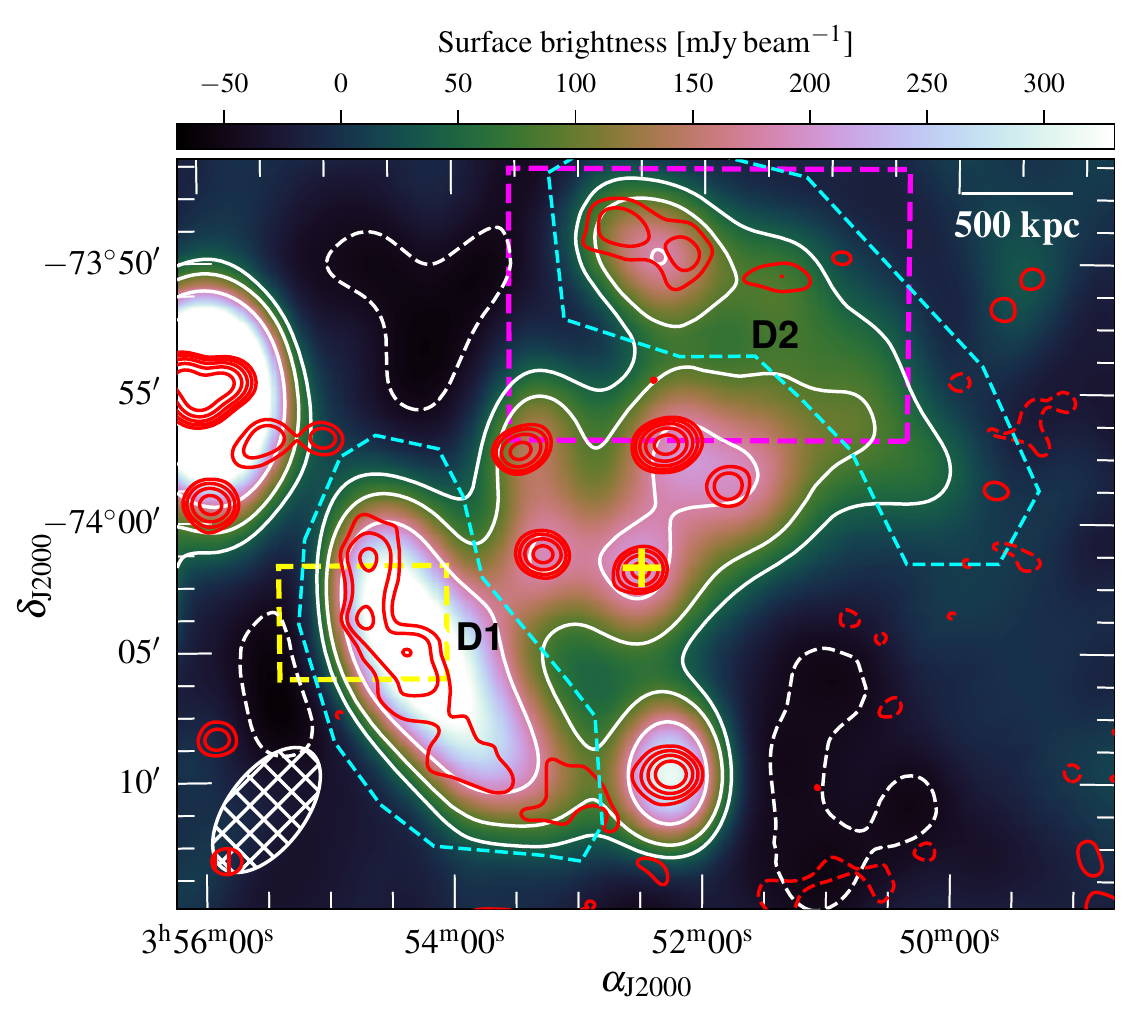}
    \caption{\label{fig:a3186:radio}}
    \end{subfigure}%
    \begin{subfigure}[b]{0.5\linewidth}
    \includegraphics[width=1\linewidth]{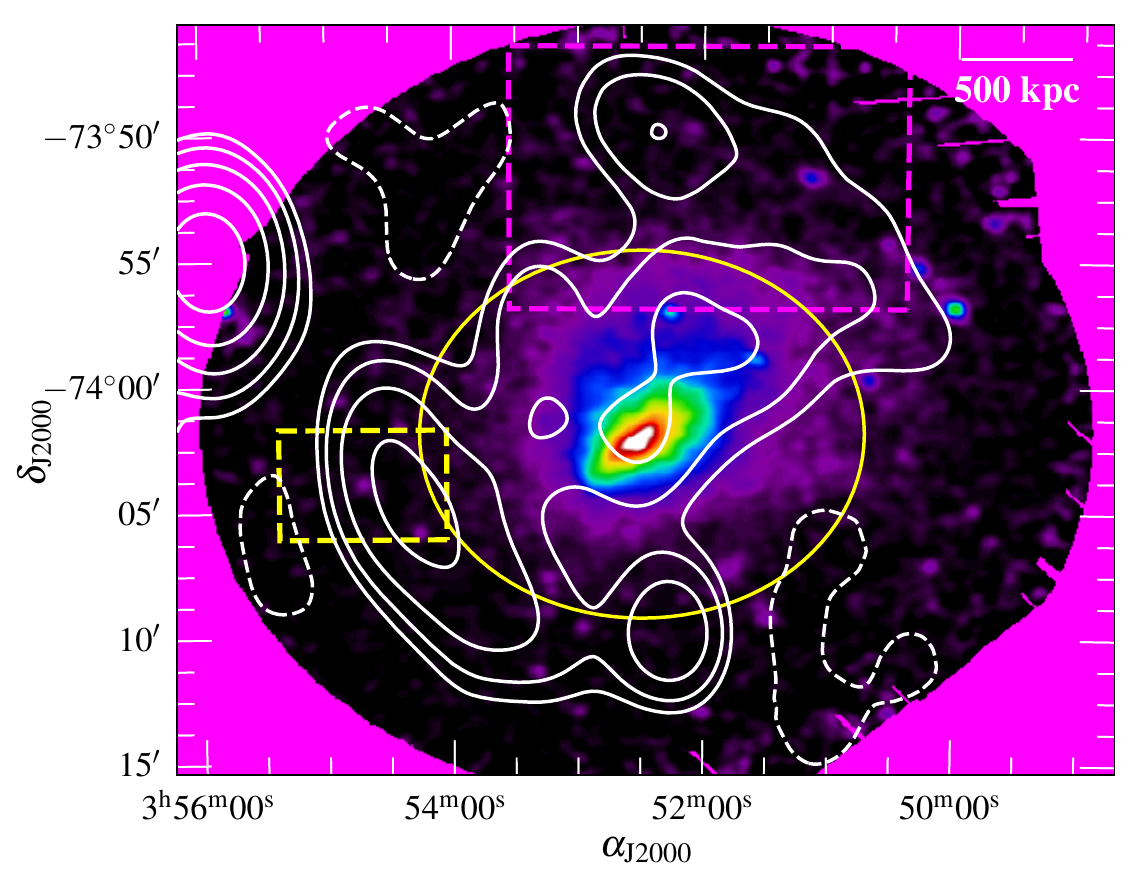}
    \caption{\label{fig:a3186:xray}}
    \end{subfigure}\\%
        \begin{subfigure}[b]{0.5\linewidth}
    \includegraphics[width=1\linewidth]{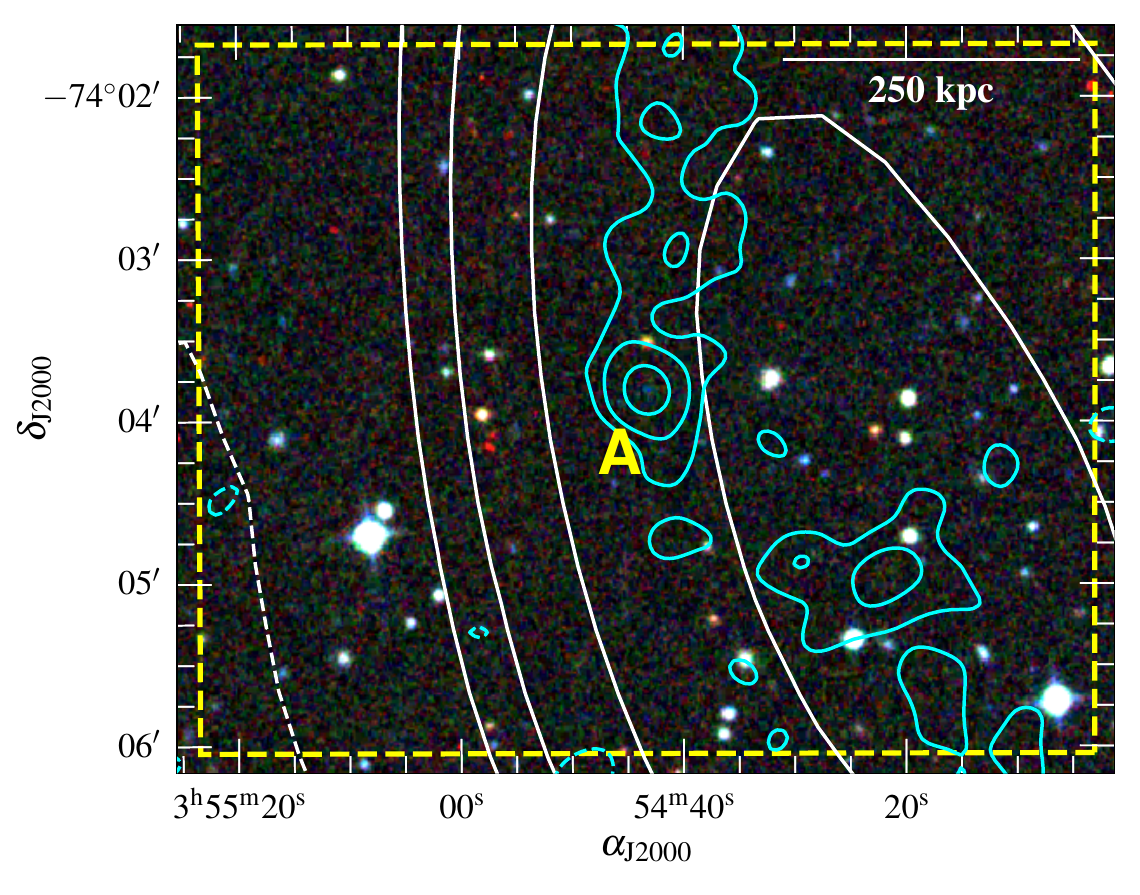}
    \caption{\label{fig:a3186:optical}}
    \end{subfigure}%
        \begin{subfigure}[b]{0.5\linewidth}
    \includegraphics[width=1\linewidth]{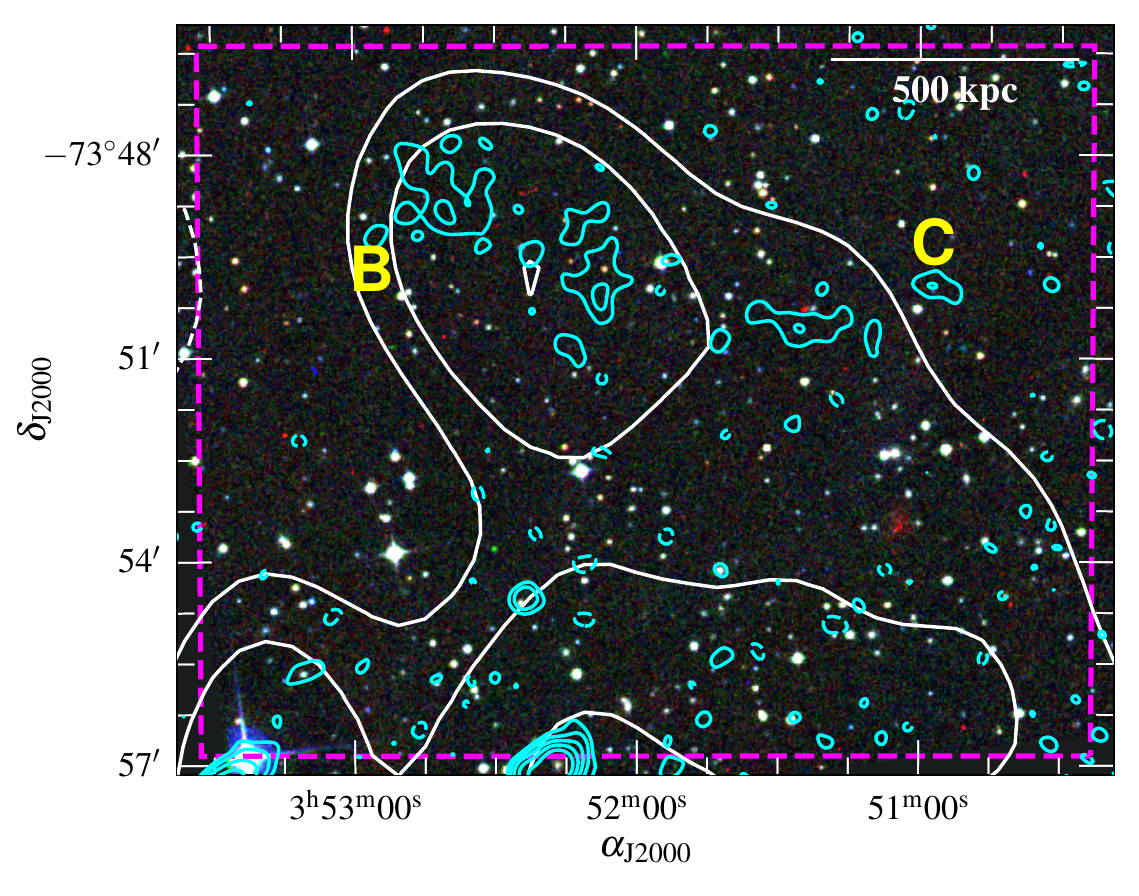}
    \caption{\label{fig:a3186:optical2}}
    \end{subfigure}%
    \caption{\label{fig:a3186} \hyperref[para:a3186]{Abell~3186}. \subref{fig:a3186:radio} Background: MWA-2, 88-MHz, robust $+2.0$ image. \subref{fig:a3186:optical} and \subref{fig:a3186:optical2} Background: RGB SSS image ($i$, $r$, $b$). \subref{fig:a3186:xray}. Background: smoothed \xmm\ EPIC image. The white contours are as in \cref{fig:a0122:radio} for the background of \subref{fig:mcxc154:radio} (with $\sigma_\text{rms} = 15$~mJy\,beam$^{-1}$). Red contours: deep ASKAP low-resolution image, $[\pm 3, 6, 12, 24, 48] \times \sigma_\text{rms}$ ($\sigma_\text{rms} = 0.70$~mJy\,beam$^{-1}$). Cyan contours: deep ASKAP robust $+0.25$, $[\pm 3, 6, 12, 24, 48] \times \sigma_\text{rms}$ ($\sigma_\text{rms} = 0.32$~mJy\,beam$^{-1}$ and $\sigma_\text{rms} = 0.25$~mJy\,beam$^{-1}$ for \subref{fig:a3186:optical} and \subref{fig:a3186:optical2}, respectively). Other image features are as in \cref{fig:a0122} and \cref{fig:a2751}, with the magenta box indicating the location of \subref{fig:a3186:optical2} on \subref{fig:a3186:radio} and \subref{fig:a3186:xray}.}
\end{figure*}

\paragraph{\hyperref[fig:mcxc154]{MCXC~J0154.2$-$5937}} \label{para:mcxc154} (\cref{fig:mcxc154}). We report peripherally located, elongated extended emission in the cluster, shown in \cref{fig:mcxc154:radio}. In \cref{fig:mcxc154:optical} a potential optical host is seen with matching emission in the RACS image (Source A, $S_{887} \sim 1$~mJy, WISEA~J015436.21$-$593929.9, no redshift). The radio SED of the whole source is uncertain, but consistent with a radio galaxy ($\alpha_{88}^{887} = -0.65 \pm 0.13$; \cref{fig:sed:mcxcj154}). If in the cluster, the source is $\sim 1$~Mpc in projected extent, classing it as a giant radio galaxy.

\subsubsection{\texttt{FIELD7}}

\begin{figure*}[h!]
    \centering
    \begin{subfigure}[b]{0.33\linewidth}
    \includegraphics[width=1\linewidth]{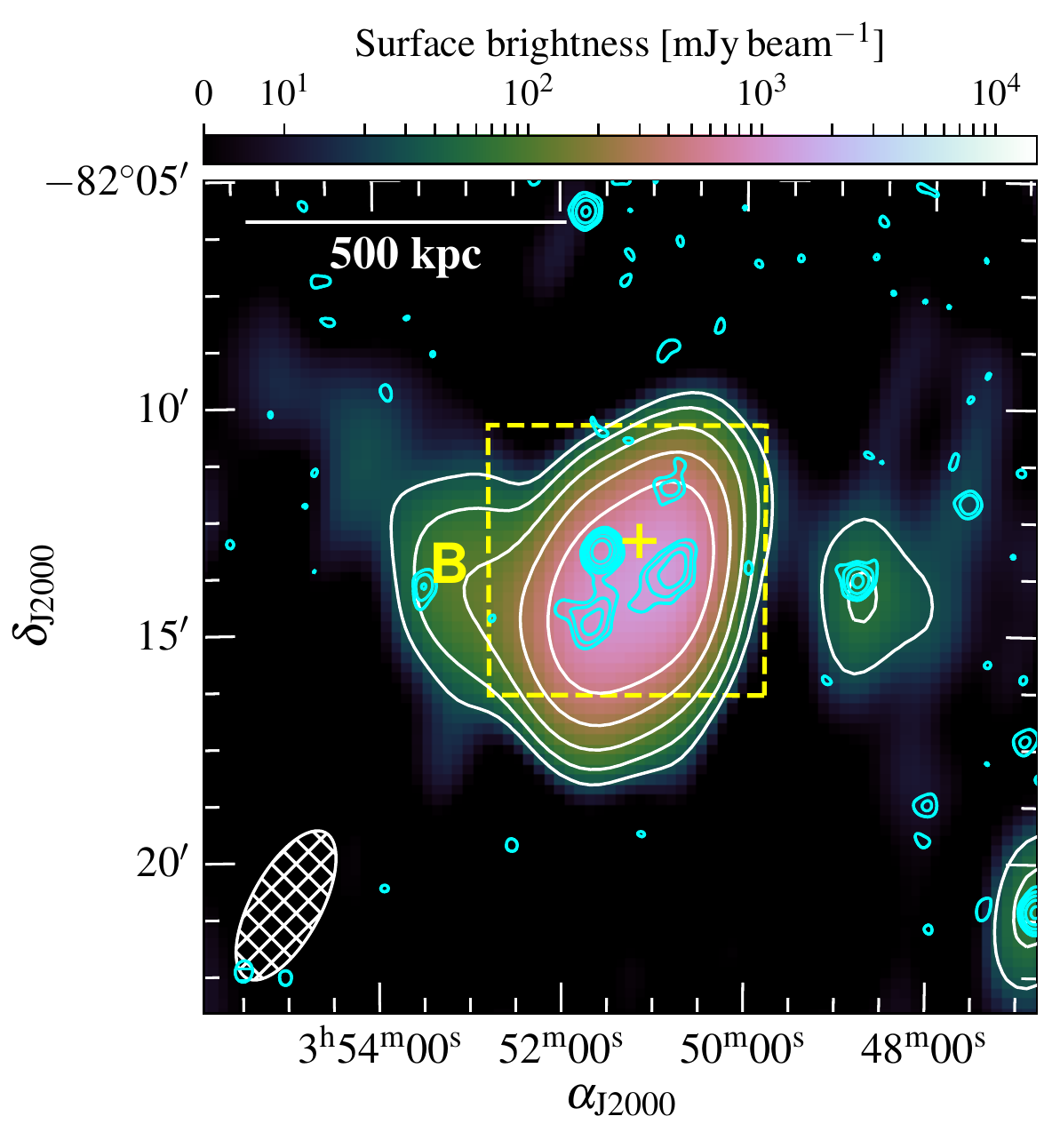}
    \caption{\label{fig:as0405:radio}}
    \end{subfigure}%
    \begin{subfigure}[b]{0.33\linewidth}
    \includegraphics[width=1\linewidth]{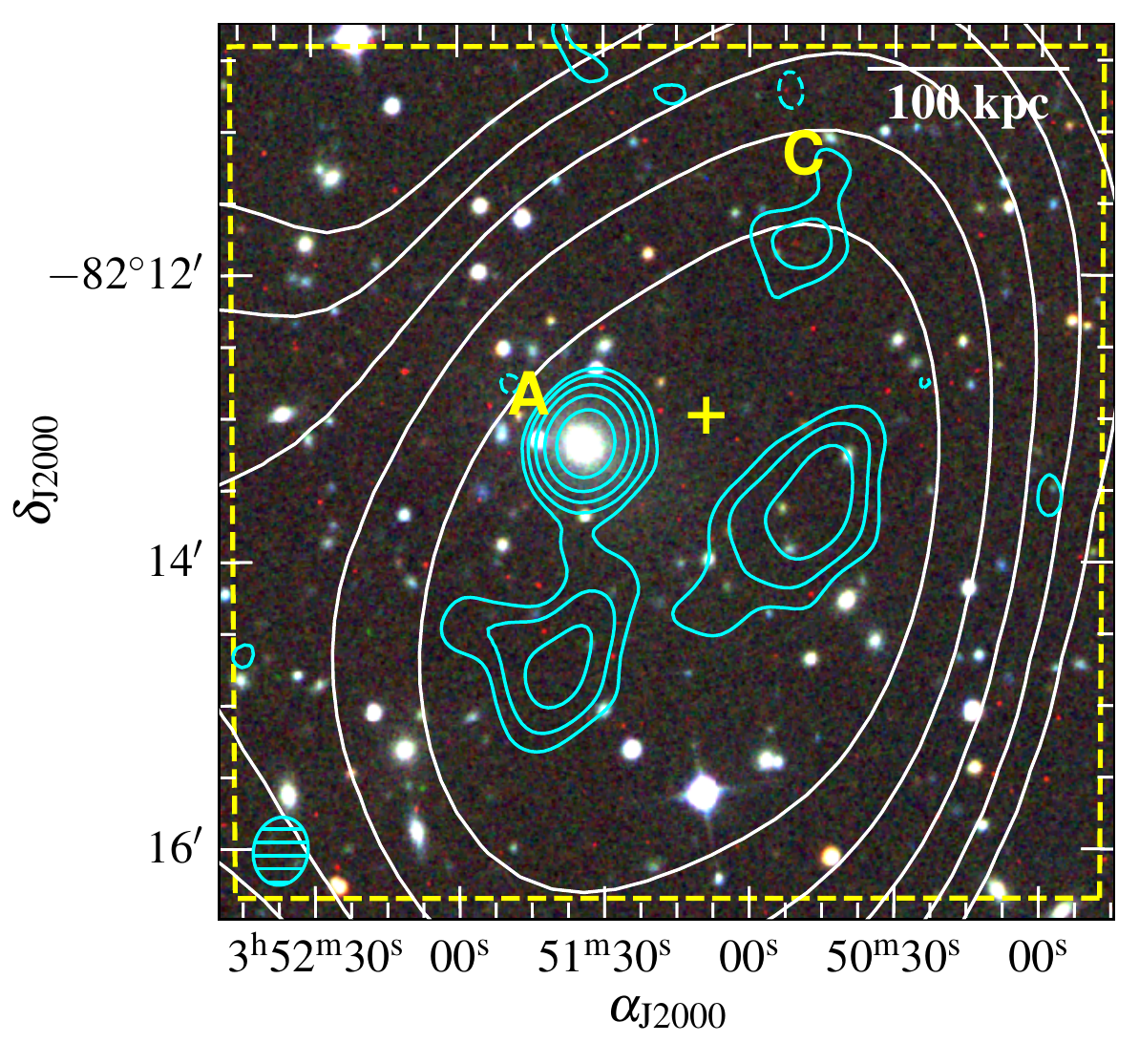}
    \caption{\label{fig:as0405:optical}}
    \end{subfigure}%
    \begin{subfigure}[b]{0.33\linewidth}
    \includegraphics[width=1\linewidth]{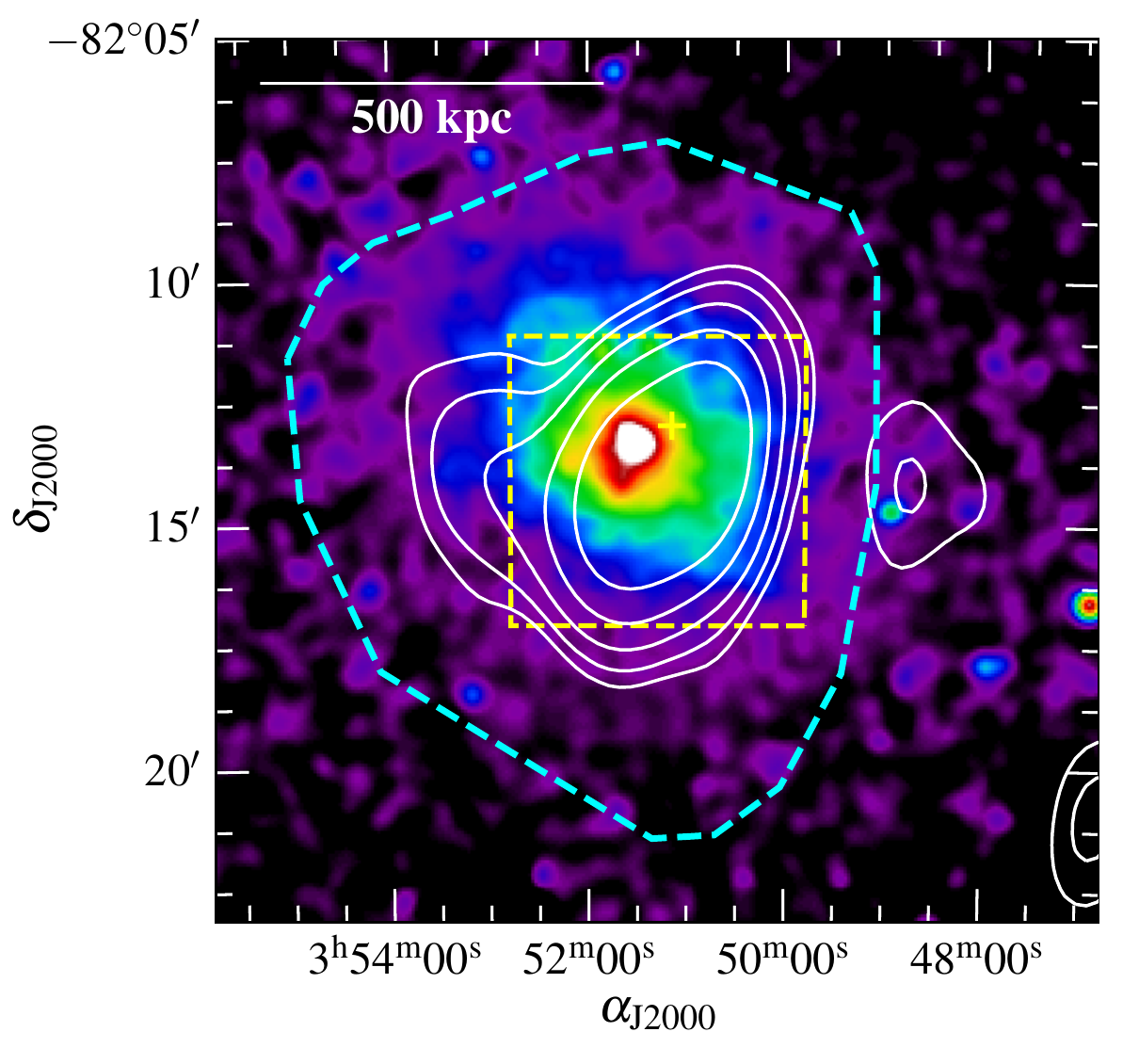}
    \caption{\label{fig:as0405:xray}}
    \end{subfigure}%
    \caption{\label{fig:as0405} \hyperref[para:as405]{Abell~S0405.} \subref{fig:as0405:radio} Background: MWA-2, 154-MHz, robust $+2.0$ image. \subref{fig:as0405:optical} Background: RGB SSS image ($i$, $r$, $b$). \subref{fig:as0405:xray}. Background: smoothed \emph{Chandra} image. The white contours are as in \cref{fig:a0122:radio} for the background of \subref{fig:as0405:radio} (with $\sigma_\text{rms} = 10$~mJy\,beam$^{-1}$). Cyan contours: RACS survey image, $[\pm 3, 6, 12, 24, 48] \times \sigma_\text{rms}$ ($\sigma_\text{rms} = 0.28$~mJy\,beam$^{-1}$). Other image features are as in \cref{fig:a0122}.}
\end{figure*}

\begin{figure*}[h!]
    \centering
    \begin{subfigure}[b]{0.33\linewidth}
    \includegraphics[width=1\linewidth]{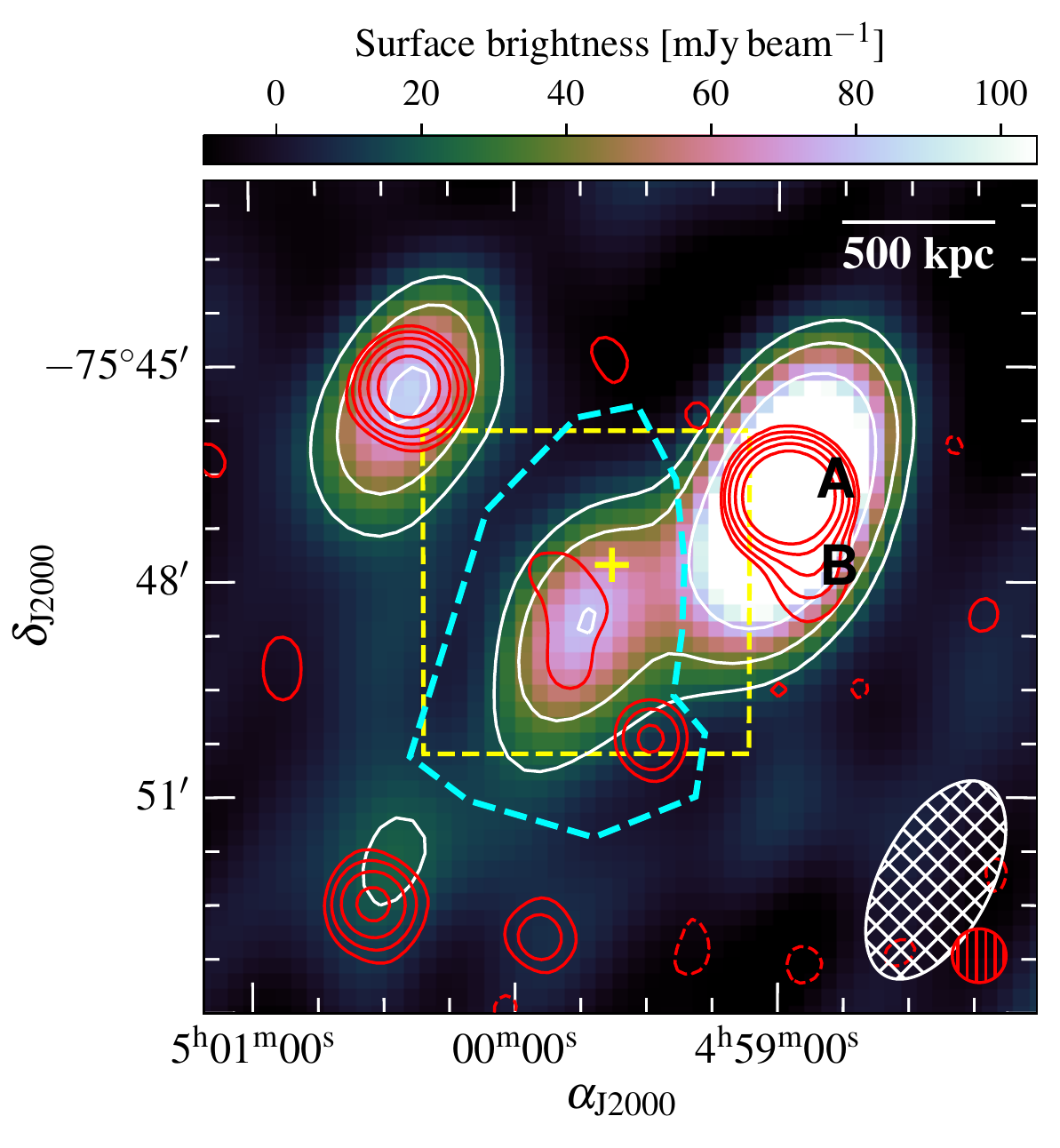}
    \caption{\label{fig:psz1g287:radio}}
    \end{subfigure}%
    \begin{subfigure}[b]{0.33\linewidth}
    \includegraphics[width=1\linewidth]{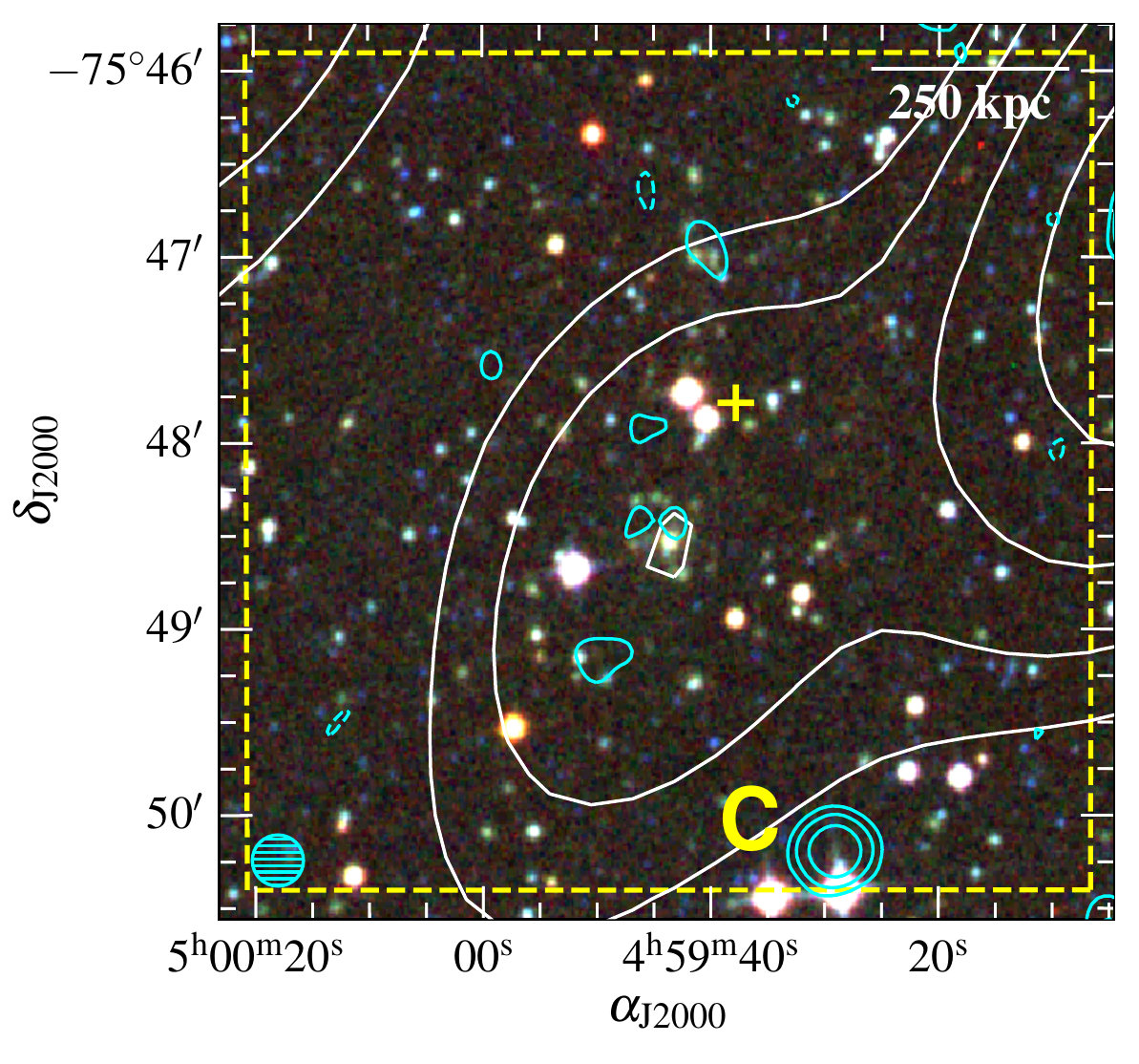}
    \caption{\label{fig:psz1g287:optical}}
    \end{subfigure}%
    \begin{subfigure}[b]{0.33\linewidth}
    \includegraphics[width=1\linewidth]{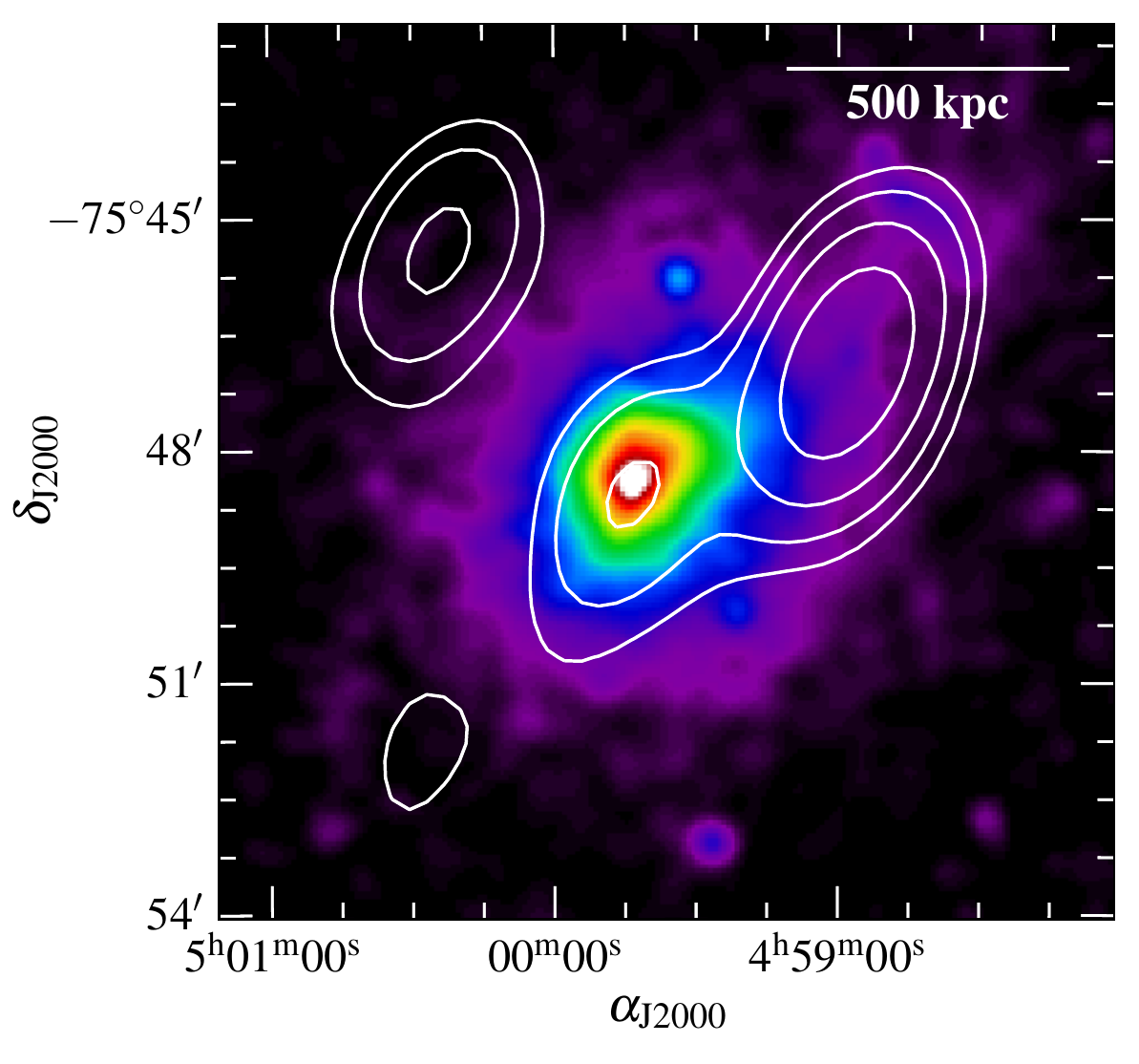}
    \caption{\label{fig:psz1g287:xray}}
    \end{subfigure}%
    \caption{\label{fig:psz1g287} \hyperref[para:psz1g287]{PSZ1~G287.95$-$32.98}. \subref{fig:psz1g287:radio} Background: MWA-2, 154-MHz, robust $+2.0$ image. \subref{fig:psz1g287:optical} Background: RGB SSS image ($i$, $r$, $b$). \subref{fig:psz1g287:xray}. Background: smoothed \xmm\ image. The white contours are as in \cref{fig:a0122:radio} for the background of \subref{fig:psz1g287:radio} (with $\sigma_\text{rms} = 6.6$~mJy\,beam$^{-1}$). Red contours: RACS low resolution image $[\pm 3, 6, 12, 24, 48] \times \sigma_\text{rms}$ ($\sigma_\text{rms} = 0.25$~mJy\,beam$^{-1}$). Cyan contours: RACS robust $+0.25$ image, $[\pm 3, 6, 12, 24, 48] \times \sigma_\text{rms}$ ($\sigma_\text{rms} = 0.15$~mJy\,beam$^{-1}$). Other image features are as in \cref{fig:a0122}.}
\end{figure*}

\paragraph{\hyperref[fig:a3186]{Abell~3186}} \label{para:a3186} (\cref{fig:a3186}). From the data presented here we report Abell~3186 to be a double relic system, potentially with central diffuse emission possible from a radio halo. The cluster shows a similar morphology to older observations of the canonical double relic cluster Abell~3667 \citep{mj-h, Hindson2014} with a larger, bright relic on one side and a smaller dimmer one on the other. 

The larger relic, D1, is detected to the SW as an elongated structure on the periphery of Abell~3186 with the MWA and ASKAP, shown in \cref{fig:a3186:radio}. The angular extent of D1 is $\sim 12$~arcmin, corresponding to $\sim 1650$~kpc. The MWA-2 images in this region at 154-, 185-, and 216-MHz, which are generally less sensitive, suffered from significant noise from a nearby bright source and the large relic is poorly detected. A somewhat compact source is detected within the emission (Source A) which has a faint, blue optical counterpart, seen in \cref{fig:a3186:optical}. After subtracting the contribution from A, we obtain a spectral index of $\alpha_{\text{D1},88}^{887} = -1.0 \pm 0.1$ (\cref{fig:sed:a3186a}), consistent with relic sources. We show the \xmm\ image in \cref{fig:a3186:xray}, and note that \citet{Lovisari2017} report the cluster has a `mixed' morphology (i.e. semi-disturbed). Additionally, the cluster hosts a second relic to the NW seen as a patchy structure, with detections in all bands, though the full extent of the emission ($\sim 14$~arcmin, $\sim 1900$~kpc), extending to the SW, is only seen at 88 and 118~MHz as with D1. After subtraction of discrete source contributions (labelled B and C in \cref{fig:a3186:optical2}), we derive a spectral index of $\alpha_{\text{D2},88}^{887} = -0.9 \pm 0.1$. The morphology and spectral properties of these relics are extremely reminiscent of Abell~3667 which as a bright, larger relic and a smaller more compact relic with average spectral indices of $-0.9$ for both across MWA bands \citep{Hindson2014}, though higher resolution, higher frequency spectral index measurements show $\alpha$ varies inside the relics from approximately $-0.8$ to $-1$ \citep{mj-h}. We consider this a classical double relic system. We note also the 88-MHz MWA-2 image (\cref{fig:a3186:radio}) appears to show an excess of diffuse flux within the central cluster region which may indicate a radio halo, though the confusion from sources in the cluster make this impossible to confirm with the present data.

\paragraph{\hyperref[fig:as0405]{Abell~S0405}} \label{para:as405} (\cref{fig:as0405}). We report a diffuse source in Abell~S0405, with RACS data revealing a double-lobed radio galaxy structure (\cref{fig:as0405:radio}) without a core or jets. No obvious optical host exists between the two lobes (\cref{fig:as0405:optical}) and we suggest this is a remnant radio galaxy. We find $\alpha_{88}^{887} = -1.99 \pm 0.08$ between the MWA and RACS data, after subtraction of Sources A and B from the MWA images (\cref{fig:sed:as405}). The source is measured to be $\sim185$~kpc if at the redshift of the cluster ($z=0.0613$; \citealt{DeGrandi1999}), and sits within the X-ray--emitting ICM (\cref{fig:as0405:xray}).

\paragraph{\hyperref[fig:psz1g287]{PSZ1~G287.95$-$32.98}} \footnote{Not to be confused with the double-relic system, PLCK~G287.0$+$32.9 reported by \citet{Bagchi2011}.} \label{para:psz1g287} (\cref{fig:psz1g287}). We report the detection of a diffuse source at the cluster centre with the MWA at 118 and 154~MHz as well as a partial detection at 887~MHz in RACS data. The projected size of the source is $\sim320$~kpc. Other MWA bands either suffer from lack of sensitivity or significant confusion with nearby Sources A and B (\cref{fig:psz1g287:radio}). We obtain a spectral index of $\alpha_{118}^{887} =-1.5 \pm 0.2$ (\cref{fig:sed:psz1g287}). \cref{fig:psz1g287:xray} shows the \xmm\ image, highlighting the central location of the candidate radio halo. Based on the \xmm\ data, \citet{Rossetti2017} use the concentration parameter \citep[see][]{Santos2008} to determine the cluster is a non-CC cluster, and the emission is therefore unlikely to be a mini-halo. Additionally, no obvious BCG with core radio emission is seen in the optical data shown in \cref{fig:psz1g287:optical}. We classify this source as a candidate radio halo.

\subsubsection{\texttt{FIELD8}}

\begin{figure*}[h!]
    \centering
    \begin{subfigure}[b]{0.33\linewidth}
    \includegraphics[width=1\linewidth]{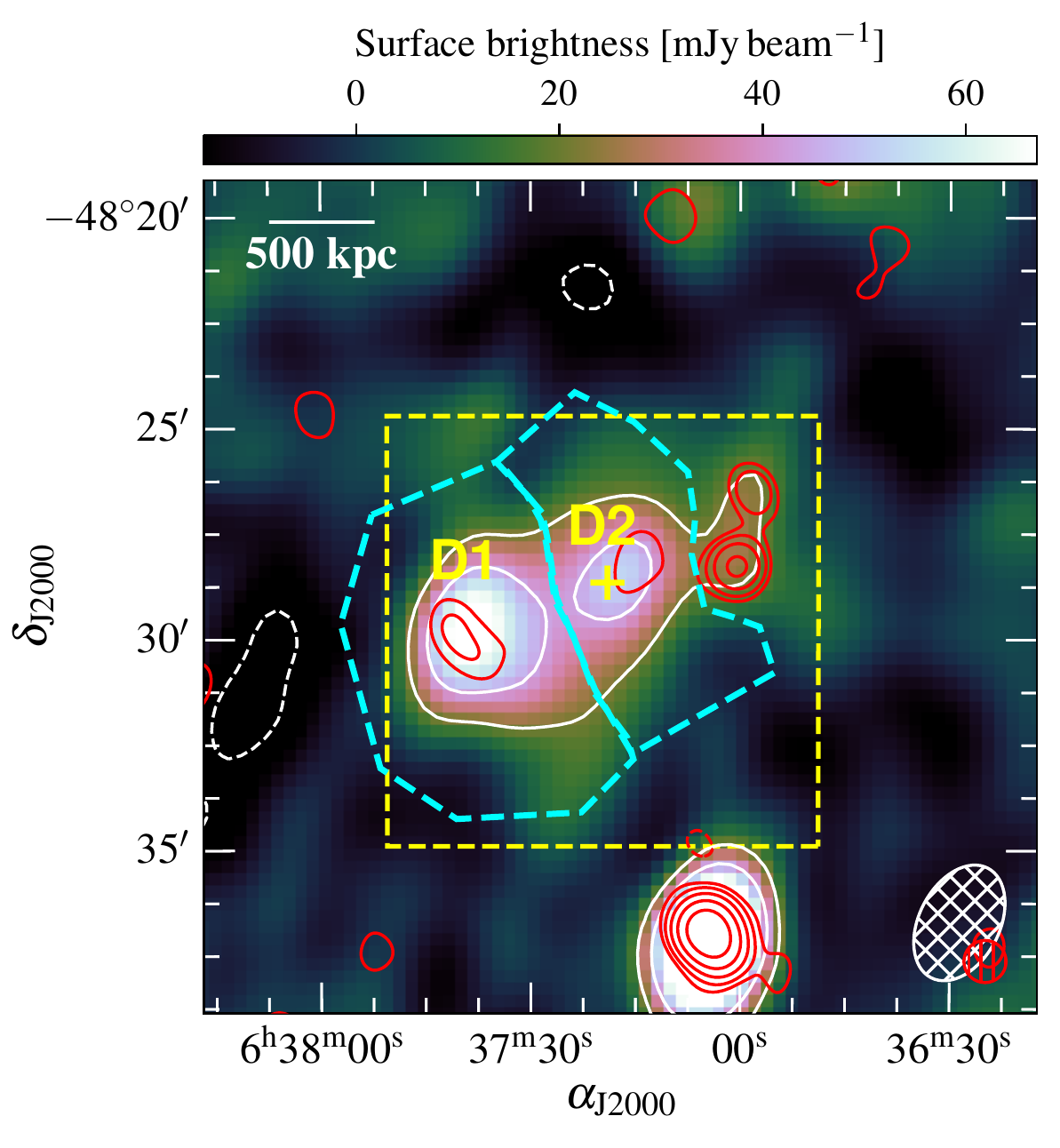}
    \caption{\label{fig:a3399:radio}}
    \end{subfigure}\hfill%
    \begin{subfigure}[b]{0.33\linewidth}
    \includegraphics[width=1\linewidth]{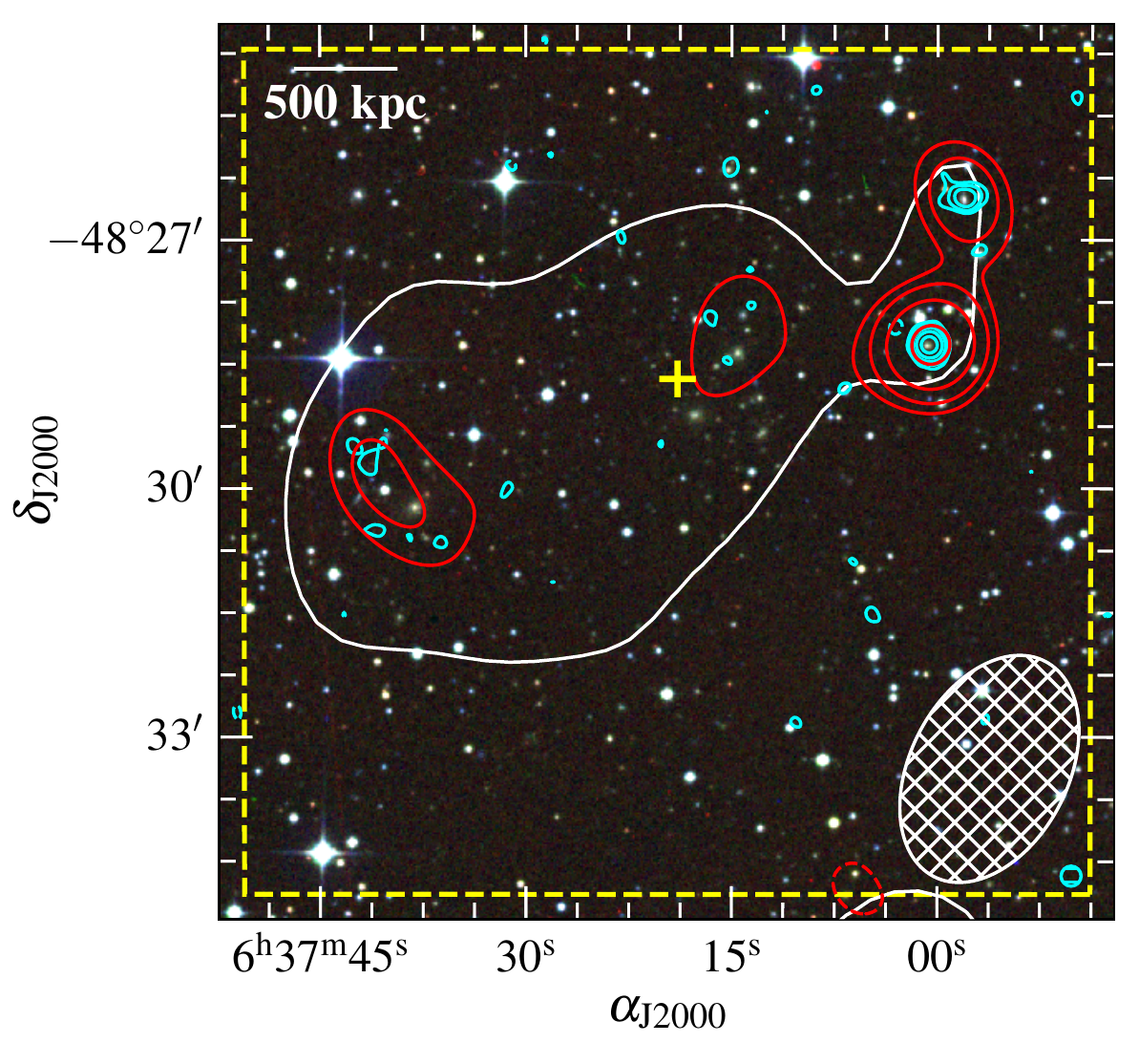}
    \caption{\label{fig:a3399:optical}}
    \end{subfigure}%
    \begin{subfigure}[b]{0.33\linewidth}
    \includegraphics[width=1\linewidth]{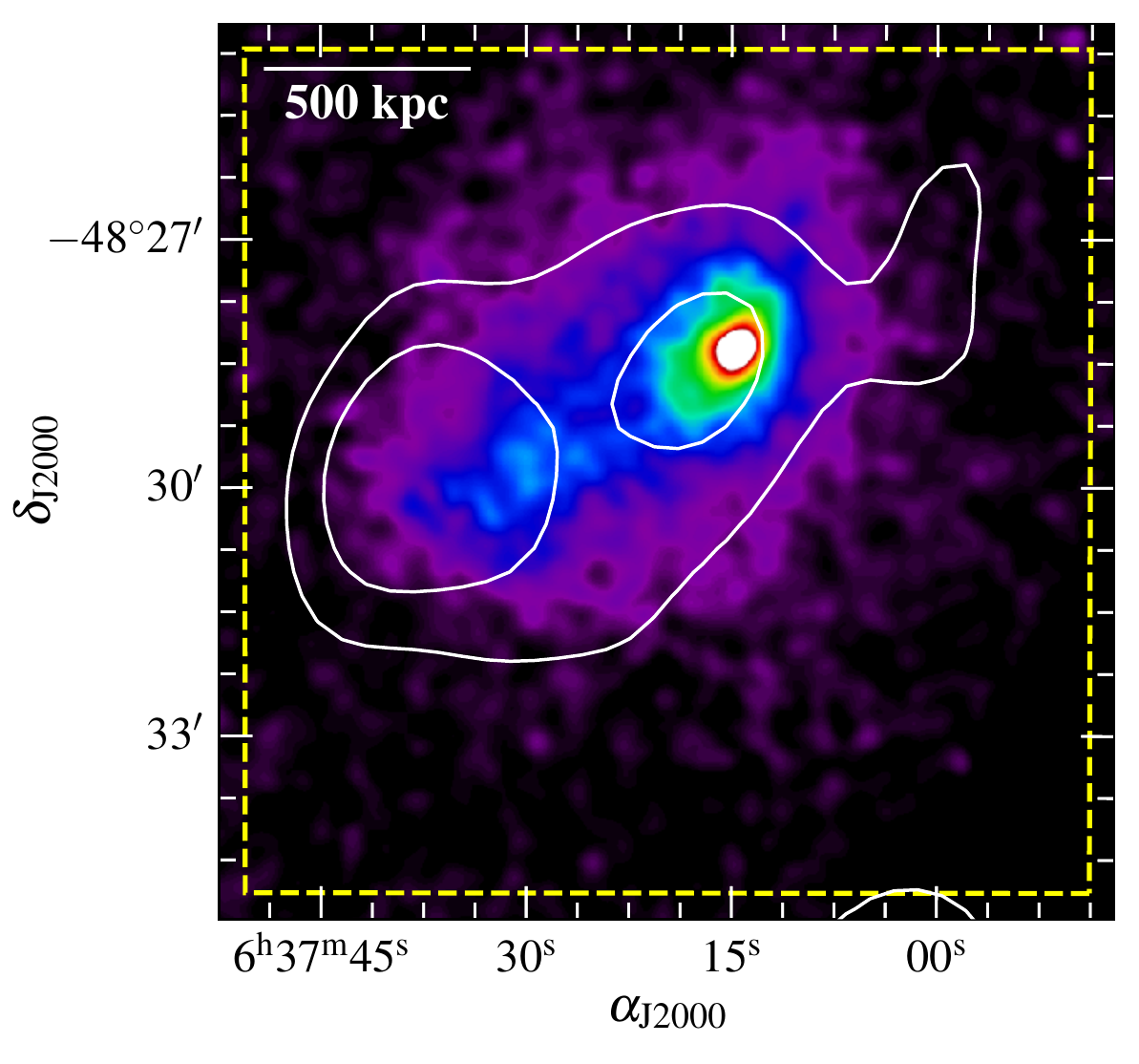}
    \caption{\label{fig:a3399:xray}}
    \end{subfigure}%
    \caption{\label{fig:a3399} \hyperref[para:a3399]{Abell~3399}. \subref{fig:a3399:radio} Background: MWA-2, 118-MHz, robust $+2.0$ image. \subref{fig:a3399:optical} Background: RGB SSS image ($i$, $r$, $b$). \subref{fig:a3399:xray}. Background: smoothed \emph{Chandra} image. The white contours are as in \cref{fig:a0122:radio} for the background of \subref{fig:a3399:radio} (with $\sigma_\text{rms} = 7.3$~mJy\,beam$^{-1}$). Red contours: RACS \corrs{discrete} source-subtracted image, $[\pm 3, 6, 12, 24, 48] \times \sigma_\text{rms}$ ($\sigma_\text{rms} = 0.45$~mJy\,beam$^{-1}$). Cyan contours: RACS survey image, $[\pm 3, 6, 12, 24, 48] \times \sigma_\text{rms}$ ($\sigma_\text{rms} = 0.17$~mJy\,beam$^{-1}$). Other image features are as in \cref{fig:a0122}.}
\end{figure*}

\paragraph{\hyperref[fig:a3399]{Abell~3399}} \label{para:a3399} (\cref{fig:a3399}). We report the detection of a candidate radio relic on the periphery of Abell~3399 (D1 in \cref{fig:a3399:radio}) {with an angular extent of $\sim 3.6$~arcmin ($\text{LLS}\approx 710$~kpc)} and a candidate radio halo at the centre of the cluster (D2 in \cref{fig:a3399:radio}) {with an angular extent of $\sim 2.8$~arcmin ($\text{LLS}\approx 570$~kpc).} As is clear from the \emph{Chandra} X-ray data shown in \cref{fig:a3399:xray}, the cluster is undergoing a merger and \citet{Lovisari2017} consider the cluster `disturbed' based on morphological analysis. Measuring flux densities of D1 at 88--154, and 887.5~MHz yields a spectral index of $\alpha_{\text{D1},88}^{887} = -1.6\pm0.3$ (\cref{fig:sed:a3399a}). For D2, the emission is is barely detected between 154--216~MHz and the 88-MHz image is too confused with D1 for a useful measurement. We instead calculate the two-point spectral index between 118 and 887~MHz, finding $\alpha_{\text{D2},118}^{887} = -1.5\pm0.2$ (\cref{fig:sed:a3399b}, though note the line shown is derived from the two-point index and not a fitted power law model). 

\subsubsection{\texttt{FIELD9}}

\begin{figure*}[h!]
    \centering
    \begin{subfigure}[b]{0.33\linewidth}
    \includegraphics[width=1\linewidth]{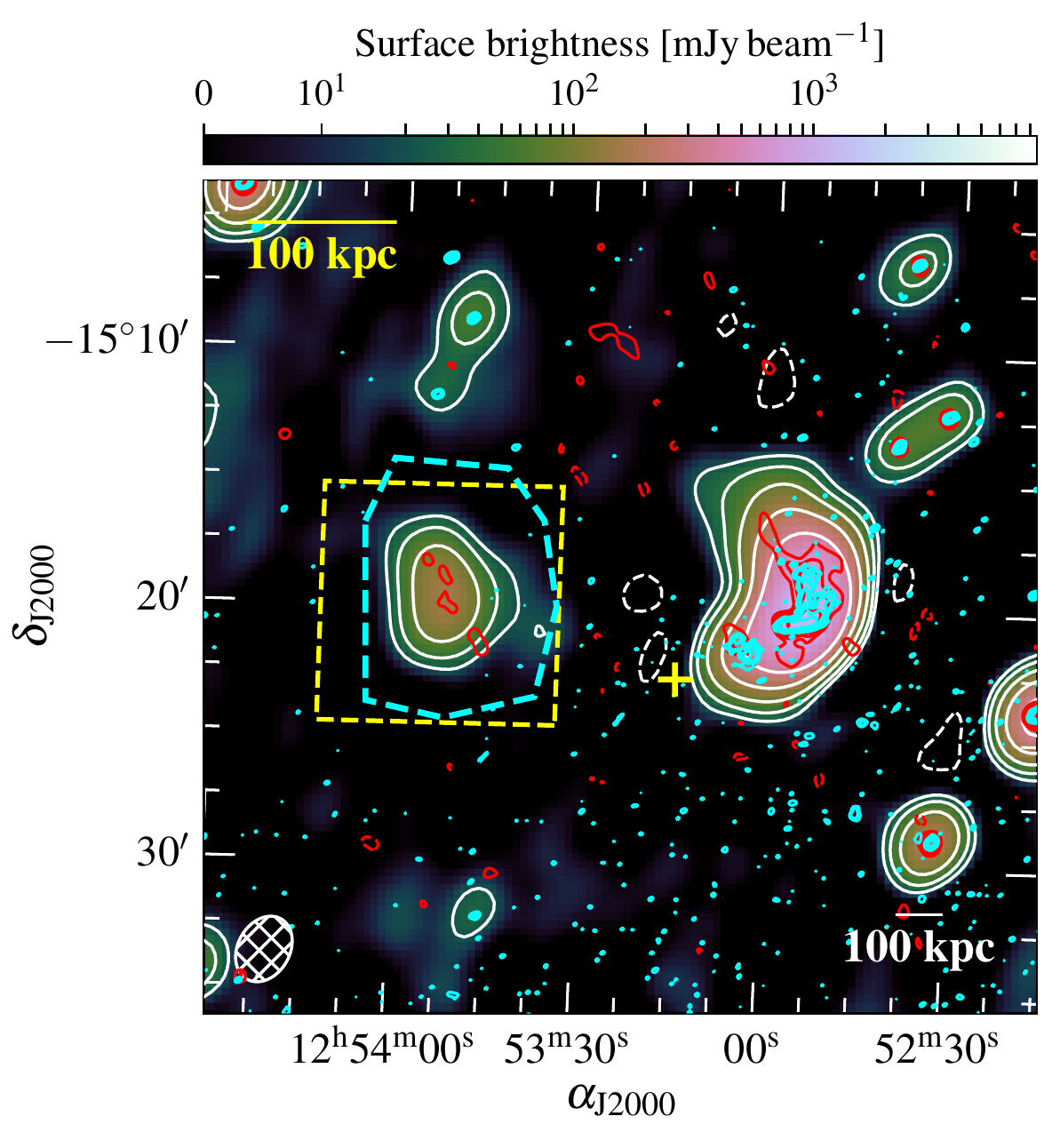}
    \caption{\label{fig:a1631:radio}}
    \end{subfigure}%
    \begin{subfigure}[b]{0.33\linewidth}
    \includegraphics[width=1\linewidth]{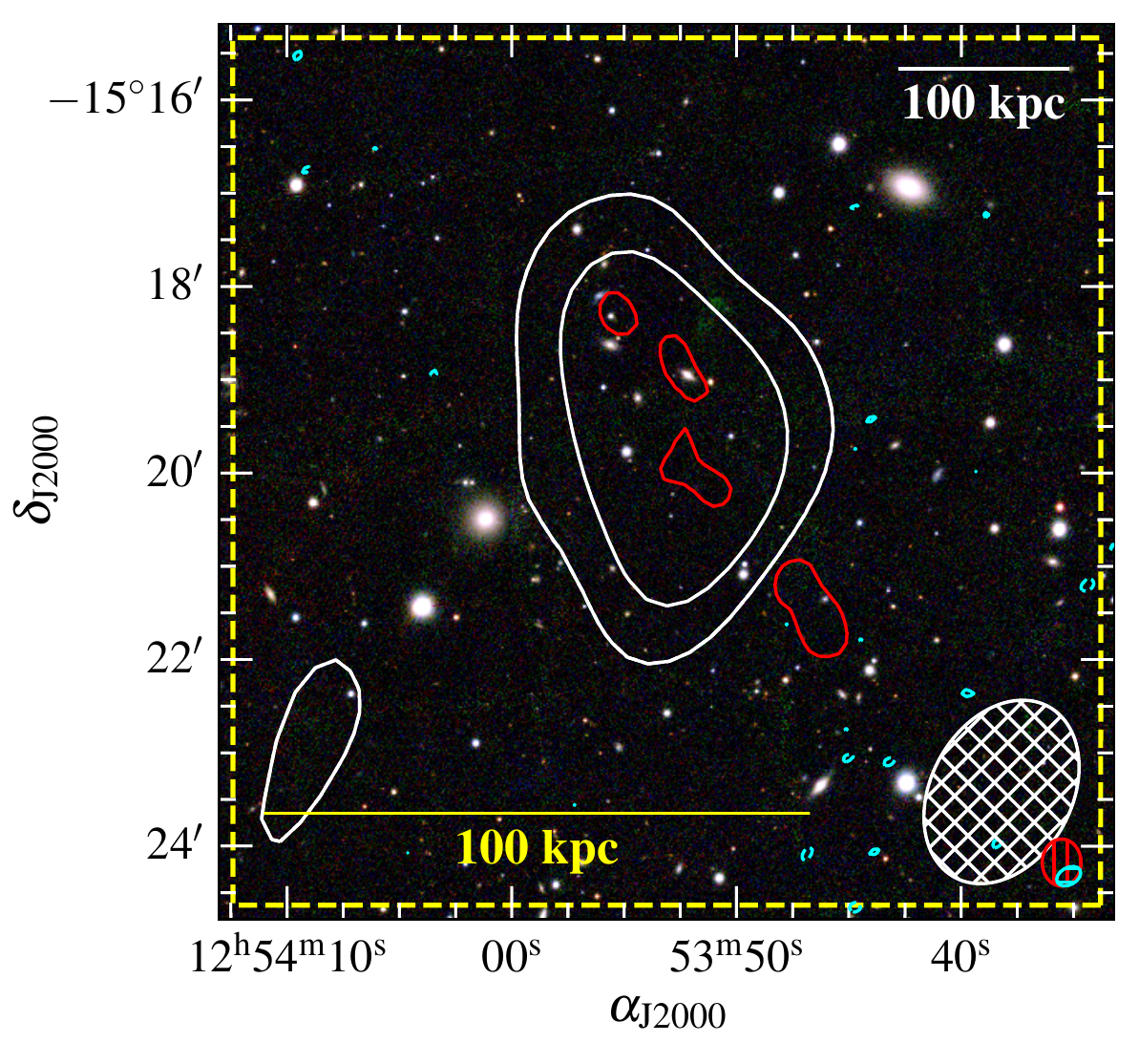}
    \caption{\label{fig:a1631:optical}}
    \end{subfigure}%
    \begin{subfigure}[b]{0.33\linewidth}
    \includegraphics[width=1\linewidth]{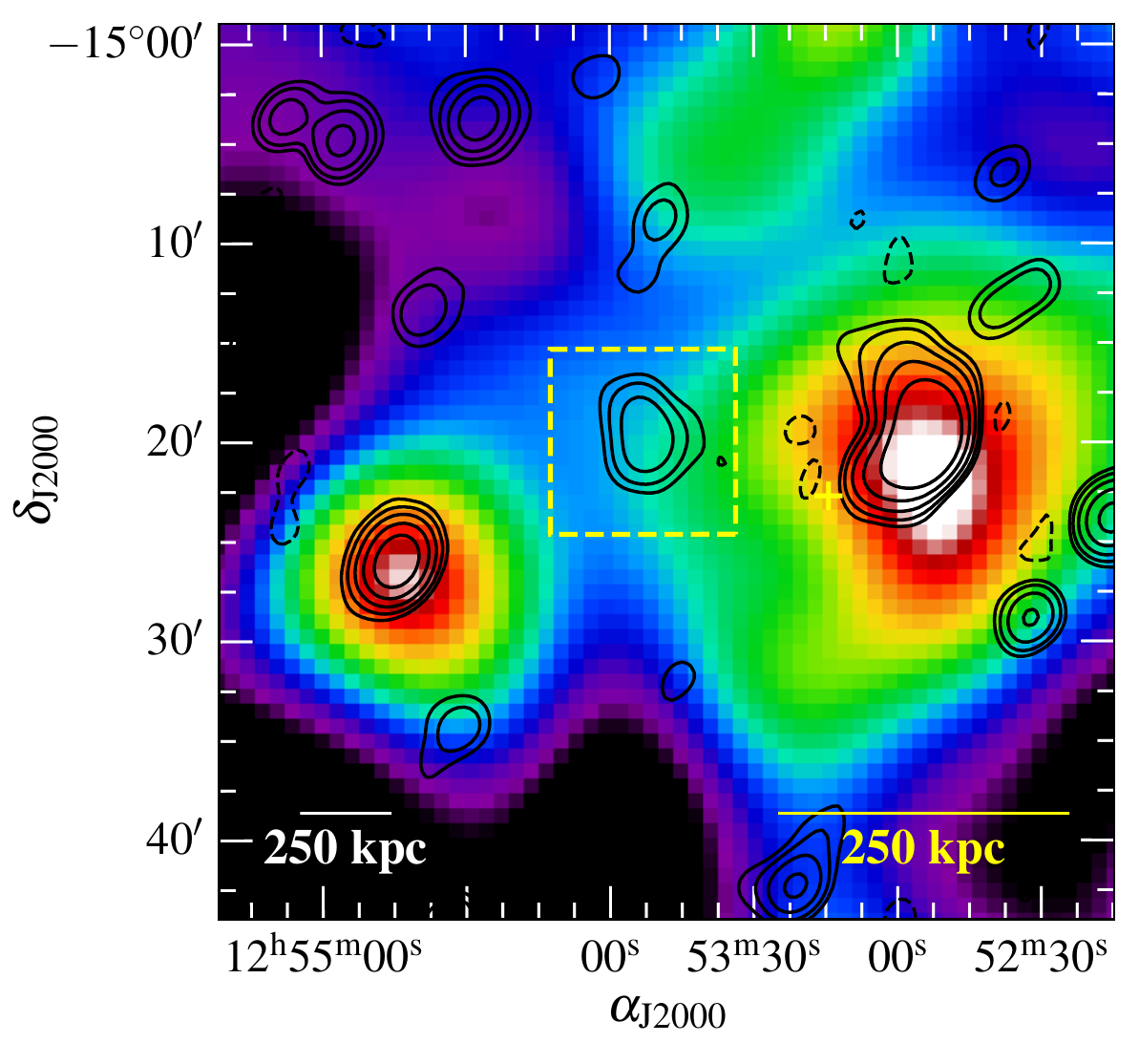}
    \caption{\label{fig:a1631:xray}}
    \end{subfigure}%
    \caption[Images of Abell~1631]{\label{fig:a1631} \hyperref[para:a1631]{MCXC~J1253.2$-$1522} (Abell~1631). \subref{fig:a1631:radio} Background: MWA-2, 118-MHz, robust $+2.0$ image. \subref{fig:a1631:optical} Background: RGB PS1 image ($i$, $r$, $g$). \subref{fig:a1631:xray}. Background: smoothed RASS image. The white (black) contours are as in \cref{fig:a0122:radio} for the background of \subref{fig:a1631:radio} (with $\sigma_\text{rms} = 7.0$~mJy\,beam$^{-1}$). Red contours: TGSS image, $[\pm 3, 6, 12, 24, 48] \times \sigma_\text{rms}$ ($\sigma_\text{rms} = 3.5$~mJy\,beam$^{-1}$). Cyan contours: RACS survey image, $[3, 6, 12, 24, 48] \times \sigma_\text{rms}$ ($\sigma_\text{rms} = 0.32$~mJy\,beam$^{-1}$). Other image features are as in \cref{fig:a0122}, with the addition of a separate, yellow linear scale at the redshift of Abell~1631.}
\end{figure*}

\paragraph{\hyperref[fig:a1631]{MCXC~J1253.2$-$1522} (Abell~1631)} \label{para:a1631} (\cref{fig:a1631}). We report a steep spectrum source projected onto MCXC~J1253.2$-$1522 in MWA data, with no counterpart in RACS, and marginal detection in the TGSS image (\cref{fig:a1631:radio}). Note the TGSS image has artefacts from the nearby complex radio galaxy that peak at the location of the diffuse source. Multiple cluster systems are reported along line of sight\corrs{: MCXC~J1253.2$-$1522 at $z=0.0462$ \citep{pap+11} and Abell~1631} at $z=0.01394$ \citep{cac+09}. Additionally, \citet{Flin2006} report that the system has complex substructure, which is clear in the smoothed RASS image shown in \cref{fig:a1631:xray}. No obvious optical host is visible in PS1 data (\cref{fig:a1631:optical}, and the projected linear extent is 200~kpc (or 70~kpc if at $z=0.01394$). While just outside of the field-of-view (FoV) of archival \xmm\ observations, RASS shows no significant X-ray emission at the location of the source, but the source sits between two X-ray clumps. We find the spectral index to be $\alpha_{88}^{200} = -1.8 \pm 0.4$ (\cref{fig:sed:a1631}), and suggest the source is fossil plasma or otherwise a remnant radio galaxy. 

\subsubsection{\texttt{FIELD10}}

\begin{figure*}[h!]
    \centering
    \begin{subfigure}[b]{0.33\linewidth}
    \includegraphics[width=1\linewidth]{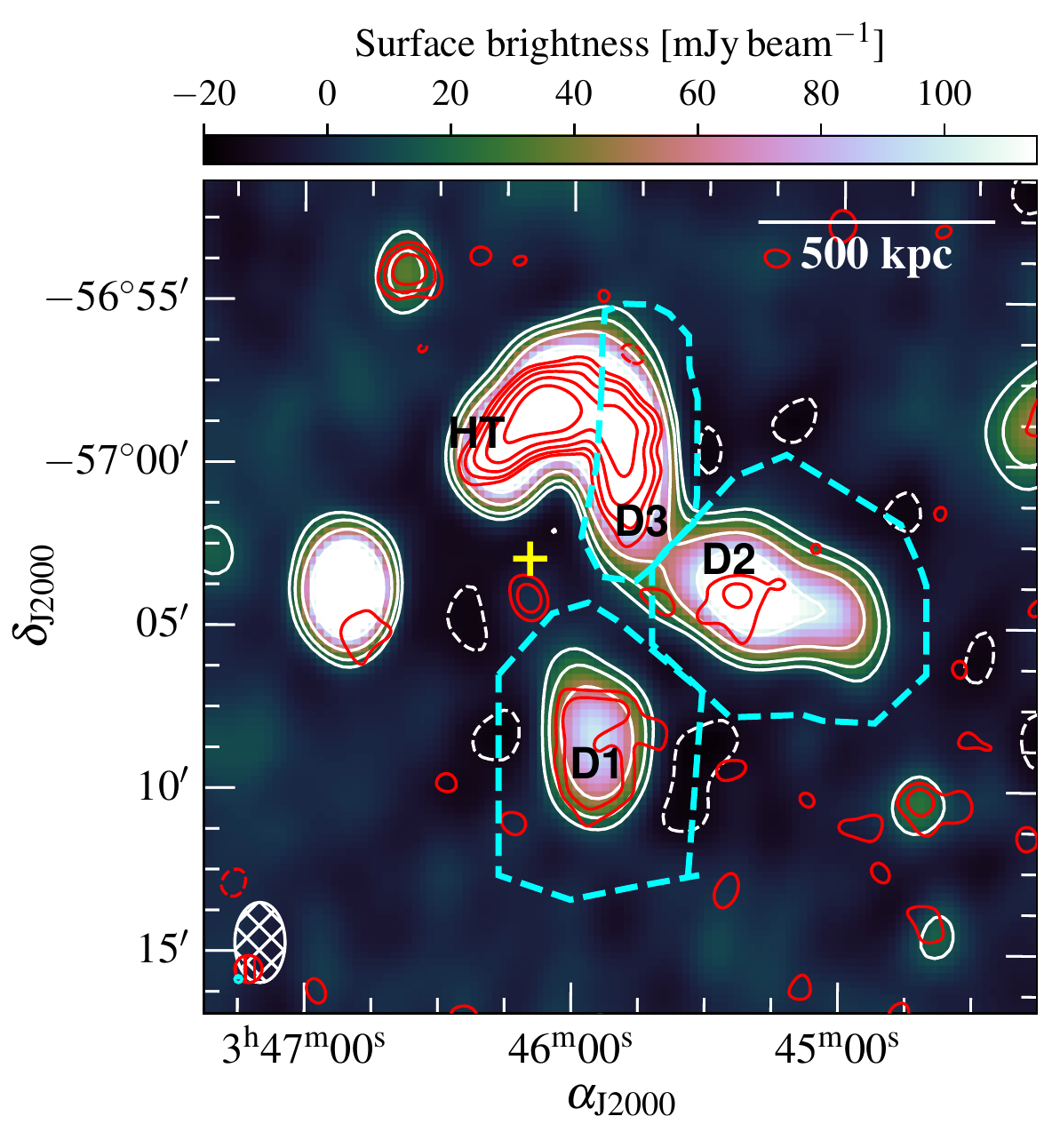}
    \caption{\label{fig:a3164:radio}}
    \end{subfigure}%
    \begin{subfigure}[b]{0.33\linewidth}
    \includegraphics[width=1\linewidth]{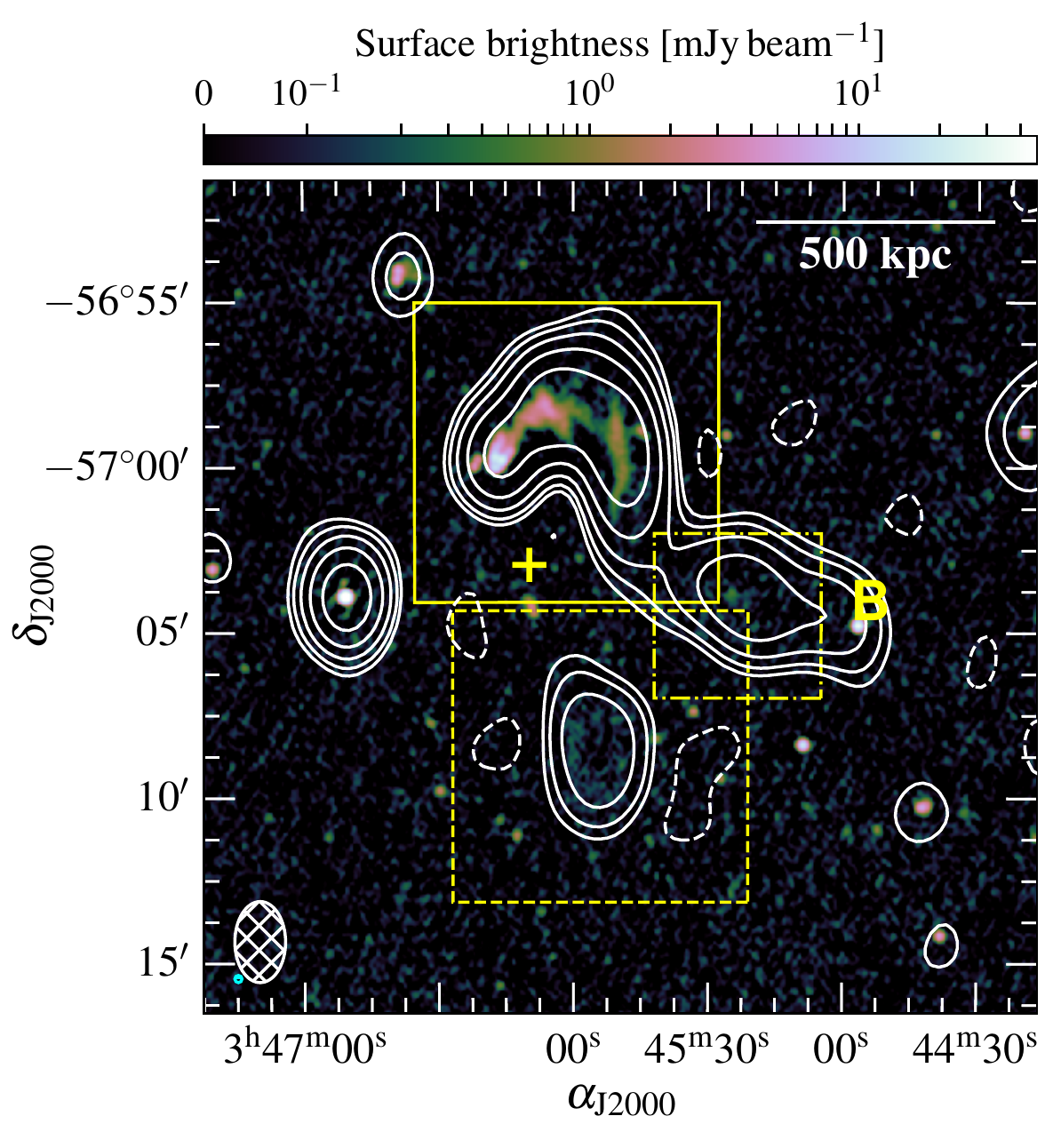}
    \caption{\label{fig:a3164:radio2}}
    \end{subfigure}%
    \begin{subfigure}[b]{0.33\linewidth}
    \includegraphics[width=1\linewidth]{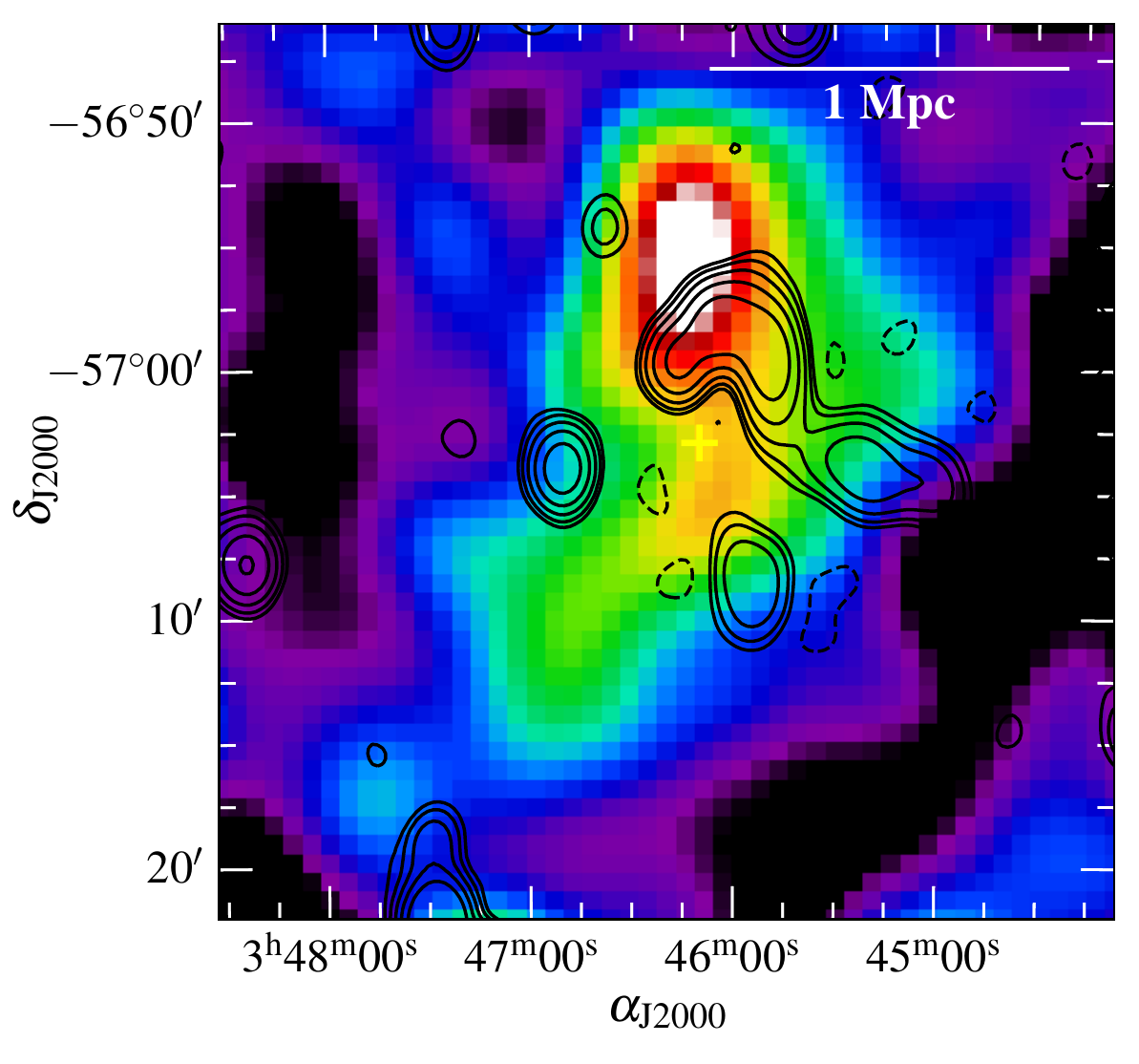}
    \caption{\label{fig:a3164:xray}}
    \end{subfigure}\\%
    \begin{subfigure}[b]{0.33\linewidth}
    \includegraphics[width=1\linewidth]{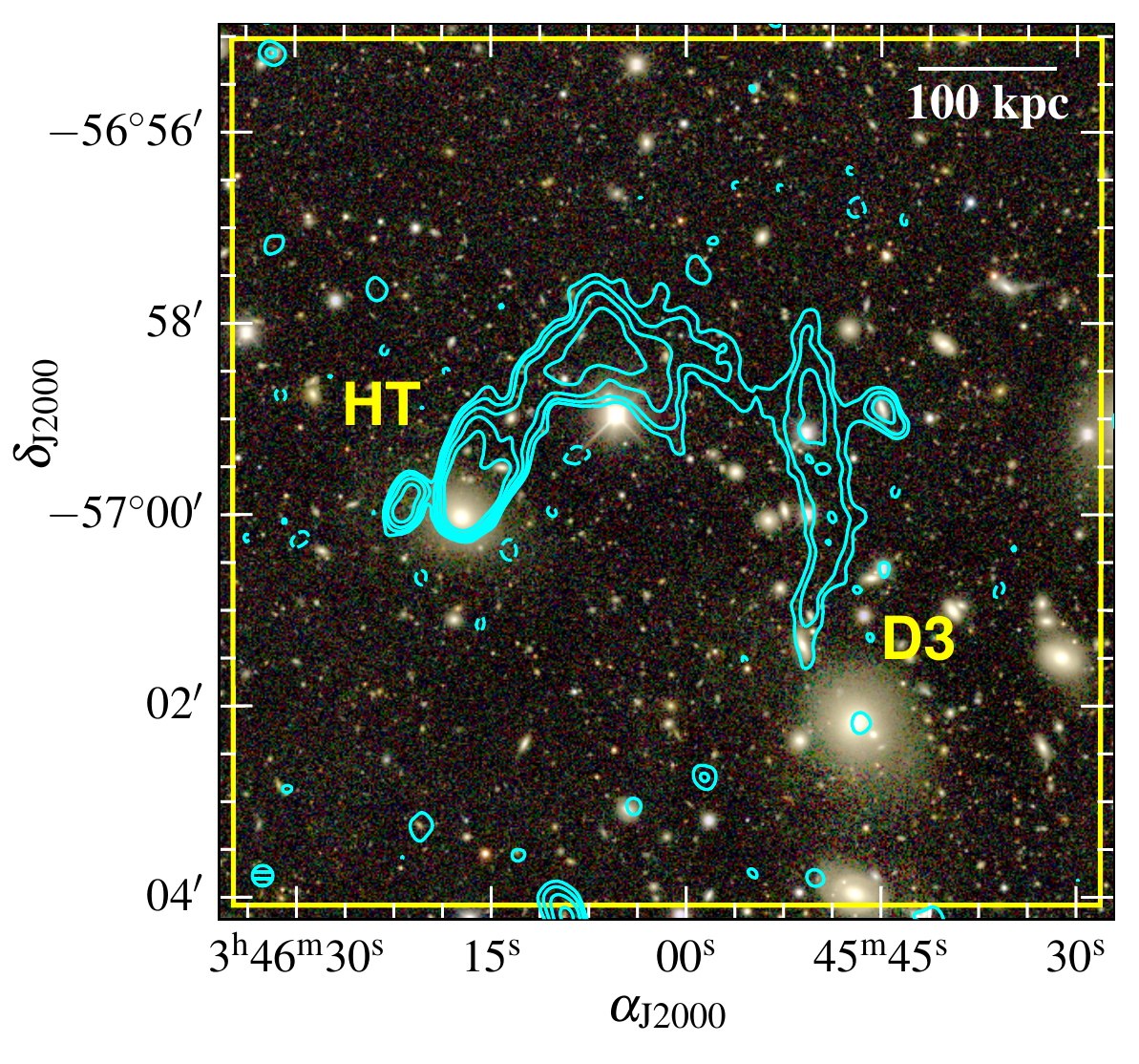}
    \caption{\label{fig:a3164:ht}}
    \end{subfigure}%
    \begin{subfigure}[b]{0.33\linewidth}
    \includegraphics[width=1\linewidth]{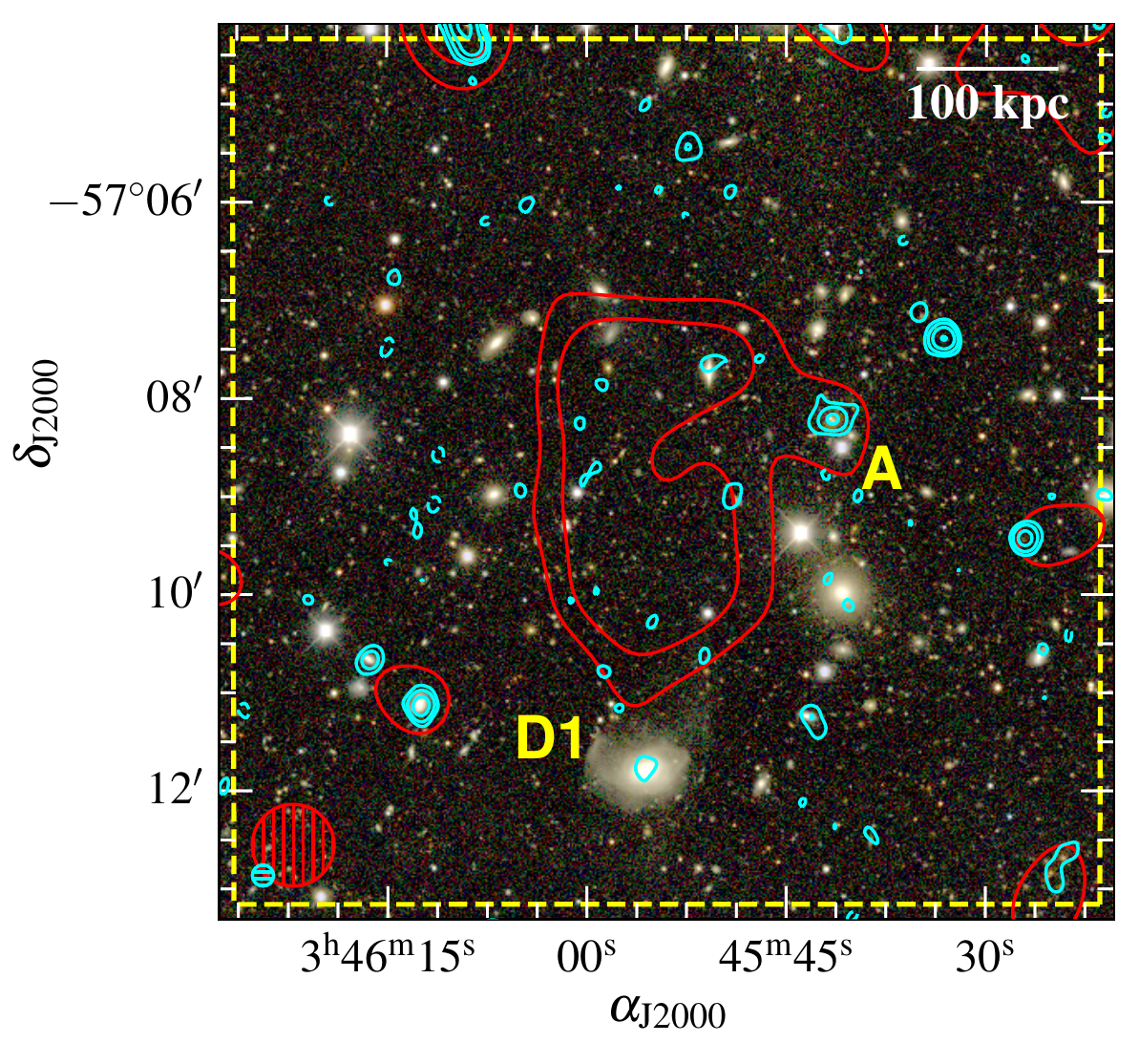}
    \caption{\label{fig:a3164:d1}}
    \end{subfigure}%
    \begin{subfigure}[b]{0.33\linewidth}
    \includegraphics[width=1\linewidth]{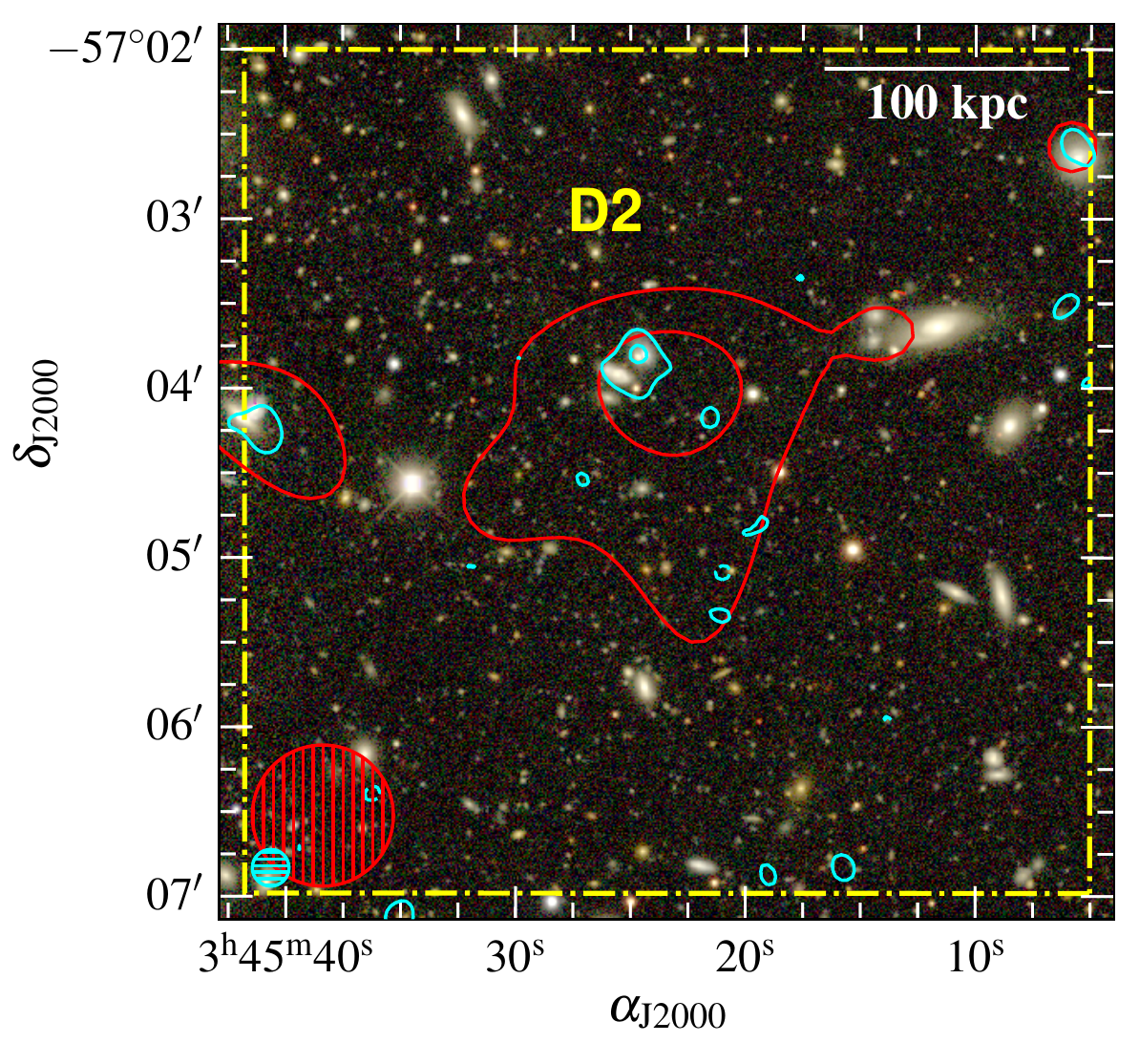}
    \caption{\label{fig:a3164:d2}}
    \end{subfigure}%
    \caption{\label{fig:a3164} \hyperref[para:a3164]{Abell~3164}. Background: MWA-2, 154-MHz, robust $+2.0$ image. \subref{fig:a3164:radio2} Background: VAST robust $0.0$ image. \subref{fig:a3164:xray} Background: smoothed, RASS image. \subref{fig:a3164:ht}--\subref{fig:a3164:d2} Background: RGB DES image ($i$, $r$, $g$). The white contours are as in \cref{fig:a0122:radio} for the background of \subref{fig:a3164:radio} (with $\sigma_\text{rms} = 4.3$~mJy\,beam$^{-1}$). Red contours: VAST source-subtracted image, $[\pm 3, 6, 12, 24, 48] \times \sigma_\text{rms}$ ($\sigma_\text{rms} = 0.17$~mJy\,beam$^{-1}$). Cyan contours: VAST robust $0.0$ image, $[\pm 3, 6, 12, 24, 48] \times \sigma_\text{rms}$ ($\sigma_\text{rms} = 0.09$~mJy\,beam$^{-1}$). Other image features are as in \cref{fig:a0122}. Yellow boxes on \subref{fig:a3164:radio2} indicate the locations of \subref{fig:a3164:ht}--\subref{fig:a3164:d2}.}
\end{figure*}


\paragraph{\hyperref[fig:a3164]{Abell~3164}} \label{para:a3164} (\cref{fig:a3164}). MWA and the VAST data reveal a complex radio galaxy with at least three distinct remnant components (D1--3 in \cref{fig:a3164:radio}) {with LASs 3.5, 3.8, and 3.1~arcmin and LLSs of 240, 260, and 210~kpc for D1--3, respectively}. The head-tail (HT) galaxy (hosted by FAIRALL~0757; \citealt{Fairall1984}) in the cluster is clearly connected to D3 and is likely responsible for at least D2, however, it is not clear whether D1 has spawned from the same galaxy. The source-subtracted VAST data is shown in \cref{fig:a3164:radio} and highlights the steepness of D2. After subtraction of Sources A and B from the relevant measurements, we obtain spectral indices of D1: $\alpha_{88}^{887} = -1.48 \pm 0.08$ (\cref{fig:sed:a3164a}), D2: $\alpha_{88}^{216} = -2.3 \pm 0.1$ (though \cref{fig:sed:a3164b} shows some curvature within the MWA band, and D2 is also fit with a generic curved model between 88--887.5~MHz), D3: $\alpha_{154}^{887} = -1.78 \pm 0.07$ (\cref{fig:sed:a3164c}). We do not believe significant flux density is missing, since D1, which is comparative in angular scale, is recovered consistently alongside the MWA. We see steepening from D3--D2, but D1 is flatter, which suggests either (1) it may not be associated with past episode from the HT or (2) it has been re-accelerated/revived by ICM motion. Steepening of the spectrum of HT radio galaxies is observed to increase with distance from the host galaxy (e.g. the HT in Abell~1132; \citealt{Wilber2018}, or in Abell~1775; \citealt{Botteon2021}).

The smoothed RASS image is shown in \cref{fig:a3164:xray} which highlights a faint X-ray structure detected as the cluster, elongated towards the SE. The X-ray emission does not elucidate the nature of the diffuse components based on location, however, the diffuse emission resides within the clearly elongated X-ray emission to the south of the cluster, thus is in a region where we can assume dynamical activity is occurring.

\subsubsection{\texttt{FIELD11}}

\begin{figure}[h!]
    \centering
    \includegraphics[width=1\linewidth]{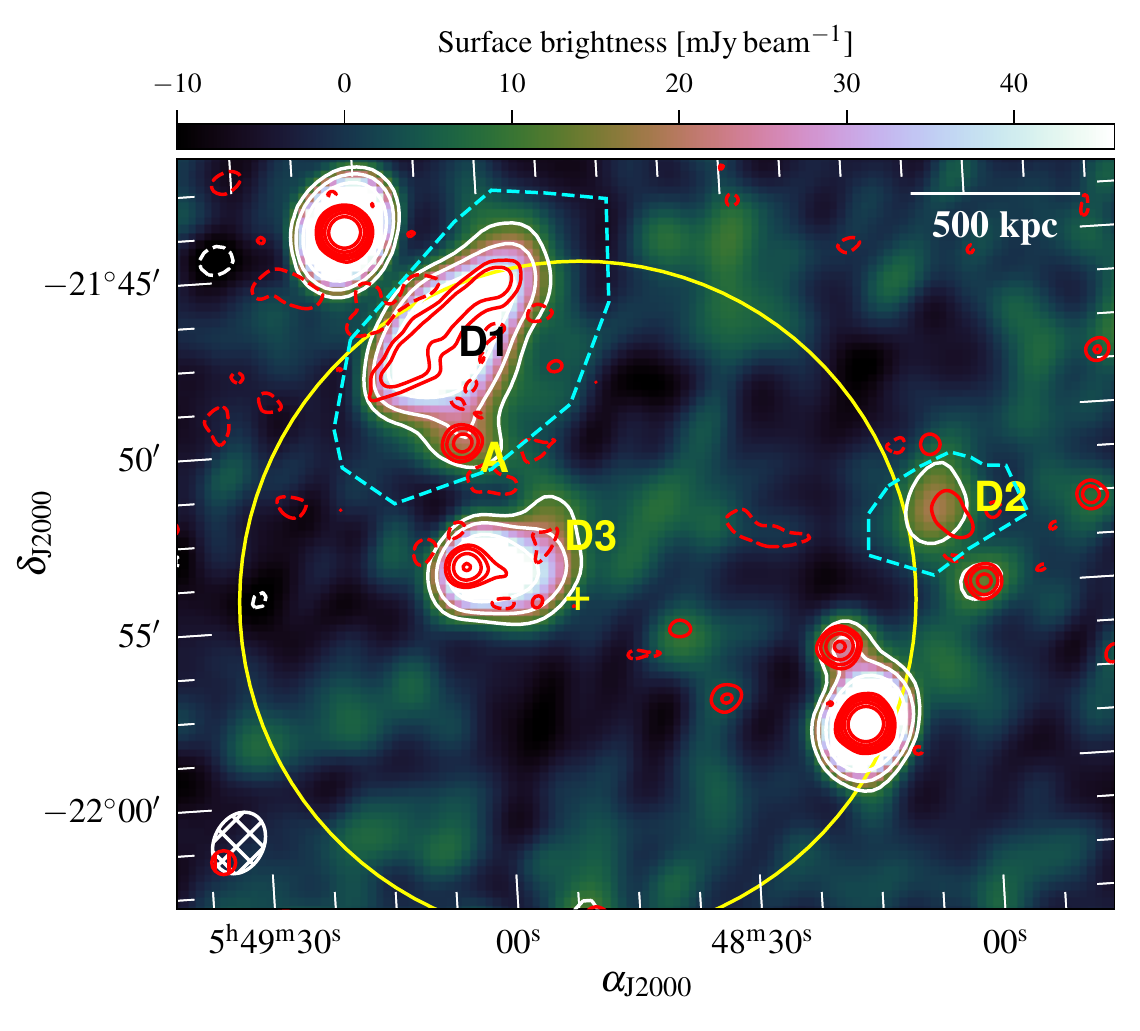}
    \caption{\label{fig:a3365} \hyperref[para:a3365]{Abell~3365}. Background: MWA-2, 154-MHz, robust $+1.0$ image. The white contours are as in \cref{fig:a0122:radio} for the background image (with $\sigma_\text{rms} = 3.5$~mJy\,beam$^{-1}$). Red contours: RACS low-resolution, $[\pm 3, 6, 12, 24, 48] \times \sigma_\text{rms}$ ($\sigma_\text{rms} = 0.47$~mJy\,beam$^{-1}$). Other image features are as in \cref{fig:a0122} and \cref{fig:a2751}. Note the white circle has a 1~Mpc radius, centered on the coordinates associated with RXC~J0548.8$-$2154 (i.e.~the X-ray component for the system).}
\end{figure}

\paragraph{\hyperref[fig:a3365]{Abell~3365}} \label{para:a3365} (\cref{fig:a3365}). Abell~3365 (RXC~J0548.8$-$2154) was reported to host a relic and candidate second relic by \citet{vanWeeren2011b}---at the time, spectral coverage was minimal. While not in our original survey scope, the cluster resides at the edge of a field that was opportunistically processed for other targets. We show the 154-MHz map in \cref{fig:a3365} with RACS contours overlaid. The NE relic is detected (D1), and the SW candidate relic source is also detected (D2). An additional extension, labelled D3, is also detected that has not been identified in other radio maps. We utilise the additional spectral coverage offered by the MWA-2 and find, after subtraction of Source A, a power law model for the NE relic, with $\alpha_{\text{D1},88}^{1420} = -0.85 \pm 0.03$ (\cref{fig:sed:a3365a}). We note a curious feature of the flux density measurements wherein the lower MWA bands separate from the higher bands---we cannot explain this feature as either instrumental or physical. While the relic is detected in the RACS data, we note significant negative bowls around the source indicating a lack of flux recovery. A lower limit is provided but not used in fitting. The resultant spectral shape is reasonably shallow for a relic. The SW relic (D2) is detected in MWA-2 data as well, but is too confused in the 88-MHz band. The resultant SED provides $\alpha_{\text{D2},118}^{1420} = -0.76 \pm 0.08$ (\cref{fig:sed:a3365b}), noting that the discrete Source A discussed by \citet{vanWeeren2011b} is not confused in these bands and is not required to be subtracted. 

\citet{Golovich2019} present deep \xmm\ observations of the cluster along with spectro-optical analysis (see their Fig.~25), highlighting the merger axis and location of the candidate SW relic with respect to three subclusters within the system. They note there is no distinct alignment of the subclusters with the W relic candidate (D2), and given the shallow spectrum we suggest this is not a relic source. 

Despite this, \citet{Urdampilleta2021} find significant temperature jumps across both relics, finding evidence for shocks with Mach numbers $\mathcal{M}_\text{west} = 3.9 \pm 0.8$ and $\mathcal{M}_\text{east} = 3.5 \pm 0.6$, which for DSA  may produce relics with flatter spectral indices as observed here. Assuming DSA for these relics, and assuming $\alpha_\text{integrated} = \alpha_\text{inj}$ as might be the case for a re-accelerated fossil electron population \citep[e.g.][]{vanWeeren2016}, we find radio Mach numbers of $\mathcal{M}_\text{RW} = 4.2 \pm 0.4$ and $\mathcal{M}_\text{RE} = 2.5 \pm 0.1$ for the western and eastern relics, respectively. If accelerated from the thermal pool, we would expect $\alpha_\text{inj} = \alpha_\text{integrated} + 0.5$, resulting in a non-physical Mach number under standard DSA. The nature of both sources (D1 and D2) is uncertain.

\begin{figure*}[h!]
    \centering
    \begin{subfigure}[b]{0.33\linewidth}
    \includegraphics[width=1\linewidth]{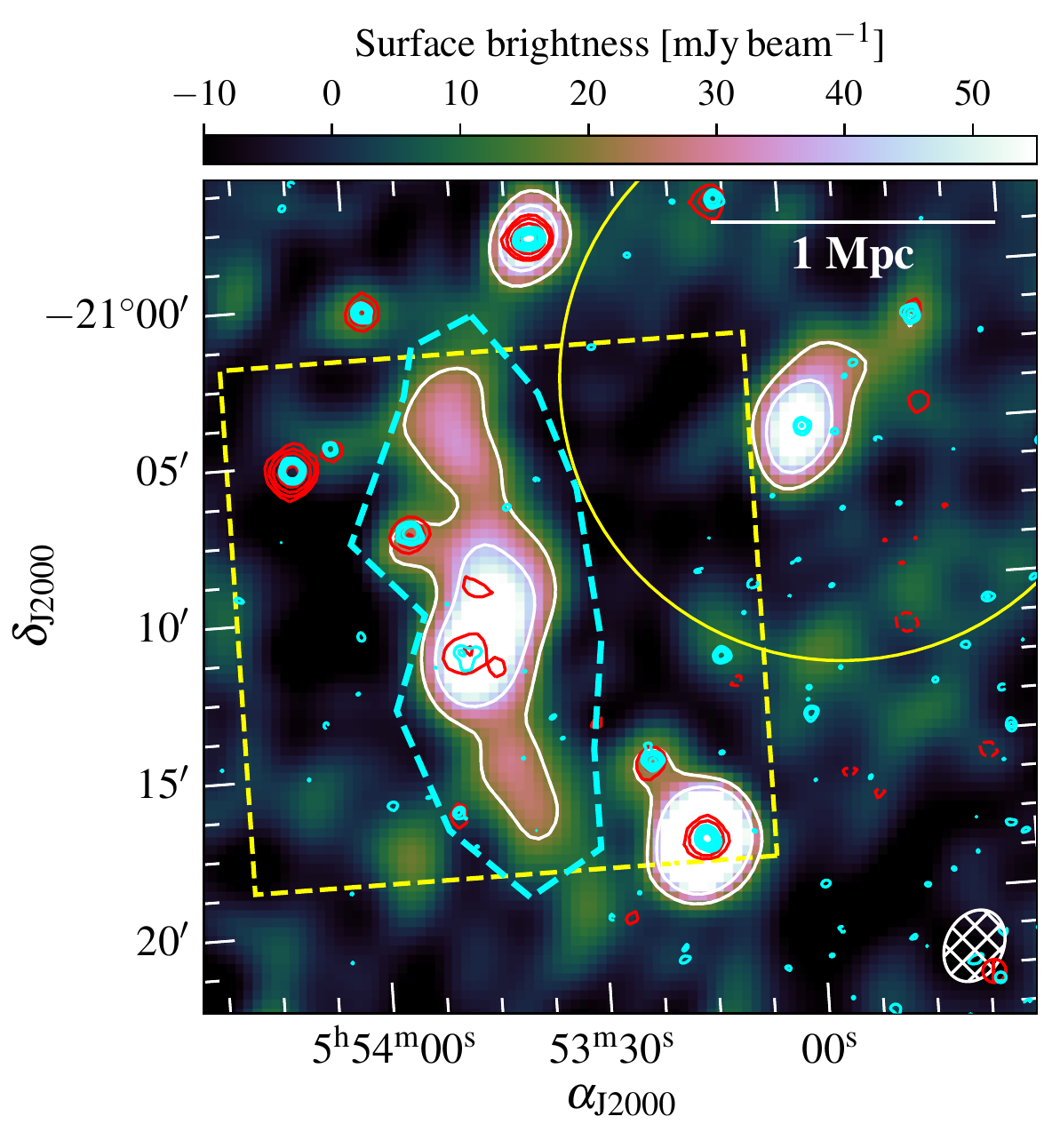}
    \caption{\label{fig:a550:radio}}
    \end{subfigure}%
    \begin{subfigure}[b]{0.33\linewidth}
    \includegraphics[width=1\linewidth]{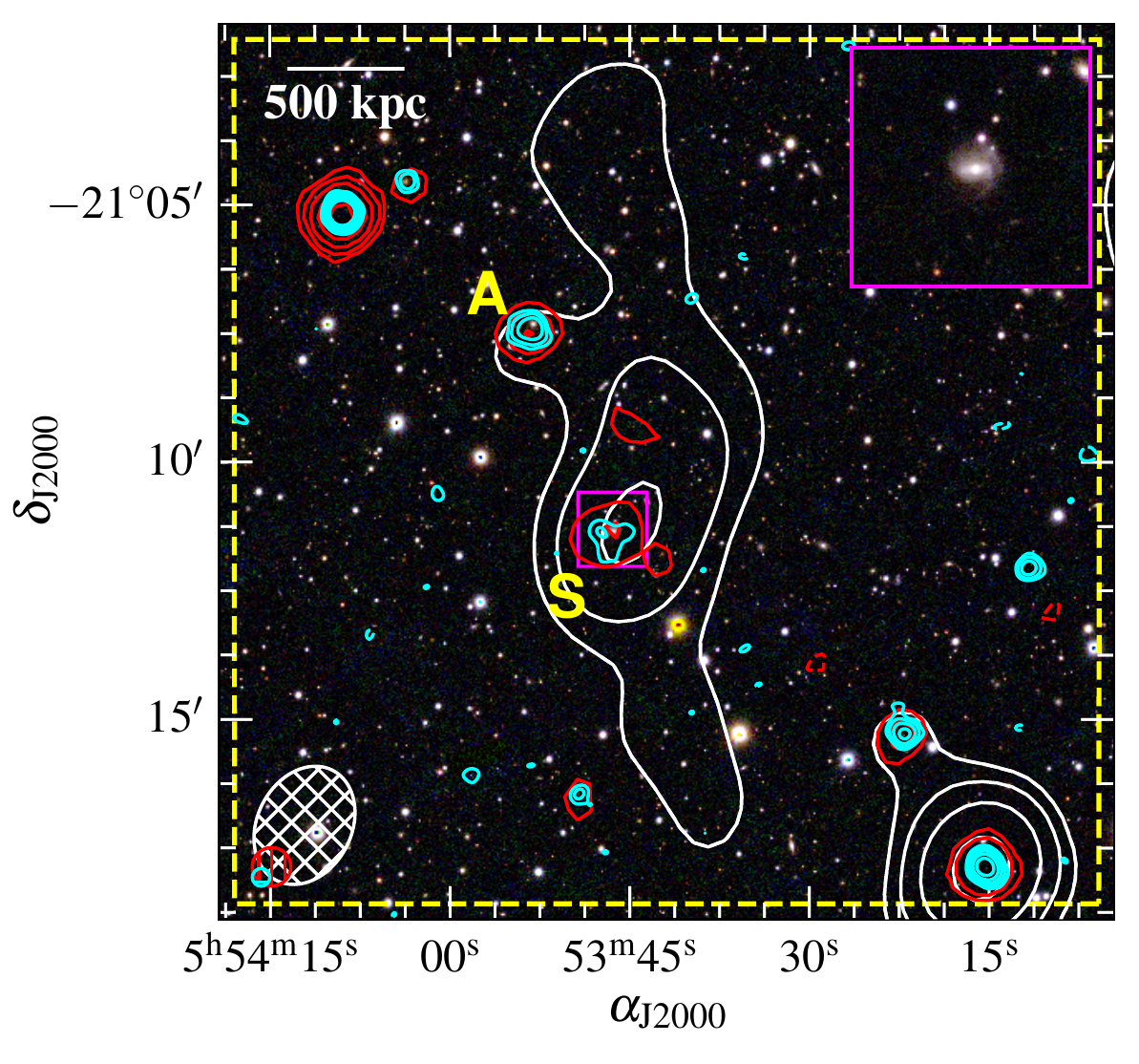}
    \caption{\label{fig:a550:optical}}
    \end{subfigure}%
    \begin{subfigure}[b]{0.33\linewidth}
    \includegraphics[width=1\linewidth]{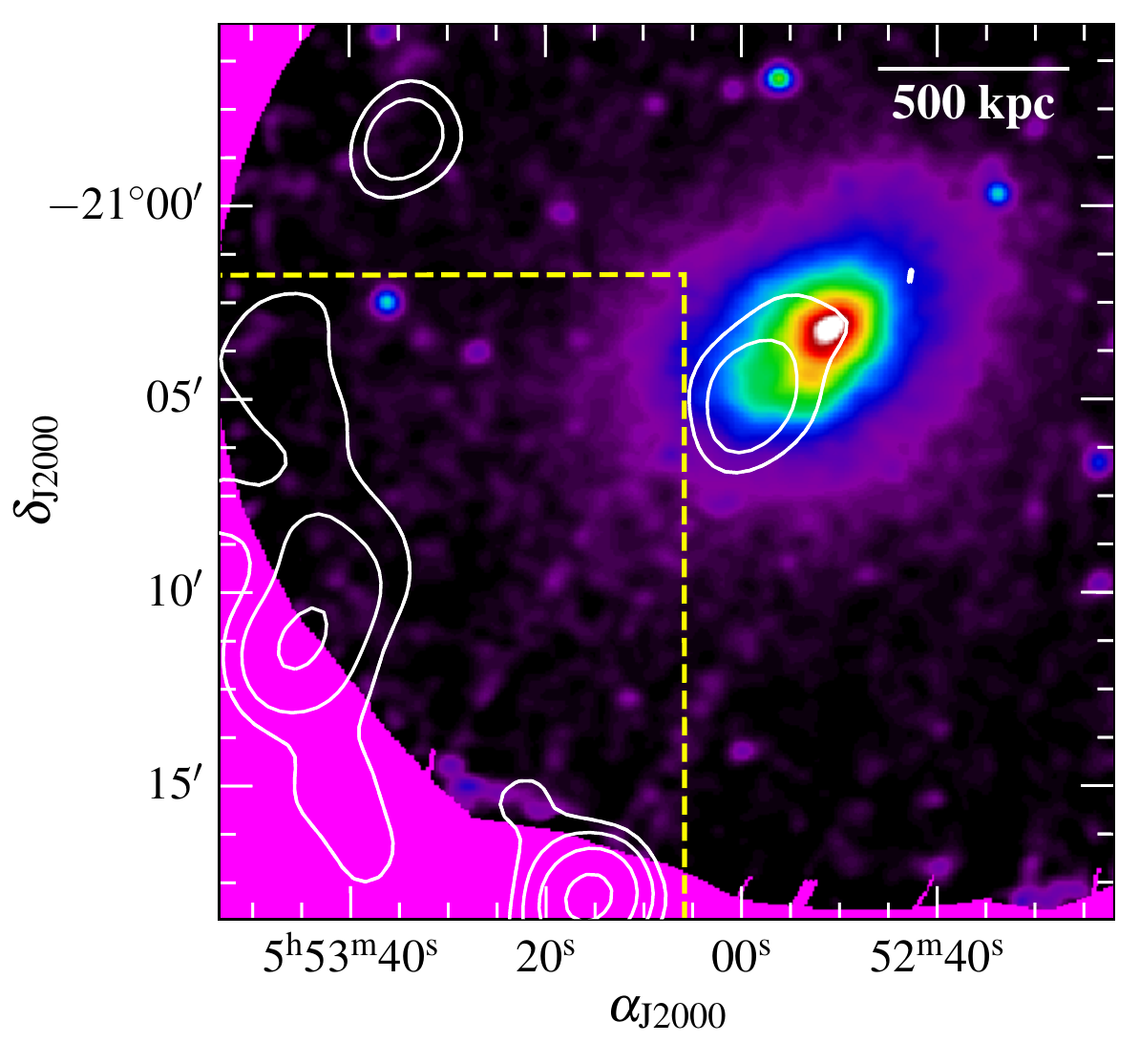}
    \caption{\label{fig:a550:xray}}
    \end{subfigure}%
    \caption{\label{fig:a550} \hyperref[para:a550]{Abell~0550}. Background: MWA-2, 118-MHz, robust $+1.0$ image. \subref{fig:a550:optical} Background: RGB PS1 image ($i$, $r$, $g$). \subref{fig:a550:xray}. Background: smoothed \xmm\ EPIC image. The white contours are as in \cref{fig:a0122:radio} for the background of \subref{fig:a550:radio} (with $\sigma_\text{rms} = 6.6$~mJy\,beam$^{-1}$). Red contours: NVSS image, $[\pm 3, 6, 12, 24, 48] \times \sigma_\text{rms}$ ($\sigma_\text{rms} = 0.57$~mJy\,beam$^{-1}$). Cyan contours: RACS robust $+0.25$ image, $[\pm 3, 6, 12, 24, 48] \times \sigma_\text{rms}$ ($\sigma_\text{rms} = 0.25$~mJy\,beam$^{-1}$). Other image features are as in \cref{fig:a0122} \corrs{and \cref{fig:a2751}}. Note \subref{fig:a550:xray} clips at the location of relic as that is the FoV of the \xmm\ observation.}
\end{figure*}

\paragraph{\hyperref[fig:a550]{Abell~0550}} \label{para:a550} (\cref{fig:a550}). We report the detection of a large-scale, elongated emission structure on the periphery of Abell~0550 (\cref{fig:a550:radio} {with an LAS of $\sim 15$~arcmin ($\text{LLS}\approx 1.6$~Mpc)}. We detect the source between 88--154~MHz in the MWA-2 data. We note that there is a spiral galaxy (WISEA~J055346.34$-$211119.6, no redshift) near the centre of the emission, shown in \cref{fig:a550:optical}, denoted with an `S'. Spiral galaxies rarely host large-scale radio lobes, though a small number have been detected, sometimes with remnant lobes \citep[e.g.][]{hso+11} and this may be such an example. With no compact emission detected in RACS from the spiral galaxy, we can rule out significant AGN contribution and suggest the emission seen in the NVSS and MWA-2 images is related to star-formation processes and that the galaxy, if the host, is not actively fuelling the lobes, consistent with a steep spectrum. The resultant spectrum (\cref{fig:sed:a550}), with the spiral and compact Source A subtracted flattens at 1.4~GHz, and is fit with a generic curved power law, though the physical interpretation of this is not clear. New kpc-scale jets from the spiral that are not subtracted may contribute to the spectral flattening, though this is unlikely to be so extreme and we suggest residual flux from the extended emission from the spiral is affecting the measurements. Across the MWA band we fit a power law model finding a spectral index of $\alpha_{88}^{154} = -2.4 \pm 0.3$. \cref{fig:a550:xray} shows the emission with respect to the X-ray--emitting core of the cluster, though the source is at the edge of the X-ray data. \citet{Bernardi2016} observed the cluster with KAT-7, searching for a central radio halo. This elongated source is outside of their observed FoV but a halo was not detected, and the data here (MWA-2 and re-processed RACS) do not suggest the presence of a halo either. 

\section{Discussion}

\subsection{Low frequency radio halos and relics}

\begin{figure*}[h!]
    \centering
    \begin{minipage}[b]{1\linewidth}
    \begin{subfigure}[b]{0.5\linewidth}
    \includegraphics[width=1\linewidth]{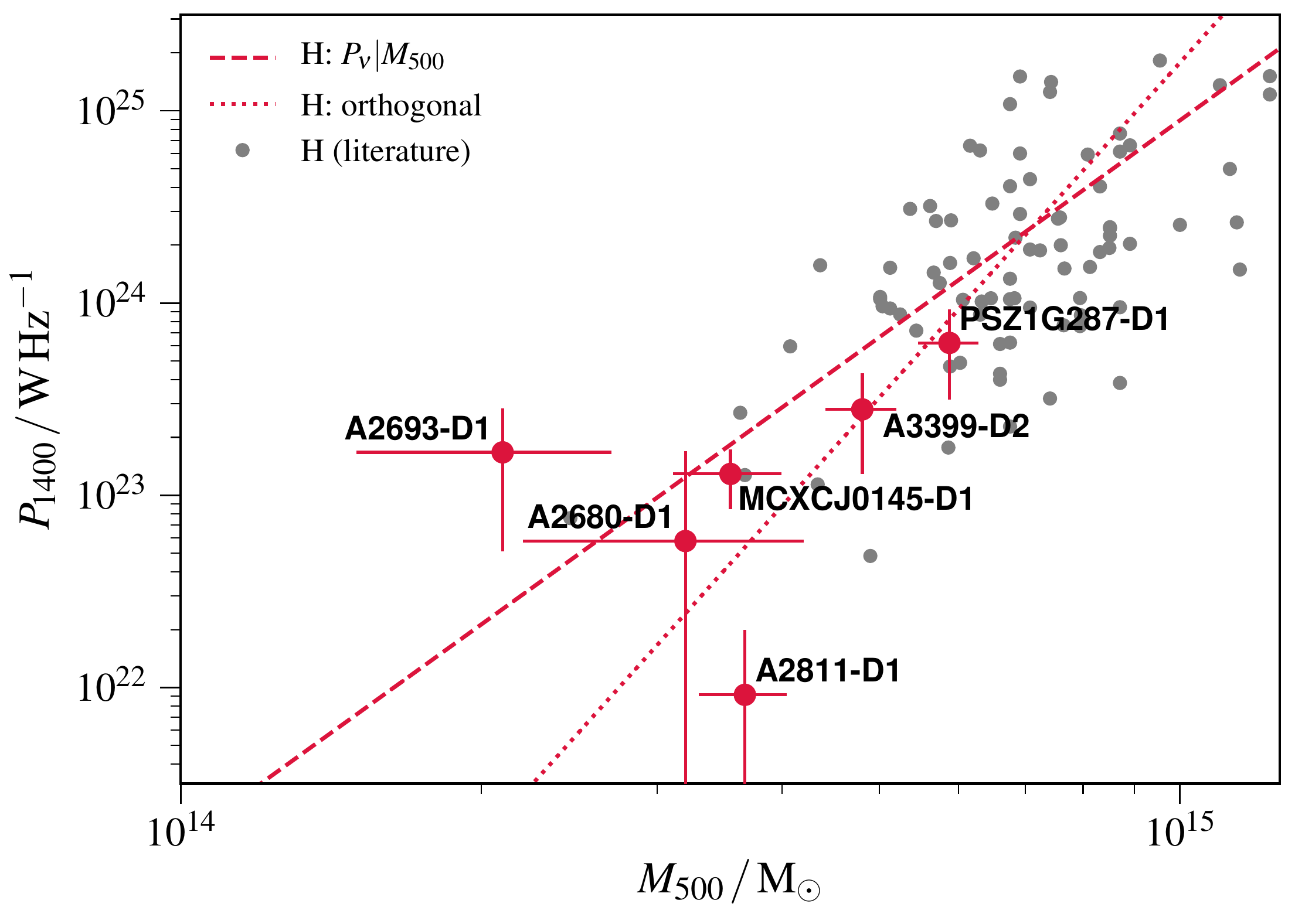}
    \caption{$P_{1400}$--$M_{500}$ halo scaling relations.\label{fig:scaling:halo:1400}}
    \end{subfigure}%
    \begin{subfigure}[b]{0.5\linewidth}
    \includegraphics[width=1\linewidth]{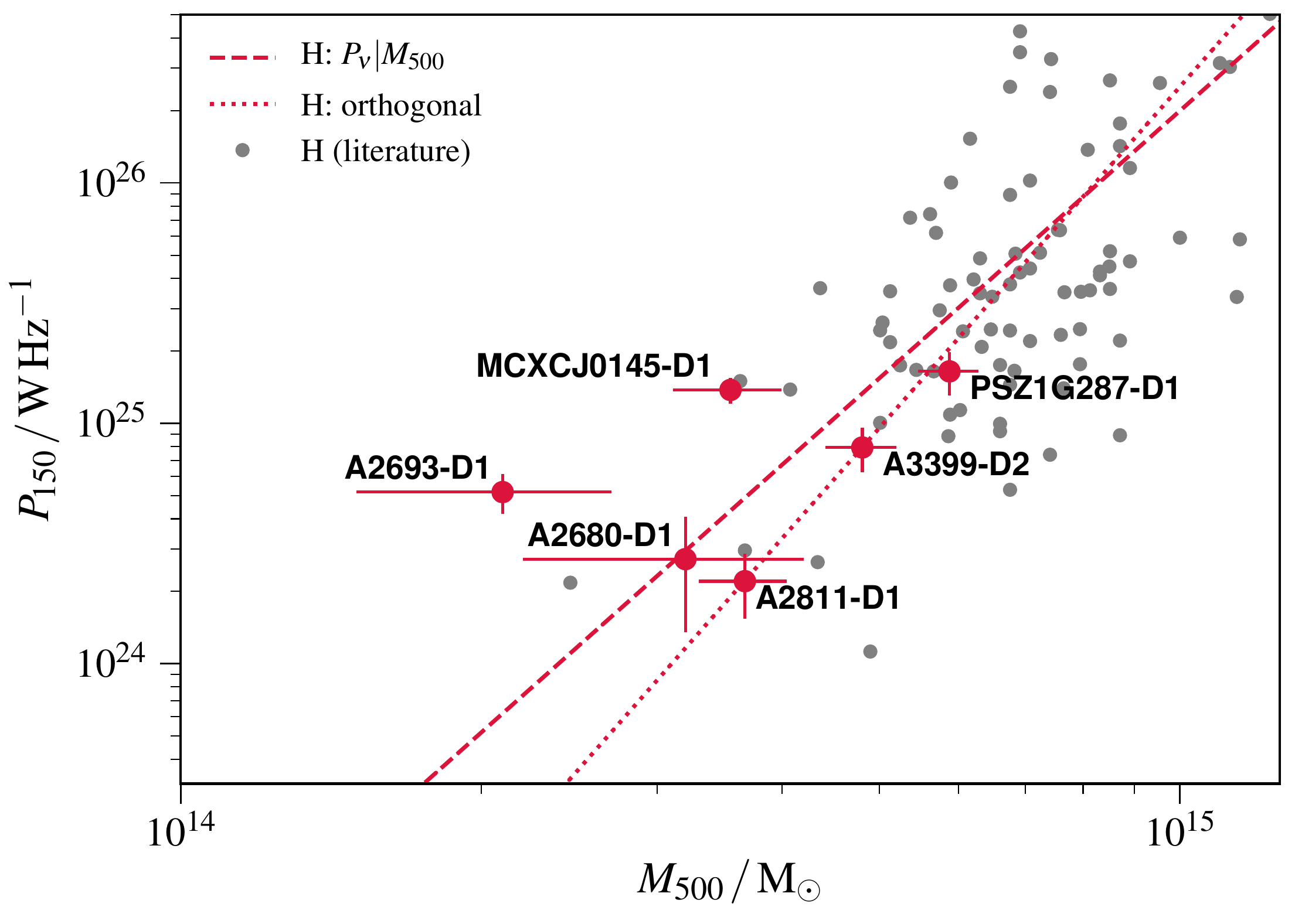}
    \caption{$P_{150}$--$M_{500}$ halo scaling relations.\label{fig:scaling:halo:150}}
    \end{subfigure}\\[0.5em]%
    \begin{subfigure}[b]{0.5\linewidth}
    \includegraphics[width=1\linewidth]{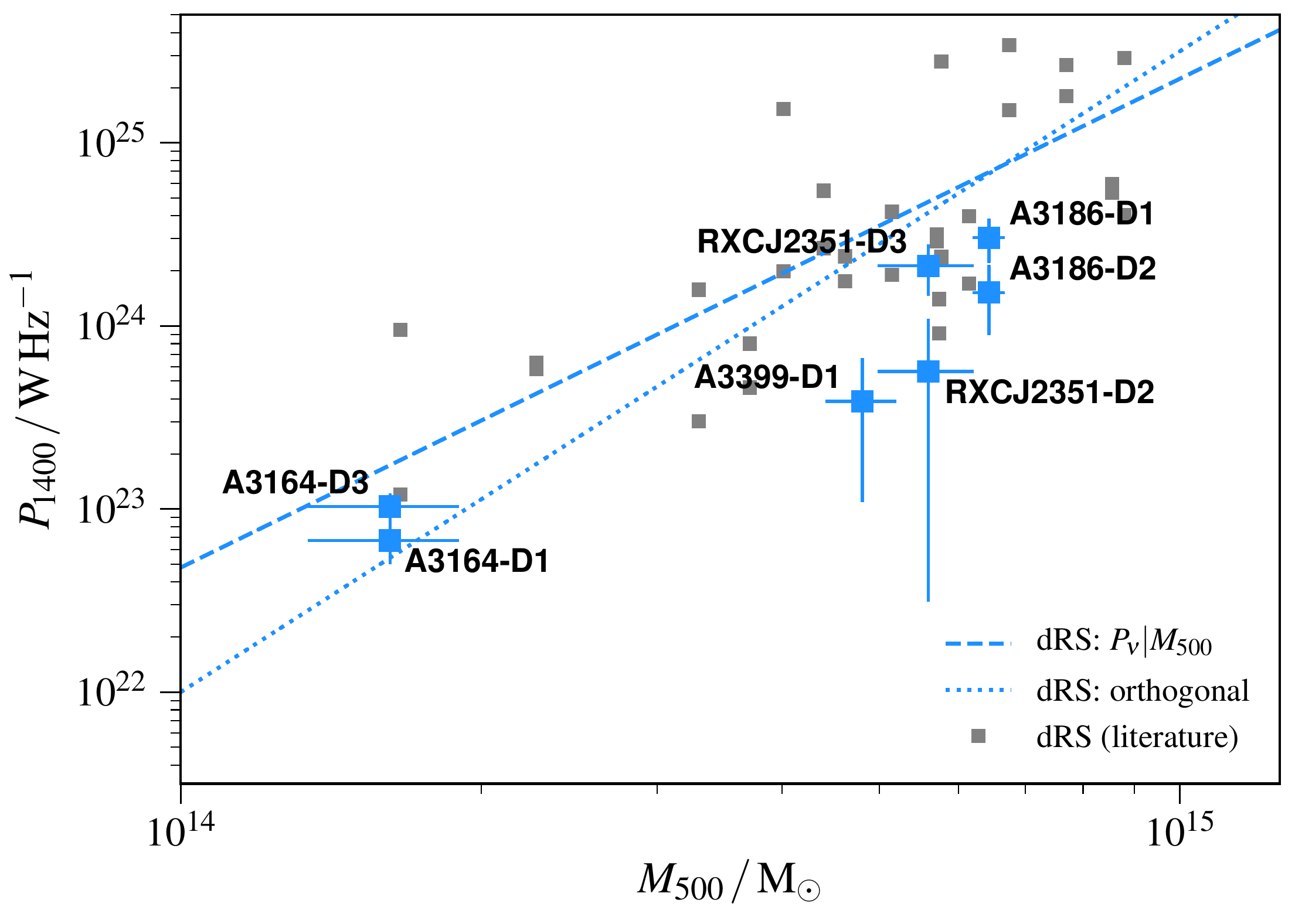}
    \caption{$P_{1400}$--$M_{500}$ double relic scaling relations.\label{fig:scaling:relic:1400}}
    \end{subfigure}%
    \begin{subfigure}[b]{0.5\linewidth}
    \includegraphics[width=1\linewidth]{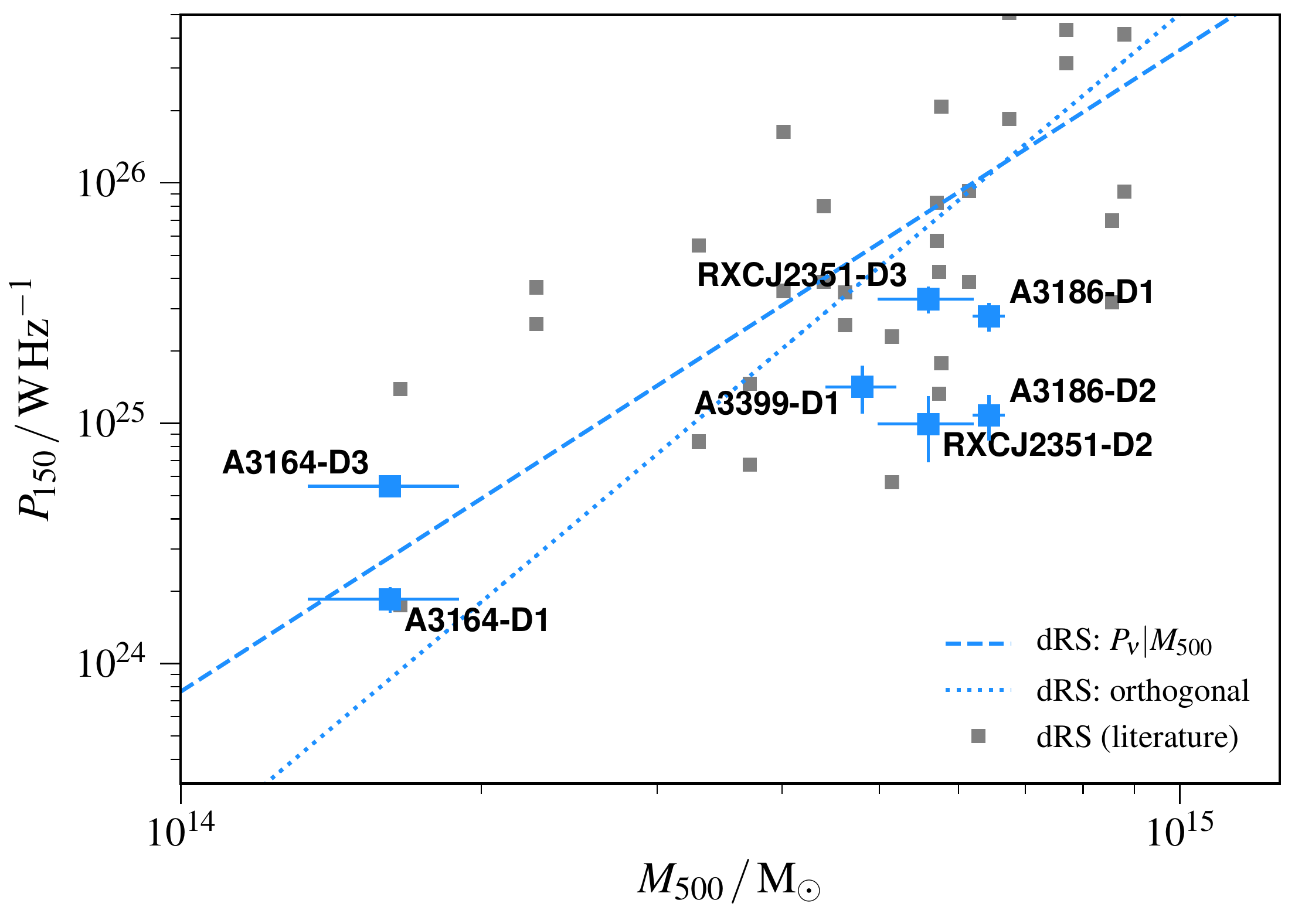}
    \caption{$P_{150}$--$M_{500}$ double relic scaling relations.\label{fig:scaling:relic:150}}
    \end{subfigure}%
    \end{minipage}\\[0.5em]%
    \begin{minipage}[b]{1\linewidth}
    \caption{\label{fig:scaling}Halo and double relic $P_{\nu}$--$M_{500}$ scaling relations with candidate sources from this work overlaid. Best-fit lines are also shown from \citet{Duchesne2020b} for relics, with scaling to 150~MHz, and \citet{Duchesne2021a} for halos. Literature data as discussed in \citet{Duchesne2020b} and \citet{Duchesne2021a} are also shown for reference. $P_{\nu}|M_{500}$ refers to fits determined assuming $P_{\nu}$ is the dependent variable and $M_{500}$ the independent variable, and `orthogonal' refers to an orthogonal regression using the Bivariate Correlated Errors and intrinsic Scatter method \citep[BCES;][]{ab96}---see \citet[][]{Duchesne2021a} for further BCES fitting details.}
    \end{minipage}%
\end{figure*}

Scaling relationships between diffuse source radio luminosity and cluster properties (e.g. $M_{500}$) have been established for halos, mini-halos, and relics. Recent works have shown that for radio halos these relationships are converging at 1.4~GHz and 150~MHz \citep{Cuciti2021b,vanWeeren2020,Duchesne2021a}. For relics this is less certain as confirmed radio relic numbers remain low, and scaling relations here typically utilise double relic detections \citep[e.g.][]{dvb+14,div+15,Duchesne2020b}. Similar relationships are found for mini-halos \citep[e.g.][]{Giacintucci2019,RichardLaferriere2020}, though the effect of the CC on cluster properties (e.g. mass) and contribution of the radio-loud BCG on the estimated mini-halo power may create additional uncertainty in the relations \citep[e.g.][]{RichardLaferriere2020}. \par 

Despite scatter in these relations, sources in each class tend to stay near the currently determined fits. As a point of comparison, we plot a selection of our sources (namely, candidate halos, relics, and some of the miscellaneous fossil sources) on the $P_{\nu}$--$M_{500}$ planes to compare against the larger samples. \cref{fig:scaling} shows these relations for halos (\cref{fig:scaling:halo:1400} and \cref{fig:scaling:halo:150}) and relics (\cref{fig:scaling:relic:1400} and \cref{fig:scaling:relic:150}) for $\nu \in \{ 150,\,1400\}$~MHz. For each, we plot the best-fit relations from \citet{Duchesne2021a} and \citet{Duchesne2020b} for halos and relics, respectively. Additionally, the full literature samples used in those works are plotted, and the $P_{150}$--$M_{500}$ relation for double relics is scaled from 1400~MHz following Equation~6 from \citet{Duchesne2021a}, assuming a mean spectral index of $\alpha = -1.2$.

As expected, at 150~MHz the candidate USSRHs in Abell~2811 and PSZ1~G287.95$-$32.98 shift to be placed on the orthogonal regression line. This is largely consistent with the USSRH population \citep[e.g.][]{ceb+13,Bruno2021,Duchesne2021a} which are normally found to be under-luminous with respect to the 1400~MHz scaling relations but shift towards the regression line when the relation is computed at 150~MHz. Conversely, the candidate halo (or point source) in Abell~2693 sits far above both relations, though the derived mass approaches the limits obtainable with RASS and carries significant uncertainty. Given how few halos are detected in low mass clusters ($\lesssim 5 \times 10^{14}$~M$_\odot$) and our lack of understanding in this low-mass regime it is difficult to rule out the radio halo classification, however, as discussed previously other characteristics (e.g.~location) are less consistent with a halo interpretation. While we suggest the sources in Abell~3164 are fossil plasma sources, their nature is still somewhat uncertain and D1 and D2 are included on \cref{fig:scaling:relic:1400} and \cref{fig:scaling:relic:150} as a reference.

\subsection{Clues from integrated spectra---seed electrons \& palaeontology}\label{sec:discussion:alphas}

\begin{figure*}[h!]
    \centering
    \begin{subfigure}[b]{0.33\linewidth}
    \includegraphics[width=1\linewidth]{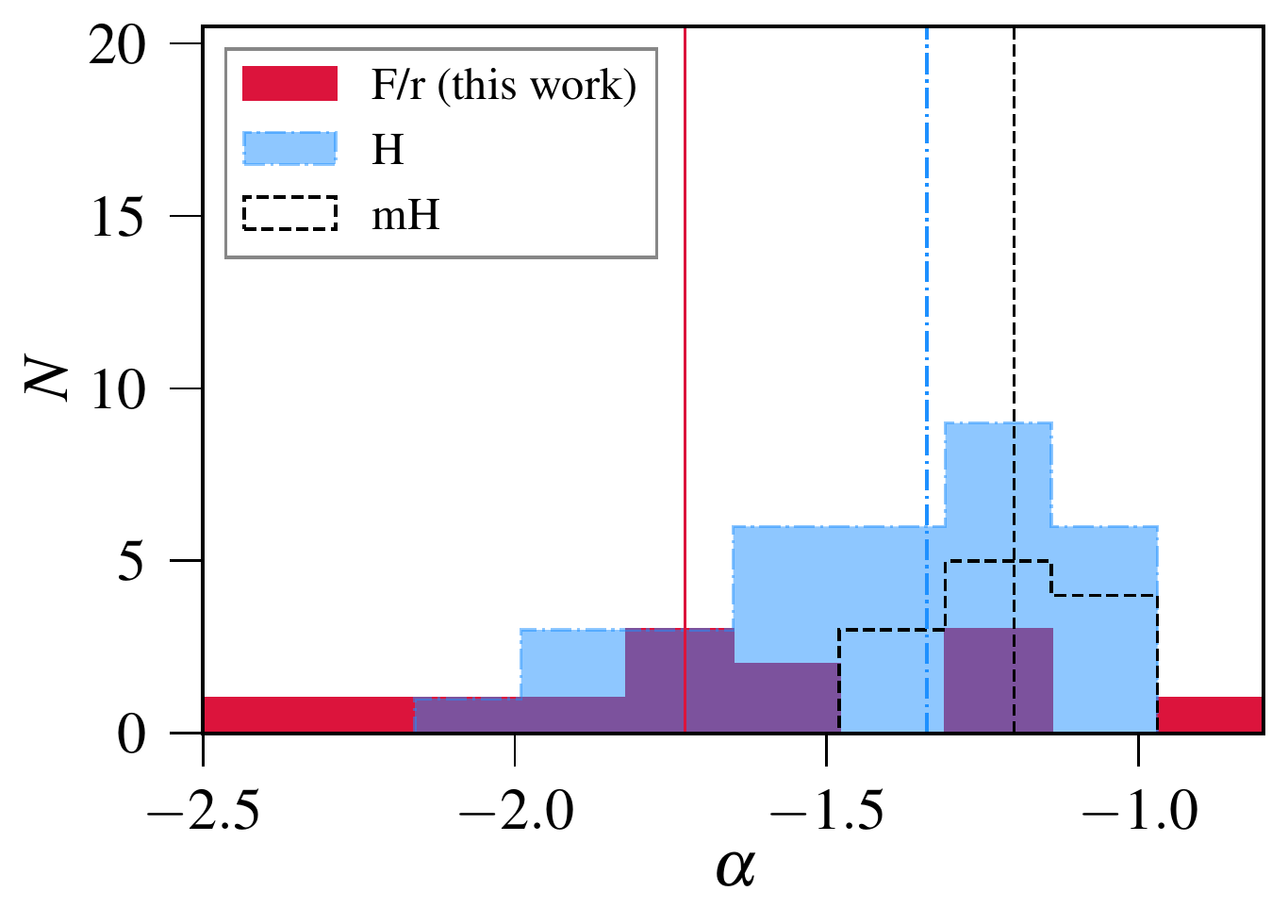}
    \caption{\label{fig:hist:halos} Halos/mini-halos}
    \end{subfigure}%
    \begin{subfigure}[b]{0.33\linewidth}
    \includegraphics[width=1\linewidth]{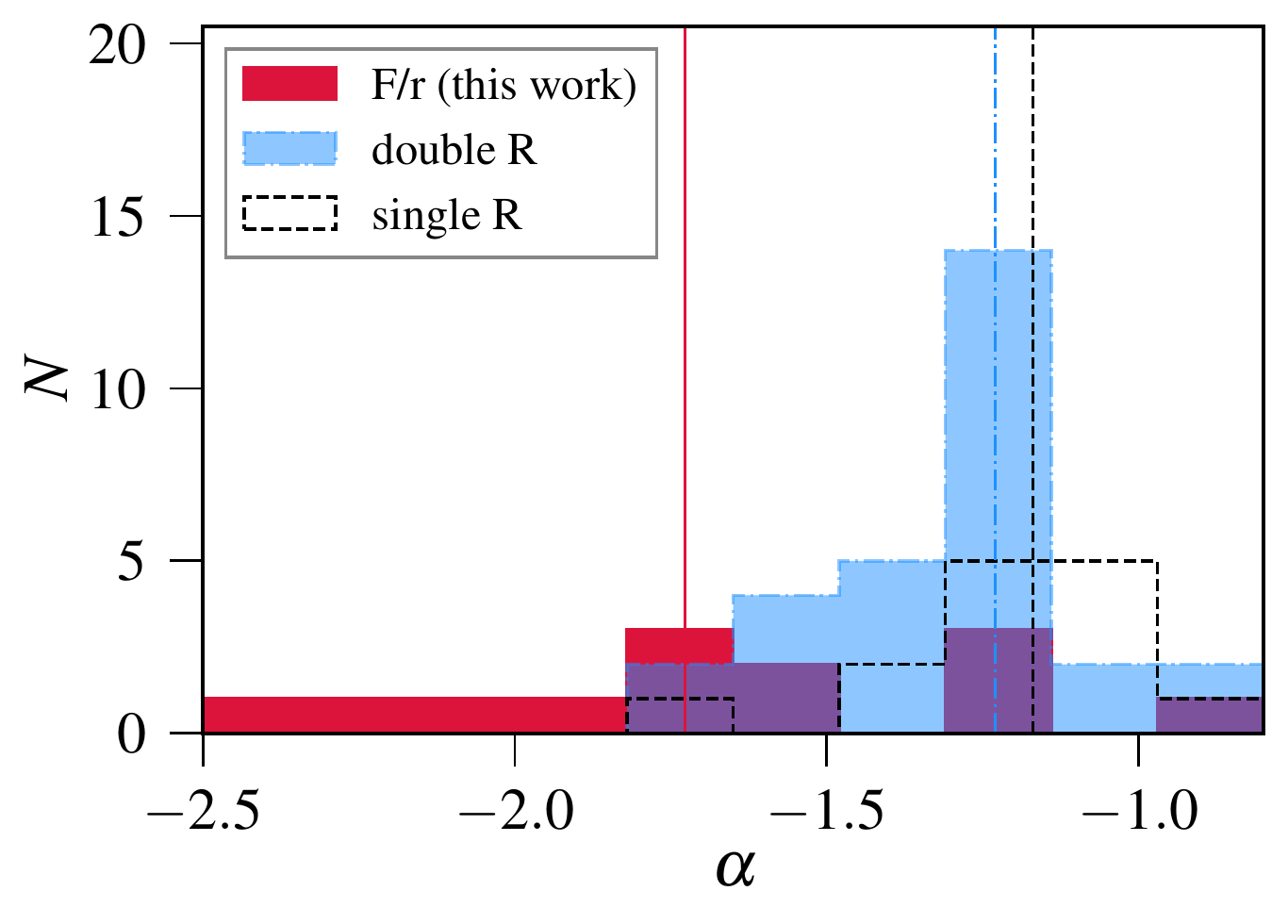}
    \caption{\label{fig:hist:relics} Relics/double relics}
    \end{subfigure}%
    \begin{subfigure}[b]{0.33\linewidth}
    \includegraphics[width=1\linewidth]{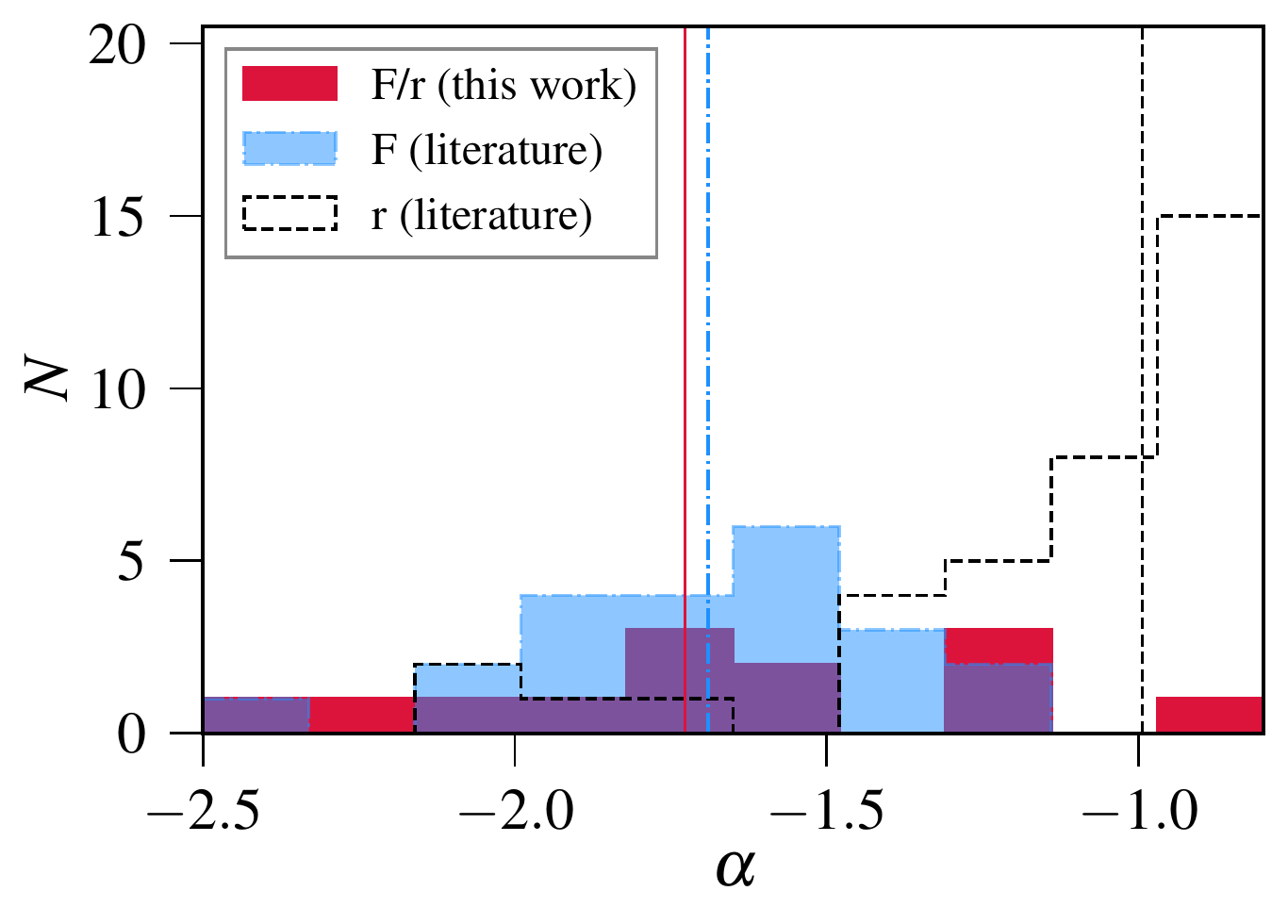}
    \caption{\label{fig:hist:fossils}Fossils/remnants}
    \end{subfigure}%
    \caption{\label{fig:hist} Spectral index distributions for \subref{fig:hist:halos} radio halos, \subref{fig:hist:relics} relics (double and single), and \subref{fig:hist:fossils} remnant radio galaxies and fossil radio plasma sources, compared to the collection of fossils plasma sources reported in this work. See \cref{sec:discussion:alphas} and \cref{tab:alphas} for population details. Vertical lines indicate median population values. Note that bins are clipped at $-2.5 \geq \alpha \geq -0.8$, and our sample is generally clipped at $\sim -0.9$ by construction due to the original search criteria. }
\end{figure*}

\subsubsection{Cluster diffuse sources samples}

We find that of the 31 sources reported in this survey \footnote{Including candidate sources, and radio galaxies/point sources discussed in \cref{sec:sources}.}, that $13$ are likely fossil radio plasma sources: remnant radio galaxies or otherwise remnant outflows from AGN at unknown stages after cessation of an episode of nuclear activity. These sources are labelled `F' and `r' in \cref{tab:clusters}. For classification purposes, in \cref{tab:clusters} remnant (`r') is used when the morphology still resembles a radio galaxy and when a possible host is identified. These sources are unlikely to be radio halos or relics based on largely morphological and other physical characteristics. 

Having established this sample we wish to consider how fossil radio plasmas relate to other diffuse cluster emission. In particular, we wish to consider the question of \corrs{whether} these sources provide a link between the source of electrons in the ICM from which relics and halos may be generated. With the MWA (and ASKAP) we are able to obtain estimates of the spectral index in the MHz regime where most emission models retain power law shapes, \corrs{and comparisons between different classes of sources (fossils, halos, relics, and remnant radio galaxies) will not be overly effected by spectral curvature.}

\cref{tab:alphas} presents details of a literature sample of integrated spectral indices for populations of halos (H), relics (R), mini halos (mH)\corrs{,} remnants (r), and cluster-based fossil sources (F), which we compare to our cluster remnant and fossil sample. Where available, we use spectral indices derived from low-frequency measurements (i.e.~within the MWA band, $\sim 150$~MHz up to $\sim 1$~GHz). 

\corrs{Spectral indices have been derived from a variety of measurements/instruments and spectra may steepen \citep[if a low-frequency observation is more sensitive; e.g.][]{Macario2010} or flatten \citep[if a high-frequency observation is more sensitive; e.g.][]{Ogrean2015} artificially due to a mismatch of $u,v$ sampling and general sensitivity between observations. This introduces further uncertainty to the spectral indices, but is not likely to bias the distributions significantly towards steeper or flatter spectra.}

We show the histograms of reported $\alpha$ in \cref{fig:hist} for our sample of cluster fossil plasma sources (r and F in \cref{tab:clusters}) against three classes of objects from the samples described in \cref{tab:alphas}: i) halos and mini halos, ii) relics (both single and double systems), and iii) cluster fossils sources and both field and cluster remnants.

\setcounter{ft}{0}
\begin{table}[t!]
    \centering
    \begin{threeparttable}
    \caption{\label{tab:alphas} Median spectral indices for the diffuse source populations.}
    \begin{tabular}{r c c c}
        \toprule
        Type & Number & med($\alpha$) & Ref.\tnote{a}\\
        \midrule
        R & 46 & $-1.2\pm0.2$ & \tabft{\ref{t:Duchesne2020b}},\tabft{\ref{t:vanWeeren2019}} \\
        H & 35 & $-1.3\pm0.3$ & \tabft{\ref{t:Duchesne2021a}} \\
        mH & 12 & $-1.2 \pm 0.1$ & \tabft{\ref{t:RichardLaferriere2020}} \\
        r (field) & 35 & $-0.9 \pm 0.2$ & \tabft{\ref{t:Brienza2017}},\tabft{\ref{t:Mahatma2018}},\tabft{\ref{t:Quici2021}} \\
        r (cluster) & 6 & $-1.8 \pm 0.4$ & \tabft{\ref{t:mpm+11}} \\
        F & 21 & $-1.7 \pm 0.4$ & \tabft{\ref{t:vanWeeren2019}},\tabft{\ref{t:Duchesne2020a}},\tabft{\ref{t:Mandal2020}},\tabft{\ref{t:Giacintucci2020}},\tabft{\ref{t:Hodgson2021}} \\
        r/F & 13 & $-1.7 \pm 0.4$ & this work \\
        \bottomrule
        \end{tabular}
        \begin{tablenotes}[flushleft]
        {\footnotesize 
        \item[a] Samples drawn from: \ft{t:Duchesne2020b} \citet{Duchesne2020b}; \ft{t:vanWeeren2019} \citet{vda+19}; \ft{t:Duchesne2021a} \citet{Duchesne2021a}; \ft{t:RichardLaferriere2020} \citet{RichardLaferriere2020}; \ft{t:Brienza2017} \citet{Brienza2017}; \ft{t:Mahatma2018} \citet{Mahatma2018}; \ft{t:Quici2021} \citet{Quici2021}; \ft{t:mpm+11} \citet{mpm+11}; \ft{t:Duchesne2020a} \citet{Duchesne2020a}; \ft{t:Mandal2020} \citet{Mandal2020}; \ft{t:Giacintucci2020} \citet{Giacintucci2020}; \ft{t:Hodgson2021} \citet{Hodgson2021}.
        }
        \end{tablenotes}
    \end{threeparttable}
\end{table}

Median $\alpha$ obtained for each distribution are reported in \cref{tab:alphas}. We note the single relic population is unlikely to be fundamentally different to the double relic population, which are thought to form through the same shock-acceleration processes. This is evident in \cref{fig:hist:relics} which shows the distribution of $\alpha$ is not significantly different between the two samples. Furthermore, the halo and relic median spectral indices are consistent, suggesting a common generation time if drawn from the same underlying electron population (though as discussed previously the generation mechanisms are thought to be different, but both associated with mergers). We note that the radio halo sample \corrs{shown in} \cref{fig:hist:halos} includes USSRHs \corrs{as expected from turbulent (re-)acceleration processes} which creates a significant tail in the distribution towards steeper spectra. Relics show a similar ultra-steep--spectrum tail but are much more densely peaked around their median value.

\subsubsection{Remnants, fossils, and revived radio plasma: seed electron populations?}

\cref{fig:hist} shows that, in general, fossil plasma sources tend towards steeper spectra than the relic and halo populations, and are significantly steeper than field remnant radio galaxies. The few remnants found in clusters \citep[e.g.][]{mpm+11} typically have steeper spectra at equivalent frequencies than their field counterparts \footnote{Note also \citet{Parma2007} reports a small number of remnants within cluster environments, though their sample was taken from sources with $\alpha < -1.3$ so by construction they are steeper than remnants found in recent searches.}. Sources such as D1 in Abell~S1099 may represent a younger cluster remnant, and with no evidence of morphological disturbance of the ICM we do not consider this a re-accelerated source. \corrs{It is likely the cluster environment plays a role in allowing these sources to remain visible for longer (e.g. by constraining diffusion of the electrons into the surrounding ICM; \citealt{mpm+11}).} Studies \citep{Brienza2017,Mahatma2018,Jurlin2020,Quici2021} suggest remnants have a short observable lifetime before surface brightness becomes prohibitively faint as the plasma both steepens in spectrum and expands. With low numbers of these remnant and fossil sources, it is difficult to draw firm conclusions about the role of the cluster environment.

Given the ubiquity of radio galaxies of complex morphologies in clusters \citep[e.g.][]{Hardcastle2004,Clarke2006,Sakelliou2008,Botteon2020b,Bruggen2020} we expect a large population of these fossil plasma sources waiting to be re-accelerated by merger-based shocks and turbulence. By construction, we have steep-spectrum (and ultra-steep--spectrum) sources in either relaxed, or `ambiguously dynamic' clusters that cannot be described as relics or halos---these are likely fossil radio plasma and highlight a population of sources in clusters that can provide seed electrons for relic and halo sources as well as smaller-scale revived fossil plasma like phoenices. 

Examples exist in the literature of re-acceleration of fossil plasmas with some connection to AGN \citep[e.g.][]{Bonafede2014,vanWeeren2017,deg17,Wilber2019}, and simulations have shown that fossil electrons from AGN outflows and long-dead radio galaxies can (1) exist diffusively within the ICM \citep{Vazza2021} and (2) be re-accelerated by normal DSA-like processes \citep[e.g.][]{Kang2018}. It is not clear that the acceleration efficiency for electrons at shocks is sufficient to produce the observed relics unless there is a mildly-relativistic population (e.g. \citealt{Botteon2020}), though many observed properties of relics do not require a population of fossil electrons \citep[e.g.][]{Rajpurohit2020,Rajpurohit2021b}. The distribution of $\alpha$ for relics and halos extends into the ultra-steep--spectrum regime which would be consistent with a population of re-accelerated fossil electrons in low-energy events (e.g. from weak mergers triggering USSRH; \citealt{bgc+08}), though it is not clear whether the density of such a fossil electron population would be sufficient to produce the observed emission. In general, such weaker events should be more common and simulations suggest a large population of faint diffuse sources should exist \citep[e.g.][]{Nuza2012}; however, these low surface brightness and ultra-steep spectrum sources are still challenging to detect and image. Surveys with e.g. LOFAR (especially combining HBA and LBA data for spectral information) should uncover a larger number of these ultra-steep--spectrum sources and indeed instruments like MWA, LOFAR, and the uGMRT are now starting to uncover them \citep[e.g.][]{Giacintucci2020,Mandal2020,Hodgson2021}.

\section{Summary}

In this work we have reported an MWA-2 follow-up survey of candidate diffuse cluster radio sources originally detected in MWA Phase I data as part of the GLEAM survey and EoR0 field survey \citepalias{Duchesne2017}. We have combined the MWA-2 data with recent ASKAP data at 887~MHz to (1) attempt to classify the sources based on their morphologies, host cluster properties, and SEDs and (2) simply ensure sources are not confused point sources---an ongoing problem with low-resolution interferometric observations. We report on 31 sources, with 6 candidate halos, 2 mini-halos (1 candidate), 3 relics (1 candidate and 1 double relic system), 13 remnant AGN or miscellaneous fossil plasmas. Some of the candidate sources are found to be point sources or radio galaxies, and we also follow-up the `twin relic' in Abell~0168 reported by \citet{Dwarakanath2018} and the double-relic system in Abell~3365 reported by \citet{vanWeeren2011b}. Specifically, we report the detection of a new double relic system associated with Abell~3186, a mini-halo associated with RXC~J0137.2$-$0912 and that the candidate halo in Abell~2811 reported by \citetalias{Duchesne2017} has a spectral index of $\alpha = -2.5\pm0.4$, though this will require follow-up high resolution, high-sensitivity observations to confirm its nature. We also observe an HT galaxy in Abell~3164 that exhibits episodic activity with a potentially re-accelerated component. 

We find that generally the relics and halos presented here sit in reasonable locations on the established $P_{1400}$--$M_{500}$ and $P_{150}$--$M_{500}$ scaling relations, with the USSRHs shifting closer towards the $P_{150}$--$M_{500}$ relation. We also compare the integrated spectra of our sample of cluster fossil and remnant sources against various samples of steep-spectrum, diffuse cluster sources as well as field remnant radio galaxies, finding that the general spectral properties are consistent with the literature fossil radio source population. This is in turn consistent with the integrated, low-frequency spectral indices of cluster-based remnant radio galaxies. We have discussed the putative link between these fossil sources and other diffuse cluster sources, noting that the distributions of spectral indices for diffuse clusters sources would be consistent with fossil radio plasmas as seed electrons for the emission. 

In this work we have just gently brushed the dust off of a small collection of old fossils and large-scale, deep, and high-resolution surveys with the SKA pathfinders---particularly ASKAP and MeerKAT with $\sim 5$--15~arcsec resolution over the Southern Sky---will enable high-fidelity follow-up of these types of sources placing constraints on the $\gtrsim 1$~GHz spectrum. In the future SKA should dig deeper, revealing myriad new ultra-steep spectrum radio fossils, providing key observational evidence for any link between these sources and cluster halos and relics. ~\\

\begin{acknowledgements}
We would like to thank an anonymous referee for their useful feedback. The authors would like to thank Dr.~Paula Tarr{\'\i}o Alonso for providing mass estimates for Abell~2680 and Abell~2693.
SWD acknowledges an Australian Government Research Training Program scholarship administered through Curtin University. 
The Australian SKA Pathfinder is part of the Australia Telescope National Facility which is managed by CSIRO. Operation of ASKAP is funded by the Australian Government with support from the National Collaborative Research Infrastructure Strategy. ASKAP uses the resources of the Pawsey Supercomputing Centre. Establishment of ASKAP, the Murchison Radio-astronomy Observatory and the Pawsey Supercomputing Centre are initiatives of the Australian Government, with support from the Government of Western Australia and the Science and Industry Endowment Fund. We acknowledge the Wajarri Yamatji people as the traditional owners of the Observatory site. Support for the operation of the MWA is provided by the Australian Government (NCRIS), under a contract to Curtin University administered by Astronomy Australia Limited.
This research has made use of data obtained from the \textit{Chandra} Data Archive and the \textit{Chandra} Source Catalog, and software provided by the \textit{Chandra} X-ray Center (CXC) in the application package \href{https://cxc.cfa.harvard.edu/ciao/}{\texttt{CIAO}}.
This research has made use of the NASA/IPAC Extragalactic Database (NED), which is operated by the Jet Propulsion Laboratory, California Institute of Technology, under contract with the National Aeronautics and Space Administration.
This research has made use of the VizieR catalogue access tool, CDS, Strasbourg, France. The original description of the VizieR service was described in \citet{vizier}. 

This research made use of a number of \texttt{python} packages not explicitly mentioned in the main text: \href{https://aplpy.github.io/}{\texttt{aplpy}} \citep{Robitaille2012}, \href{https://www.astropy.org/}{\texttt{astropy}} \citep{Astropy2013}, \texttt{matplotlib} \citep{Hunter2007}, \href{https://numpy.org/}{\texttt{numpy}} \citep{Numpy2011} and \texttt{scipy} \citep{Jones2001}.
\end{acknowledgements}

{\footnotesize
\bibliographystyle{pasa-mnras}
\bibliography{references}

\begin{thebibliography}{}
\makeatletter
\relax
\def\mn@urlcharsother{\let\do\@makeother \do\$\do\&\do\#\do\^\do\_\do\%\do\~}
\definecolor{darkblue}{RGB}{30, 144, 255}
\def\mndoi{\begingroup\mn@urlcharsother \@ifnextchar [ {\mndoi@} {\mndoi@[]}}
\def\mndoi@[#1]#2{\def\@tempa{#1}\ifx\@tempa\@empty \href
  {http://dx.doi.org/#2} {\textcolor{darkblue}{doi:#2}}\else \href
  {http://dx.doi.org/#2} {\textcolor{darkblue}{#1}}\fi \endgroup}
\def\mn@eprint#1#2{\mn@eprint@#1:#2::\@nil}
\def\mn@eprint@arXiv#1{\href {http://arxiv.org/abs/#1} {{\tt arXiv:#1}}}
\def\mn@eprint@dblp#1{\href {http://dblp.uni-trier.de/rec/bibtex/#1.xml}
  {dblp:#1}}
\def\mn@eprint@#1:#2:#3:#4\@nil{\def\@tempa {#1}\def\@tempb {#2}\def\@tempc
  {#3}\ifx \@tempc \@empty \let \@tempc \@tempb \let \@tempb \@tempa \fi \ifx
  \@tempb \@empty \def\@tempb {arXiv}\fi \@ifundefined
  {mn@eprint@\@tempb}{\@tempb:\@tempc}{\expandafter \expandafter \csname
  mn@eprint@\@tempb\endcsname \expandafter{\@tempc}}}

\bibitem[\protect\citeauthoryear{{Abbott} et~al.,}{{Abbott}
  et~al.}{2018}]{des1}
{Abbott} T.~M.~C.,  et~al., 2018, \mndoi [\apjs] {10.3847/1538-4365/aae9f0},
  \href {https://ui.adsabs.harvard.edu/abs/2018ApJS..239...18A} {239, 18}

\bibitem[\protect\citeauthoryear{{Abell}}{{Abell}}{1958}]{abell}
{Abell} G.~O.,  1958, \mndoi [\apjs] {10.1086/190036}, \href
  {http://adsabs.harvard.edu/abs/1958ApJS....3..211A} {3, 211}

\bibitem[\protect\citeauthoryear{{Abell}, {Corwin}  \& {Olowin}}{{Abell}
  et~al.}{1989}]{aco89}
{Abell} G.~O.,  {Corwin} Jr. H.~G.,   {Olowin} R.~P.,  1989, \mndoi [\apjs]
  {10.1086/191333}, \href {http://adsabs.harvard.edu/abs/1989ApJS...70....1A}
  {70, 1}

\bibitem[\protect\citeauthoryear{{Akritas} \& {Bershady}}{{Akritas} \&
  {Bershady}}{1996}]{ab96}
{Akritas} M.~G.,  {Bershady} M.~A.,  1996, \mndoi [\apj] {10.1086/177901},
  \href {http://adsabs.harvard.edu/abs/1996ApJ...470..706A} {470, 706}

\bibitem[\protect\citeauthoryear{{Astropy Collaboration} et~al.,}{{Astropy
  Collaboration} et~al.}{2013}]{Astropy2013}
{Astropy Collaboration} et~al., 2013, \mndoi [\aap]
  {10.1051/0004-6361/201322068}, \href
  {http://adsabs.harvard.edu/abs/2013A%26A...558A..33A} {558, A33}

\bibitem[\protect\citeauthoryear{{Bagchi} et~al.,}{{Bagchi}
  et~al.}{2011}]{Bagchi2011}
{Bagchi} J.,  et~al., 2011, \mndoi [\apjl] {10.1088/2041-8205/736/1/L8}, \href
  {https://ui.adsabs.harvard.edu/abs/2011ApJ...736L...8B} {736, L8}

\bibitem[\protect\citeauthoryear{{Bartalucci} et~al.,}{{Bartalucci}
  et~al.}{2017}]{bartalucci2017}
{Bartalucci} I.,  et~al., 2017, \mndoi [\aap] {10.1051/0004-6361/201731689},
  \href {https://ui.adsabs.harvard.edu/abs/2017A&A...608A..88B} {608, A88}

\bibitem[\protect\citeauthoryear{{Bernardi} et~al.,}{{Bernardi}
  et~al.}{2016}]{Bernardi2016}
{Bernardi} G.,  et~al., 2016, \mndoi [\mnras] {10.1093/mnras/stv2589}, \href
  {http://adsabs.harvard.edu/abs/2016MNRAS.456.1259B} {456, 1259}

\bibitem[\protect\citeauthoryear{{Bock}, {Large}  \& {Sadler}}{{Bock}
  et~al.}{1999}]{bls99}
{Bock} D.~C.-J.,  {Large} M.~I.,   {Sadler} E.~M.,  1999, \mndoi [\aj]
  {10.1086/300786}, \href {http://adsabs.harvard.edu/abs/1999AJ....117.1578B}
  {117, 1578}

\bibitem[\protect\citeauthoryear{{Bogd{\'a}n} et~al.,}{{Bogd{\'a}n}
  et~al.}{2013}]{bogdan13}
{Bogd{\'a}n} {\'A}.,  et~al., 2013, \mndoi [\apj] {10.1088/0004-637X/772/2/97},
  \href {https://ui.adsabs.harvard.edu/abs/2013ApJ...772...97B} {772, 97}

\bibitem[\protect\citeauthoryear{{B{\"o}hringer} et~al.,}{{B{\"o}hringer}
  et~al.}{2004}]{bsg+04}
{B{\"o}hringer} H.,  et~al., 2004, \mndoi [\aap] {10.1051/0004-6361:20034484},
  \href {http://adsabs.harvard.edu/abs/2004A%26A...425..367B} {425, 367}

\bibitem[\protect\citeauthoryear{{B{\"o}hringer} et~al.,}{{B{\"o}hringer}
  et~al.}{2010}]{Bohringer2010}
{B{\"o}hringer} H.,  et~al., 2010, \mndoi [\aap] {10.1051/0004-6361/200913911},
  \href {https://ui.adsabs.harvard.edu/abs/2010A&A...514A..32B} {514, A32}

\bibitem[\protect\citeauthoryear{{Bonafede} et~al.,}{{Bonafede}
  et~al.}{2012}]{Bonafede2012}
{Bonafede} A.,  et~al., 2012, \mndoi [\mnras]
  {10.1111/j.1365-2966.2012.21570.x}, \href
  {https://ui.adsabs.harvard.edu/abs/2012MNRAS.426...40B} {426, 40}

\bibitem[\protect\citeauthoryear{{Bonafede}, {Intema}, {Br{\"u}ggen},
  {Girardi}, {Nonino}, {Kantharia}, {van Weeren}  \&
  {R{\"o}ttgering}}{{Bonafede} et~al.}{2014}]{Bonafede2014}
{Bonafede} A.,  {Intema} H.~T.,  {Br{\"u}ggen} M.,  {Girardi} M.,  {Nonino} M.,
   {Kantharia} N.,  {van Weeren} R.~J.,   {R{\"o}ttgering} H.~J.~A.,  2014,
  \mndoi [\apj] {10.1088/0004-637X/785/1/1}, \href
  {https://ui.adsabs.harvard.edu/abs/2014ApJ...785....1B} {785, 1}

\bibitem[\protect\citeauthoryear{{Botteon}, {Gastaldello}, {Brunetti}  \&
  {Dallacasa}}{{Botteon} et~al.}{2016a}]{botteon_shock_2016}
{Botteon} A.,  {Gastaldello} F.,  {Brunetti} G.,   {Dallacasa} D.,  2016a,
  \mndoi [\mnras] {10.1093/mnrasl/slw082}, \href
  {https://ui.adsabs.harvard.edu/abs/2016MNRAS.460L..84B} {460, L84}

\bibitem[\protect\citeauthoryear{Botteon, Gastaldello, Brunetti  \&
  Kale}{Botteon et~al.}{2016b}]{botteon_mathcal_2016}
Botteon A.,  Gastaldello F.,  Brunetti G.,   Kale R.,  2016b, \mndoi [\mnras]
  {10.1093/mnras/stw2089}, 463, 1534

\bibitem[\protect\citeauthoryear{{Botteon} et~al.,}{{Botteon}
  et~al.}{2018}]{Botteon2018}
{Botteon} A.,  et~al., 2018, \mndoi [\mnras] {10.1093/mnras/sty1102}, \href
  {https://ui.adsabs.harvard.edu/abs/2018MNRAS.478..885B} {478, 885}

\bibitem[\protect\citeauthoryear{{Botteon} et~al.,}{{Botteon}
  et~al.}{2020a}]{Botteon2020c}
{Botteon} A.,  et~al., 2020a, \mndoi [\mnras] {10.1093/mnrasl/slaa142}, \href
  {https://ui.adsabs.harvard.edu/abs/2020MNRAS.499L..11B} {499, L11}

\bibitem[\protect\citeauthoryear{{Botteon}, {Brunetti}, {Ryu}  \&
  {Roh}}{{Botteon} et~al.}{2020b}]{Botteon2020}
{Botteon} A.,  {Brunetti} G.,  {Ryu} D.,   {Roh} S.,  2020b, \mndoi [\aap]
  {10.1051/0004-6361/201936216}, \href
  {https://ui.adsabs.harvard.edu/abs/2020A&A...634A..64B} {634, A64}

\bibitem[\protect\citeauthoryear{{Botteon} et~al.,}{{Botteon}
  et~al.}{2020c}]{Botteon2020b}
{Botteon} A.,  et~al., 2020c, \mndoi [\apj] {10.3847/1538-4357/ab9a2f}, \href
  {https://ui.adsabs.harvard.edu/abs/2020ApJ...897...93B} {897, 93}

\bibitem[\protect\citeauthoryear{{Botteon} et~al.,}{{Botteon}
  et~al.}{2021a}]{Botteon2021}
{Botteon} A.,  et~al., 2021a, \mndoi [\aap] {10.1051/0004-6361/202040083},
  \href {https://ui.adsabs.harvard.edu/abs/2021A&A...649A..37B} {649, A37}

\bibitem[\protect\citeauthoryear{{Botteon} et~al.,}{{Botteon}
  et~al.}{2021b}]{Botteon2021b}
{Botteon} A.,  et~al., 2021b, \mndoi [\apjl] {10.3847/2041-8213/ac0636}, \href
  {https://ui.adsabs.harvard.edu/abs/2021ApJ...914L..29B} {914, L29}

\bibitem[\protect\citeauthoryear{{Bourdin} \& {Mazzotta}}{{Bourdin} \&
  {Mazzotta}}{2008}]{bourdin08}
{Bourdin} H.,  {Mazzotta} P.,  2008, \mndoi [\aap]
  {10.1051/0004-6361:20065758}, \href
  {https://ui.adsabs.harvard.edu/abs/2008A&A...479..307B} {479, 307}

\bibitem[\protect\citeauthoryear{{Bourdin}, {Mazzotta}, {Markevitch},
  {Giacintucci}  \& {Brunetti}}{{Bourdin} et~al.}{2013}]{bourdin13}
{Bourdin} H.,  {Mazzotta} P.,  {Markevitch} M.,  {Giacintucci} S.,   {Brunetti}
  G.,  2013, \mndoi [\apj] {10.1088/0004-637X/764/1/82}, \href
  {https://ui.adsabs.harvard.edu/abs/2013ApJ...764...82B} {764, 82}

\bibitem[\protect\citeauthoryear{{Bravi}, {Gitti}  \& {Brunetti}}{{Bravi}
  et~al.}{2016}]{bgb16}
{Bravi} L.,  {Gitti} M.,   {Brunetti} G.,  2016, \mndoi [\mnras]
  {10.1093/mnrasl/slv137}, \href
  {http://adsabs.harvard.edu/abs/2016MNRAS.455L..41B} {455, L41}

\bibitem[\protect\citeauthoryear{{Brienza} et~al.,}{{Brienza}
  et~al.}{2017}]{Brienza2017}
{Brienza} M.,  et~al., 2017, \mndoi [\aap] {10.1051/0004-6361/201730932}, \href
  {https://ui.adsabs.harvard.edu/abs/2017A&A...606A..98B} {606, A98}

\bibitem[\protect\citeauthoryear{{Briggs}}{{Briggs}}{1995}]{bri95}
{Briggs} D.~S.,  1995, PhD thesis, The New Mexico Institute of Mining and
  Technology, Socorro, New Mexico, \url
  {http://www.aoc.nrao.edu/dissertations/dbriggs/}

\bibitem[\protect\citeauthoryear{{Br{\"u}ggen} et~al.,}{{Br{\"u}ggen}
  et~al.}{2021}]{Bruggen2020}
{Br{\"u}ggen} M.,  et~al., 2021, \mndoi [\aap] {10.1051/0004-6361/202039533},
  \href {https://ui.adsabs.harvard.edu/abs/2021A&A...647A...3B} {647, A3}

\bibitem[\protect\citeauthoryear{{Brunetti} \& {Jones}}{{Brunetti} \&
  {Jones}}{2014}]{bj14}
{Brunetti} G.,  {Jones} T.~W.,  2014, \mndoi [International Journal of Modern
  Physics D] {10.1142/S0218271814300079}, \href
  {http://adsabs.harvard.edu/abs/2014IJMPD..2330007B} {23, 1430007}

\bibitem[\protect\citeauthoryear{{Brunetti} \& {Vazza}}{{Brunetti} \&
  {Vazza}}{2020}]{Brunetti2020}
{Brunetti} G.,  {Vazza} F.,  2020, \mndoi [\prl]
  {10.1103/PhysRevLett.124.051101}, \href
  {https://ui.adsabs.harvard.edu/abs/2020PhRvL.124e1101B} {124, 051101}

\bibitem[\protect\citeauthoryear{{Brunetti}, {Setti}, {Feretti}  \&
  {Giovannini}}{{Brunetti} et~al.}{2001}]{bsfg01}
{Brunetti} G.,  {Setti} G.,  {Feretti} L.,   {Giovannini} G.,  2001, \mndoi
  [\mnras] {10.1046/j.1365-8711.2001.03978.x}, \href
  {http://adsabs.harvard.edu/abs/2001MNRAS.320..365B} {320, 365}

\bibitem[\protect\citeauthoryear{{Brunetti} et~al.,}{{Brunetti}
  et~al.}{2008}]{bgc+08}
{Brunetti} G.,  et~al., 2008, \mndoi [\nat] {10.1038/nature07379}, \href
  {http://adsabs.harvard.edu/abs/2008Natur.455..944B} {455, 944}

\bibitem[\protect\citeauthoryear{{Brunetti}, {Cassano}, {Dolag}  \&
  {Setti}}{{Brunetti} et~al.}{2009}]{Brunetti2009}
{Brunetti} G.,  {Cassano} R.,  {Dolag} K.,   {Setti} G.,  2009, \mndoi [\aap]
  {10.1051/0004-6361/200912751}, \href
  {https://ui.adsabs.harvard.edu/abs/2009A&A...507..661B} {507, 661}

\bibitem[\protect\citeauthoryear{{Bruno} et~al.,}{{Bruno}
  et~al.}{2021}]{Bruno2021}
{Bruno} L.,  et~al., 2021, \mndoi [\aap] {10.1051/0004-6361/202039877}, \href
  {https://ui.adsabs.harvard.edu/abs/2021A&A...650A..44B} {650, A44}

\bibitem[\protect\citeauthoryear{{Buote}}{{Buote}}{2001}]{Buote2001}
{Buote} D.~A.,  2001, \mndoi [\apjl] {10.1086/320500}, \href
  {https://ui.adsabs.harvard.edu/abs/2001ApJ...553L..15B} {553, L15}

\bibitem[\protect\citeauthoryear{{Caretta}, {Maia}, {Kawasaki}  \&
  {Willmer}}{{Caretta} et~al.}{2002}]{cmkw02}
{Caretta} C.~A.,  {Maia} M.~A.~G.,  {Kawasaki} W.,   {Willmer} C.~N.~A.,  2002,
  \mndoi [\aj] {10.1086/338894}, \href
  {http://adsabs.harvard.edu/abs/2002AJ....123.1200C} {123, 1200}

\bibitem[\protect\citeauthoryear{{Cassano}, {Ettori}, {Giacintucci},
  {Brunetti}, {Markevitch}, {Venturi}  \& {Gitti}}{{Cassano}
  et~al.}{2010}]{Cassano2010}
{Cassano} R.,  {Ettori} S.,  {Giacintucci} S.,  {Brunetti} G.,  {Markevitch}
  M.,  {Venturi} T.,   {Gitti} M.,  2010, \mndoi [\apjl]
  {10.1088/2041-8205/721/2/L82}, \href
  {https://ui.adsabs.harvard.edu/abs/2010ApJ...721L..82C} {721, L82}

\bibitem[\protect\citeauthoryear{{Cassano} et~al.,}{{Cassano}
  et~al.}{2013}]{ceb+13}
{Cassano} R.,  et~al., 2013, \mndoi [\apj] {10.1088/0004-637X/777/2/141}, \href
  {http://adsabs.harvard.edu/abs/2013ApJ...777..141C} {777, 141}

\bibitem[\protect\citeauthoryear{{Cavagnolo}, {Donahue}, {Voit}  \&
  {Sun}}{{Cavagnolo} et~al.}{2008}]{Cavagnolo2008}
{Cavagnolo} K.~W.,  {Donahue} M.,  {Voit} G.~M.,   {Sun} M.,  2008, \mndoi
  [\apj] {10.1086/588630}, \href
  {https://ui.adsabs.harvard.edu/abs/2008ApJ...682..821C} {682, 821}

\bibitem[\protect\citeauthoryear{{Cavaliere} \& {Fusco-Femiano}}{{Cavaliere} \&
  {Fusco-Femiano}}{1976}]{Cavaliere1976}
{Cavaliere} A.,  {Fusco-Femiano} R.,  1976, \aap, \href
  {https://ui.adsabs.harvard.edu/abs/1976A%26A....49..137C} {49, 137}

\bibitem[\protect\citeauthoryear{{Chambers} et~al.,}{{Chambers}
  et~al.}{2016}]{cmm+16}
{Chambers} K.~C.,  et~al., 2016, preprint, \href
  {http://adsabs.harvard.edu/abs/2016arXiv161205560C} {} (\mn@eprint {arXiv}
  {1612.05560})

\bibitem[\protect\citeauthoryear{{Chapman}, {Dempsey}, {Miller}, {Heywood},
  {Pritchard}, {Sangster}, {Whiting}  \& {Dart}}{{Chapman}
  et~al.}{2017}]{casda}
{Chapman} J.~M.,  {Dempsey} J.,  {Miller} D.,  {Heywood} I.,  {Pritchard} J.,
  {Sangster} E.,  {Whiting} M.,   {Dart} M.,  2017, {CASDA: The CSIRO ASKAP
  Science Data Archive}.
p.~73

\bibitem[\protect\citeauthoryear{{Chon} \& {B{\"o}hringer}}{{Chon} \&
  {B{\"o}hringer}}{2012}]{cb12}
{Chon} G.,  {B{\"o}hringer} H.,  2012, \mndoi [\aap]
  {10.1051/0004-6361/201117996}, 538, A35

\bibitem[\protect\citeauthoryear{{Clarke} \& {Ensslin}}{{Clarke} \&
  {Ensslin}}{2006}]{Clarke2006}
{Clarke} T.~E.,  {Ensslin} T.,  2006, \mndoi [Astronomische Nachrichten]
  {10.1002/asna.200610586}, \href
  {https://ui.adsabs.harvard.edu/abs/2006AN....327..553C} {327, 553}

\bibitem[\protect\citeauthoryear{{Clarke}, {Kronberg}  \&
  {B{\"o}hringer}}{{Clarke} et~al.}{2001}]{Clarke2001}
{Clarke} T.~E.,  {Kronberg} P.~P.,   {B{\"o}hringer} H.,  2001, \mndoi [\apjl]
  {10.1086/318896}, \href
  {https://ui.adsabs.harvard.edu/abs/2001ApJ...547L.111C} {547, L111}

\bibitem[\protect\citeauthoryear{{Cohen} \& {Clarke}}{{Cohen} \&
  {Clarke}}{2011}]{cc11}
{Cohen} A.~S.,  {Clarke} T.~E.,  2011, \mndoi [\aj]
  {10.1088/0004-6256/141/5/149}, \href
  {http://adsabs.harvard.edu/abs/2011AJ....141..149C} {141, 149}

\bibitem[\protect\citeauthoryear{{Condon}, {Cotton}, {Greisen}, {Yin},
  {Perley}, {Taylor}  \& {Broderick}}{{Condon} et~al.}{1998}]{ccg+98}
{Condon} J.~J.,  {Cotton} W.~D.,  {Greisen} E.~W.,  {Yin} Q.~F.,  {Perley}
  R.~A.,  {Taylor} G.~B.,   {Broderick} J.~J.,  1998, \mndoi [\aj]
  {10.1086/300337}, \href {http://adsabs.harvard.edu/abs/1998AJ....115.1693C}
  {115, 1693}

\bibitem[\protect\citeauthoryear{{Coziol}, {Andernach}, {Caretta},
  {Alamo-Mart{\'{\i}}nez}  \& {Tago}}{{Coziol} et~al.}{2009}]{cac+09}
{Coziol} R.,  {Andernach} H.,  {Caretta} C.~A.,  {Alamo-Mart{\'{\i}}nez} K.~A.,
    {Tago} E.,  2009, \mndoi [\aj] {10.1088/0004-6256/137/6/4795}, \href
  {http://adsabs.harvard.edu/abs/2009AJ....137.4795C} {137, 4795}

\bibitem[\protect\citeauthoryear{{Cruddace} et~al.,}{{Cruddace}
  et~al.}{2002}]{Cruddace2002}
{Cruddace} R.,  et~al., 2002, \mndoi [\apjs] {10.1086/324519}, \href
  {https://ui.adsabs.harvard.edu/abs/2002ApJS..140..239C} {140, 239}

\bibitem[\protect\citeauthoryear{{Cuciti} et~al.,}{{Cuciti}
  et~al.}{2021}]{Cuciti2021b}
{Cuciti} V.,  et~al., 2021, \mndoi [\aap] {10.1051/0004-6361/202039208}, \href
  {https://ui.adsabs.harvard.edu/abs/2021A&A...647A..51C} {647, A51}

\bibitem[\protect\citeauthoryear{{De Grandi} et~al.,}{{De Grandi}
  et~al.}{1999}]{DeGrandi1999}
{De Grandi} S.,  et~al., 1999, \mndoi [\apj] {10.1086/306939}, \href
  {https://ui.adsabs.harvard.edu/abs/1999ApJ...514..148D} {514, 148}

\bibitem[\protect\citeauthoryear{{Di Gennaro} et~al.,}{{Di Gennaro}
  et~al.}{2021}]{DiGennaro2020}
{Di Gennaro} G.,  et~al., 2021, \mndoi [Nature Astronomy]
  {10.1038/s41550-020-01244-5}, \href
  {https://ui.adsabs.harvard.edu/abs/2021NatAs...5..268D} {5, 268}

\bibitem[\protect\citeauthoryear{{Duchesne}}{{Duchesne}}{2021}]{mwa:clusters}
{Duchesne} S.,  2021, Data for an MWA-2 survey of diffuse radio emission in
  galaxy clusters, \mndoi{10.26185/611f33b774e96}, \url
  {https://dx.doi.org/10.26185/611f33b774e96}

\bibitem[\protect\citeauthoryear{{Duchesne}, {Johnston-Hollitt}, {Zhu}, {Wayth}
   \& {Line}}{{Duchesne} et~al.}{2020}]{Duchesne2020a}
{Duchesne} S.~W.,  {Johnston-Hollitt} M.,  {Zhu} Z.,  {Wayth} R.~B.,   {Line}
  J.~L.~B.,  2020, \mndoi [\pasa] {10.1017/pasa.2020.29}, \href
  {https://ui.adsabs.harvard.edu/abs/2020PASA...37...37D} {37, e037}

\bibitem[\protect\citeauthoryear{{Duchesne}, {Johnston-Hollitt}, {Bartalucci},
  {Hodgson}  \& {Pratt}}{{Duchesne} et~al.}{2021a}]{Duchesne2020b}
{Duchesne} S.~W.,  {Johnston-Hollitt} M.,  {Bartalucci} I.,  {Hodgson} T.,
  {Pratt} G.~W.,  2021a, \mndoi [\pasa] {10.1017/pasa.2020.51}, \href
  {https://ui.adsabs.harvard.edu/abs/2021PASA...38....5D} {38, e005}

\bibitem[\protect\citeauthoryear{{Duchesne}, {Johnston-Hollitt}, {Offringa},
  {Pratt}, {Zheng}  \& {Dehghan}}{{Duchesne} et~al.}{2021b}]{Duchesne2017}
{Duchesne} S.~W.,  {Johnston-Hollitt} M.,  {Offringa} A.~R.,  {Pratt} G.~W.,
  {Zheng} Q.,   {Dehghan} S.,  2021b, \mndoi [\pasa] {10.1017/pasa.2021.7},
  \href {https://ui.adsabs.harvard.edu/abs/2021PASA...38...10D} {38, e010}

\bibitem[\protect\citeauthoryear{{Duchesne}, {Johnston-Hollitt}  \&
  {Wilber}}{{Duchesne} et~al.}{2021c}]{Duchesne2021a}
{Duchesne} S.~W.,  {Johnston-Hollitt} M.,   {Wilber} A.~G.,  2021c, \mndoi
  [\pasa] {10.1017/pasa.2021.24}, \href
  {https://ui.adsabs.harvard.edu/abs/2021PASA...38...31D} {38, e031}

\bibitem[\protect\citeauthoryear{{Duffy} \& {Blundell}}{{Duffy} \&
  {Blundell}}{2012}]{db12}
{Duffy} P.,  {Blundell} K.~M.,  2012, \mndoi [\mnras]
  {10.1111/j.1365-2966.2011.20239.x}, \href
  {http://adsabs.harvard.edu/abs/2012MNRAS.421..108D} {421, 108}

\bibitem[\protect\citeauthoryear{{Dwarakanath}, {Parekh}, {Kale}  \&
  {George}}{{Dwarakanath} et~al.}{2018}]{Dwarakanath2018}
{Dwarakanath} K.~S.,  {Parekh} V.,  {Kale} R.,   {George} L.~T.,  2018, \mndoi
  [\mnras] {10.1093/mnras/sty744}, \href
  {https://ui.adsabs.harvard.edu/abs/2018MNRAS.477..957D} {477, 957}

\bibitem[\protect\citeauthoryear{{En{\ss}lin} \& {Br{\"u}ggen}}{{En{\ss}lin} \&
  {Br{\"u}ggen}}{2002}]{eb02}
{En{\ss}lin} T.~A.,  {Br{\"u}ggen} M.,  2002, \mndoi [\mnras]
  {10.1046/j.1365-8711.2002.05261.x}, \href
  {http://adsabs.harvard.edu/abs/2002MNRAS.331.1011E} {331, 1011}

\bibitem[\protect\citeauthoryear{{En{\ss}lin} \& {Gopal-Krishna}}{{En{\ss}lin}
  \& {Gopal-Krishna}}{2001}]{eg01}
{En{\ss}lin} T.~A.,  {Gopal-Krishna} 2001, \mndoi [\aap]
  {10.1051/0004-6361:20000198}, \href
  {http://adsabs.harvard.edu/abs/2001A%26A...366...26E} {366, 26}

\bibitem[\protect\citeauthoryear{{En{\ss}lin}, {Biermann}, {Klein}  \&
  {Kohle}}{{En{\ss}lin} et~al.}{1998}]{Ensslin1998}
{En{\ss}lin} T.~A.,  {Biermann} P.~L.,  {Klein} U.,   {Kohle} S.,  1998, \aap,
  \href {https://ui.adsabs.harvard.edu/abs/1998A&A...332..395E} {332, 395}

\bibitem[\protect\citeauthoryear{{Fairall}}{{Fairall}}{1984}]{Fairall1984}
{Fairall} A.~P.,  1984, \mndoi [\mnras] {10.1093/mnras/210.1.69}, \href
  {https://ui.adsabs.harvard.edu/abs/1984MNRAS.210...69F} {210, 69}

\bibitem[\protect\citeauthoryear{{Finoguenov}, {Sarazin}, {Nakazawa}, {Wik}  \&
  {Clarke}}{{Finoguenov} et~al.}{2010}]{Finoguenov2010}
{Finoguenov} A.,  {Sarazin} C.~L.,  {Nakazawa} K.,  {Wik} D.~R.,   {Clarke}
  T.~E.,  2010, \mndoi [\apj] {10.1088/0004-637X/715/2/1143}, \href
  {https://ui.adsabs.harvard.edu/abs/2010ApJ...715.1143F} {715, 1143}

\bibitem[\protect\citeauthoryear{{Flaugher} et~al.,}{{Flaugher}
  et~al.}{2015}]{decam}
{Flaugher} B.,  et~al., 2015, \mndoi [\aj] {10.1088/0004-6256/150/5/150}, \href
  {https://ui.adsabs.harvard.edu/abs/2015AJ....150..150F} {150, 150}

\bibitem[\protect\citeauthoryear{{Fleenor}, {Rose}, {Christiansen},
  {Johnston-Hollitt}, {Hunstead}, {Drinkwater}  \& {Saunders}}{{Fleenor}
  et~al.}{2006}]{Fleenor2006}
{Fleenor} M.~C.,  {Rose} J.~A.,  {Christiansen} W.~A.,  {Johnston-Hollitt} M.,
  {Hunstead} R.~W.,  {Drinkwater} M.~J.,   {Saunders} W.,  2006, \mndoi [\aj]
  {10.1086/500197}, \href
  {https://ui.adsabs.harvard.edu/abs/2006AJ....131.1280F} {131, 1280}

\bibitem[\protect\citeauthoryear{{Flin} \& {Krywult}}{{Flin} \&
  {Krywult}}{2006}]{Flin2006}
{Flin} P.,  {Krywult} J.,  2006, \mndoi [\aap] {10.1051/0004-6361:20041635},
  \href {https://ui.adsabs.harvard.edu/abs/2006A&A...450....9F} {450, 9}

\bibitem[\protect\citeauthoryear{{Fruscione} et~al.,}{{Fruscione}
  et~al.}{2006}]{Fuscione2006}
{Fruscione} A.,  et~al., 2006, in \procspie. p. 62701V,
  \mndoi{10.1117/12.671760}

\bibitem[\protect\citeauthoryear{{Garilli}, {Maccagni}  \&
  {Tarenghi}}{{Garilli} et~al.}{1993}]{Garilli1993}
{Garilli} B.,  {Maccagni} D.,   {Tarenghi} M.,  1993, \aaps, \href
  {https://ui.adsabs.harvard.edu/abs/1993A&AS..100...33G} {100, 33}

\bibitem[\protect\citeauthoryear{{Garmire}, {Bautz}, {Ford}, {Nousek}  \&
  {Ricker}}{{Garmire} et~al.}{2003}]{garmire2003}
{Garmire} G.~P.,  {Bautz} M.~W.,  {Ford} P.~G.,  {Nousek} J.~A.,   {Ricker}
  George~R. J.,  2003, in {Truemper} J.~E.,  {Tananbaum} H.~D.,  eds,  Society
  of Photo-Optical Instrumentation Engineers (SPIE) Conference Series Vol.
  4851, X-Ray and Gamma-Ray Telescopes and Instruments for Astronomy.. pp
  28--44, \mndoi{10.1117/12.461599}

\bibitem[\protect\citeauthoryear{{Giacintucci}, {Markevitch}, {Cassano},
  {Venturi}, {Clarke}, {Kale}  \& {Cuciti}}{{Giacintucci}
  et~al.}{2019}]{Giacintucci2019}
{Giacintucci} S.,  {Markevitch} M.,  {Cassano} R.,  {Venturi} T.,  {Clarke}
  T.~E.,  {Kale} R.,   {Cuciti} V.,  2019, \mndoi [\apj]
  {10.3847/1538-4357/ab29f1}, \href
  {https://ui.adsabs.harvard.edu/abs/2019ApJ...880...70G} {880, 70}

\bibitem[\protect\citeauthoryear{{Giacintucci}, {Markevitch},
  {Johnston-Hollitt}, {Wik}, {Wang}  \& {Clarke}}{{Giacintucci}
  et~al.}{2020}]{Giacintucci2020}
{Giacintucci} S.,  {Markevitch} M.,  {Johnston-Hollitt} M.,  {Wik} D.~R.,
  {Wang} Q.~H.~S.,   {Clarke} T.~E.,  2020, \mndoi [\apj]
  {10.3847/1538-4357/ab6a9d}, \href
  {https://ui.adsabs.harvard.edu/abs/2020ApJ...891....1G} {891, 1}

\bibitem[\protect\citeauthoryear{{Gitti}, {Brunetti}  \& {Setti}}{{Gitti}
  et~al.}{2002}]{Gitti2002}
{Gitti} M.,  {Brunetti} G.,   {Setti} G.,  2002, \mndoi [\aap]
  {10.1051/0004-6361:20020284}, \href
  {https://ui.adsabs.harvard.edu/abs/2002A&A...386..456G} {386, 456}

\bibitem[\protect\citeauthoryear{{Golovich} et~al.,}{{Golovich}
  et~al.}{2019}]{Golovich2019}
{Golovich} N.,  et~al., 2019, \mndoi [\apj] {10.3847/1538-4357/ab2f90}, \href
  {https://ui.adsabs.harvard.edu/abs/2019ApJ...882...69G} {882, 69}

\bibitem[\protect\citeauthoryear{{Govoni} et~al.,}{{Govoni}
  et~al.}{2019}]{Govoni2019}
{Govoni} F.,  et~al., 2019, \mndoi [Science] {10.1126/science.aat7500}, \href
  {https://ui.adsabs.harvard.edu/abs/2019Sci...364..981G} {364, 981}

\bibitem[\protect\citeauthoryear{{Hambly} et~al.,}{{Hambly}
  et~al.}{2001a}]{supercosmos1}
{Hambly} N.~C.,  et~al., 2001a, \mndoi [\mnras]
  {10.1111/j.1365-2966.2001.04660.x}, \href
  {http://adsabs.harvard.edu/abs/2001MNRAS.326.1279H} {326, 1279}

\bibitem[\protect\citeauthoryear{{Hambly}, {Irwin}  \& {MacGillivray}}{{Hambly}
  et~al.}{2001b}]{supercosmos2}
{Hambly} N.~C.,  {Irwin} M.~J.,   {MacGillivray} H.~T.,  2001b, \mndoi [\mnras]
  {10.1111/j.1365-2966.2001.04661.x}, \href
  {http://adsabs.harvard.edu/abs/2001MNRAS.326.1295H} {326, 1295}

\bibitem[\protect\citeauthoryear{{Hambly}, {Davenhall}, {Irwin}  \&
  {MacGillivray}}{{Hambly} et~al.}{2001c}]{supercosmos3}
{Hambly} N.~C.,  {Davenhall} A.~C.,  {Irwin} M.~J.,   {MacGillivray} H.~T.,
  2001c, \mndoi [\mnras] {10.1111/j.1365-2966.2001.04662.x}, \href
  {http://adsabs.harvard.edu/abs/2001MNRAS.326.1315H} {326, 1315}

\bibitem[\protect\citeauthoryear{{Hancock}, {Trott}  \&
  {Hurley-Walker}}{{Hancock} et~al.}{2018}]{hth18}
{Hancock} P.~J.,  {Trott} C.~M.,   {Hurley-Walker} N.,  2018, \mndoi [\pasa]
  {10.1017/pasa.2018.3}, \href
  {http://adsabs.harvard.edu/abs/2018PASA...35...11H} {35, e011}

\bibitem[\protect\citeauthoryear{{Hardcastle} \& {Sakelliou}}{{Hardcastle} \&
  {Sakelliou}}{2004}]{Hardcastle2004}
{Hardcastle} M.~J.,  {Sakelliou} I.,  2004, \mndoi [\mnras]
  {10.1111/j.1365-2966.2004.07522.x}, \href
  {https://ui.adsabs.harvard.edu/abs/2004MNRAS.349..560H} {349, 560}

\bibitem[\protect\citeauthoryear{{Hindson} et~al.,}{{Hindson}
  et~al.}{2014}]{Hindson2014}
{Hindson} L.,  et~al., 2014, \mndoi [\mnras] {10.1093/mnras/stu1669}, \href
  {https://ui.adsabs.harvard.edu/abs/2014MNRAS.445..330H} {445, 330}

\bibitem[\protect\citeauthoryear{{Hodgson}, {Johnston-Hollitt}, {McKinley},
  {Vernstrom}  \& {Vacca}}{{Hodgson} et~al.}{2020}]{Hodgson2020}
{Hodgson} T.,  {Johnston-Hollitt} M.,  {McKinley} B.,  {Vernstrom} T.,
  {Vacca} V.,  2020, \mndoi [\pasa] {10.1017/pasa.2020.26}, \href
  {https://ui.adsabs.harvard.edu/abs/2020PASA...37...32H} {37, e032}

\bibitem[\protect\citeauthoryear{{Hodgson}, {Bartalucci}, {Johnston-Hollitt},
  {McKinley}, {Vazza}  \& {Wittor}}{{Hodgson} et~al.}{2021}]{Hodgson2021}
{Hodgson} T.,  {Bartalucci} I.,  {Johnston-Hollitt} M.,  {McKinley} B.,
  {Vazza} F.,   {Wittor} D.,  2021, \mndoi [\apj] {10.3847/1538-4357/abe384},
  \href {https://ui.adsabs.harvard.edu/abs/2021ApJ...909..198H} {909, 198}

\bibitem[\protect\citeauthoryear{{Hoeft} \& {Br{\"u}ggen}}{{Hoeft} \&
  {Br{\"u}ggen}}{2007}]{Hoeft2007}
{Hoeft} M.,  {Br{\"u}ggen} M.,  2007, \mndoi [\mnras]
  {10.1111/j.1365-2966.2006.11111.x}, \href
  {https://ui.adsabs.harvard.edu/abs/2007MNRAS.375...77H} {375, 77}

\bibitem[\protect\citeauthoryear{{Hota} et~al.,}{{Hota} et~al.}{2011}]{hso+11}
{Hota} A.,  et~al., 2011, \mndoi [\mnras] {10.1111/j.1745-3933.2011.01115.x},
  \href {http://adsabs.harvard.edu/abs/2011MNRAS.417L..36H} {417, L36}

\bibitem[\protect\citeauthoryear{{Hotan}, {McConnell}, {Whiting}  \&
  {Huynh}}{{Hotan} et~al.}{2020b}]{askap:sb25035}
{Hotan} A.,  {McConnell} D.,  {Whiting} M.,   {Huynh} M.,  2020b, ASKAP Data
  Products for Project AS110 (The Rapid ASKAP Continuum Survey): images and
  visibilities. v1. CSIRO. Data Collection, \url
  {http://hdl.handle.net/102.100.100/374841?index=1}

\bibitem[\protect\citeauthoryear{{Hotan}, {Whiting}, {Huynh}  \&
  {Moss}}{{Hotan} et~al.}{2020a}]{askap:racs}
{Hotan} A.,  {Whiting} M.,  {Huynh} M.,   {Moss} V.,  2020a, ASKAP Data
  Products for Project AS113 (Other ASKAP pilot science including tests, TOOs
  or guest observations): images and visibilities. v1. CSIRO. Data Collection.,
  \url {http://hdl.handle.net/102.100.100/348894?index=1}

\bibitem[\protect\citeauthoryear{{Hotan} et~al.,}{{Hotan}
  et~al.}{2021}]{Hotan2021}
{Hotan} A.~W.,  et~al., 2021, \mndoi [\pasa] {10.1017/pasa.2021.1}, \href
  {https://ui.adsabs.harvard.edu/abs/2021PASA...38....9H} {38, e009}

\bibitem[\protect\citeauthoryear{{Hunter}}{{Hunter}}{2007}]{Hunter2007}
{Hunter} J.~D.,  2007, \mndoi [Computing in Science and Engineering]
  {10.1109/MCSE.2007.55}, \href
  {http://adsabs.harvard.edu/abs/2007CSE.....9...90H} {9, 90}

\bibitem[\protect\citeauthoryear{{Hurley-Walker} et~al.,}{{Hurley-Walker}
  et~al.}{2017}]{gleamegc}
{Hurley-Walker} N.,  et~al., 2017, \mndoi [\mnras] {10.1093/mnras/stw2337},
  \href {http://adsabs.harvard.edu/abs/2017MNRAS.464.1146H} {464, 1146}

\bibitem[\protect\citeauthoryear{{Huynh}, {Dempsey}, {Whiting}  \&
  {Ophel}}{{Huynh} et~al.}{2020}]{Huynh2020}
{Huynh} M.,  {Dempsey} J.,  {Whiting} M.~T.,   {Ophel} M.,  2020, in
  {Ballester} P.,  {Ibsen} J.,  {Solar} M.,   {Shortridge} K.,  eds,
  Astronomical Society of the Pacific Conference Series Vol. 522, Astronomical
  Data Analysis Software and Systems XXVII. p.~263

\bibitem[\protect\citeauthoryear{{Intema}, {Jagannathan}, {Mooley}  \&
  {Frail}}{{Intema} et~al.}{2017}]{ijmf16}
{Intema} H.~T.,  {Jagannathan} P.,  {Mooley} K.~P.,   {Frail} D.~A.,  2017,
  \mndoi [\aap] {10.1051/0004-6361/201628536}, \href
  {http://adsabs.harvard.edu/abs/2017A%26A...598A..78I} {598, A78}

\bibitem[\protect\citeauthoryear{{Jaffe}}{{Jaffe}}{1977}]{Jaffe1977}
{Jaffe} W.~J.,  1977, \mndoi [\apj] {10.1086/155011}, \href
  {https://ui.adsabs.harvard.edu/abs/1977ApJ...212....1J} {212, 1}

\bibitem[\protect\citeauthoryear{{Johnston-Hollitt}}{{Johnston-Hollitt}}{2003}]{mj-h}
{Johnston-Hollitt} M.,  2003, PhD thesis, University of Adelaide

\bibitem[\protect\citeauthoryear{{Jonas} \& {MeerKAT Team}}{{Jonas} \& {MeerKAT
  Team}}{2016}]{Jonas2016}
{Jonas} J.,  {MeerKAT Team} 2016, in MeerKAT Science: On the Pathway to the
  SKA. p.~1

\bibitem[\protect\citeauthoryear{Jones, Oliphant, Peterson  et~al.}{Jones
  et~al.}{2001}]{Jones2001}
Jones E.,  Oliphant T.,  Peterson P.,   et~al., 2001, {SciPy}: Open source
  scientific tools for {Python}, \url {http://www.scipy.org/}

\bibitem[\protect\citeauthoryear{{Jones} et~al.,}{{Jones}
  et~al.}{2009}]{Jones2009}
{Jones} D.~H.,  et~al., 2009, \mndoi [\mnras]
  {10.1111/j.1365-2966.2009.15338.x}, \href
  {https://ui.adsabs.harvard.edu/abs/2009MNRAS.399..683J} {399, 683}

\bibitem[\protect\citeauthoryear{{Jurlin} et~al.,}{{Jurlin}
  et~al.}{2020}]{Jurlin2020}
{Jurlin} N.,  et~al., 2020, \mndoi [\aap] {10.1051/0004-6361/201936955}, \href
  {https://ui.adsabs.harvard.edu/abs/2020A&A...638A..34J} {638, A34}

\bibitem[\protect\citeauthoryear{{Kang}}{{Kang}}{2015}]{Kang2015}
{Kang} H.,  2015, \mndoi [Journal of Korean Astronomical Society]
  {10.5303/JKAS.2015.48.2.155}, \href
  {https://ui.adsabs.harvard.edu/abs/2015JKAS...48..155K} {48, 155}

\bibitem[\protect\citeauthoryear{{Kang}}{{Kang}}{2018}]{Kang2018}
{Kang} H.,  2018, \mndoi [Journal of Korean Astronomical Society]
  {10.5303/JKAS.2018.51.6.185}, \href
  {https://ui.adsabs.harvard.edu/abs/2018JKAS...51..185K} {51, 185}

\bibitem[\protect\citeauthoryear{{Kang} \& {Ryu}}{{Kang} \&
  {Ryu}}{2011}]{Kang2011}
{Kang} H.,  {Ryu} D.,  2011, \mndoi [\apj] {10.1088/0004-637X/734/1/18}, \href
  {https://ui.adsabs.harvard.edu/abs/2011ApJ...734...18K} {734, 18}

\bibitem[\protect\citeauthoryear{{Kang} \& {Ryu}}{{Kang} \&
  {Ryu}}{2016}]{Kang2016}
{Kang} H.,  {Ryu} D.,  2016, \mndoi [\apj] {10.3847/0004-637X/823/1/13}, \href
  {https://ui.adsabs.harvard.edu/abs/2016ApJ...823...13K} {823, 13}

\bibitem[\protect\citeauthoryear{{Kempner} \& {Sarazin}}{{Kempner} \&
  {Sarazin}}{2001}]{Kempner2001}
{Kempner} J.~C.,  {Sarazin} C.~L.,  2001, \mndoi [\apj] {10.1086/319024}, \href
  {https://ui.adsabs.harvard.edu/abs/2001ApJ...548..639K} {548, 639}

\bibitem[\protect\citeauthoryear{{Knowles} et~al.,}{{Knowles}
  et~al.}{2021}]{Knowles2020}
{Knowles} K.,  et~al., 2021, \mndoi [\mnras] {10.1093/mnras/stab939}, \href
  {https://ui.adsabs.harvard.edu/abs/2021MNRAS.504.1749K} {504, 1749}

\bibitem[\protect\citeauthoryear{{Line}, {Webster}, {Pindor}, {Mitchell}  \&
  {Trott}}{{Line} et~al.}{2017}]{puma}
{Line} J.~L.~B.,  {Webster} R.~L.,  {Pindor} B.,  {Mitchell} D.~A.,   {Trott}
  C.~M.,  2017, \mndoi [\pasa] {10.1017/pasa.2016.58}, \href
  {https://ui.adsabs.harvard.edu/abs/2017PASA...34....3L} {34, e003}

\bibitem[\protect\citeauthoryear{{Loi} et~al.,}{{Loi} et~al.}{2017}]{Loi2017}
{Loi} F.,  et~al., 2017, \mndoi [\mnras] {10.1093/mnras/stx2197}, \href
  {https://ui.adsabs.harvard.edu/abs/2017MNRAS.472.3605L} {472, 3605}

\bibitem[\protect\citeauthoryear{{Lovisari} et~al.,}{{Lovisari}
  et~al.}{2017}]{Lovisari2017}
{Lovisari} L.,  et~al., 2017, \mndoi [\apj] {10.3847/1538-4357/aa855f}, \href
  {https://ui.adsabs.harvard.edu/abs/2017ApJ...846...51L} {846, 51}

\bibitem[\protect\citeauthoryear{{Macario}, {Venturi}, {Brunetti}, {Dallacasa},
  {Giacintucci}, {Cassano}, {Bardelli}  \& {Athreya}}{{Macario}
  et~al.}{2010}]{Macario2010}
{Macario} G.,  {Venturi} T.,  {Brunetti} G.,  {Dallacasa} D.,  {Giacintucci}
  S.,  {Cassano} R.,  {Bardelli} S.,   {Athreya} R.,  2010, \mndoi [\aap]
  {10.1051/0004-6361/201014109}, \href
  {https://ui.adsabs.harvard.edu/abs/2010A&A...517A..43M} {517, A43}

\bibitem[\protect\citeauthoryear{{Macario} et~al.,}{{Macario}
  et~al.}{2013}]{Macario2013}
{Macario} G.,  et~al., 2013, \mndoi [\aap] {10.1051/0004-6361/201220667}, \href
  {https://ui.adsabs.harvard.edu/abs/2013A&A...551A.141M} {551, A141}

\bibitem[\protect\citeauthoryear{{Mahatma} et~al.,}{{Mahatma}
  et~al.}{2018}]{Mahatma2018}
{Mahatma} V.~H.,  et~al., 2018, \mndoi [\mnras] {10.1093/mnras/sty025}, \href
  {https://ui.adsabs.harvard.edu/abs/2018MNRAS.475.4557M} {475, 4557}

\bibitem[\protect\citeauthoryear{{Mandal} et~al.,}{{Mandal}
  et~al.}{2020}]{Mandal2020}
{Mandal} S.,  et~al., 2020, \mndoi [\aap] {10.1051/0004-6361/201936560}, \href
  {https://ui.adsabs.harvard.edu/abs/2020A&A...634A...4M} {634, A4}

\bibitem[\protect\citeauthoryear{{Markevitch}, {Govoni}, {Brunetti}  \&
  {Jerius}}{{Markevitch} et~al.}{2005}]{Markevitch2005}
{Markevitch} M.,  {Govoni} F.,  {Brunetti} G.,   {Jerius} D.,  2005, \mndoi
  [\apj] {10.1086/430695}, \href
  {https://ui.adsabs.harvard.edu/abs/2005ApJ...627..733M} {627, 733}

\bibitem[\protect\citeauthoryear{{Mauch}, {Murphy}, {Buttery}, {Curran},
  {Hunstead}, {Piestrzynski}, {Robertson}  \& {Sadler}}{{Mauch}
  et~al.}{2003}]{mmb+03}
{Mauch} T.,  {Murphy} T.,  {Buttery} H.~J.,  {Curran} J.,  {Hunstead} R.~W.,
  {Piestrzynski} B.,  {Robertson} J.~G.,   {Sadler} E.~M.,  2003, \mndoi
  [\mnras] {10.1046/j.1365-8711.2003.06605.x}, \href
  {http://adsabs.harvard.edu/abs/2003MNRAS.342.1117M} {342, 1117}

\bibitem[\protect\citeauthoryear{{McConnell} et~al.,}{{McConnell}
  et~al.}{2020}]{racs1}
{McConnell} D.,  et~al., 2020, \mndoi [\pasa] {10.1017/pasa.2020.41}, \href
  {https://ui.adsabs.harvard.edu/abs/2020PASA...37...48M} {37, e048}

\bibitem[\protect\citeauthoryear{{Morganson} et~al.,}{{Morganson}
  et~al.}{2018}]{des2}
{Morganson} E.,  et~al., 2018, \mndoi [\pasp] {10.1088/1538-3873/aab4ef}, \href
  {https://ui.adsabs.harvard.edu/abs/2018PASP..130g4501M} {130, 074501}

\bibitem[\protect\citeauthoryear{{Murgia}, {Govoni}, {Markevitch}, {Feretti},
  {Giovannini}, {Taylor}  \& {Carretti}}{{Murgia} et~al.}{2009}]{Murgia2009}
{Murgia} M.,  {Govoni} F.,  {Markevitch} M.,  {Feretti} L.,  {Giovannini} G.,
  {Taylor} G.~B.,   {Carretti} E.,  2009, \mndoi [\aap]
  {10.1051/0004-6361/200911659}, \href
  {https://ui.adsabs.harvard.edu/abs/2009A&A...499..679M} {499, 679}

\bibitem[\protect\citeauthoryear{{Murgia} et~al.,}{{Murgia}
  et~al.}{2011}]{mpm+11}
{Murgia} M.,  et~al., 2011, \mndoi [\aap] {10.1051/0004-6361/201015302}, \href
  {http://adsabs.harvard.edu/abs/2011A%26A...526A.148M} {526, A148}

\bibitem[\protect\citeauthoryear{{Murphy}, {Mauch}, {Green}, {Hunstead},
  {Piestrzynska}, {Kels}  \& {Sztajer}}{{Murphy} et~al.}{2007}]{mmg+07}
{Murphy} T.,  {Mauch} T.,  {Green} A.,  {Hunstead} R.~W.,  {Piestrzynska} B.,
  {Kels} A.~P.,   {Sztajer} P.,  2007, \mndoi [\mnras]
  {10.1111/j.1365-2966.2007.12379.x}, \href
  {http://adsabs.harvard.edu/abs/2007MNRAS.382..382M} {382, 382}

\bibitem[\protect\citeauthoryear{{Murphy} et~al.,}{{Murphy}
  et~al.}{2013}]{Murphy2013}
{Murphy} T.,  et~al., 2013, \mndoi [\pasa] {10.1017/pasa.2012.006}, \href
  {https://ui.adsabs.harvard.edu/abs/2013PASA...30....6M} {30, e006}

\bibitem[\protect\citeauthoryear{{Murphy}, {Lenc}, {Whiting}, {Huynh}  \&
  {Hotan}}{{Murphy} et~al.}{2019}]{askap:a141}
{Murphy} T.,  {Lenc} E.,  {Whiting} M.,  {Huynh} M.,   {Hotan} A.,  2019, ASKAP
  Data Products for Project AS111 (ASKAP Pilot Survey for Gravitational Wave
  Counterparts): images and visibilities. v1. CSIRO. Data Collection, \url
  {http://hdl.handle.net/102.100.100/175570?index=1}

\bibitem[\protect\citeauthoryear{{Murphy} et~al.,}{{Murphy}
  et~al.}{2020}]{askap:vast}
{Murphy} T.,  et~al., 2020, ASKAP Data Products for Project AS107 (ASKAP Pilot
  Survey for VAST): images and visibilities. v1. CSIRO. Data Collection, \url
  {http://hdl.handle.net/102.100.100/340961?index=1}

\bibitem[\protect\citeauthoryear{{Nuza}, {Hoeft}, {van Weeren}, {Gottl{\"o}ber}
   \& {Yepes}}{{Nuza} et~al.}{2012}]{Nuza2012}
{Nuza} S.~E.,  {Hoeft} M.,  {van Weeren} R.~J.,  {Gottl{\"o}ber} S.,   {Yepes}
  G.,  2012, \mndoi [\mnras] {10.1111/j.1365-2966.2011.20118.x}, \href
  {https://ui.adsabs.harvard.edu/abs/2012MNRAS.420.2006N} {420, 2006}

\bibitem[\protect\citeauthoryear{{Ochsenbein}, {Bauer}  \&
  {Marcout}}{{Ochsenbein} et~al.}{2000}]{vizier}
{Ochsenbein} F.,  {Bauer} P.,   {Marcout} J.,  2000, \mndoi [\aaps]
  {10.1051/aas:2000169}, \href
  {http://adsabs.harvard.edu/abs/2000A%26AS..143...23O} {143, 23}

\bibitem[\protect\citeauthoryear{{Offringa} \& {Smirnov}}{{Offringa} \&
  {Smirnov}}{2017}]{wsclean2}
{Offringa} A.~R.,  {Smirnov} O.,  2017, \mndoi [\mnras]
  {10.1093/mnras/stx1547}, \href
  {https://ui.adsabs.harvard.edu/abs/2017MNRAS.471..301O} {471, 301}

\bibitem[\protect\citeauthoryear{{Offringa} et~al.,}{{Offringa}
  et~al.}{2014}]{wsclean1}
{Offringa} A.~R.,  et~al., 2014, \mndoi [\mnras] {10.1093/mnras/stu1368}, \href
  {https://ui.adsabs.harvard.edu/abs/2014MNRAS.444..606O} {444, 606}

\bibitem[\protect\citeauthoryear{{Offringa} et~al.,}{{Offringa}
  et~al.}{2015}]{owh+15}
{Offringa} A.~R.,  et~al., 2015, \mndoi [\pasa] {10.1017/pasa.2015.7}, \href
  {http://adsabs.harvard.edu/abs/2015PASA...32....8O} {32, e008}

\bibitem[\protect\citeauthoryear{{Offringa} et~al.,}{{Offringa}
  et~al.}{2016}]{oth+16}
{Offringa} A.~R.,  et~al., 2016, \mndoi [\mnras] {10.1093/mnras/stw310}, \href
  {http://adsabs.harvard.edu/abs/2016MNRAS.458.1057O} {458, 1057}

\bibitem[\protect\citeauthoryear{{Ogrean} et~al.,}{{Ogrean}
  et~al.}{2015}]{Ogrean2015}
{Ogrean} G.~A.,  et~al., 2015, \mndoi [\apj] {10.1088/0004-637X/812/2/153},
  \href {https://ui.adsabs.harvard.edu/abs/2015ApJ...812..153O} {812, 153}

\bibitem[\protect\citeauthoryear{{Orr{\'u}}, {Murgia}, {Feretti}, {Govoni},
  {Brunetti}, {Giovannini}, {Girardi}  \& {Setti}}{{Orr{\'u}}
  et~al.}{2007}]{Orru2007}
{Orr{\'u}} E.,  {Murgia} M.,  {Feretti} L.,  {Govoni} F.,  {Brunetti} G.,
  {Giovannini} G.,  {Girardi} M.,   {Setti} G.,  2007, \mndoi [\aap]
  {10.1051/0004-6361:20066118}, \href
  {https://ui.adsabs.harvard.edu/abs/2007A&A...467..943O} {467, 943}

\bibitem[\protect\citeauthoryear{{Parma}, {Murgia}, {de Ruiter}, {Fanti},
  {Mack}  \& {Govoni}}{{Parma} et~al.}{2007}]{Parma2007}
{Parma} P.,  {Murgia} M.,  {de Ruiter} H.~R.,  {Fanti} R.,  {Mack} K.~H.,
  {Govoni} F.,  2007, \mndoi [\aap] {10.1051/0004-6361:20077592}, \href
  {https://ui.adsabs.harvard.edu/abs/2007A&A...470..875P} {470, 875}

\bibitem[\protect\citeauthoryear{{Pearce} et~al.,}{{Pearce}
  et~al.}{2017}]{Pearce2017}
{Pearce} C.~J.~J.,  et~al., 2017, \mndoi [\apj] {10.3847/1538-4357/aa7e2f},
  \href {https://ui.adsabs.harvard.edu/abs/2017ApJ...845...81P} {845, 81}

\bibitem[\protect\citeauthoryear{{Piffaretti}, {Arnaud}, {Pratt},
  {Pointecouteau}  \& {Melin}}{{Piffaretti} et~al.}{2011}]{pap+11}
{Piffaretti} R.,  {Arnaud} M.,  {Pratt} G.~W.,  {Pointecouteau} E.,   {Melin}
  J.-B.,  2011, \mndoi [\aap] {10.1051/0004-6361/201015377}, \href
  {http://adsabs.harvard.edu/abs/2011A%26A...534A.109P} {534, A109}

\bibitem[\protect\citeauthoryear{{Planck Collaboration} et~al.,}{{Planck
  Collaboration} et~al.}{2014}]{planck14}
{Planck Collaboration} et~al., 2014, \mndoi [\aap]
  {10.1051/0004-6361/201321523}, \href
  {https://ui.adsabs.harvard.edu/abs/2014A&A...571A..29P} {571, A29}

\bibitem[\protect\citeauthoryear{{Planck Collaboration} et~al.,}{{Planck
  Collaboration} et~al.}{2015}]{planck15}
{Planck Collaboration} et~al., 2015, \mndoi [\aap]
  {10.1051/0004-6361/201525787}, \href
  {http://cdsads.u-strasbg.fr/abs/2015A%26A...581A..14P} {581, A14}

\bibitem[\protect\citeauthoryear{{Planck Collaboration} et~al.,}{{Planck
  Collaboration} et~al.}{2016}]{psz2}
{Planck Collaboration} et~al., 2016, \mndoi [\aap]
  {10.1051/0004-6361/201525823}, \href
  {https://ui.adsabs.harvard.edu/abs/2016A&A...594A..27P} {594, A27}

\bibitem[\protect\citeauthoryear{{Poole}, {Fardal}, {Babul}, {McCarthy},
  {Quinn}  \& {Wadsley}}{{Poole} et~al.}{2006}]{Poole2006}
{Poole} G.~B.,  {Fardal} M.~A.,  {Babul} A.,  {McCarthy} I.~G.,  {Quinn} T.,
  {Wadsley} J.,  2006, \mndoi [\mnras] {10.1111/j.1365-2966.2006.10916.x},
  \href {https://ui.adsabs.harvard.edu/abs/2006MNRAS.373..881P} {373, 881}

\bibitem[\protect\citeauthoryear{{Pratt}, {Croston}, {Arnaud}  \&
  {B{\"o}hringer}}{{Pratt} et~al.}{2009}]{pcab09}
{Pratt} G.~W.,  {Croston} J.~H.,  {Arnaud} M.,   {B{\"o}hringer} H.,  2009,
  \mndoi [\aap] {10.1051/0004-6361/200810994}, \href
  {http://adsabs.harvard.edu/abs/2009A%26A...498..361P} {498, 361}

\bibitem[\protect\citeauthoryear{{Quici} et~al.,}{{Quici}
  et~al.}{2021}]{Quici2021}
{Quici} B.,  et~al., 2021, \mndoi [\pasa] {10.1017/pasa.2020.49}, \href
  {https://ui.adsabs.harvard.edu/abs/2021PASA...38....8Q} {38, e008}

\bibitem[\protect\citeauthoryear{{Rajpurohit} et~al.,}{{Rajpurohit}
  et~al.}{2020}]{Rajpurohit2020}
{Rajpurohit} K.,  et~al., 2020, \mndoi [\aap] {10.1051/0004-6361/202039165},
  \href {https://ui.adsabs.harvard.edu/abs/2020A&A...642L..13R} {642, L13}

\bibitem[\protect\citeauthoryear{{Rajpurohit} et~al.,}{{Rajpurohit}
  et~al.}{2021a}]{Rajpurohit2021c}
{Rajpurohit} K.,  et~al., 2021a, arXiv e-prints, \href
  {https://ui.adsabs.harvard.edu/abs/2021arXiv210405690R} {p. arXiv:2104.05690}

\bibitem[\protect\citeauthoryear{{Rajpurohit} et~al.,}{{Rajpurohit}
  et~al.}{2021b}]{Rajpurohit2021b}
{Rajpurohit} K.,  et~al., 2021b, \mndoi [\aap] {10.1051/0004-6361/202039428},
  \href {https://ui.adsabs.harvard.edu/abs/2021A&A...646A..56R} {646, A56}

\bibitem[\protect\citeauthoryear{{Rajpurohit} et~al.,}{{Rajpurohit}
  et~al.}{2021c}]{Rajpurohit2021a}
{Rajpurohit} K.,  et~al., 2021c, \mndoi [\aap] {10.1051/0004-6361/202039591},
  \href {https://ui.adsabs.harvard.edu/abs/2021A&A...646A.135R} {646, A135}

\bibitem[\protect\citeauthoryear{{Richard-Laferri{\`e}re}
  et~al.,}{{Richard-Laferri{\`e}re} et~al.}{2020}]{RichardLaferriere2020}
{Richard-Laferri{\`e}re} A.,  et~al., 2020, \mndoi [\mnras]
  {10.1093/mnras/staa2877}, \href
  {https://ui.adsabs.harvard.edu/abs/2020MNRAS.499.2934R} {499, 2934}

\bibitem[\protect\citeauthoryear{{Robitaille} \& {Bressert}}{{Robitaille} \&
  {Bressert}}{2012}]{Robitaille2012}
{Robitaille} T.,  {Bressert} E.,  2012, {APLpy: Astronomical Plotting Library
  in Python}, Astrophysics Source Code Library (\mn@eprint {ascl} {1208.017})

\bibitem[\protect\citeauthoryear{{Rossetti}, {Gastaldello}, {Eckert}, {Della
  Torre}, {Pantiri}, {Cazzoletti}  \& {Molendi}}{{Rossetti}
  et~al.}{2017}]{Rossetti2017}
{Rossetti} M.,  {Gastaldello} F.,  {Eckert} D.,  {Della Torre} M.,  {Pantiri}
  G.,  {Cazzoletti} P.,   {Molendi} S.,  2017, \mndoi [\mnras]
  {10.1093/mnras/stx493}, \href
  {https://ui.adsabs.harvard.edu/abs/2017MNRAS.468.1917R} {468, 1917}

\bibitem[\protect\citeauthoryear{{Sakelliou}, {Hardcastle}  \&
  {Jetha}}{{Sakelliou} et~al.}{2008}]{Sakelliou2008}
{Sakelliou} I.,  {Hardcastle} M.~J.,   {Jetha} N.~N.,  2008, \mndoi [\mnras]
  {10.1111/j.1365-2966.2007.12465.x}, \href
  {https://ui.adsabs.harvard.edu/abs/2008MNRAS.384...87S} {384, 87}

\bibitem[\protect\citeauthoryear{{Santos}, {Rosati}, {Tozzi}, {B{\"o}hringer},
  {Ettori}  \& {Bignamini}}{{Santos} et~al.}{2008}]{Santos2008}
{Santos} J.~S.,  {Rosati} P.,  {Tozzi} P.,  {B{\"o}hringer} H.,  {Ettori} S.,
  {Bignamini} A.,  2008, \mndoi [\aap] {10.1051/0004-6361:20078815}, \href
  {https://ui.adsabs.harvard.edu/abs/2008A%26A...483...35S} {483, 35}

\bibitem[\protect\citeauthoryear{{Schwope} et~al.,}{{Schwope}
  et~al.}{2000}]{shl+00}
{Schwope} A.,  et~al., 2000, \mndoi [Astronomische Nachrichten]
  {10.1002/(SICI)1521-3994(200003)321:1<1::AID-ASNA1>3.0.CO;2-C}, \href
  {http://adsabs.harvard.edu/abs/2000AN....321....1S} {321, 1}

\bibitem[\protect\citeauthoryear{{Slee}, {Roy}, {Murgia}, {Andernach}  \&
  {Ehle}}{{Slee} et~al.}{2001}]{srm+01}
{Slee} O.~B.,  {Roy} A.~L.,  {Murgia} M.,  {Andernach} H.,   {Ehle} M.,  2001,
  \mndoi [\aj] {10.1086/322105}, \href
  {http://adsabs.harvard.edu/abs/2001AJ....122.1172S} {122, 1172}

\bibitem[\protect\citeauthoryear{{Struble} \& {Rood}}{{Struble} \&
  {Rood}}{1999}]{sr99}
{Struble} M.~F.,  {Rood} H.~J.,  1999, \mndoi [\apjs] {10.1086/313274}, \href
  {http://adsabs.harvard.edu/abs/1999ApJS..125...35S} {125, 35}

\bibitem[\protect\citeauthoryear{{Str{\"u}der} et~al.,}{{Str{\"u}der}
  et~al.}{2001}]{struder2001}
{Str{\"u}der} L.,  et~al., 2001, \mndoi [\aap] {10.1051/0004-6361:20000066},
  \href {http://adsabs.harvard.edu/abs/2001A%26A...365L..18S} {365, L18}

\bibitem[\protect\citeauthoryear{{Subrahmanyan}, {Ekers}, {Saripalli}  \&
  {Sadler}}{{Subrahmanyan} et~al.}{2010}]{Subrahmanyan2010}
{Subrahmanyan} R.,  {Ekers} R.~D.,  {Saripalli} L.,   {Sadler} E.~M.,  2010,
  \mndoi [\mnras] {10.1111/j.1365-2966.2009.16105.x}, \href
  {https://ui.adsabs.harvard.edu/abs/2010MNRAS.402.2792S} {402, 2792}

\bibitem[\protect\citeauthoryear{{Tarr{\'\i}o}, {Melin}, {Arnaud}  \&
  {Pratt}}{{Tarr{\'\i}o} et~al.}{2016}]{Tarrio2016}
{Tarr{\'\i}o} P.,  {Melin} J.~B.,  {Arnaud} M.,   {Pratt} G.~W.,  2016, \mndoi
  [\aap] {10.1051/0004-6361/201628366}, \href
  {https://ui.adsabs.harvard.edu/abs/2016A&A...591A..39T} {591, A39}

\bibitem[\protect\citeauthoryear{{Tarr{\'\i}o}, {Melin}  \&
  {Arnaud}}{{Tarr{\'\i}o} et~al.}{2018}]{Tarrio2018}
{Tarr{\'\i}o} P.,  {Melin} J.~B.,   {Arnaud} M.,  2018, \mndoi [\aap]
  {10.1051/0004-6361/201731984}, \href
  {https://ui.adsabs.harvard.edu/abs/2018A&A...614A..82T} {614, A82}

\bibitem[\protect\citeauthoryear{{Thierbach}, {Klein}  \&
  {Wielebinski}}{{Thierbach} et~al.}{2003}]{tkw03}
{Thierbach} M.,  {Klein} U.,   {Wielebinski} R.,  2003, \mndoi [\aap]
  {10.1051/0004-6361:20021474}, \href
  {http://adsabs.harvard.edu/abs/2003A%26A...397...53T} {397, 53}

\bibitem[\protect\citeauthoryear{{Thorat}, {Subrahmanyan}, {Saripalli}  \&
  {Ekers}}{{Thorat} et~al.}{2013}]{Thorat2013}
{Thorat} K.,  {Subrahmanyan} R.,  {Saripalli} L.,   {Ekers} R.~D.,  2013,
  \mndoi [\apj] {10.1088/0004-637X/762/1/16}, \href
  {https://ui.adsabs.harvard.edu/abs/2013ApJ...762...16T} {762, 16}

\bibitem[\protect\citeauthoryear{{Tingay} et~al.,}{{Tingay}
  et~al.}{2013}]{tgb+13}
{Tingay} S.~J.,  et~al., 2013, \mndoi [\pasa] {10.1017/pasa.2012.007}, \href
  {http://adsabs.harvard.edu/abs/2013PASA...30....7T} {30, 7}

\bibitem[\protect\citeauthoryear{{Tonry} et~al.,}{{Tonry}
  et~al.}{2012}]{tsl+12}
{Tonry} J.~L.,  et~al., 2012, \mndoi [\apj] {10.1088/0004-637X/750/2/99}, \href
  {http://adsabs.harvard.edu/abs/2012ApJ...750...99T} {750, 99}

\bibitem[\protect\citeauthoryear{{Trasatti}, {Akamatsu}, {Lovisari}, {Klein},
  {Bonafede}, {Br{\"u}ggen}, {Dallacasa}  \& {Clarke}}{{Trasatti}
  et~al.}{2015}]{Trasatti2015}
{Trasatti} M.,  {Akamatsu} H.,  {Lovisari} L.,  {Klein} U.,  {Bonafede} A.,
  {Br{\"u}ggen} M.,  {Dallacasa} D.,   {Clarke} T.,  2015, \mndoi [\aap]
  {10.1051/0004-6361/201423972}, \href
  {https://ui.adsabs.harvard.edu/abs/2015A&A...575A..45T} {575, A45}

\bibitem[\protect\citeauthoryear{{Turner} et~al.,}{{Turner}
  et~al.}{2001}]{turner2001}
{Turner} M.~J.~L.,  et~al., 2001, \mndoi [\aap] {10.1051/0004-6361:20000087},
  \href {http://adsabs.harvard.edu/abs/2001A%26A...365L..27T} {365, L27}

\bibitem[\protect\citeauthoryear{{Urdampilleta}, {Simionescu}, {Kaastra},
  {Zhang}, {Di Gennaro}, {Mernier}, {de Plaa}  \& {Brunetti}}{{Urdampilleta}
  et~al.}{2021}]{Urdampilleta2021}
{Urdampilleta} I.,  {Simionescu} A.,  {Kaastra} J.~S.,  {Zhang} X.,  {Di
  Gennaro} G.,  {Mernier} F.,  {de Plaa} J.,   {Brunetti} G.,  2021, \mndoi
  [\aap] {10.1051/0004-6361/201937160}, \href
  {https://ui.adsabs.harvard.edu/abs/2021A&A...646A..95U} {646, A95}

\bibitem[\protect\citeauthoryear{{Vazza}, {Wittor}, {Brunetti}  \&
  {Br{\"u}ggen}}{{Vazza} et~al.}{2021}]{Vazza2021}
{Vazza} F.,  {Wittor} D.,  {Brunetti} G.,   {Br{\"u}ggen} M.,  2021, arXiv
  e-prints, \href {https://ui.adsabs.harvard.edu/abs/2021arXiv210204193V} {p.
  arXiv:2102.04193}

\bibitem[\protect\citeauthoryear{{Venturi}, {Giacintucci}, {Dallacasa},
  {Cassano}, {Brunetti}, {Bardelli}  \& {Setti}}{{Venturi}
  et~al.}{2008}]{vgd+08}
{Venturi} T.,  {Giacintucci} S.,  {Dallacasa} D.,  {Cassano} R.,  {Brunetti}
  G.,  {Bardelli} S.,   {Setti} G.,  2008, \mndoi [\aap]
  {10.1051/0004-6361:200809622}, \href
  {http://adsabs.harvard.edu/abs/2008A%26A...484..327V} {484, 327}

\bibitem[\protect\citeauthoryear{{Vikhlinin}, {McNamara}, {Forman}, {Jones},
  {Quintana}  \& {Hornstrup}}{{Vikhlinin} et~al.}{1998}]{Vikhlinin1998}
{Vikhlinin} A.,  {McNamara} B.~R.,  {Forman} W.,  {Jones} C.,  {Quintana} H.,
  {Hornstrup} A.,  1998, \mndoi [\apj] {10.1086/305951}, \href
  {https://ui.adsabs.harvard.edu/abs/1998ApJ...502..558V} {502, 558}

\bibitem[\protect\citeauthoryear{{Voges} et~al.,}{{Voges}
  et~al.}{1999}]{vab+99}
{Voges} W.,  et~al., 1999, \aap, \href
  {http://adsabs.harvard.edu/abs/1999A%26A...349..389V} {349, 389}

\bibitem[\protect\citeauthoryear{{Wayth} et~al.,}{{Wayth}
  et~al.}{2015}]{wlb+15}
{Wayth} R.~B.,  et~al., 2015, \mndoi [\pasa] {10.1017/pasa.2015.26}, \href
  {http://adsabs.harvard.edu/abs/2015PASA...32...25W} {32, 25}

\bibitem[\protect\citeauthoryear{{Wayth} et~al.,}{{Wayth}
  et~al.}{2018}]{wtt+18}
{Wayth} R.~B.,  et~al., 2018, \mndoi [\pasa] {10.1017/pasa.2018.37}, \href
  {http://adsabs.harvard.edu/abs/2018PASA...35...33W} {35}

\bibitem[\protect\citeauthoryear{{Wen} \& {Han}}{{Wen} \&
  {Han}}{2015}]{Wen2015}
{Wen} Z.~L.,  {Han} J.~L.,  2015, \mndoi [\apj] {10.1088/0004-637X/807/2/178},
  \href {https://ui.adsabs.harvard.edu/abs/2015ApJ...807..178W} {807, 178}

\bibitem[\protect\citeauthoryear{{Wen}, {Han}  \& {Liu}}{{Wen}
  et~al.}{2012}]{whl12}
{Wen} Z.~L.,  {Han} J.~L.,   {Liu} F.~S.,  2012, \mndoi [\apjs]
  {10.1088/0067-0049/199/2/34}, \href
  {http://adsabs.harvard.edu/abs/2012ApJS..199...34W} {199, 34}

\bibitem[\protect\citeauthoryear{{Wilber} et~al.,}{{Wilber}
  et~al.}{2018}]{Wilber2018}
{Wilber} A.,  et~al., 2018, \mndoi [\mnras] {10.1093/mnras/stx2568}, \href
  {https://ui.adsabs.harvard.edu/abs/2018MNRAS.473.3536W} {473, 3536}

\bibitem[\protect\citeauthoryear{{Wilber} et~al.,}{{Wilber}
  et~al.}{2019}]{Wilber2019}
{Wilber} A.,  et~al., 2019, \mndoi [\aap] {10.1051/0004-6361/201833884}, \href
  {https://ui.adsabs.harvard.edu/abs/2019A&A...622A..25W} {622, A25}

\bibitem[\protect\citeauthoryear{{Wilber}, {Johnston-Hollitt}, {Duchesne},
  {Tasse}, {Akamatsu}, {Intema}  \& {Hodgson}}{{Wilber}
  et~al.}{2020}]{Wilber2020}
{Wilber} A.~G.,  {Johnston-Hollitt} M.,  {Duchesne} S.~W.,  {Tasse} C.,
  {Akamatsu} H.,  {Intema} H.,   {Hodgson} T.,  2020, \mndoi [\pasa]
  {10.1017/pasa.2020.34}, \href
  {https://ui.adsabs.harvard.edu/abs/2020PASA...37...40W} {37, e040}

\bibitem[\protect\citeauthoryear{{Xie} et~al.,}{{Xie} et~al.}{2020}]{Xie2020}
{Xie} C.,  et~al., 2020, \mndoi [\aap] {10.1051/0004-6361/201936953}, \href
  {https://ui.adsabs.harvard.edu/abs/2020A&A...636A...3X} {636, A3}

\bibitem[\protect\citeauthoryear{{Zaritsky}, {Gonzalez}  \&
  {Zabludoff}}{{Zaritsky} et~al.}{2006}]{zgz06}
{Zaritsky} D.,  {Gonzalez} A.~H.,   {Zabludoff} A.~I.,  2006, \mndoi [\apj]
  {10.1086/498672}, \href {http://adsabs.harvard.edu/abs/2006ApJ...638..725Z}
  {638, 725}

\bibitem[\protect\citeauthoryear{{de Gasperin}}{{de Gasperin}}{2017}]{deg17}
{de Gasperin} F.,  2017, \mndoi [\mnras] {10.1093/mnras/stx210}, \href
  {http://adsabs.harvard.edu/abs/2017MNRAS.467.2234D} {467, 2234}

\bibitem[\protect\citeauthoryear{{de Gasperin}, {van Weeren}, {Br{\"u}ggen},
  {Vazza}, {Bonafede}  \& {Intema}}{{de Gasperin} et~al.}{2014}]{dvb+14}
{de Gasperin} F.,  {van Weeren} R.~J.,  {Br{\"u}ggen} M.,  {Vazza} F.,
  {Bonafede} A.,   {Intema} H.~T.,  2014, \mndoi [\mnras]
  {10.1093/mnras/stu1658}, \href
  {http://adsabs.harvard.edu/abs/2014MNRAS.444.3130D} {444, 3130}

\bibitem[\protect\citeauthoryear{{de Gasperin}, {Intema}, {van Weeren},
  {Dawson}, {Golovich}, {Wittman}, {Bonafede}  \& {Br{\"u}ggen}}{{de Gasperin}
  et~al.}{2015}]{div+15}
{de Gasperin} F.,  {Intema} H.~T.,  {van Weeren} R.~J.,  {Dawson} W.~A.,
  {Golovich} N.,  {Wittman} D.,  {Bonafede} A.,   {Br{\"u}ggen} M.,  2015,
  \mndoi [\mnras] {10.1093/mnras/stv1873}, \href
  {http://adsabs.harvard.edu/abs/2015MNRAS.453.3483D} {453, 3483}

\bibitem[\protect\citeauthoryear{{de Gasperin} et~al.,}{{de Gasperin}
  et~al.}{2017}]{deGasperin2017a}
{de Gasperin} F.,  et~al., 2017, \mndoi [Science Advances]
  {10.1126/sciadv.1701634}, \href
  {https://ui.adsabs.harvard.edu/abs/2017SciA....3E1634D} {3, e1701634}

\bibitem[\protect\citeauthoryear{{van Haarlem} et~al.,}{{van Haarlem}
  et~al.}{2013}]{lofar}
{van Haarlem} M.~P.,  et~al., 2013, \mndoi [\aap]
  {10.1051/0004-6361/201220873}, \href
  {http://adsabs.harvard.edu/abs/2013A%26A...556A...2V} {556, A2}

\bibitem[\protect\citeauthoryear{{van Weeren}, {R{\"o}ttgering}, {Br{\"u}ggen}
  \& {Hoeft}}{{van Weeren} et~al.}{2010}]{vanWeeren2010}
{van Weeren} R.~J.,  {R{\"o}ttgering} H. J.~A.,  {Br{\"u}ggen} M.,   {Hoeft}
  M.,  2010, \mndoi [Science] {10.1126/science.1194293}, \href
  {https://ui.adsabs.harvard.edu/abs/2010Sci...330..347V} {330, 347}

\bibitem[\protect\citeauthoryear{{van Weeren}, {Br{\"u}ggen}, {R{\"o}ttgering},
  {Hoeft}, {Nuza}  \& {Intema}}{{van Weeren} et~al.}{2011}]{vanWeeren2011b}
{van Weeren} R.~J.,  {Br{\"u}ggen} M.,  {R{\"o}ttgering} H.~J.~A.,  {Hoeft} M.,
   {Nuza} S.~E.,   {Intema} H.~T.,  2011, \mndoi [\aap]
  {10.1051/0004-6361/201117149}, \href
  {https://ui.adsabs.harvard.edu/abs/2011A&A...533A..35V} {533, A35}

\bibitem[\protect\citeauthoryear{{van Weeren}, {R{\"o}ttgering}, {Intema},
  {Rudnick}, {Br{\"u}ggen}, {Hoeft}  \& {Oonk}}{{van Weeren}
  et~al.}{2012}]{vanWeeren2012}
{van Weeren} R.~J.,  {R{\"o}ttgering} H.~J.~A.,  {Intema} H.~T.,  {Rudnick} L.,
   {Br{\"u}ggen} M.,  {Hoeft} M.,   {Oonk} J.~B.~R.,  2012, \mndoi [\aap]
  {10.1051/0004-6361/201219000}, \href
  {https://ui.adsabs.harvard.edu/abs/2012A&A...546A.124V} {546, A124}

\bibitem[\protect\citeauthoryear{{van Weeren} et~al.,}{{van Weeren}
  et~al.}{2016}]{vanWeeren2016}
{van Weeren} R.~J.,  et~al., 2016, \mndoi [\apj] {10.3847/0004-637X/818/2/204},
  \href {https://ui.adsabs.harvard.edu/abs/2016ApJ...818..204V} {818, 204}

\bibitem[\protect\citeauthoryear{{van Weeren} et~al.,}{{van Weeren}
  et~al.}{2017}]{vanWeeren2017}
{van Weeren} R.~J.,  et~al., 2017, \mndoi [Nature Astronomy]
  {10.1038/s41550-016-0005}, \href
  {https://ui.adsabs.harvard.edu/abs/2017NatAs...1E...5V} {1, 0005}

\bibitem[\protect\citeauthoryear{{van Weeren}, {de Gasperin}, {Akamatsu},
  {Br{\"u}ggen}, {Feretti}, {Kang}, {Stroe}  \& {Zandanel}}{{van Weeren}
  et~al.}{2019}]{vda+19}
{van Weeren} R.~J.,  {de Gasperin} F.,  {Akamatsu} H.,  {Br{\"u}ggen} M.,
  {Feretti} L.,  {Kang} H.,  {Stroe} A.,   {Zandanel} F.,  2019, \mndoi [\ssr]
  {10.1007/s11214-019-0584-z}, \href
  {http://adsabs.harvard.edu/abs/2019SSRv..215...16V} {215, 16}

\bibitem[\protect\citeauthoryear{{van Weeren} et~al.,}{{van Weeren}
  et~al.}{2021}]{vanWeeren2020}
{van Weeren} R.~J.,  et~al., 2021, \mndoi [\aap] {10.1051/0004-6361/202039826},
  \href {https://ui.adsabs.harvard.edu/abs/2021A&A...651A.115V} {651, A115}

\bibitem[\protect\citeauthoryear{van~der Walt, Colbert  \& Varoquaux}{van~der
  Walt et~al.}{2011}]{Numpy2011}
van~der Walt S.,  Colbert S.~C.,   Varoquaux G.,  2011, \mndoi [Computing in
  Science Engineering] {10.1109/MCSE.2011.37}, 13, 22

\makeatother
\end{thebibliography}
}

\begin{appendix}

\section{Non-candidates}\label{app:nondetections}
In this section we will record clusters previously reported as having candidate diffuse emission by \citetalias{Duchesne2017} but are shown by the MWA-2 data to be discrete sources. Note that while \verb|FIELD1| and \verb|FIELD2| overlap with a majority image used by \citealt{Duchesne2017}, for some clusters towards the edges of the fields we are unable to confirm the presence of either diffuse emission of discrete sources---these clusters are not mentioned here.

\paragraph{Abell~0022} Edge of image. Inconclusive.
\paragraph{Abell~0033} Edge of image. Inconclusive.
\paragraph{Abell~2798} Resolution still not sufficient. Inconclusive.
\paragraph{Abell~S1136} To be discussed in Macgregor et al. (in prepatation).
\paragraph{Abell~S1063} Halo reported by \citet{Xie2020}.
\paragraph{Abell~2556} The MWA-2 216-MHz, robust $+0.5$ image breaks up into three discrete point sources. RACS data confirm the source is a number of point sources and a double-lobed radio galaxy, with a spectral index $\alpha \sim -0.8$. The steeper spectral index reported by \citetalias{Duchesne2017} is a result of point sources contributing to the 168-MHz measurement but not the 1.4-GHz measurement. 
\paragraph{Abell~S0084} MWA-2 data breaks into two discrete components and RACS confirms this. We cannot confirm the candidate halo/mini-halo.
\paragraph{Abell~S1121} Out of images.
\paragraph{PSZ~G082.31$-$67.01} Out of images.


\section{Measured source properties}\label{sec:app:table}
A table of measured and derived source properties are provided for all frequencies (88, 118, 154, 169, 185, 200, 216, 887, 943, and 1400~MHz, though note not all sources are measured at every frequency) as supplementary online material available at the data store for the Publications of the Astronomical Society of Australia, hosted at \url{https://data-portal.hpc.swin.edu.au/institute/pasa} \citep{mwa:clusters} \footnote{\url{ https://dx.doi.org/10.26185/611f33b774e96}}. The following details the columns available for each source:

\begin{enumerate}[label=C\arabic*]
    \addtocounter{enumi}{-1}
    \item \texttt{cluster\_name} \\ Name of cluster as it appears in \cref{tab:clusters}.
    \item \texttt{source\_id} \\ ID of source as used in \cref{sec:sources} and in figures.
    \item \texttt{flux\_\$nu} (mJy) \\ Final flux density of source at frequency \texttt{\$nu} MHz.
    \item \texttt{err\_flux\_\$nu} (mJy) \\ Uncertainty on the final flux density of source at frequency \texttt{\$nu} MHz.
    \item \texttt{conf\_\$nu} (mJy) \\ Total confusing flux density subtracted from initial measurement at frequency \texttt{\$nu} MHz.
    \item \texttt{err\_conf\_\$nu} (mJy) \\ Uncertainty on total confusing flux density subtracted from initial measurement at frequency \texttt{\$nu} MHz. This is added in quadrature to the initial measurement. 
    \item \texttt{psf\_a\_\$nu} (arcsec) \\ FWHM of PSF major axis in \texttt{\$nu}-MHz image at the source location.
    \item \texttt{psf\_b\_\$nu} (arcsec) \\ FWHM of PSF minor axis in \texttt{\$nu}-MHz image at the source location. 
    \\~\\ \vdots \\
    \setItemnumber{56}
    \item \texttt{alpha\_mwa} \\ Spectral index across the MWA-2 images as shown in \cref{fig:sed:a0122}--\subref{fig:sed:a550}.
    \item \texttt{alpha} \\ Spectral index across all data for power law models as shown in \cref{fig:sed:a0122}--\subref{fig:sed:a550}.
    \item \texttt{q} \\ Curvature parameter for curved power law model fits \corrs{(\autoref{eq:cpowerlaw})}, as shown in \cref{fig:sed:a0122}--\subref{fig:sed:a550}.
\end{enumerate}

\section{Integrated spectra}\label{app:seds}
\cref{fig:sed:a0122}--\subref{fig:sed:a550} show the SEDs for all sources reported in \cref{sec:sources}. For each source with MWA-2 and additional data, we \corrs{additionally} provide a power law fit \corrs{to} the MWA-2 data only. 

\begin{figure*}
\centering
\begin{subfigure}[b]{0.25\linewidth}
\includegraphics[width=1\linewidth]{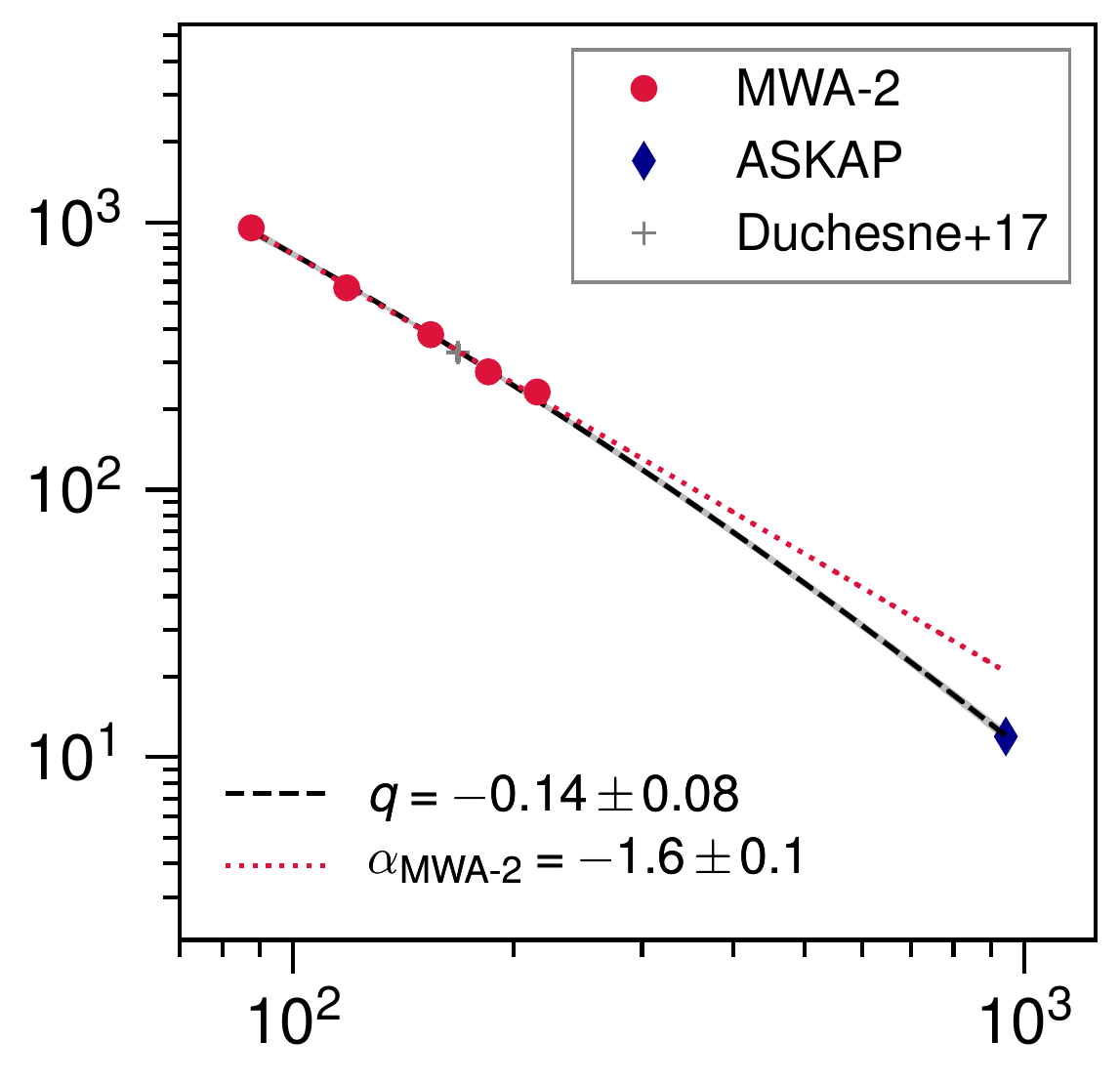}
\cprotect\subcaption{\label{fig:sed:a0122}  \hyperref[para:a0122]{Abell~0122}, D1.}
\end{subfigure}%
\begin{subfigure}[b]{0.25\linewidth}
\includegraphics[width=1\linewidth]{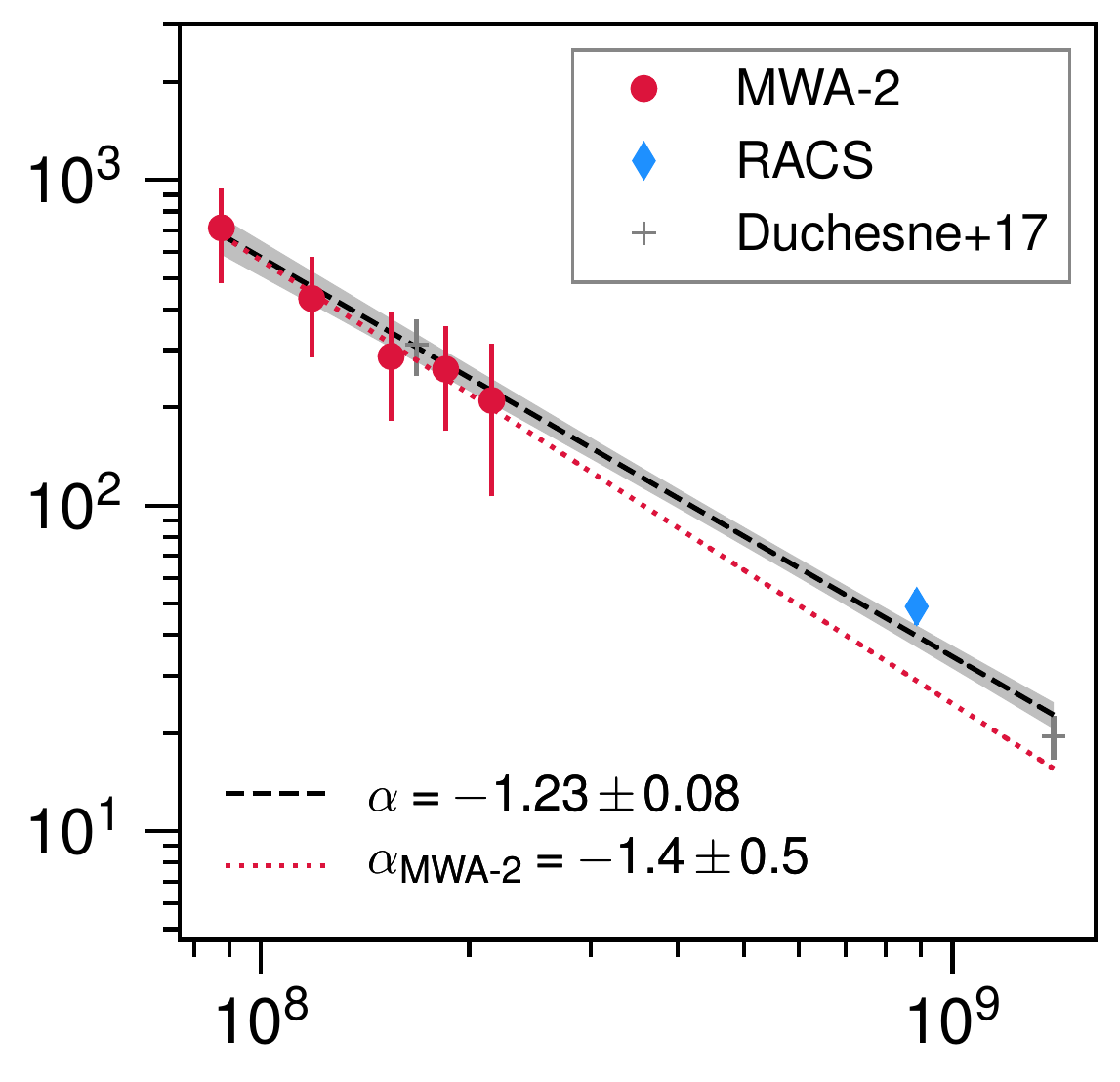}
\cprotect\subcaption{\label{fig:sed:a2751}  \hyperref[para:a2751]{Abell~2751}, D1.}
\end{subfigure}%
\begin{subfigure}[b]{0.25\linewidth}
\includegraphics[width=1\linewidth]{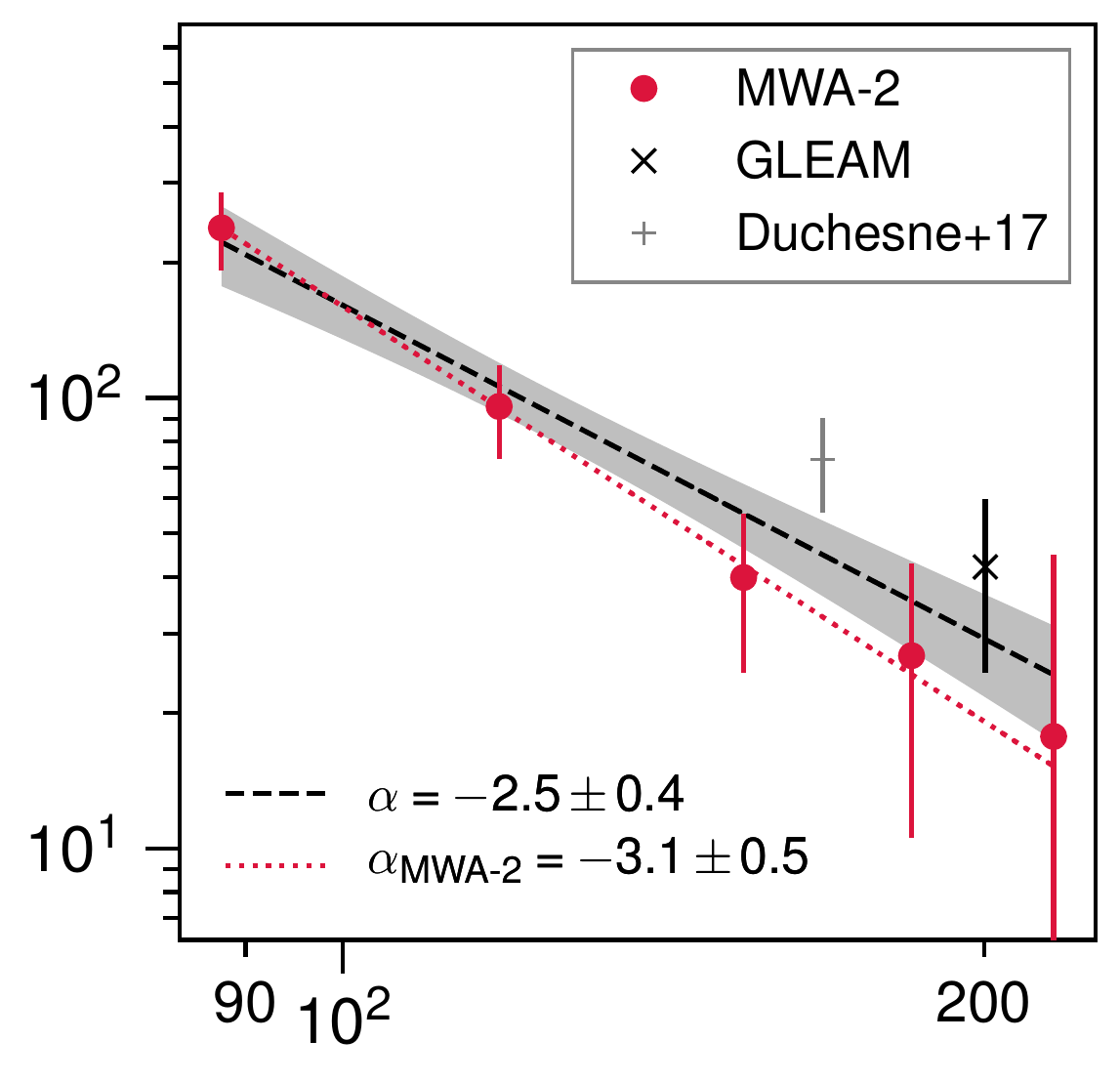}
\subcaption{\label{fig:sed:a2811} \hyperref[para:a2811]{Abell~2811}, D1.}
\end{subfigure}\\%
\begin{subfigure}[b]{0.25\linewidth}
\includegraphics[width=1\linewidth]{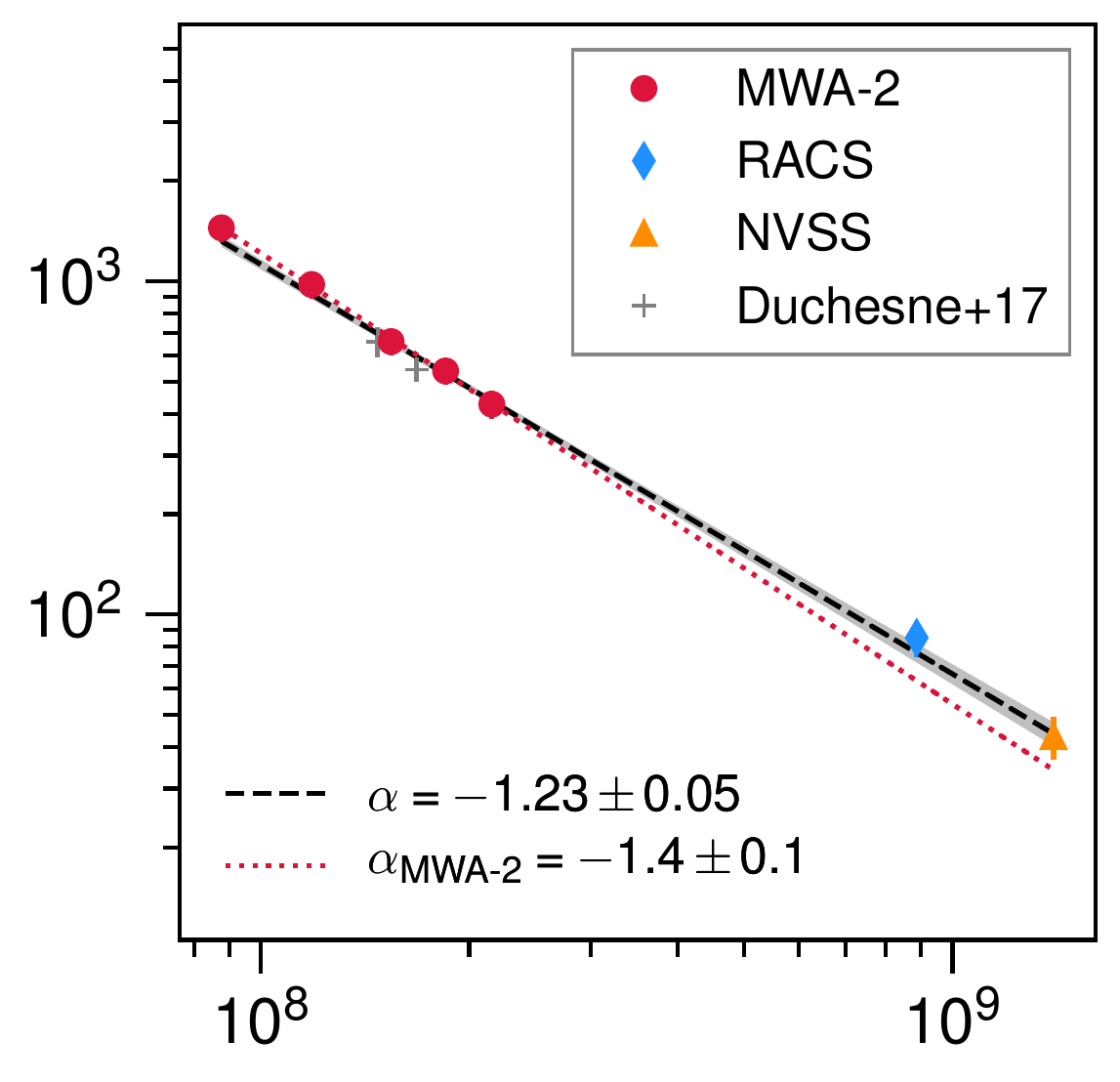}
\subcaption{\label{fig:sed:a2496} \hyperref[para:a2496]{Abell~2496}, D1.}
\end{subfigure}%
\begin{subfigure}[b]{0.25\linewidth}
\includegraphics[width=1\linewidth]{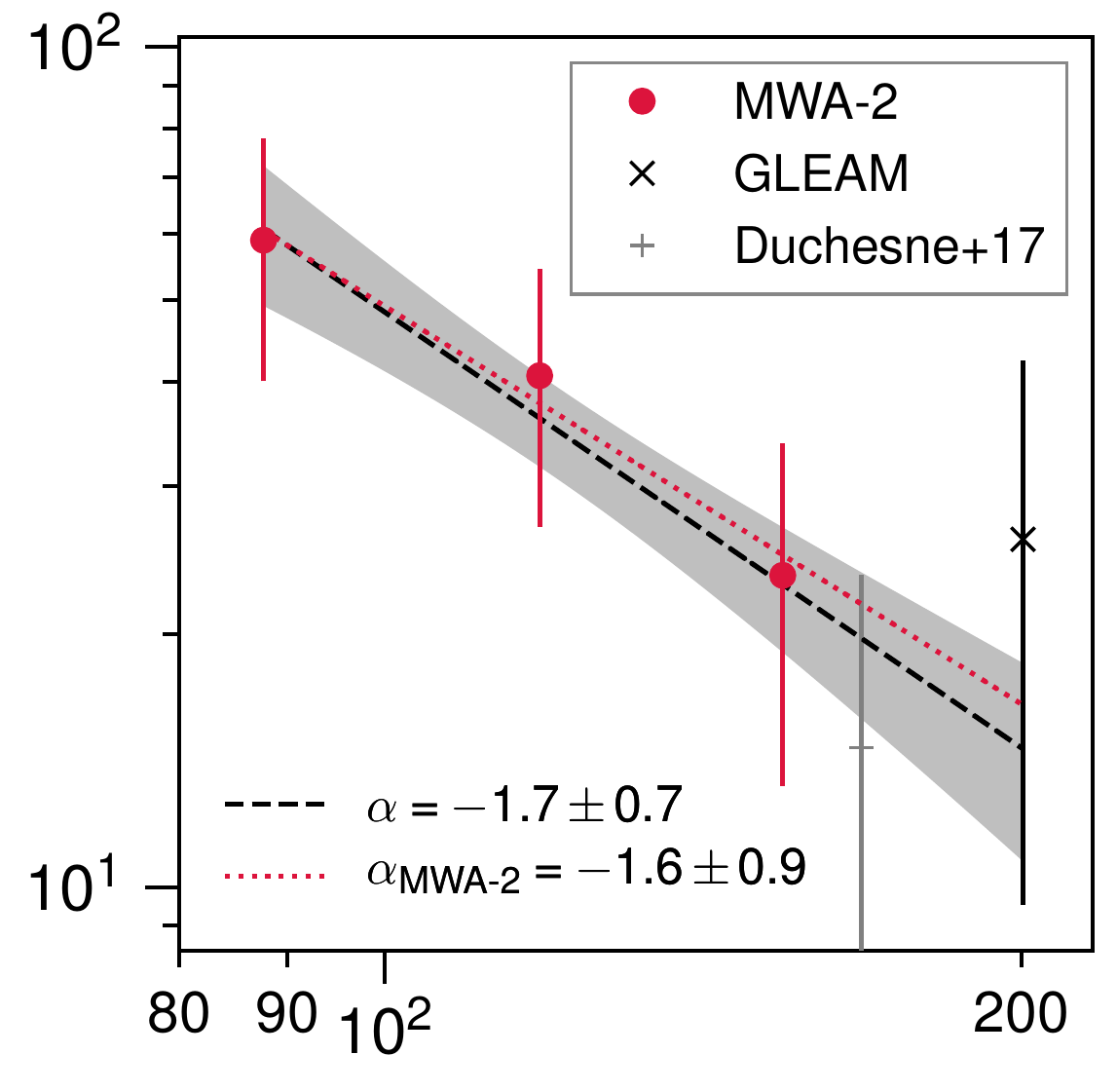}
\subcaption{\label{fig:sed:a2680} \hyperref[para:a2680]{Abell~2680}, D1.}
\end{subfigure}%
\begin{subfigure}[b]{0.25\linewidth}
\includegraphics[width=1\linewidth]{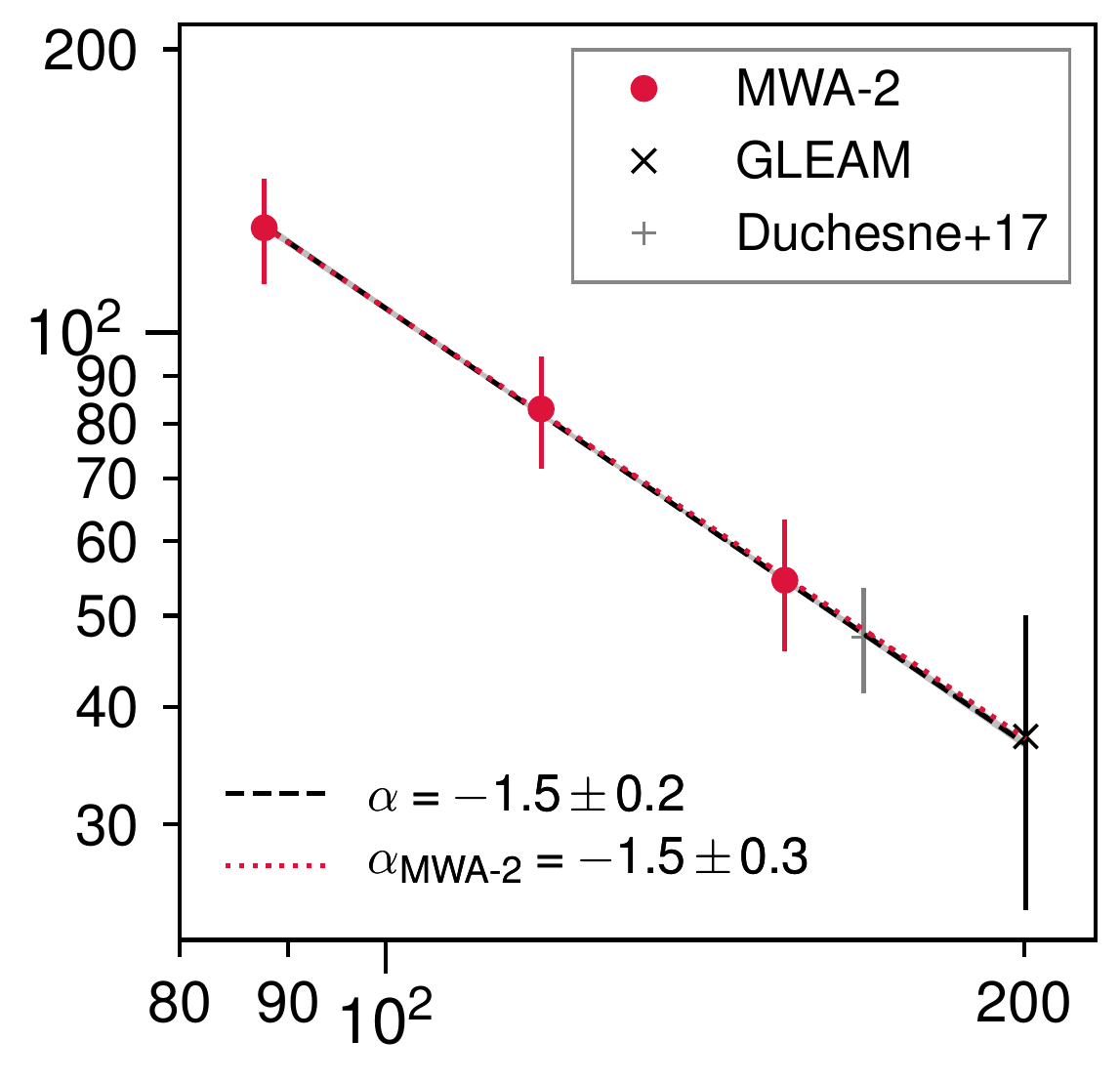}
\subcaption{\label{fig:sed:a2693} \hyperref[para:a2693]{Abell~2693}, D1.}
\end{subfigure}%
\begin{subfigure}[b]{0.25\linewidth}
\includegraphics[width=1\linewidth]{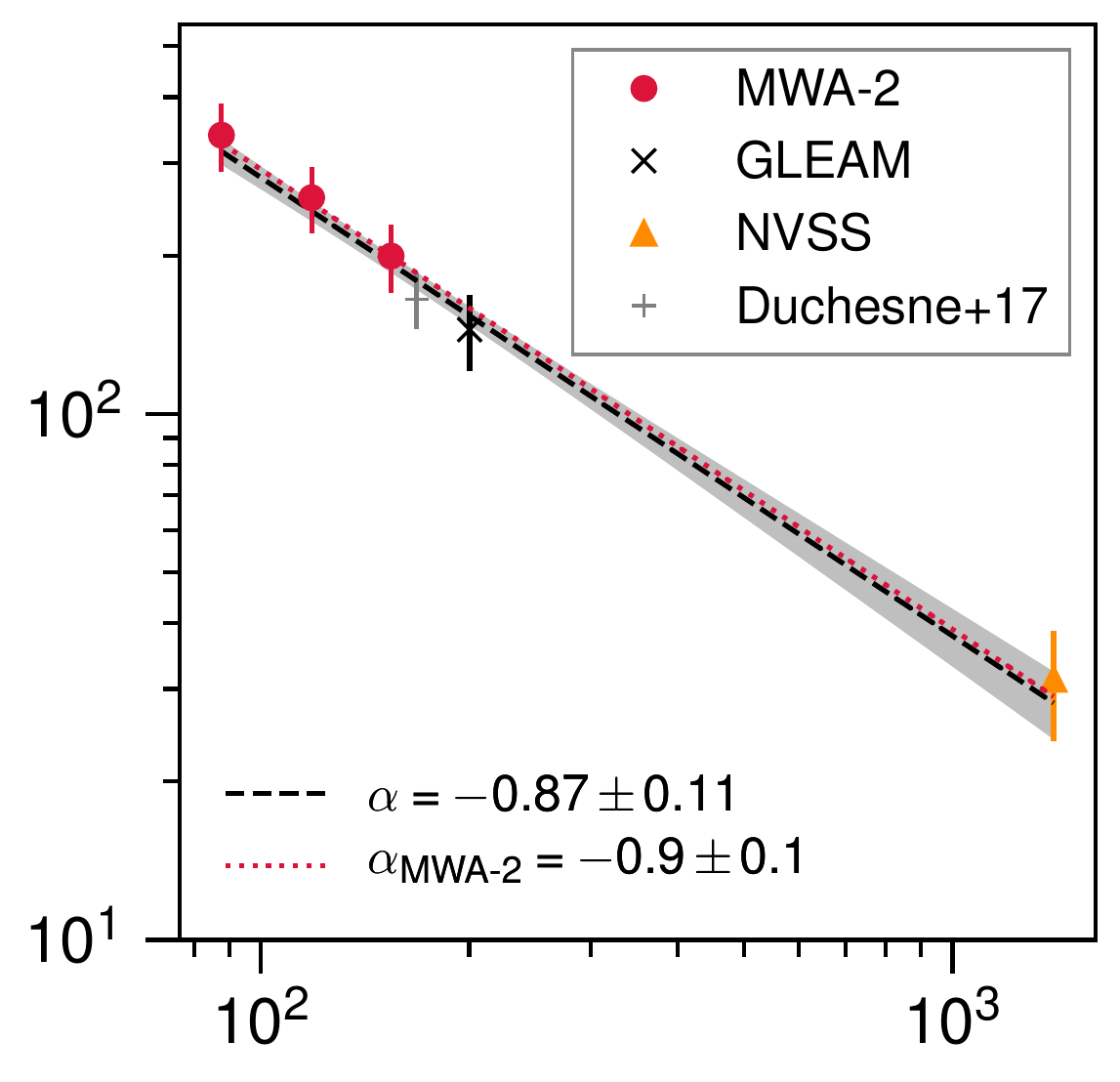}
\subcaption{\label{fig:sed:as1099} \hyperref[para:as1099]{Abell~S1099}, D1.}
\end{subfigure}\\[0.5em]%
\begin{subfigure}[b]{0.25\linewidth}
\includegraphics[width=1\linewidth]{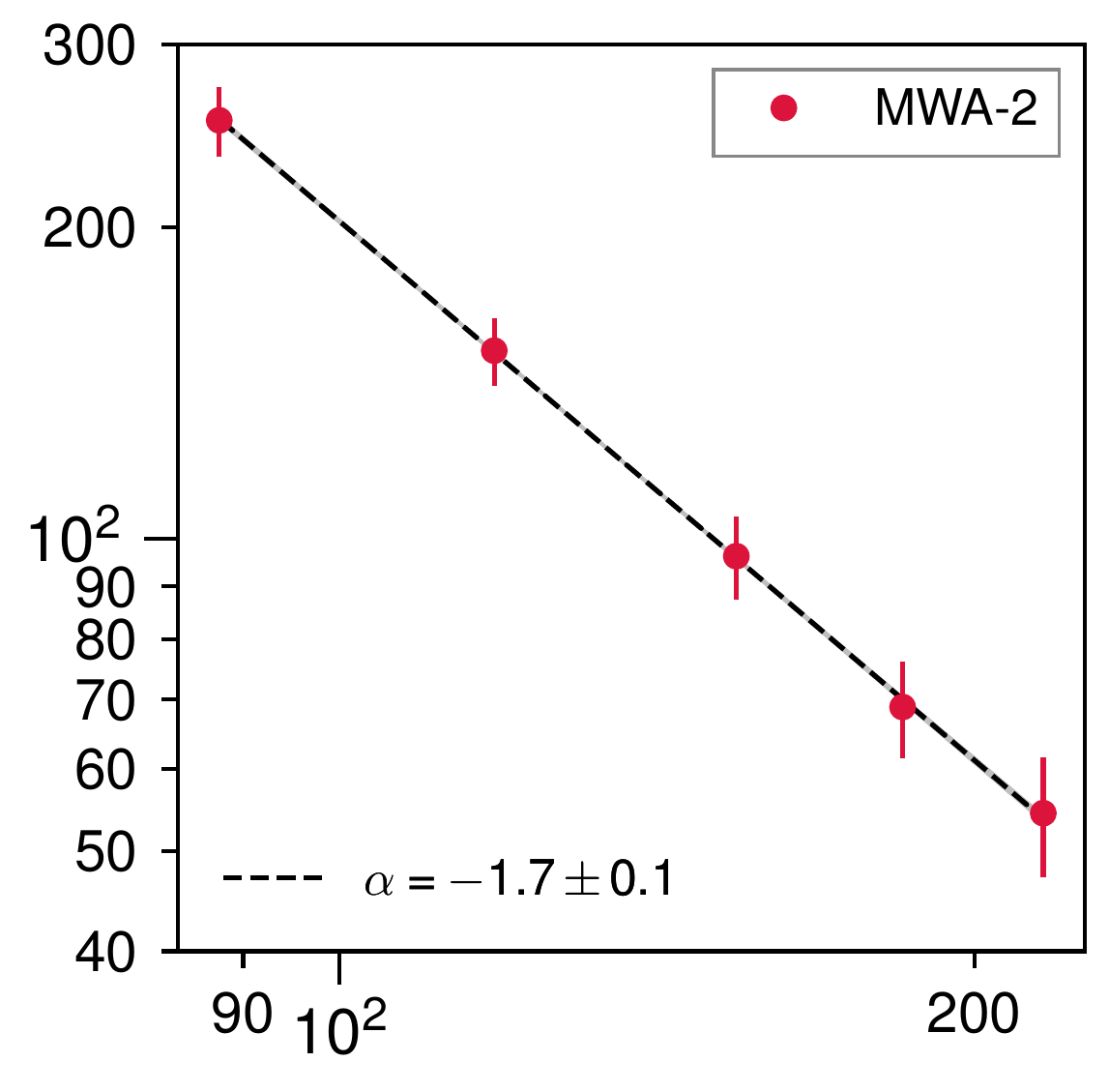}
\subcaption{\label{fig:sed:aqrcc} \hyperref[para:aqrcc]{AqrCC~087}, D1}
\end{subfigure}%
\begin{subfigure}[b]{0.25\linewidth}
\includegraphics[width=1\linewidth]{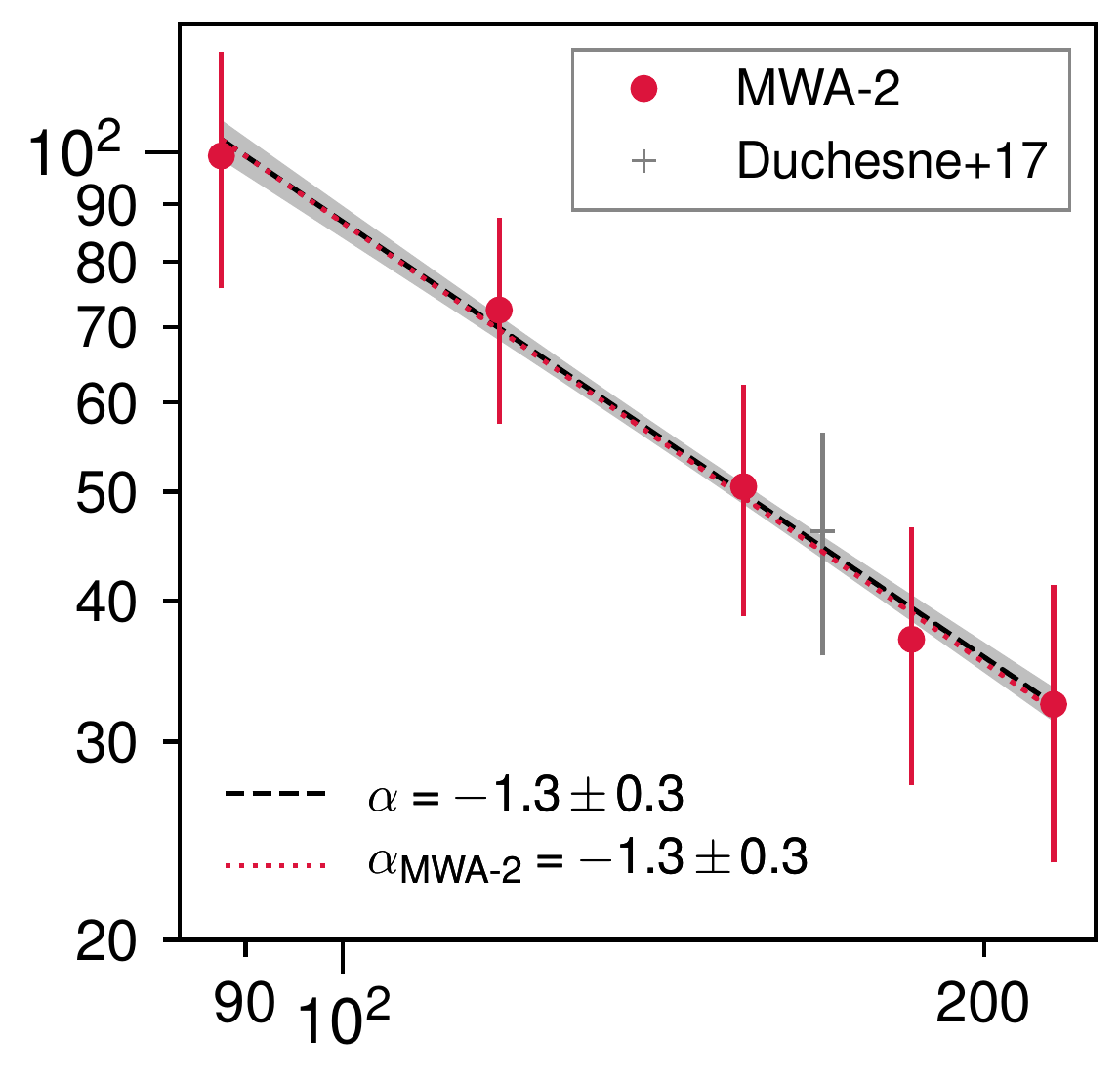}
\subcaption{\label{fig:sed:rxcja} \hyperref[para:rxcj]{RXC~J2351.0$-$1954}, D2}
\end{subfigure}%
\begin{subfigure}[b]{0.25\linewidth}
\includegraphics[width=1\linewidth]{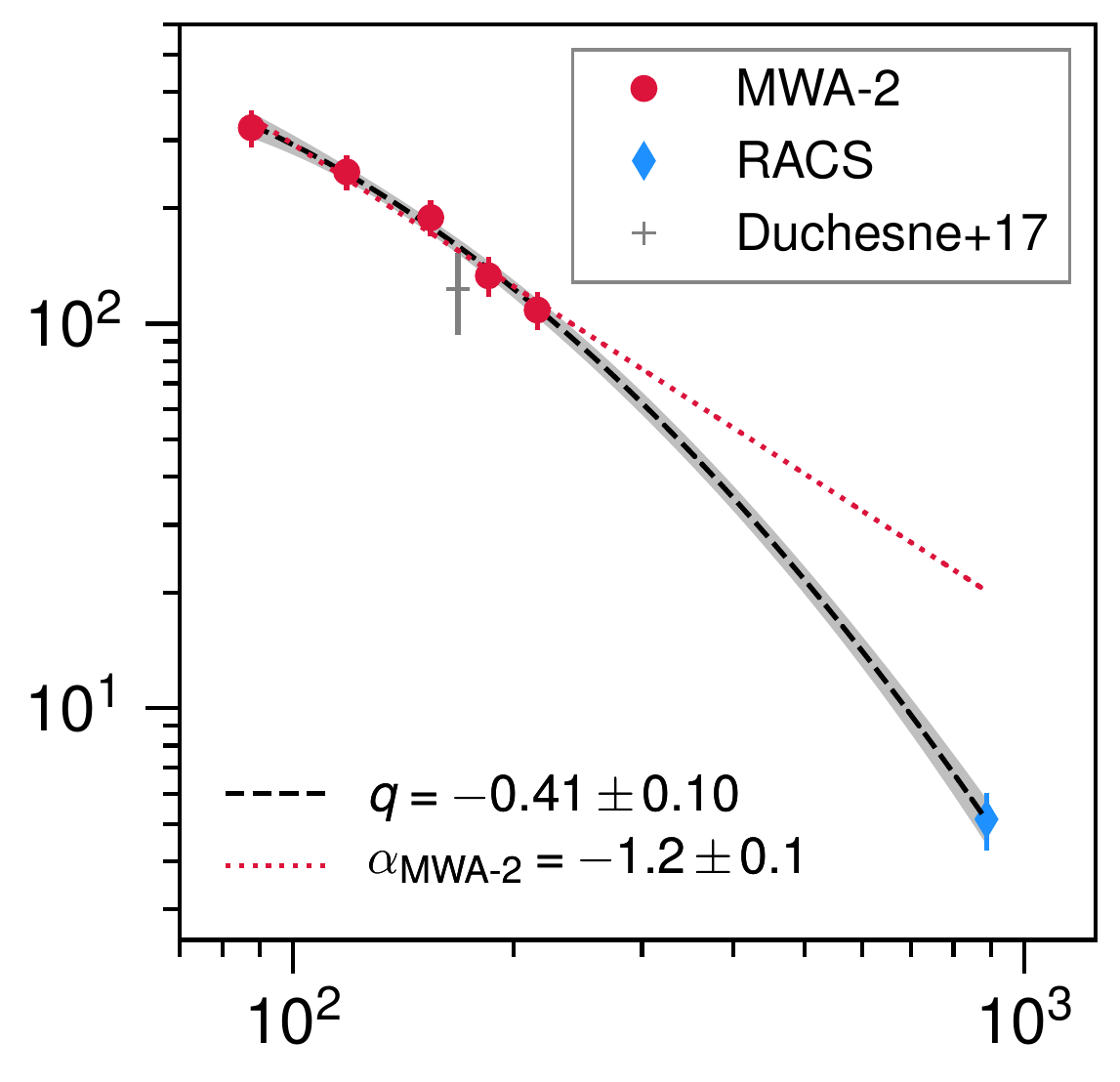}
\subcaption{\label{fig:sed:rxcjb} \hyperref[para:rxcj]{RXC~J2351.0$-$1954}, D3}
\end{subfigure}%
\begin{subfigure}[b]{0.25\linewidth}
\includegraphics[width=1\linewidth]{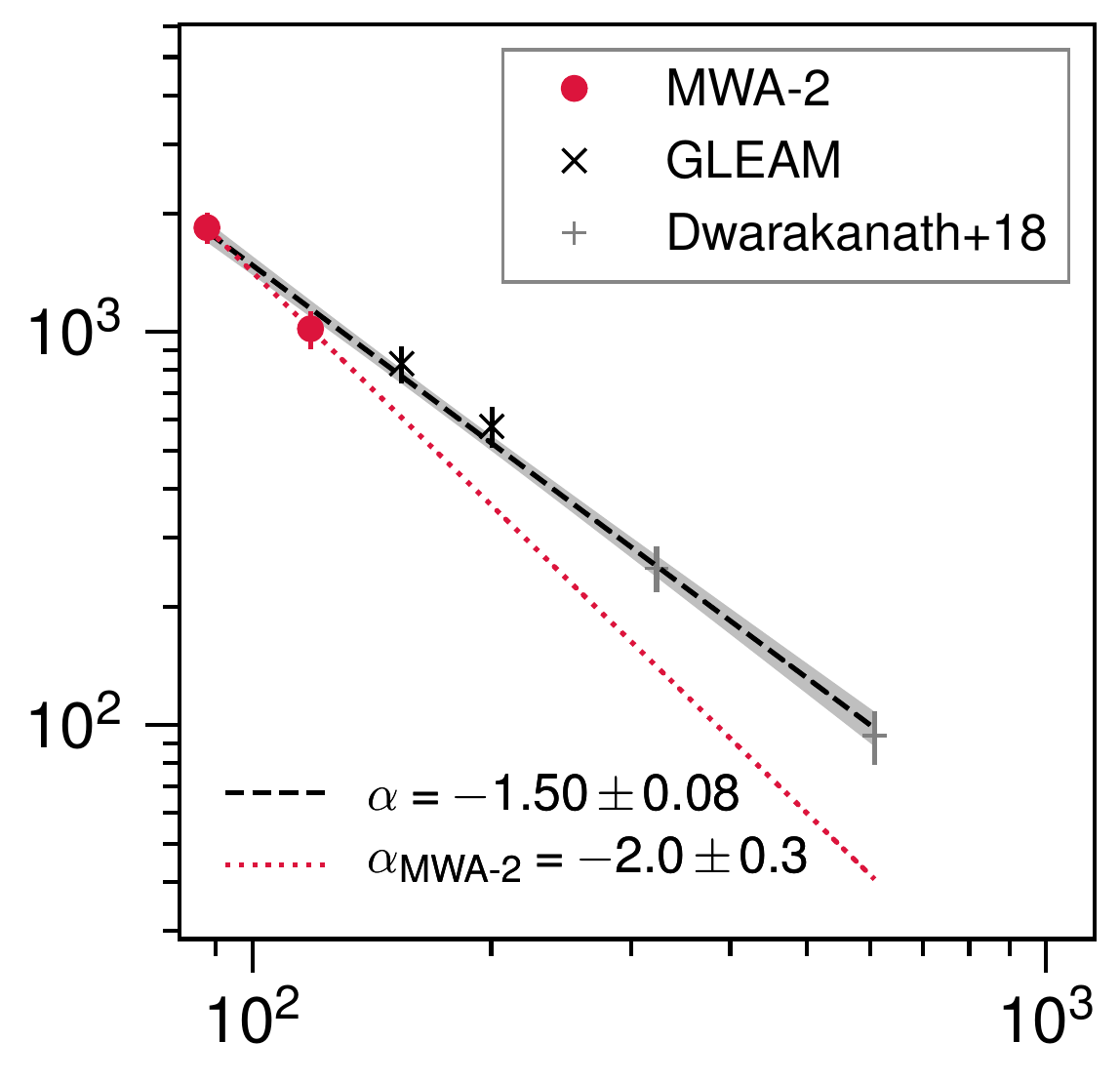}
\subcaption{\label{fig:sed:a168} \hyperref[para:a168]{Abell~0168}, D1.}
\end{subfigure}\\[0.5em]%
\begin{subfigure}[b]{0.25\linewidth}
\includegraphics[width=1\linewidth]{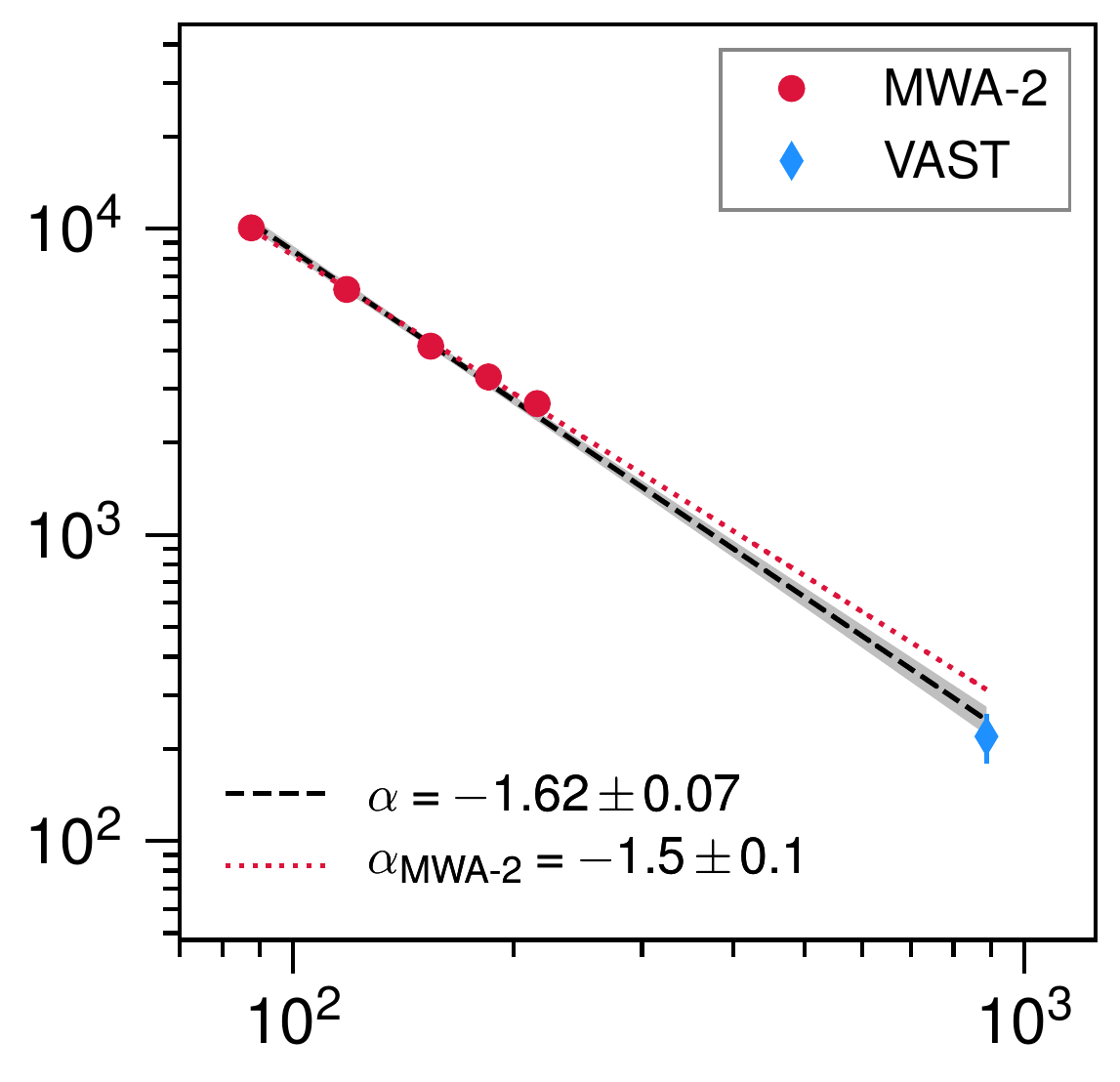}
\subcaption{\label{fig:sed:mcxcj137:1} \hyperref[para:mcxc137]{RXC~J0137.2$-$0912}, D1.}
\end{subfigure}%
\begin{subfigure}[b]{0.25\linewidth}
\includegraphics[width=1\linewidth]{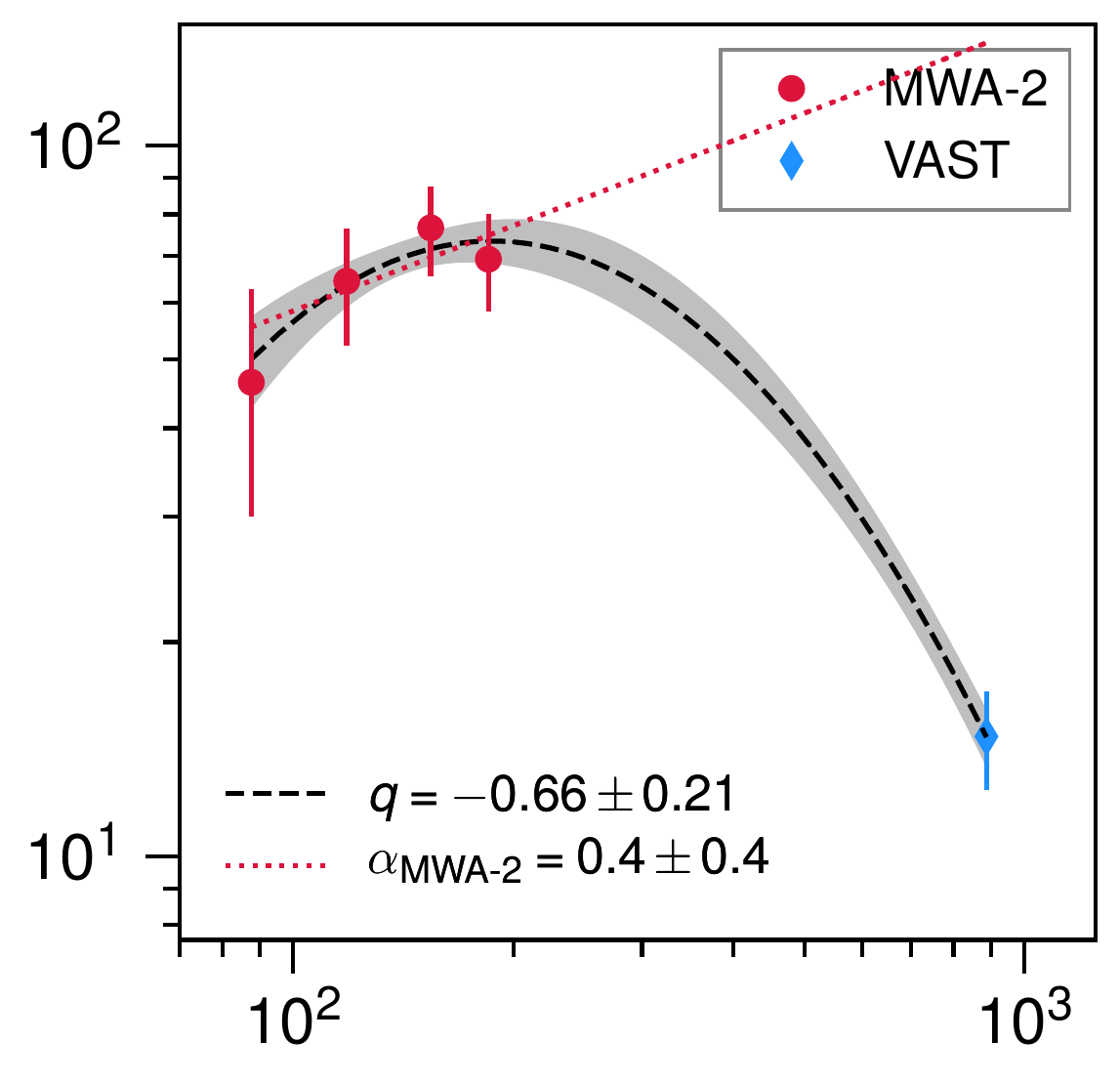}
\subcaption{\label{fig:sed:mcxcj137:2} \hyperref[para:mcxc137]{RXC~J0137.2$-$0912}, D2.}
\end{subfigure}%
\caption{\label{fig:sed} Integrated spectra of diffuse sources described in \cref{sec:sources} in the order they are reported. The ordinate and abscissa are integrated flux density (mJy) and frequency \corrs{(MHz)}, respectively. Individual measurements are reported in the online table described in \cref{sec:app:table}. The dashed, black lines are the fits for the full set of measurements (with grey, shaded regions corresponding to 95\% confidence intervals) and the red, dotted lines are fits for only the MWA-2 data. Note black arrows represent limits. Note if only two data points are available, a two-point spectral index was calculated and the resultant line is drawn based on that spectral index. For curved power law spectra, we report the curvature, $q$, rather than the equivalent spectral index.}

\end{figure*}%

\begin{figure*}\ContinuedFloat
\centering
\begin{subfigure}[b]{0.25\linewidth}
\includegraphics[width=1\linewidth]{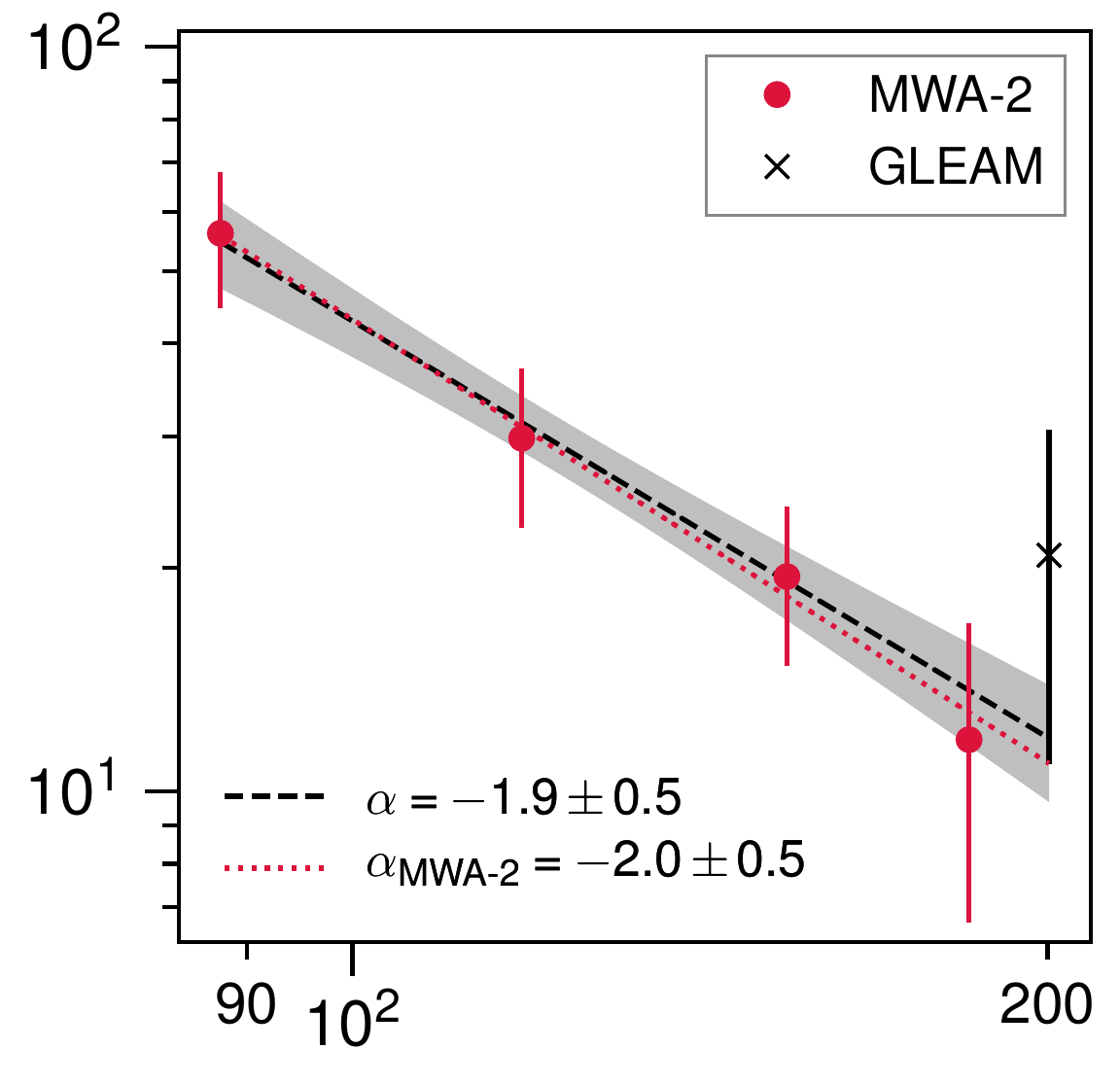}
\subcaption{\label{fig:sed:as112} \hyperref[para:as112]{Abell~S0112}, D1}
\end{subfigure}%
\begin{subfigure}[b]{0.25\linewidth}
\includegraphics[width=1\linewidth]{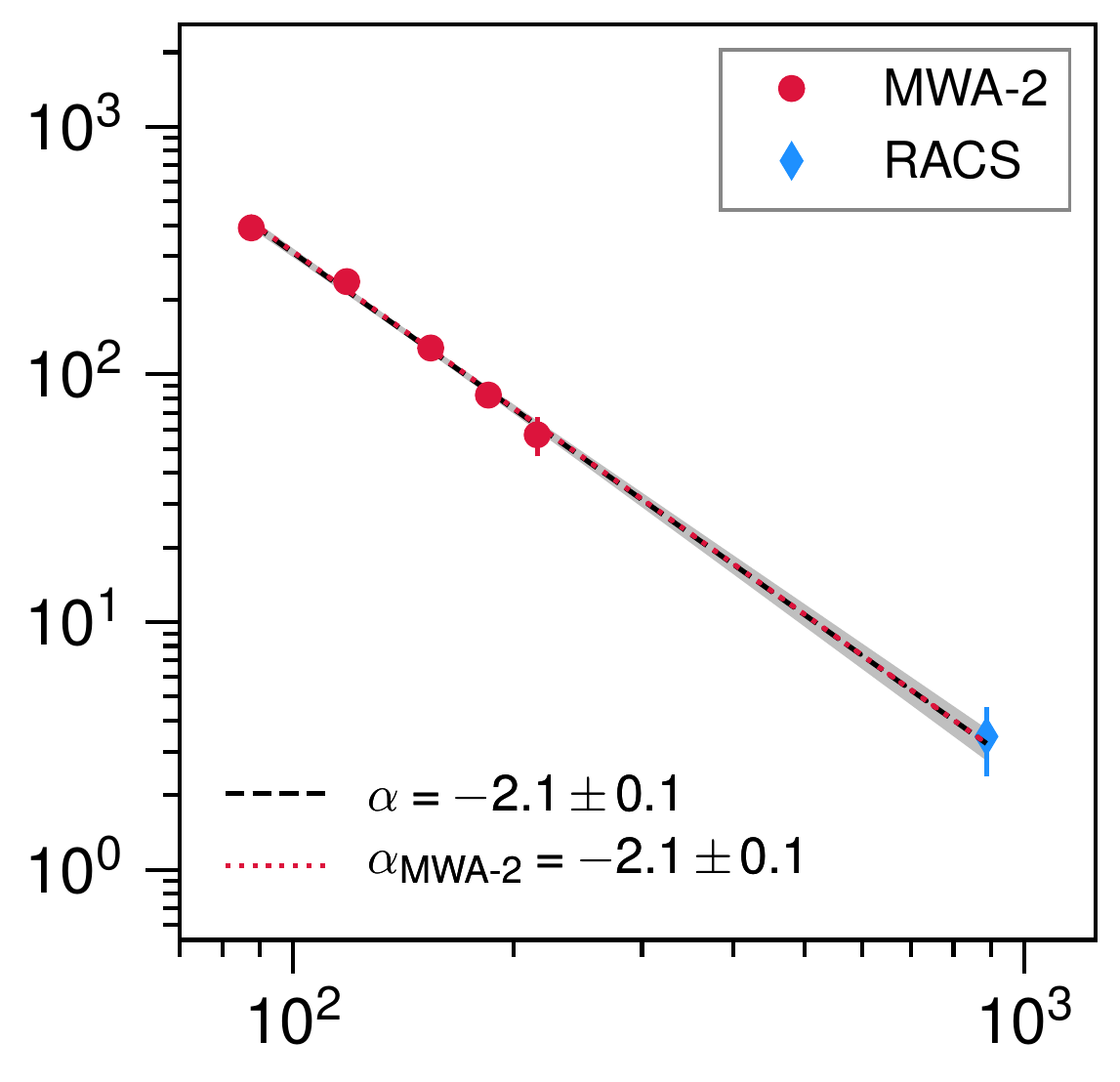}
\subcaption{\label{fig:sed:mcxcj145} \hyperref[para:mcxc145]{MCXC~J0145.2$-$6033}, D1.}
\end{subfigure}%
\begin{subfigure}[b]{0.25\linewidth}
\includegraphics[width=1\linewidth]{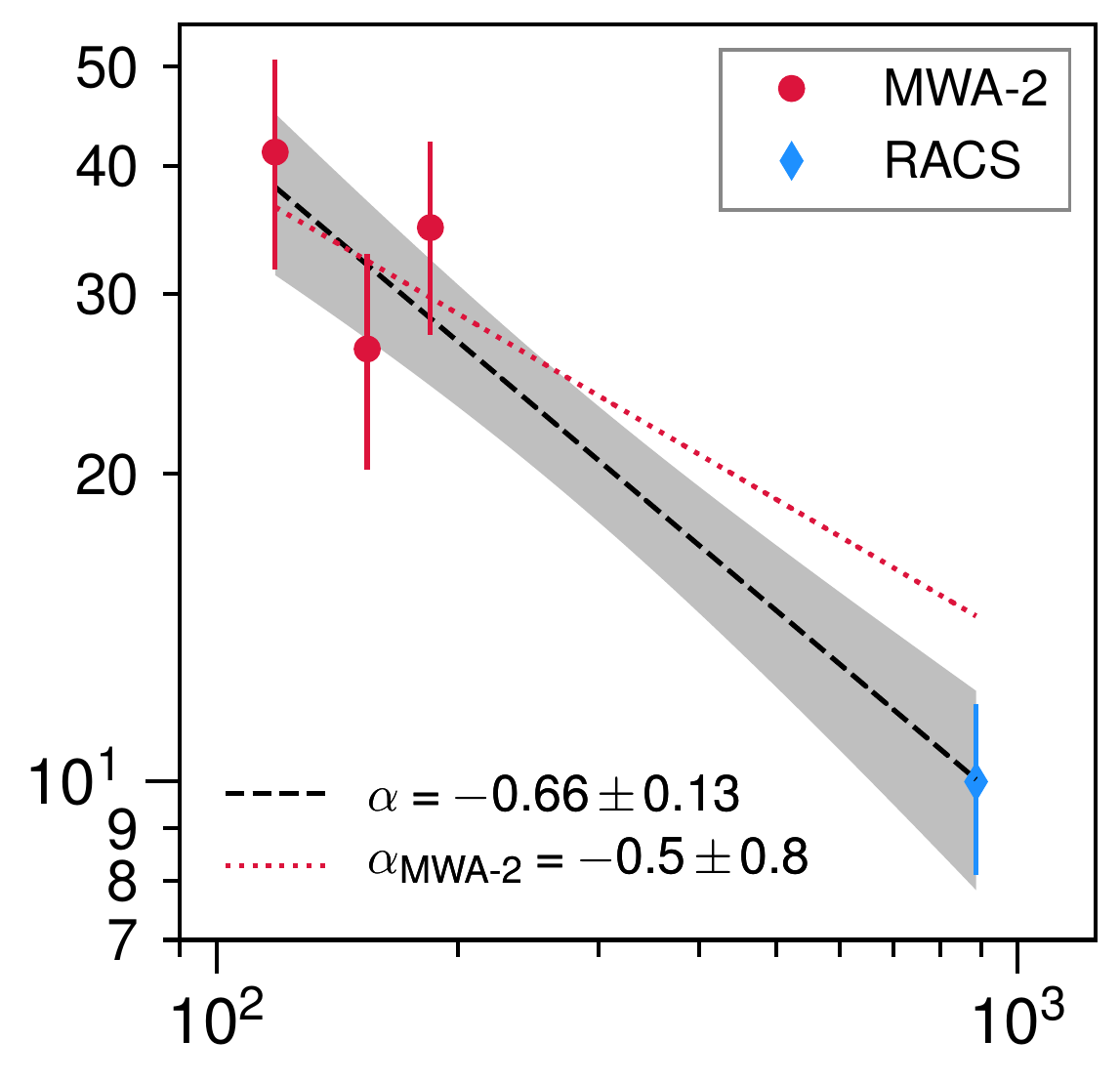}
\subcaption{\label{fig:sed:mcxcj154} \hyperref[para:mcxc154]{MCXC~J0154.2$-$5937}, D1.}
\end{subfigure}%
\begin{subfigure}[b]{0.25\linewidth}
\includegraphics[width=1\linewidth]{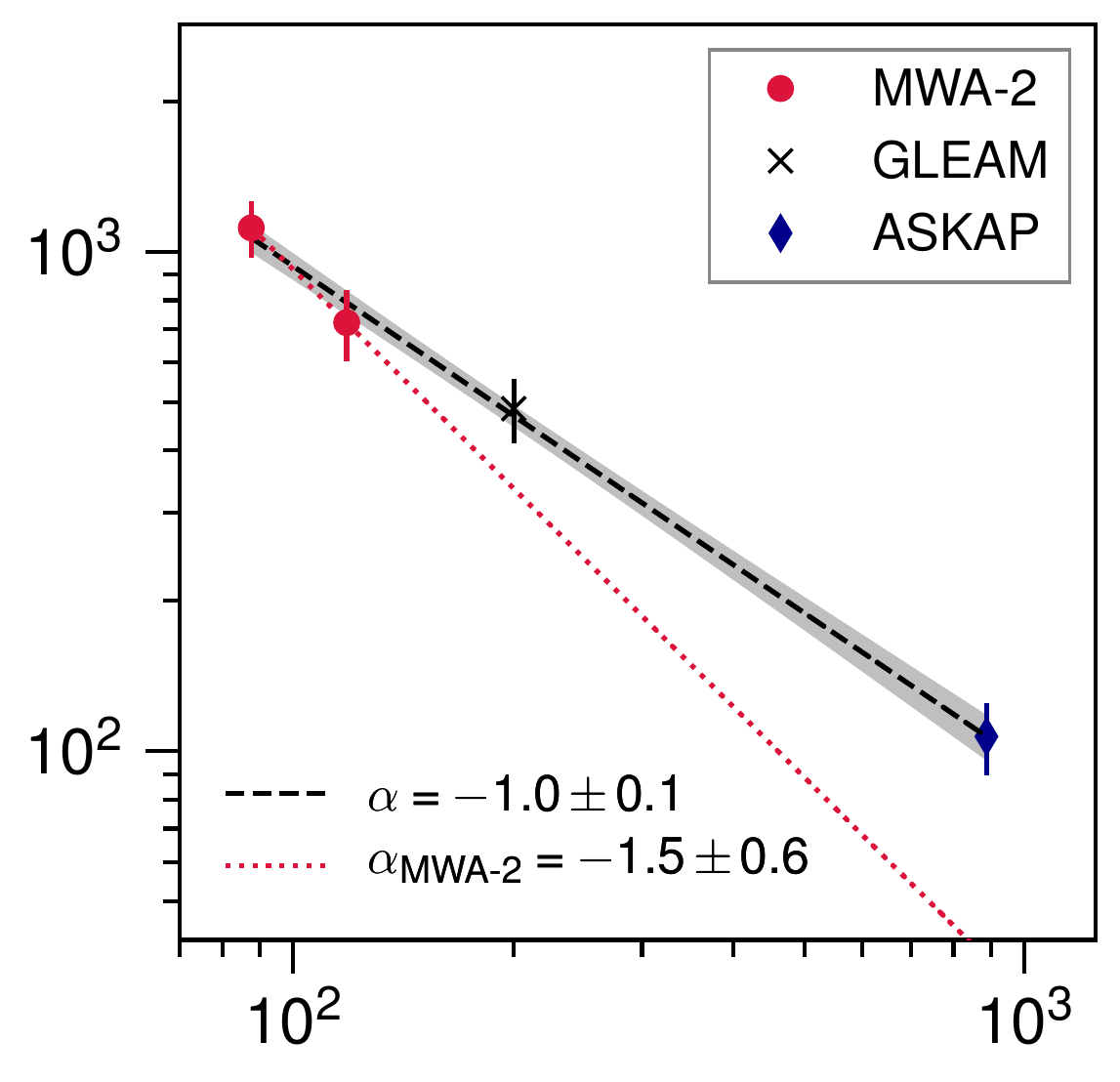}
\subcaption{\label{fig:sed:a3186a} \hyperref[para:a3186]{Abell~3186}, D1.}
\end{subfigure}\\[0.5em]%
\begin{subfigure}[b]{0.25\linewidth}
\includegraphics[width=1\linewidth]{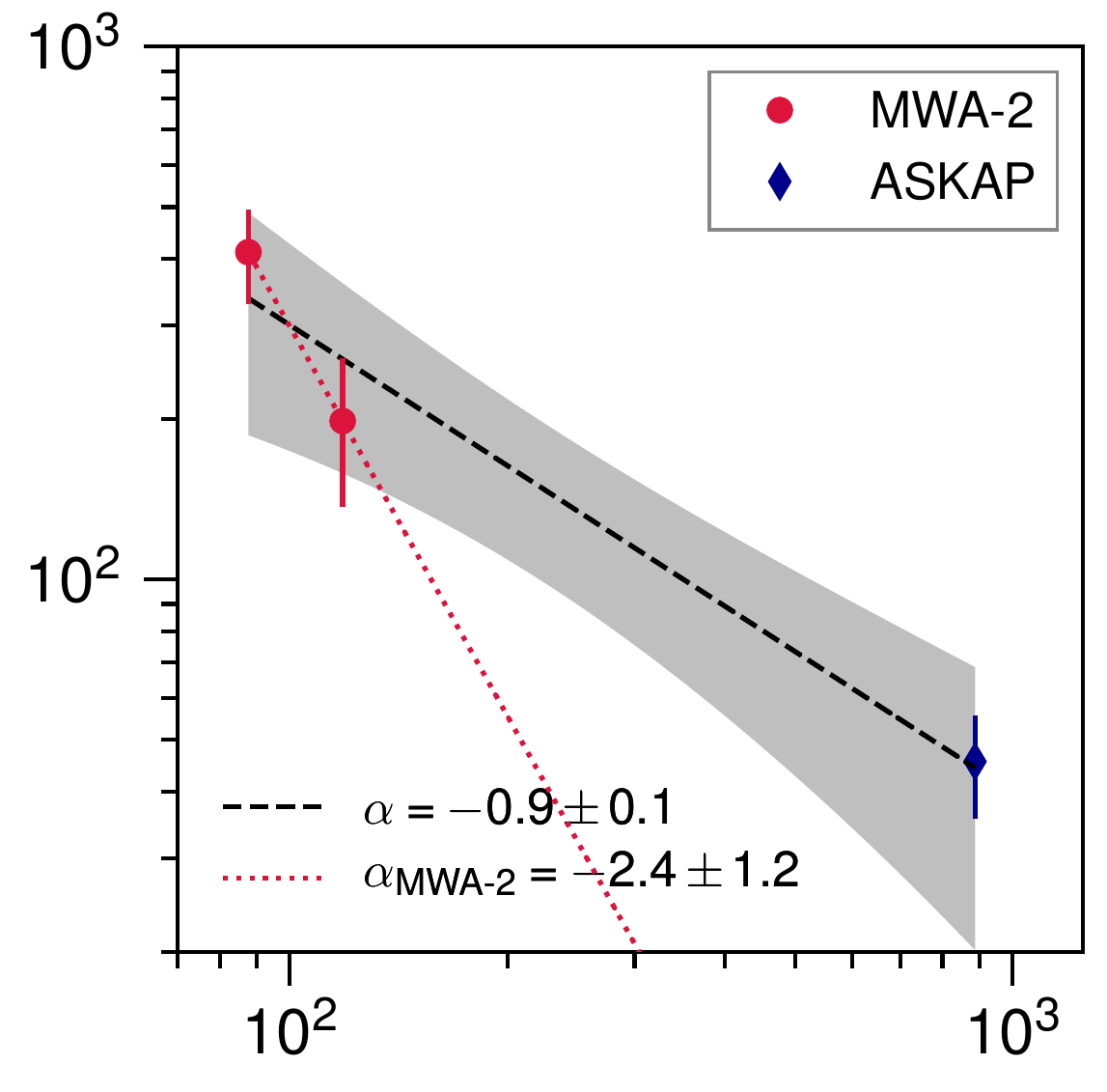}
\caption{\label{fig:sed:a3186b} \hyperref[para:a3186]{Abell~3186}, D2.}
\end{subfigure}%
\begin{subfigure}[b]{0.25\linewidth}
\includegraphics[width=1\linewidth]{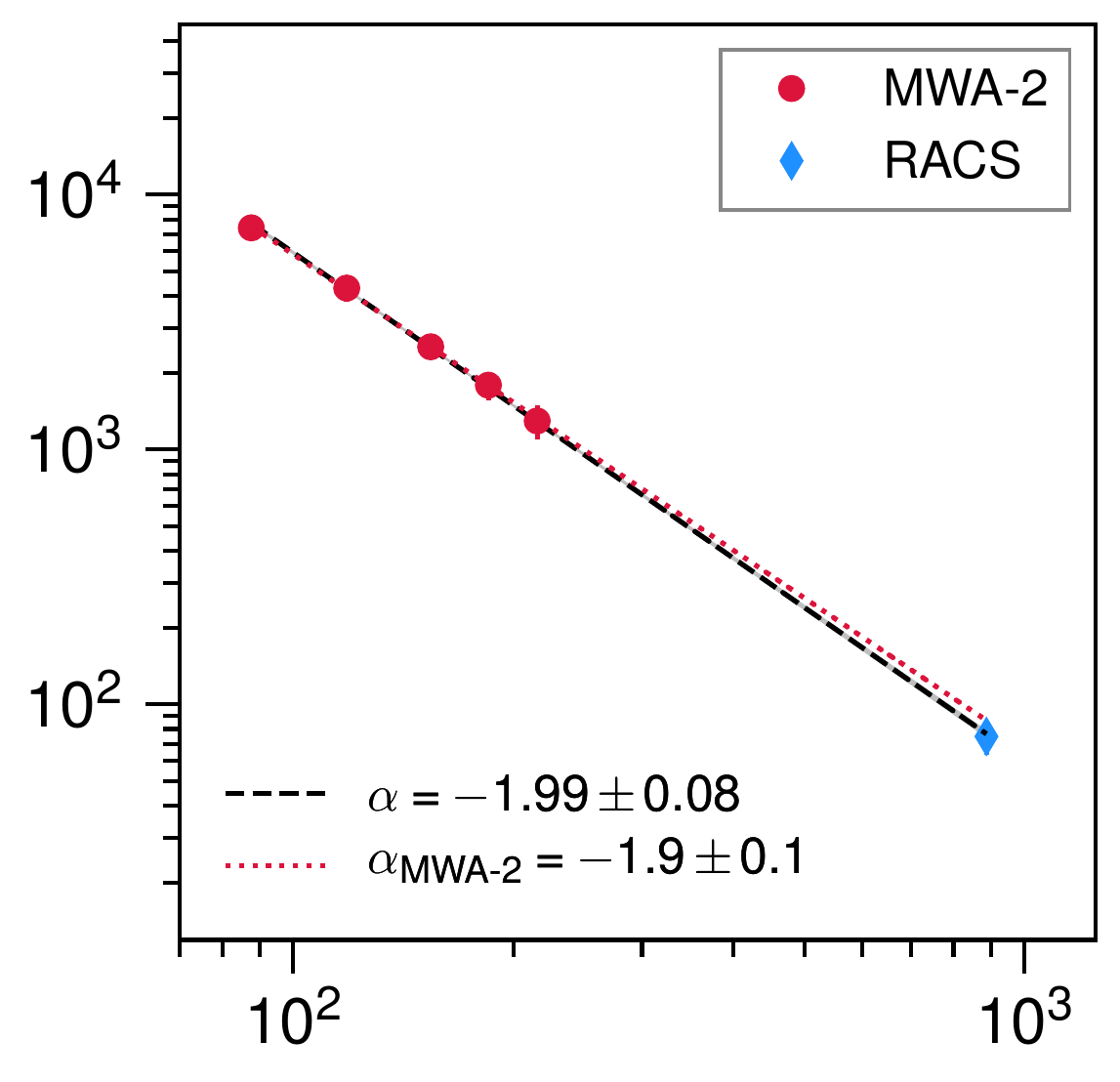}
\subcaption{\label{fig:sed:as405} \hyperref[para:as405]{Abell~S0405}, D1.}
\end{subfigure}%
\begin{subfigure}[b]{0.25\linewidth}
\includegraphics[width=1\linewidth]{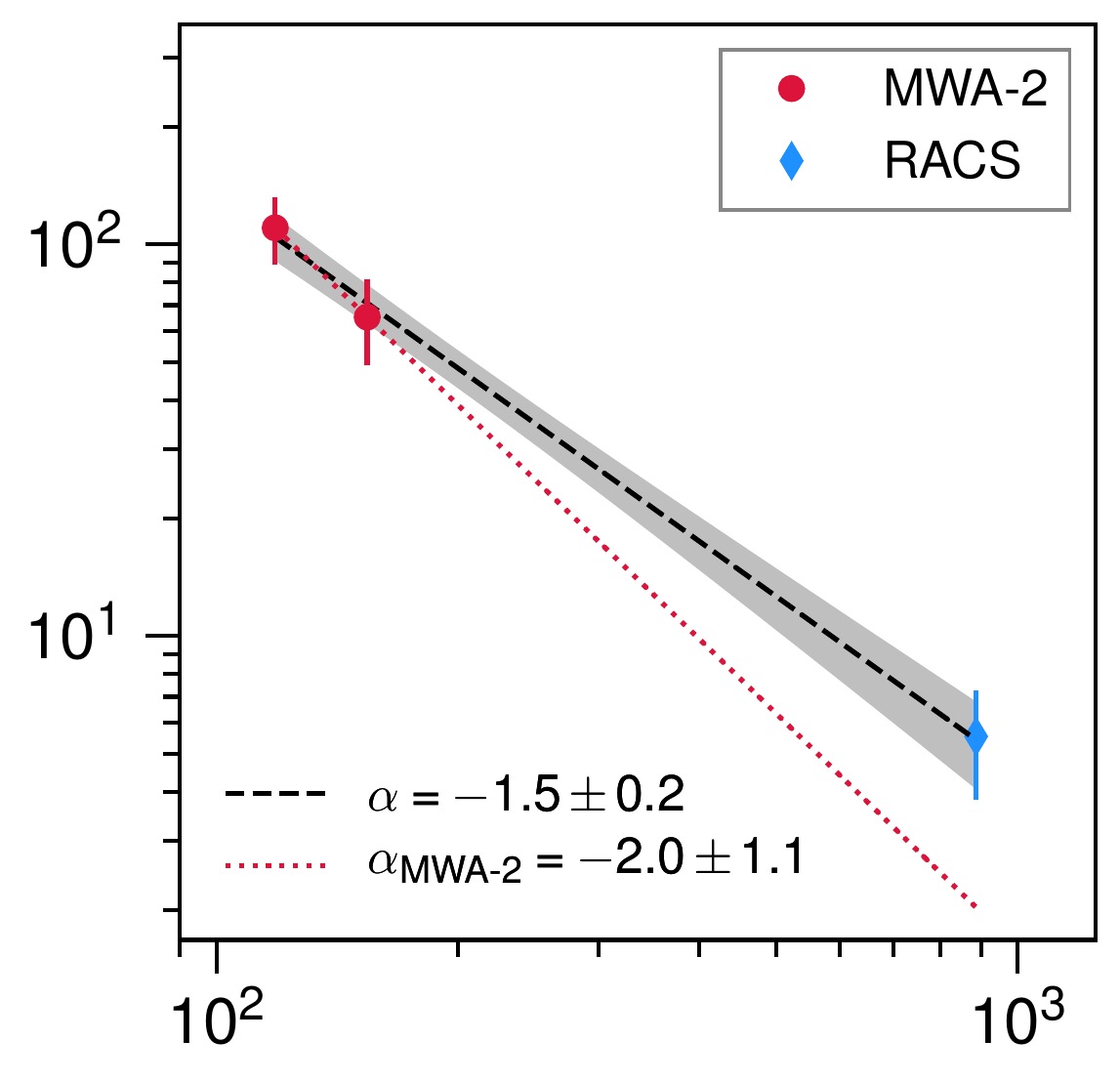}
\subcaption{\label{fig:sed:psz1g287} \hyperref[para:psz1g287]{PSZ1 G287.95$-$32.98}, D1.}
\end{subfigure}%
\begin{subfigure}[b]{0.25\linewidth}
\includegraphics[width=1\linewidth]{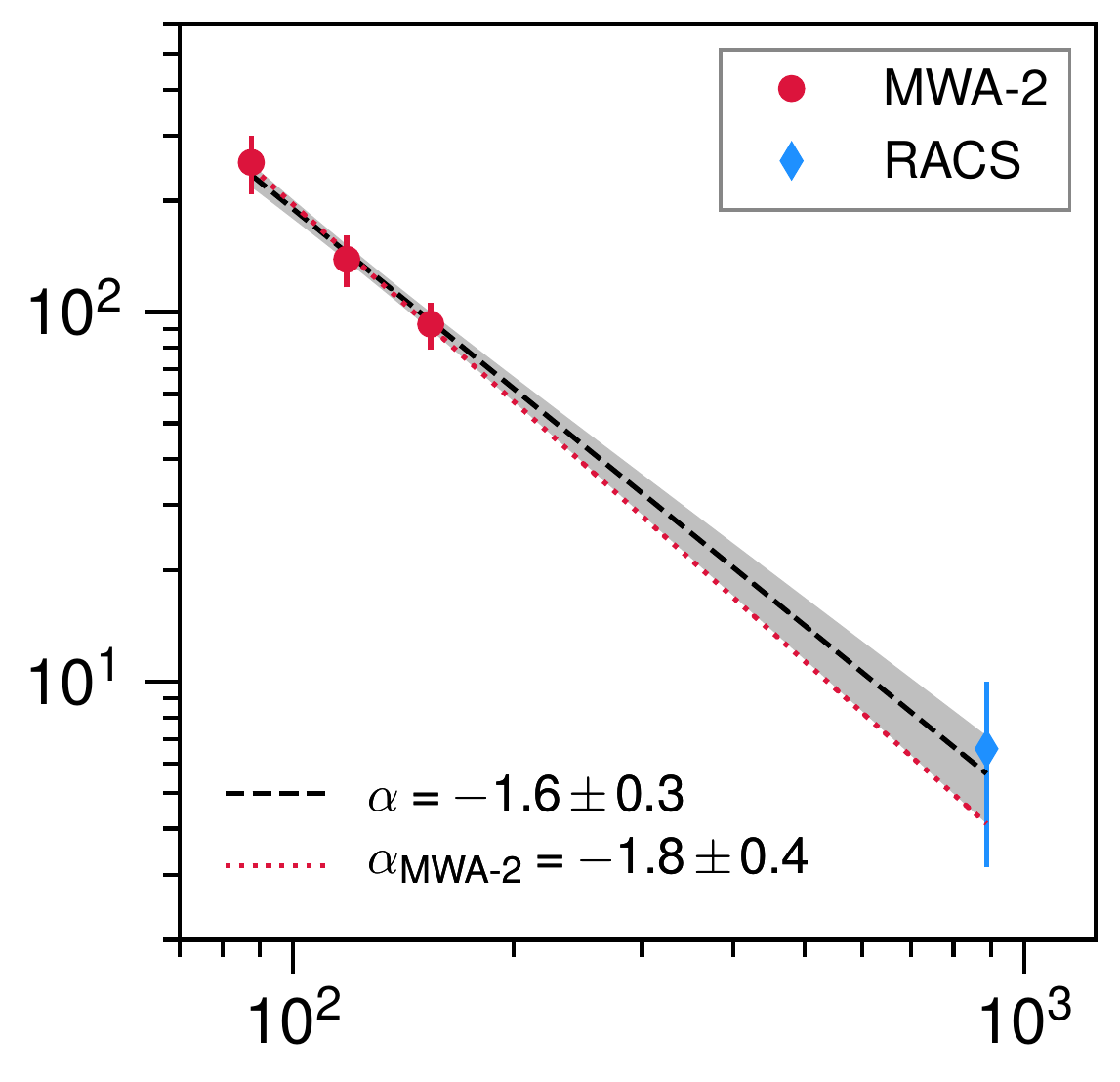}
\subcaption{\label{fig:sed:a3399a} \hyperref[para:a3399]{Abell~3399}, D1.}
\end{subfigure}\\[0.5em]%
\begin{subfigure}[b]{0.25\linewidth}
\includegraphics[width=1\linewidth]{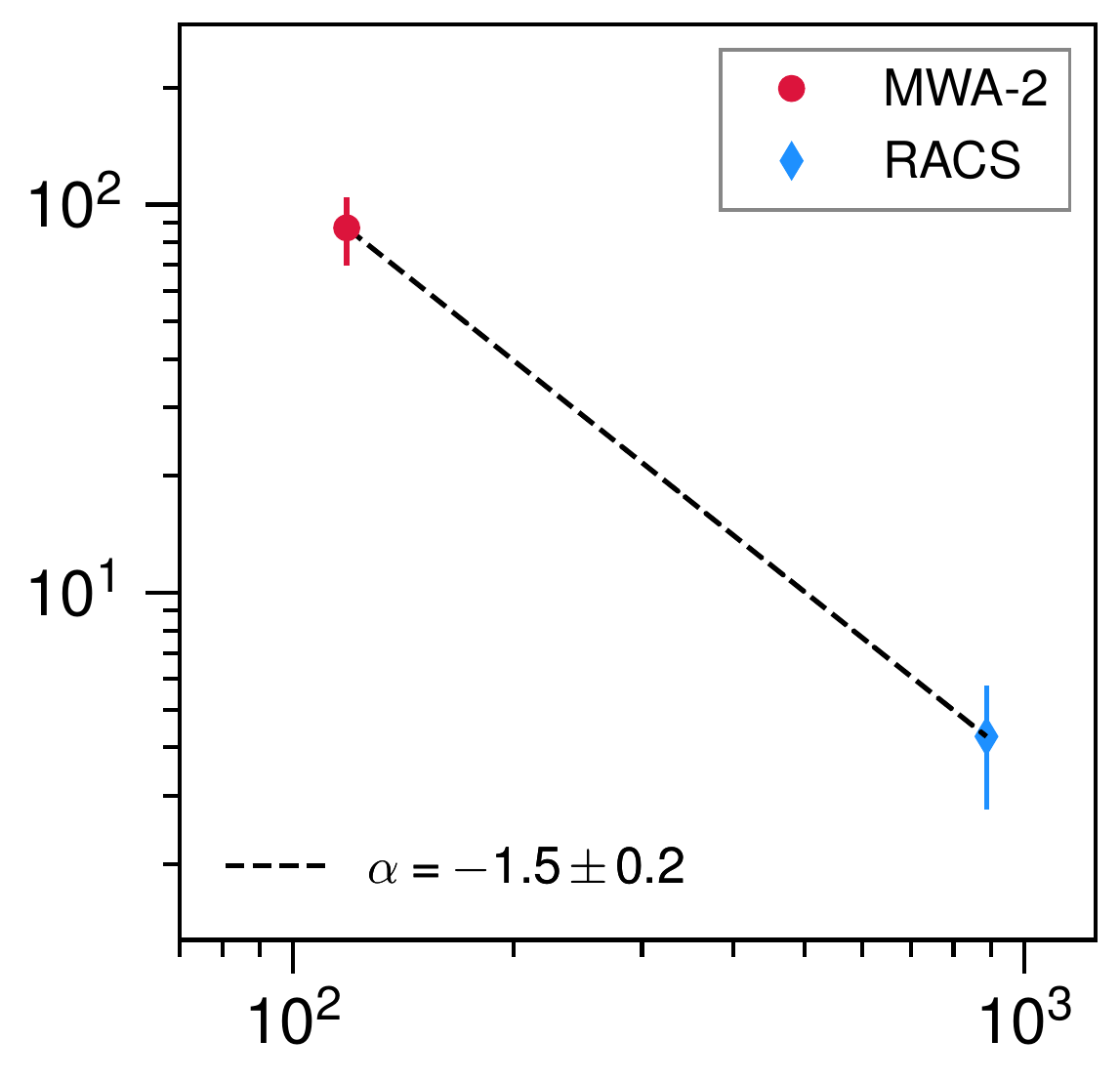}
\subcaption{\label{fig:sed:a3399b} \hyperref[para:a3399]{Abell~3399}, D2}
\end{subfigure}%
\begin{subfigure}[b]{0.25\linewidth}
\includegraphics[width=1\linewidth]{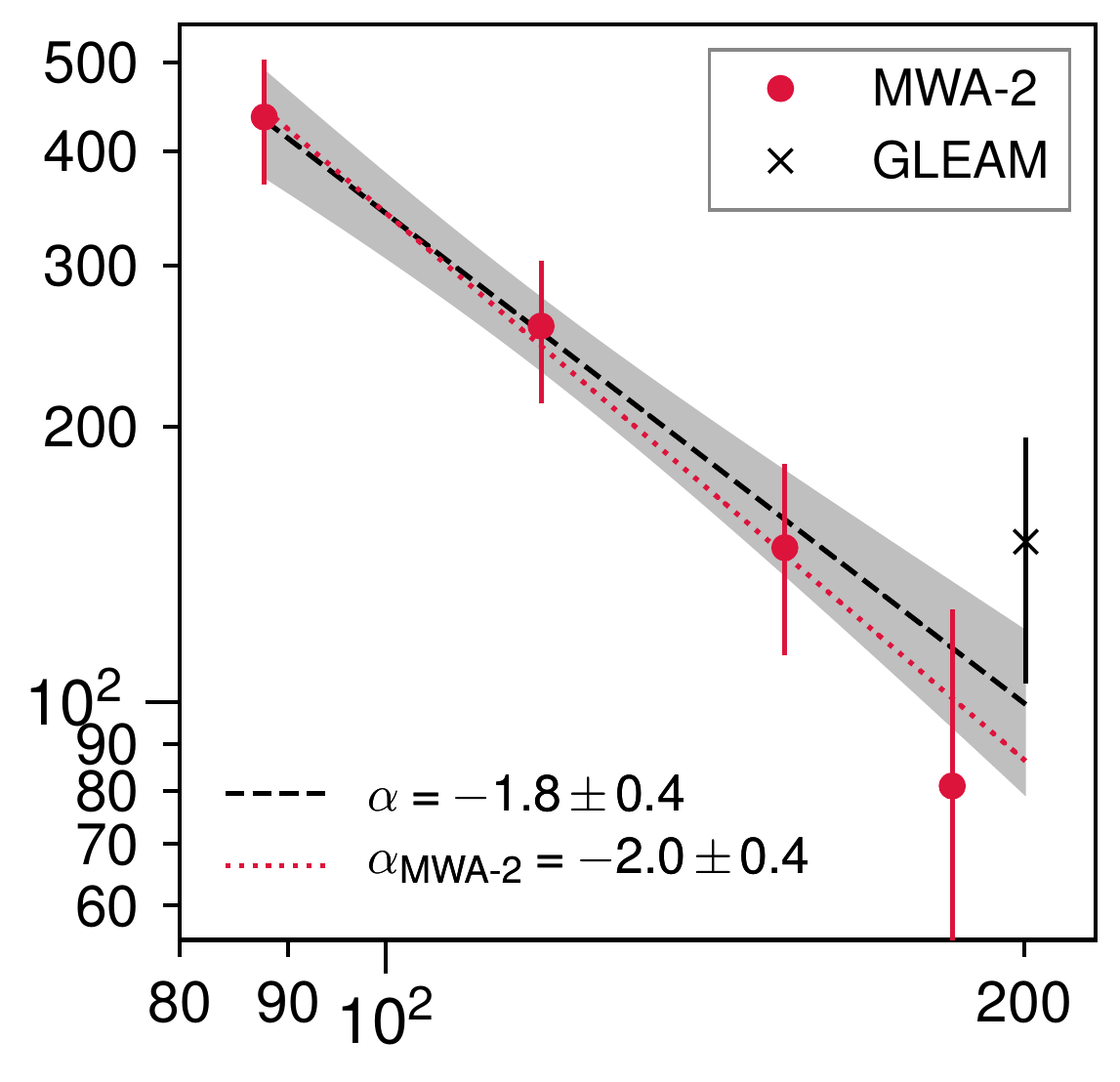}
\subcaption{\label{fig:sed:a1631} \hyperref[para:a1631]{Abell~1631}, D1.}
\end{subfigure}%
\begin{subfigure}[b]{0.25\linewidth}
\includegraphics[width=1\linewidth]{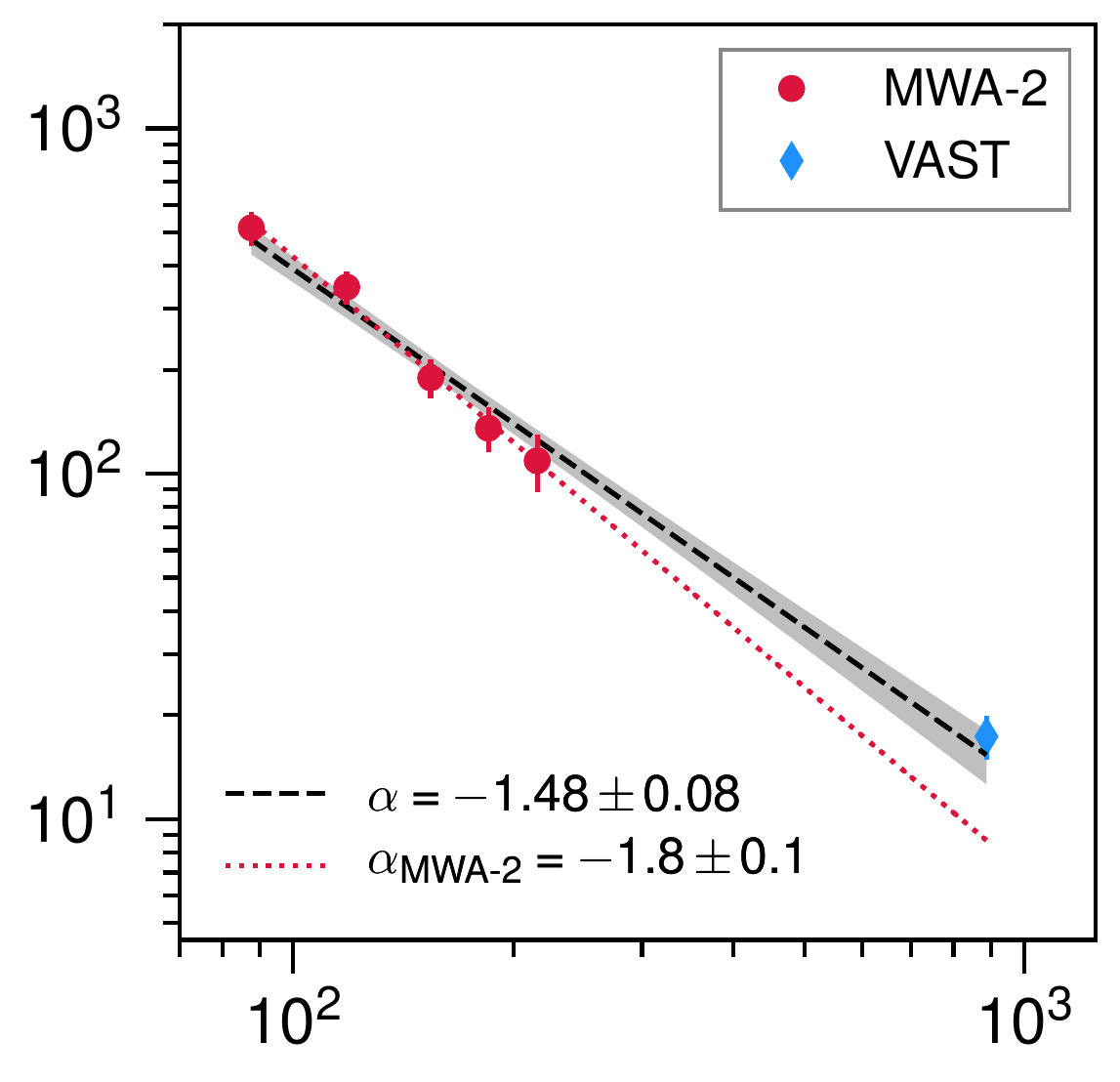}
\subcaption{\label{fig:sed:a3164a} \hyperref[para:a3164]{Abell~3164}, D1.}
\end{subfigure}%
\begin{subfigure}[b]{0.25\linewidth}
\includegraphics[width=1\linewidth]{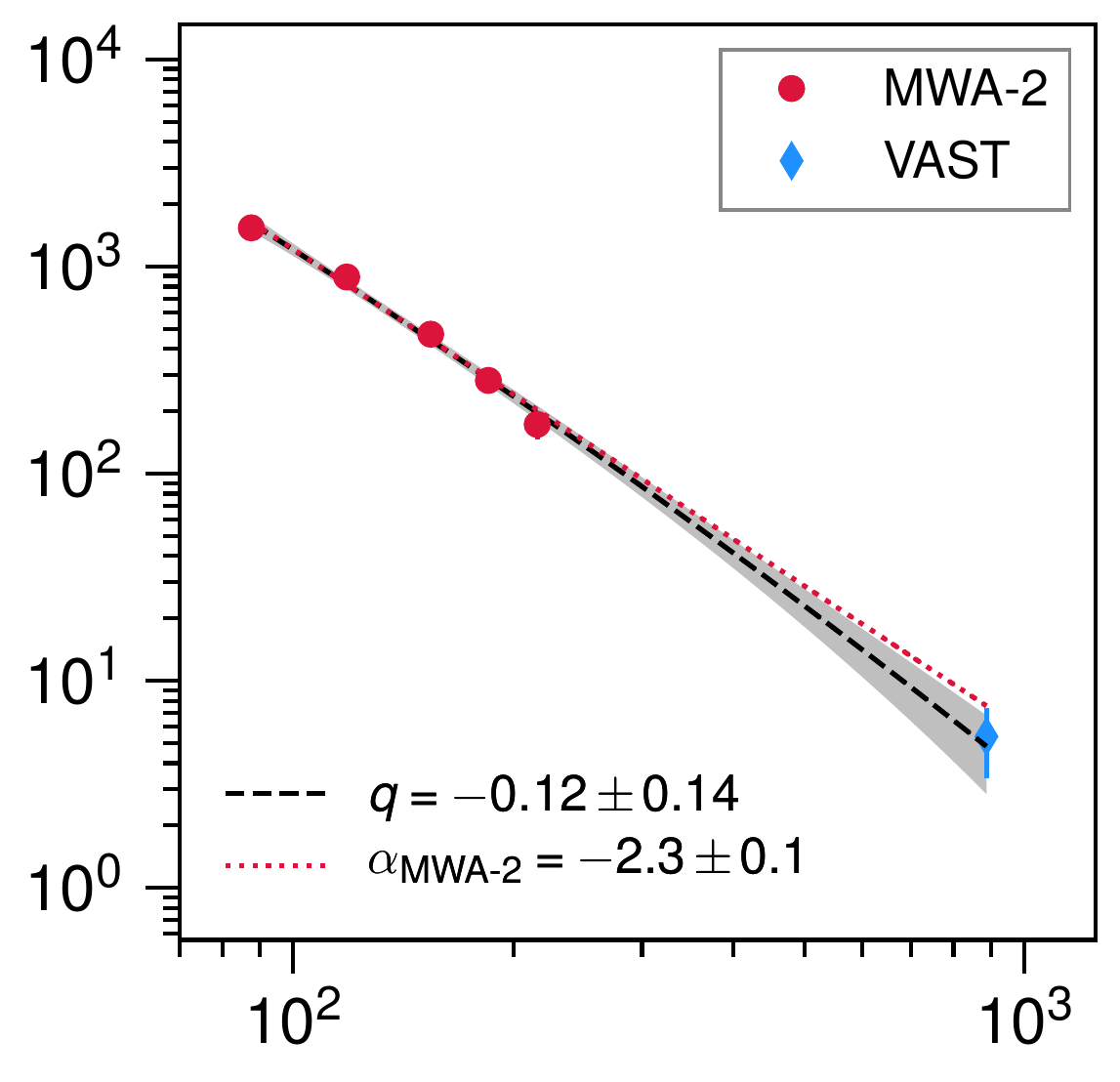}
\subcaption{\label{fig:sed:a3164b} \hyperref[para:a3164]{Abell~3164}, D2.}
\end{subfigure}\\[0.5em]%
\begin{subfigure}[b]{0.25\linewidth}
\includegraphics[width=1\linewidth]{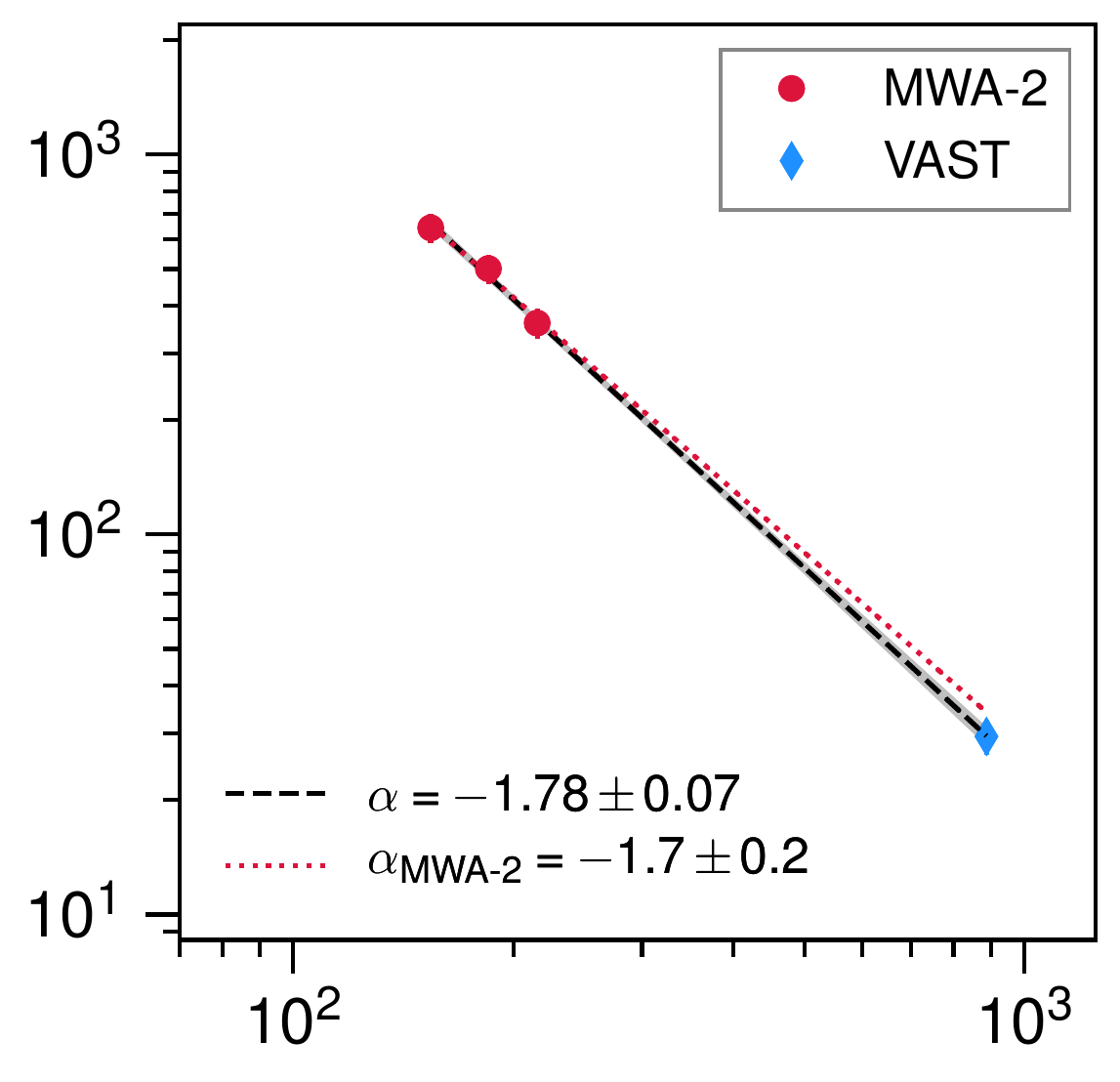}
\subcaption{\label{fig:sed:a3164c} \hyperref[para:a3164]{Abell~3164}, D3.}
\end{subfigure}%
\begin{subfigure}[b]{0.25\linewidth}
\includegraphics[width=1\linewidth]{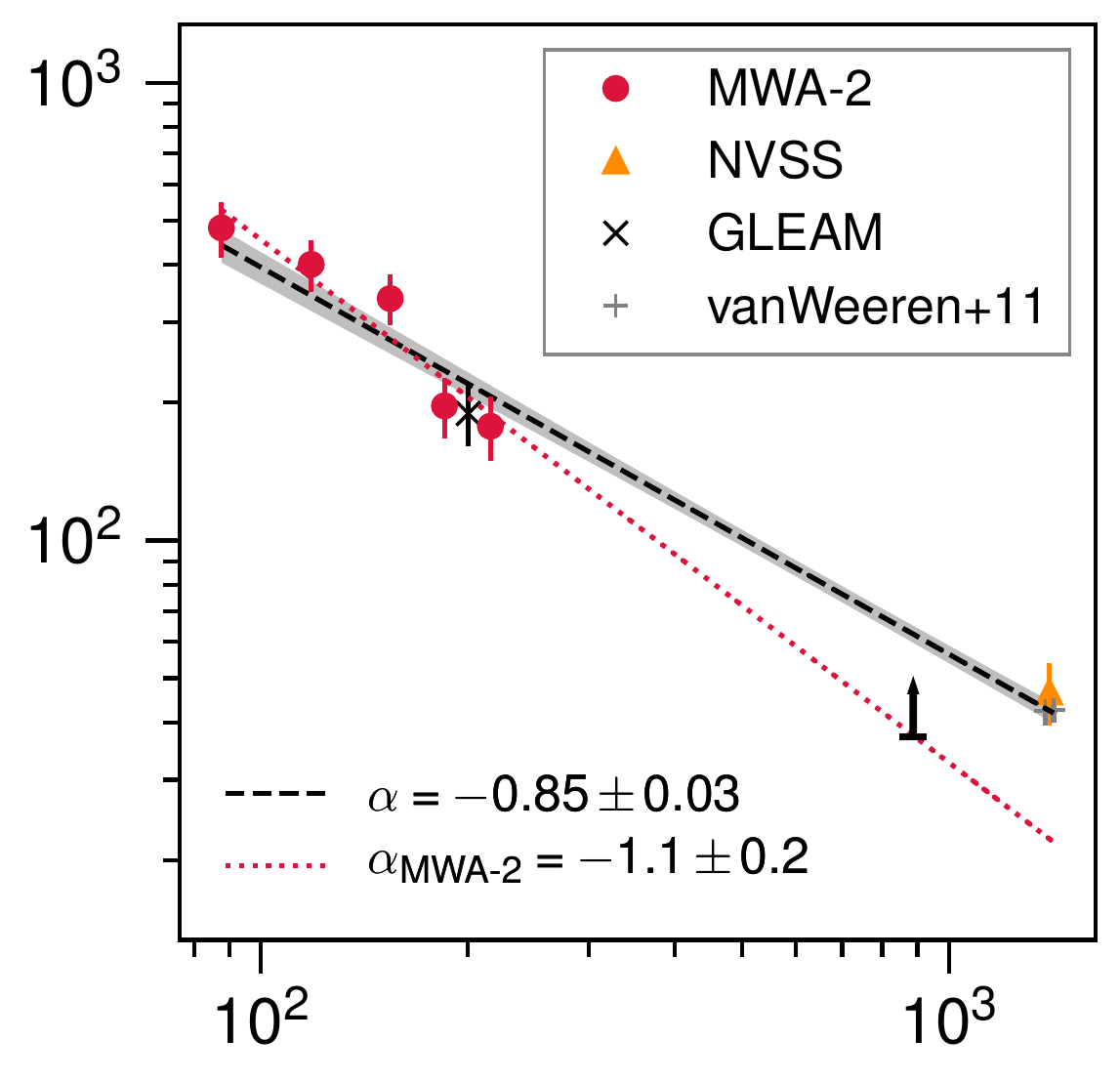}
\subcaption{\label{fig:sed:a3365a} \hyperref[para:a3365]{Abell~3365}, D1.}
\end{subfigure}%
\begin{subfigure}[b]{0.25\linewidth}
\includegraphics[width=1\linewidth]{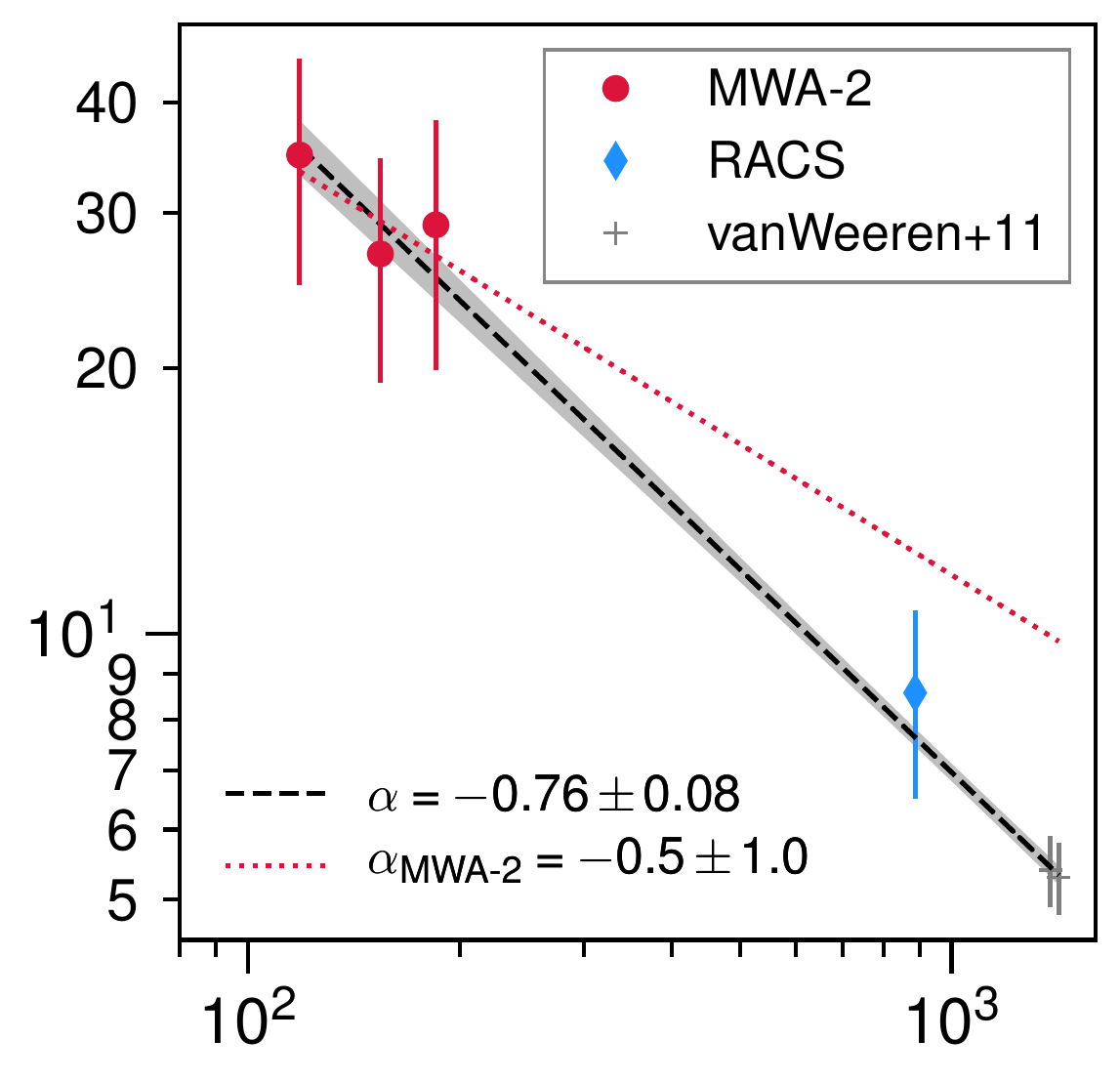}
\subcaption{\label{fig:sed:a3365b} \hyperref[para:a3365]{Abell~3365}, D2.}
\end{subfigure}%
\begin{subfigure}[b]{0.25\linewidth}
\includegraphics[width=1\linewidth]{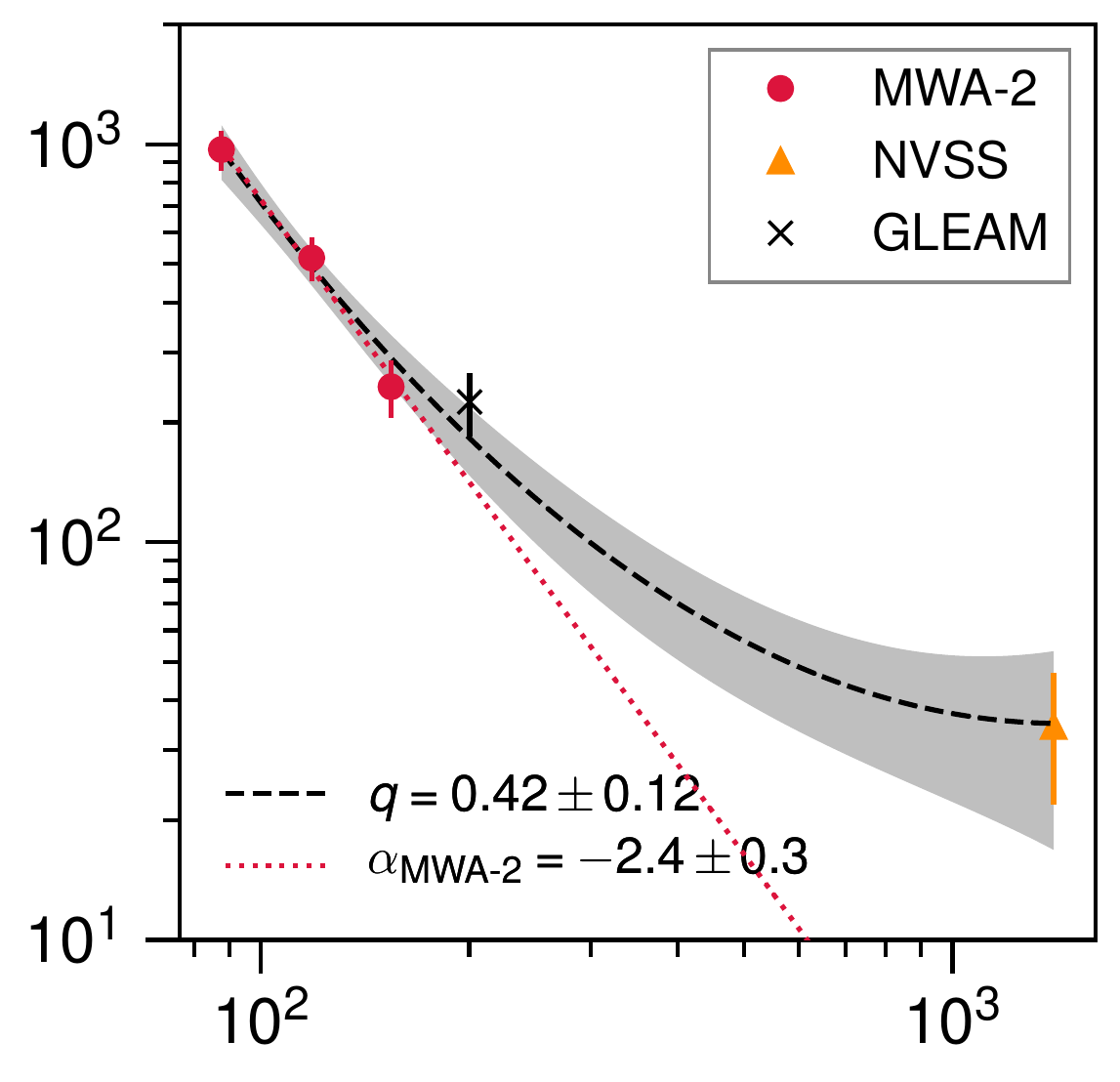}
\subcaption{\label{fig:sed:a550} \hyperref[para:a550]{Abell~0550}, D1.}
\end{subfigure}%
\caption{\emph{continued.}}
\end{figure*}

\section{Flux recovery in MWA-2 data}\label{app:dirty}
\begin{figure*}[b!]
\centering
\begin{subfigure}[b]{0.5\linewidth}
\includegraphics[width=1\linewidth]{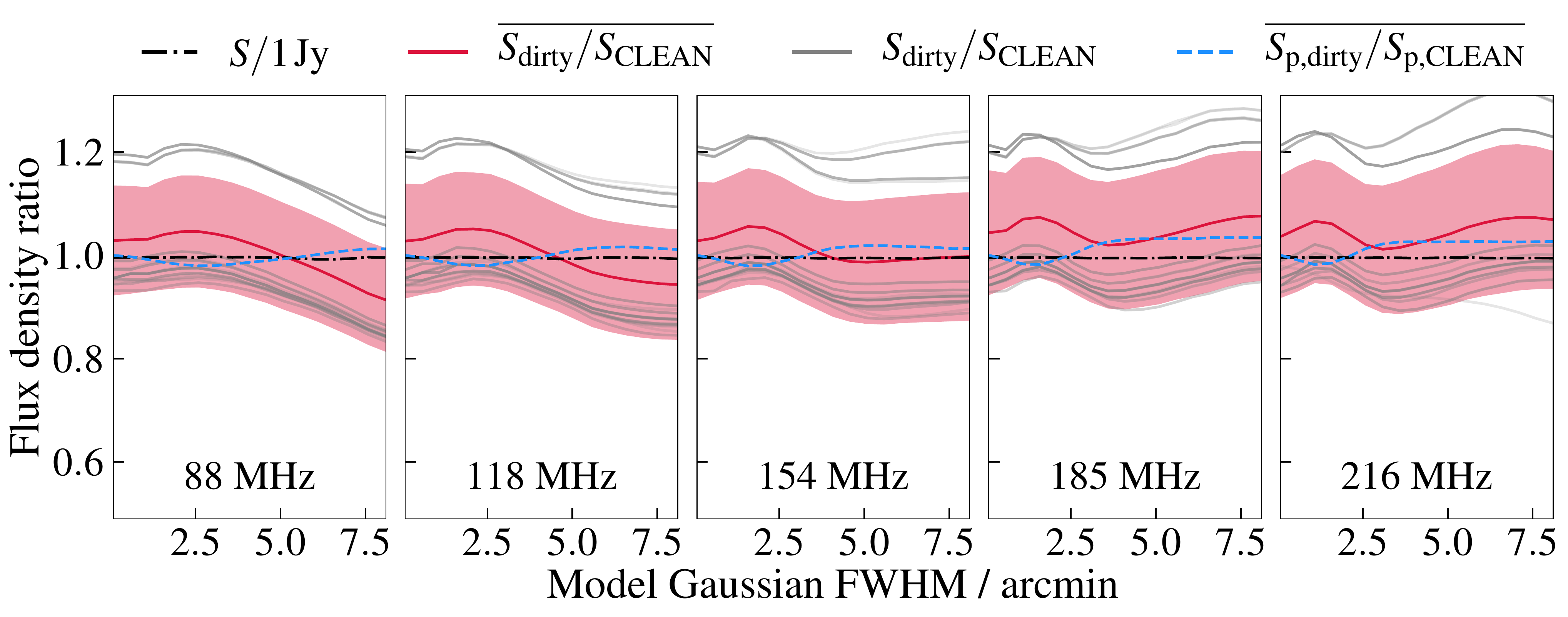}
\subcaption{\label{fig:dirty:f1} \texttt{FIELD1} robust $+2.0$.}
\end{subfigure}\hfill%
\begin{subfigure}[b]{0.5\linewidth}
\includegraphics[width=1\linewidth]{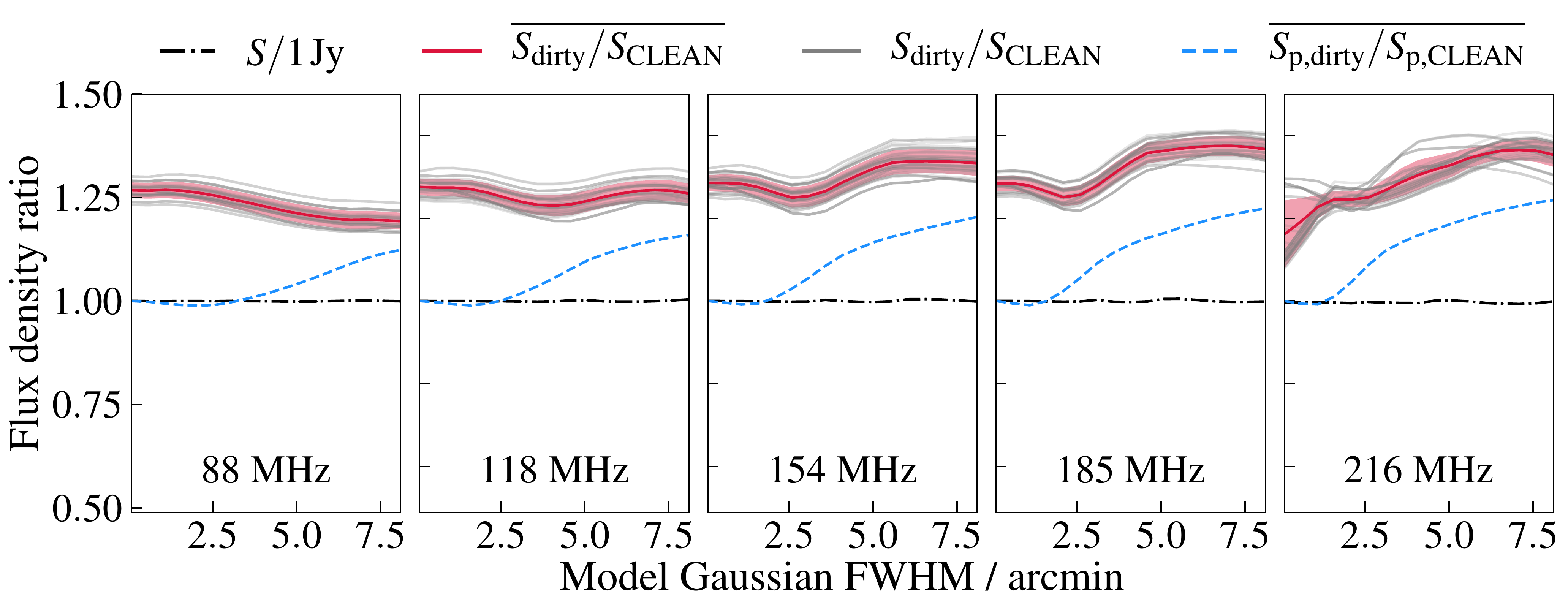}
\subcaption{\label{fig:dirty:f2} \texttt{FIELD2} robust $+2.0$.}
\end{subfigure}\\%
\begin{subfigure}[b]{0.5\linewidth}
\includegraphics[width=1\linewidth]{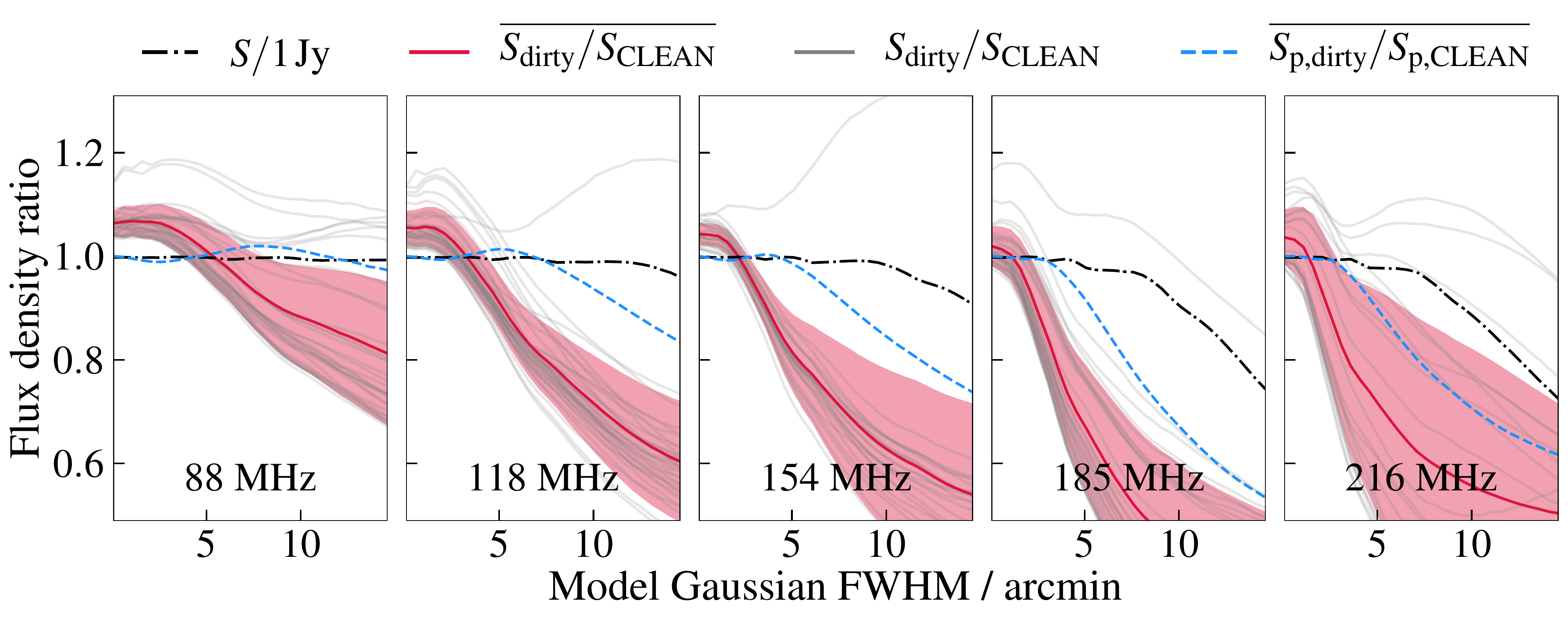}
\subcaption{\label{fig:dirty:f3} \texttt{FIELD3} robust $+2.0$.}
\end{subfigure}\hfill%
\begin{subfigure}[b]{0.5\linewidth}
\includegraphics[width=1\linewidth]{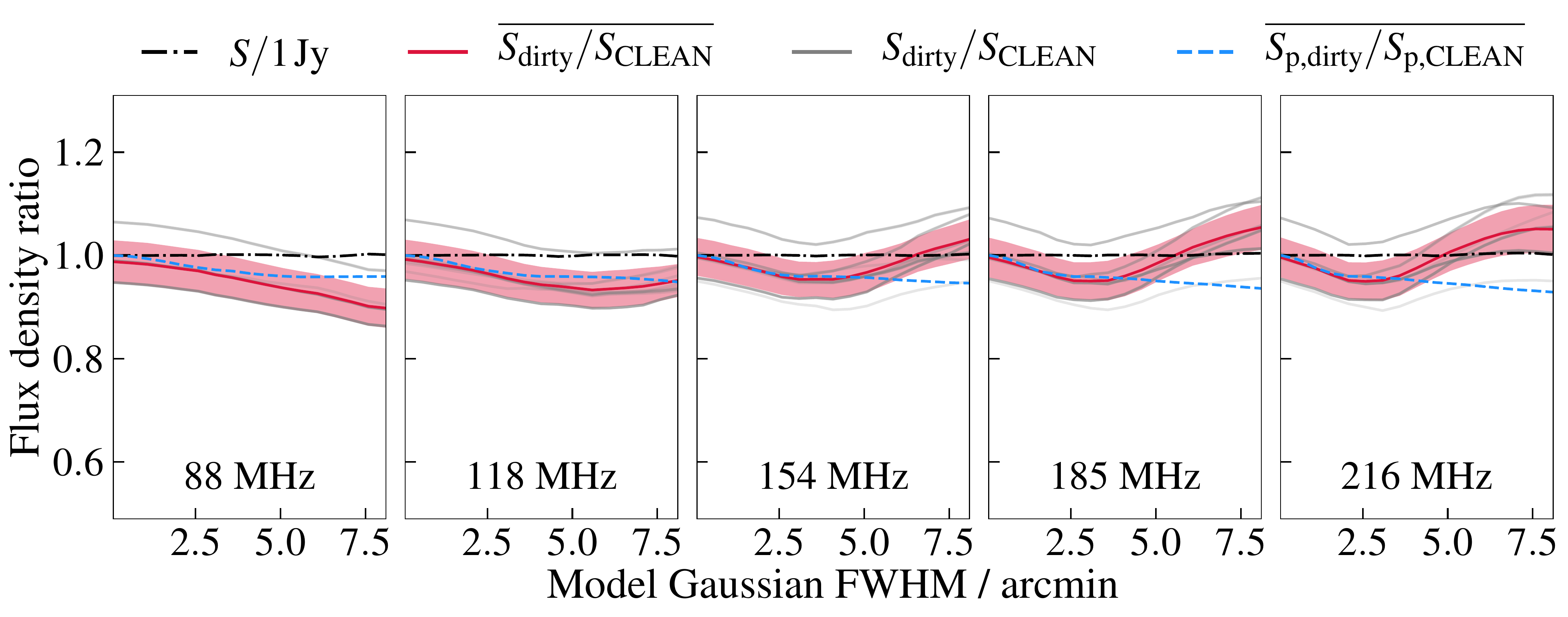}
\subcaption{\label{fig:dirty:f6} \texttt{FIELD6} robust $+2.0$.}
\end{subfigure}\\%
\begin{subfigure}[b]{0.5\linewidth}
\includegraphics[width=1\linewidth]{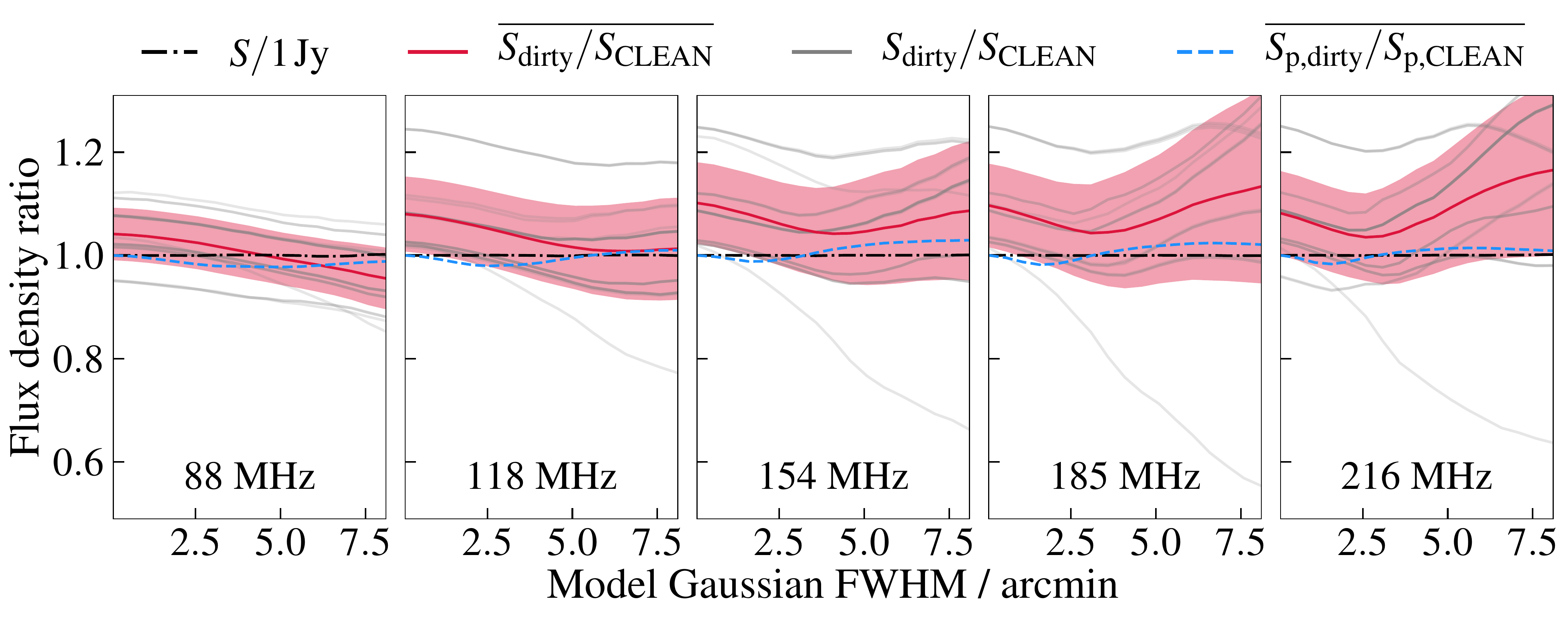}
\subcaption{\label{fig:dirty:f7} \texttt{FIELD7} robust $+2.0$.}
\end{subfigure}\hfill%
\begin{subfigure}[b]{0.5\linewidth}
\includegraphics[width=1\linewidth]{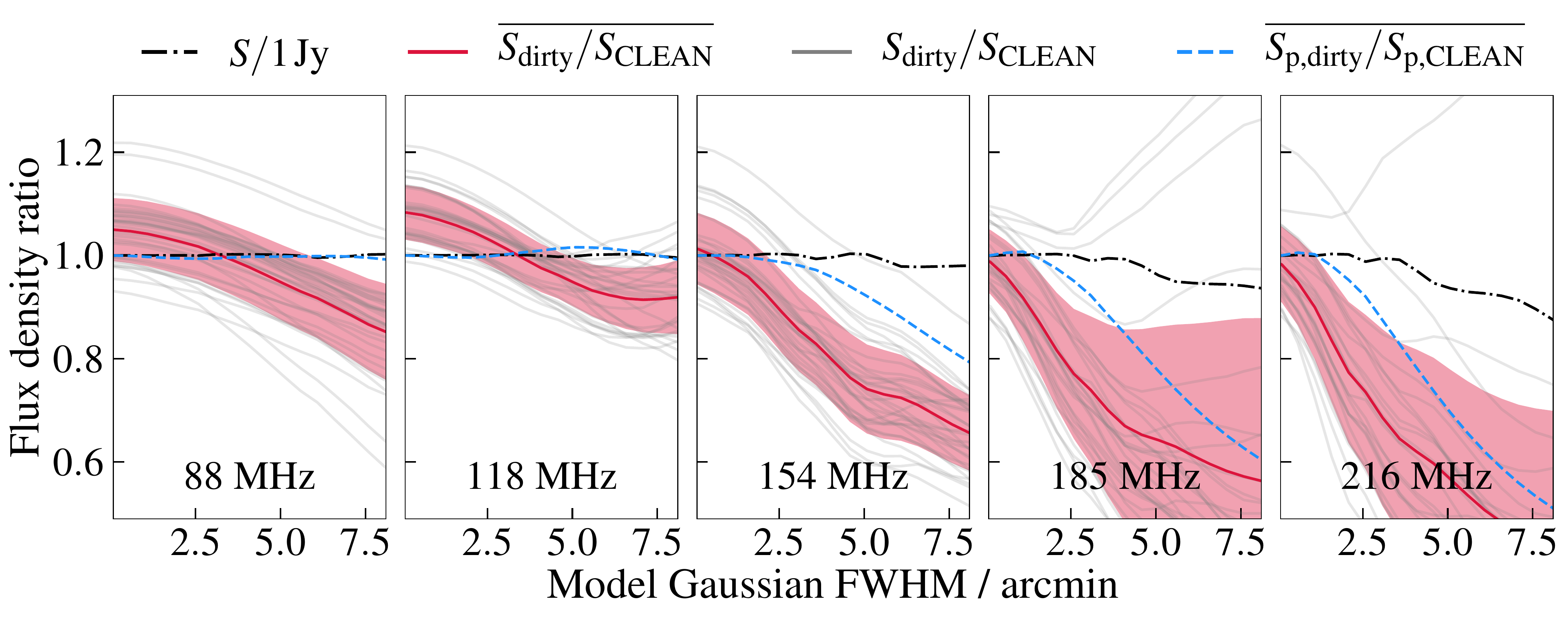}
\subcaption{\label{fig:dirty:f8} \texttt{FIELD8} robust $+2.0$.}
\end{subfigure}\\%
\begin{subfigure}[b]{0.5\linewidth}
\includegraphics[width=1\linewidth]{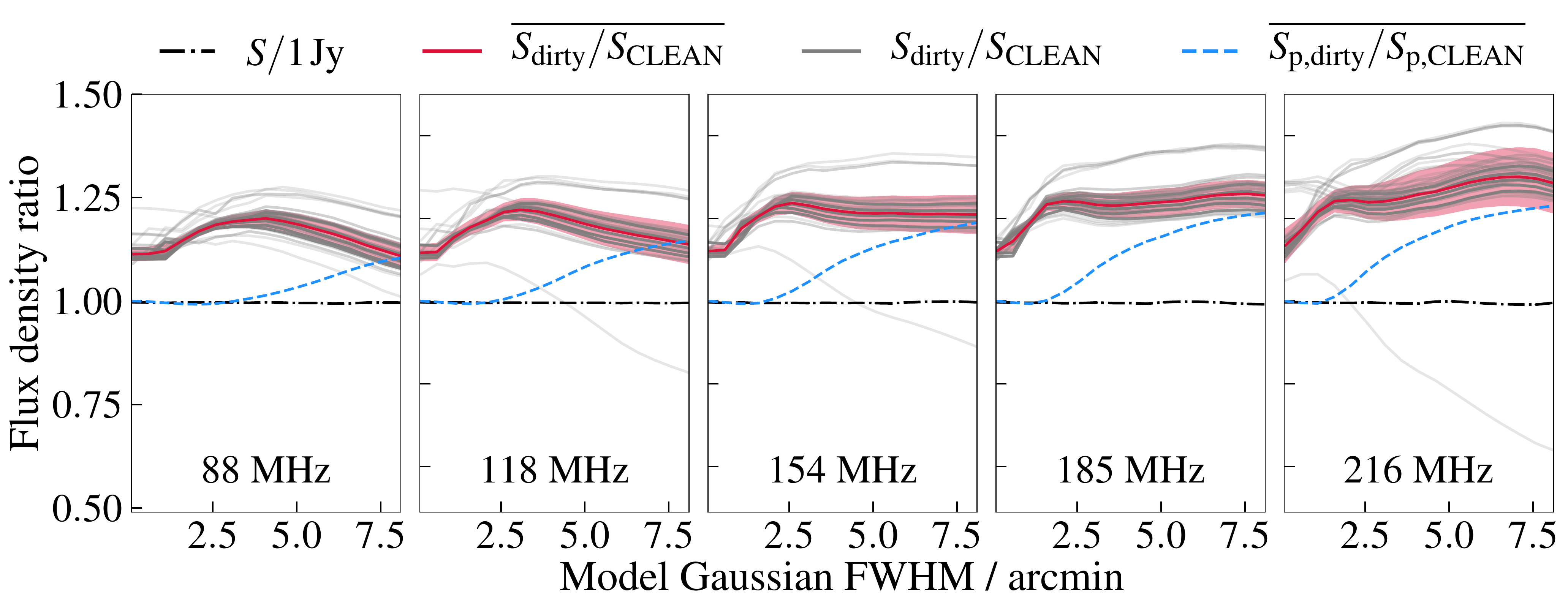}
\subcaption{\label{fig:dirty:f9} \texttt{FIELD9} robust $+2.0$.}
\end{subfigure}\hfill%
\begin{subfigure}[b]{0.5\linewidth}
\includegraphics[width=1\linewidth]{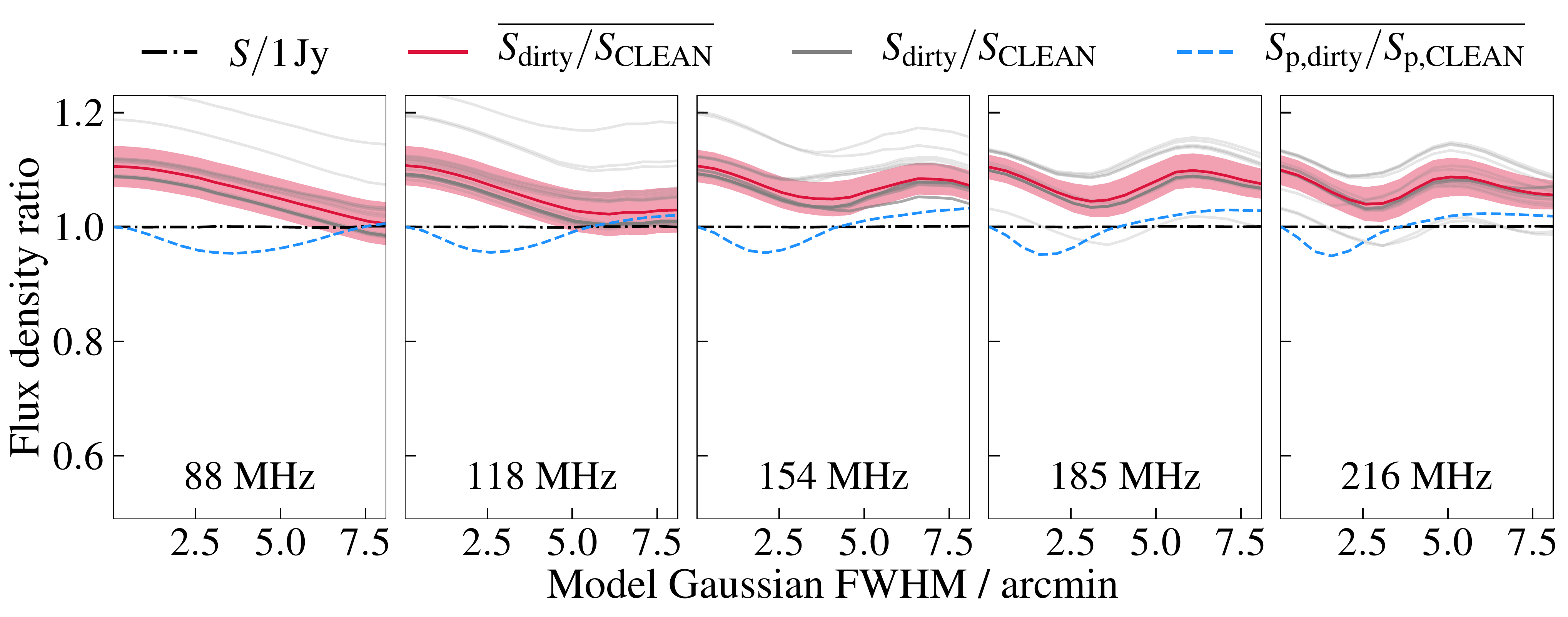}
\subcaption{\label{fig:dirty:f10} \texttt{FIELD10} robust $+2.0$.}
\end{subfigure}\\%
\begin{subfigure}[b]{0.5\linewidth}
\includegraphics[width=1\linewidth]{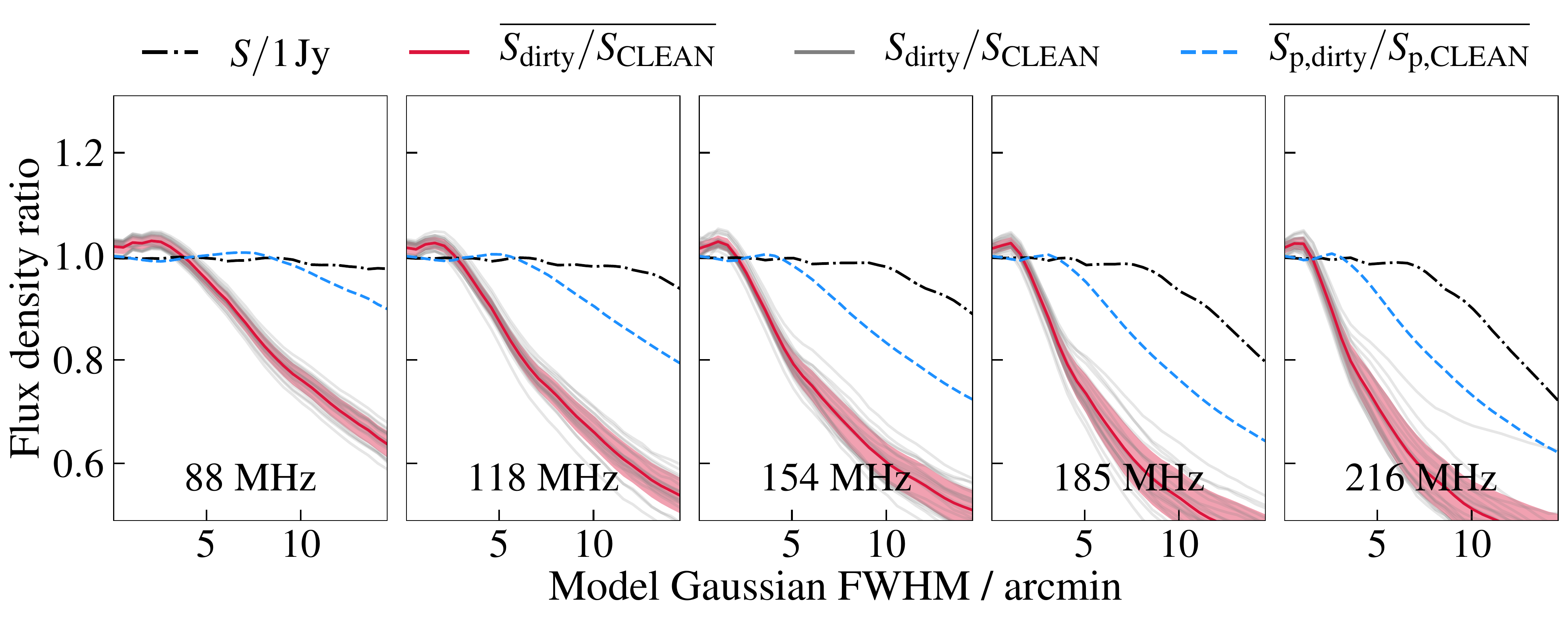}
\subcaption{\label{fig:dirty:f422} \corrs{\texttt{FIELD11}} robust $+1.0$.}
\end{subfigure}\\%
\cprotect\caption{Flux recovery and ratio of deconvolved (`CLEAN') to un-deconvolved (`dirty') integrated flux density for individual snapshots (grey lines). The angular scale on the abscissa correspond to FWHM of the simulated Gaussian (sampled every 30~arcsec). The mean profile, $\overline{S_\text{dirty} / S_\text{CLEAN}}$, is plotted with the standard deviation plotted as a red shaded region. The mean peak flux profile, $\overline{S_\text{p,dirty} / S_\text{p,CLEAN}}$, is also shown. \subref{fig:dirty:f1} and \subref{fig:dirty:f8} are re-produced from \citet{Duchesne2021a} for completeness.}
\end{figure*}

\cref{fig:dirty:f1}--\subref{fig:dirty:f422} show the ratios of dirty to CLEAN flux density for Gaussian models of varying FWHM. For each source the residual flux density is integrated and multiplied by the factor $S_\text{CLEAN} / S_\text{dirty}$ to account for this. \corrs{Additional detail of this process is provided in \citet{Duchesne2021a}.}

\end{appendix}

\end{document}